\documentclass[10pt,a4paper,twoside,notitlepage,openany]{report}

\PassOptionsToPackage{bookmarks=false}{hyperref} %

\usepackage{jheppub} %

\usepackage[version=3]{mhchem}	%

\usepackage{booktabs} %

\usepackage{datetime} %

\usepackage{mathrsfs}

\usepackage{subfig} 

\usepackage{xspace} %
\usepackage{amsfonts} %
\usepackage{amsmath} %
\usepackage{amssymb} %
\usepackage{bbm} %
\usepackage{bm} %
\usepackage{cancel} %
\usepackage{ctable} %
\usepackage{fancybox} %
\cornersize{.2} %
\usepackage[displaymath]{lineno} %
\usepackage{longtable} %
\usepackage{paralist} %
\usepackage{pdfpages} %
\usepackage[section]{placeins}%
\usepackage{setspace} %
\usepackage{slashed} %
\usepackage[draft,textsize=footnotesize,color=ForestGreen!15]{todonotes}%
\usepackage{units} %
\usepackage[normalem]{ulem}%
\usepackage{varioref} %
\usepackage{xcolor}%
\definecolor{ForestGreen}{rgb}{0.13, 0.55, 0.13} %

\newcommand{\ship}{SHiP\xspace} %
\newcommand{\lsim}{\mathrel{\hbox{\rlap{\lower.55ex \hbox{$\sim$}} \kern-.3em
    \raise.4ex \hbox{$<$}}}} %
\newcommand{\gsim}{\mathrel{\hbox{\rlap{\lower.55ex \hbox{$\sim$}} \kern-.3em
    \raise.4ex \hbox{$>$}}}} %

\newcommand{\be}{\begin{equation}}%
\newcommand{\ee}{\end{equation}}

\newcommand{\beq}{\begin{equation}}%
\newcommand{\eeq}{\end{equation}}

\newcommand{\bea}{\begin{eqnarray}}
\newcommand{\eea}{\end{eqnarray}}

\newcommand{\contrib}[1]{}

\newcommand{\InsertSection}[1]{\IfFileExists{#1}{\subfile{#1}}{}}

\usepackage{subfiles} %

\toccontinuoustrue %
\begin{document}
\hypersetup{pageanchor=false}

\title{A facility to Search for Hidden Particles at the CERN SPS: the SHiP
  physics case}

\abstract{This paper describes the physics case for a new fixed target
  facility at CERN SPS. The SHiP (\emph{Search for Hidden Particles})
  experiment is intended to hunt for new physics in the largely unexplored
  domain of very weakly interacting particles with masses below the Fermi
  scale, inaccessible to the LHC experiments, and to study tau neutrino
  physics. The same proton beam setup can be used later to look for decays of
  tau-leptons with lepton flavour number non-conservation, $\tau\to 3\mu$ and
  to search for weakly-interacting sub-GeV dark matter candidates.  We discuss
  the evidence for physics beyond the Standard Model and describe interactions
  between new particles and four different \emph{portals} --- scalars,
  vectors, fermions or axion-like particles. We discuss motivations for
  different models, manifesting themselves via these interactions, and how
  they can be probed with the SHiP experiment and present several case
  studies. The prospects to search for relatively light SUSY and composite
  particles at SHiP are also discussed. We demonstrate that the SHiP
  experiment has a unique potential to discover new physics and can directly
  probe a number of solutions of beyond the Standard Model puzzles, such as
  neutrino masses, baryon asymmetry of the Universe, dark matter, and
  inflation.}

--------------------------------------------------------------------------------
\author[1,2]{Sergey~Alekhin,}
\author[3]{Wolfgang~Altmannshofer,}
\author[4]{Takehiko~Asaka,}
\author[5]{Brian~Batell,}
\author[6,7]{Fedor~Bezrukov,}
\author[8]{Kyrylo~Bondarenko,}
\author[8]{Alexey~Boyarsky$^{\bm{\displaystyle  \star}}$,}
\author[9]{Nathaniel~Craig,}
\author[10]{Ki-Young~Choi,}
\author[11]{Crist\'obal~Corral,}
\author[12]{David~Curtin,}
\author[13,14]{Sacha~Davidson,}
\author[15]{Andr\'e~de Gouv\^ea,}
\author[16]{Stefano~Dell'Oro,}
\author[17]{Patrick~deNiverville,}
\author[18]{P.~S. Bhupal Dev,}
\author[19]{Herbi~Dreiner,}
\author[20]{Marco~Drewes,}
\author[21]{Shintaro~Eijima,}
\author[22]{Rouven~Essig,}
\author[17]{Anthony~Fradette,}
\author[20]{Bj\"orn~Garbrecht,}
\author[23]{Belen~Gavela,}
\author[5]{Gian~F.~Giudice,}
\author[24,25]{Dmitry~Gorbunov,}
\author[3]{Stefania~Gori,}
\author[26,27]{Christophe~Grojean$^{\bm{\displaystyle \S}}$,}
\author[28,29]{Mark~D.~Goodsell,}
\author[30]{Alberto~Guffanti,}
\author[31]{Thomas~Hambye,}
\author[32]{Steen~H.~Hansen,}
\author[11]{Juan~Carlos~Helo,}
\author[33]{Pilar~Hernandez,}
\author[20]{Alejandro~Ibarra,}
\author[8,34]{Artem~Ivashko,}
\author[3]{Eder~Izaguirre,}
\author[35]{Joerg~Jaeckel$^{\bm{\displaystyle \S}}$,}
\author[36]{Yu~Seon Jeong,}
\author[27]{Felix~Kahlhoefer,}
\author[37]{Yonatan~Kahn,}
\author[5,38,39]{Andrey~Katz,}
\author[36]{Choong~Sun~Kim,}
\author[11]{Sergey~Kovalenko,}
\author[3]{Gordan~Krnjaic,}
\author[40,41,42]{Valery~E. Lyubovitskij,}
\author[16]{Simone~Marcocci,}
\author[5]{Matthew~Mccullough,}
\author[43]{~David McKeen,}
\author[44]{Guenakh~Mitselmakher~,}
\author[45]{Sven-Olaf~Moch,}
\author[46]{Rabindra~N. Mohapatra,}
\author[47]{David~E.~Morrissey,}
\author[34]{Maksym~Ovchynnikov,}
\author[48]{Emmanuel~Paschos,}
\author[18]{Apostolos~Pilaftsis,}
\author[3,17]{Maxim~Pospelov$^{\bm{\displaystyle \S}}$,}
\author[49]{Mary~Hall~Reno,}
\author[27]{Andreas~Ringwald,}
\author[17]{Adam~Ritz,}
\author[50]{Leszek~Roszkowski,}
\author[24]{Valery~Rubakov,}
\author[21]{Oleg~Ruchayskiy$^{\bm{\displaystyle  \star}}$,}
\author[51]{Jessie~Shelton,}
\author[52]{Ingo~Schienbein,}
\author[19]{Daniel~Schmeier,}
\author[27]{Kai~Schmidt-Hoberg,}
\author[5]{Pedro~Schwaller,}
\author[53,54]{Goran~Senjanovic,}
\author[55]{Osamu~Seto,}
\author[21]{Mikhail~Shaposhnikov$^{\bm{\displaystyle  \star, \S}}$,}
\author[3]{Brian~Shuve,}
\author[56]{Robert~Shrock,}
\author[44]{Lesya~Shchutska$^{\bm{\displaystyle \S}}$,}
\author[57]{Michael~Spannowsky,}
\author[58]{Andy~Spray,}
\author[5]{Florian~Staub,}
\author[5]{Daniel~Stolarski,}
\author[39]{Matt~Strassler,}
\author[53]{Vladimir~Tello,}
\author[59,60]{Francesco~Tramontano$^{\bm{\displaystyle \S}}$,}
\author[59]{Anurag~Tripathi,}
\author[61]{Sean~Tulin,}
\author[16,62]{Francesco~Vissani,}
\author[63]{Martin~W. Winkler,}
\author[64,65]{Kathryn~M.~Zurek}

\affiliation[1]{Deutsches Elektronensynchrotron DESY, Platanenallee 6, D--15738 Zeuthen, Germany}
\affiliation[2]{Institute for High Energy Physics,  142281 Protvino, Moscow region, Russia}
\affiliation[3]{Perimeter Institute for Theoretical Physics, 31 Caroline St. N, Waterloo, Ontario, Canada}
\affiliation[4]{Department of Physics, Niigata University, Niigata 950-2181, Japan}
\affiliation[5]{Theory Division, Physics Department, CERN, CH-1211 Geneva 23, Switzerland}
\affiliation[6]{Physics Department, University of Connecticut, Storrs, CT 06269-3046, USA}
\affiliation[7]{RIKEN-BNL Research Center, Brookhaven National Laboratory, Upton, NY 11973, USA}
\affiliation[8]{Instituut-Lorentz for Theoretical Physics, Universiteit Leiden, Niels Bohrweg 2, Leiden, The Netherlands}
\affiliation[9]{Department of Physics, University of California, Santa Barbara, CA 93106, USA}
\affiliation[10]{Korea Astronomy and Space Science Institute,Daejon 305-348,  Republic of Korea}
\affiliation[11]{Departamento de F\'isica, Universidad T\'ecnica Federico Santa Mar\'ia and Centro Cient\'ifico Tecnol\'ogico de Valpara\'iso, Casilla 110-V, Valpara\'iso, Chile}
\affiliation[12]{Maryland Center for Fundamental Physics, University of Maryland, College Park, MD 20742, USA}
\affiliation[13]{IPNL, CNRS/IN2P3,  4 rue E. Fermi, Universit\'e Lyon 1,69622 Villeurbanne cedex, France}
\affiliation[14]{Universit\'e de Lyon, F-69622, Lyon, France}
\affiliation[15]{Northwestern University, Department of Physics \& Astronomy, Evanston, IL~60208-3112, USA}
\affiliation[16]{Gran Sasso Science Institute, Viale Crispi 7, 67100 L'Aquila, Italy}
\affiliation[17]{Department of Physics and Astronomy, University of Victoria, Victoria, BC, V8P 5C2, Canada}
\affiliation[18]{Consortium for Fundamental Physics, School of Physics and Astronomy, University of Manchester, Manchester M13 9PL, United Kingdom}
\affiliation[19]{Bethe Center for Theoretical Physics \& Physikalisches Institut der Universit\"at Bonn, 53115 Bonn, Germany}
\affiliation[20]{Physik-Department, Technische Universit\"at M\"unchen, James-Franck-Stra\ss{}e, 85748 Garching, Germany}
\affiliation[21]{Ecole Polytechnique F\'ed\'erale de Lausanne, FSB/ITP/LPPC, BSP, CH-1015, Lausanne, Switzerland}
\affiliation[22]{C. N. Yang Institute for Theoretical Physics, Stony Brook University, Stony Brook, NY 11794, USA}
\affiliation[23]{Departamento de F\'isica Te\'orica and Instituto de F\'isica Te\'orica, IFT-UAM/CSIC, Universidad Aut\'onoma de Madrid, Cantoblanco, 28049, Madrid, Spain}
\affiliation[24]{Institute for Nuclear Research of the Russian Academy of Sciences, Moscow 117312, Russia}
\affiliation[25]{Moscow Institute of Physics and Technology, Dolgoprudny 141700, Russia}
\affiliation[26]{ICREA at IFAE, Universitat Aut\`onoma de Barcelona, E-08193 Bellaterra, Spain}
\affiliation[27]{Deutsches Elektronen-Synchroton (DESY), Notkestrasse 85, D-22607 Hamburg, Germany}
\affiliation[28]{Sorbonne Universit\'es, UPMC Univ Paris 06, UMR 7589, LPTHE, F-75005, Paris, France}
\affiliation[29]{CNRS, UMR 7589, LPTHE, F-75005, Paris, France}
\affiliation[30]{Niels Bohr International Academy and Discovery Center,  Niels Bohr Institute, University of Copenhagen,  Blegdamsvej 17, DK-2100 Copenhagen, Denmark}
\affiliation[31]{Service de Physique Th\'eorique, Universit\'e Libre de Bruxelles,Bld du Triomphe, CP225, 1050 Brussels, Belgium}
\affiliation[32]{Dark Cosmology Centre, Niels Bohr Institute, University of Copenhagen, Juliane Maries Vej 30, 2100 Copenhagen, Denmark}
\affiliation[33]{Instituto de F\'isica Corpuscular (IFIC), CSIC-Universitat de Val\`encia Apartado de Correos 22085,E-46071 Valencia, Spain}
\affiliation[34]{Department of Physics, Kiev National Taras Shevchenko University, Glushkov str. 2 building 6, Kiev, 03022, Ukraine}
\affiliation[35]{Institut f\"ur theoretische Physik, Universit\"at Heidelberg, Philosophenweg 16, 69120 Heidelberg, Germany}
\affiliation[36]{Department of Physics and IPAP, Yonsei University, Seoul 120-749, Korea}
\affiliation[37]{Massachusetts Institute of Technology, Cambridge, MA 02139, USA}
\affiliation[38]{Universit\'e de Gen\`eve, Department of Theoretical Physics and Center for Astroparticle Physics (CAP), 24 quai E. Ansermet, CH-1211 Geneva 4, Switzerland}
\affiliation[39]{Department of Physics, Harvard University, Cambridge, MA 02138, USA}
\affiliation[40]{Institut f\"ur Theoretische Physik, Universit\"at T\"ubingen, Kepler Center for Astro and Particle Physics, Auf der Morgenstelle 14, D-72076 T\"ubingen, Germany}
\affiliation[41]{Department of Physics, Tomsk State University, 634050 Tomsk, Russia}
\affiliation[42]{Mathematical Physics Department, Tomsk Polytechnic University, Lenin avenue 30, 634050 Tomsk, Russia}
\affiliation[43]{Department of Physics, University of Washington, Seattle, Washington 98195, USA}
\affiliation[44]{University of Florida, Gainesville, USA}
\affiliation[45]{II. Institut f\"ur Theoretische Physik, Universit\"at Hamburg, Luruper Chaussee 149, D--22761 Hamburg, Germany}
\affiliation[46]{Maryland Center for Fundamental Physics and Department of Physics,  University of Maryland, College Park, Maryland 20742, USA}
\affiliation[47]{TRIUMF, 4004 Wesbrook Mall, Vancouver, BC V6T 2A3, Canada}
\affiliation[48]{Department of Physics, Technical University of Dortmund, D-44221, Dortmund, Germany}
\affiliation[49]{University of Iowa, Iowa City, Iowa, 52242, USA}
\affiliation[50]{National Centre for Nuclear Research, Hoza 69, 00-681 Warsaw, Poland}
\affiliation[51]{1110 West Green Street Urbana, IL 61801, Dept of Physics, University of Illinois at Urbana-Champaign, USA}
\affiliation[52]{LPSC, Universit\'e Grenoble-Alpes, CNRS/IN2P3, 53 avenue des Martyrs, 38026 Grenoble, France}
\affiliation[53]{Theory Group, Gran Sasso Science Institute, Viale Crispi 7, 67100 L'Aquila, Italy}
\affiliation[54]{ICTP, Trieste, Italy}
\affiliation[55]{Department of Life Science and Technology,Hokkai-Gakuen University,Sapporo 062-8605, Japan}
\affiliation[56]{C. N. Yang Institute for Theoretical Physics, Stony Brook University   Stony Brook, NY 11794 USA}
\affiliation[57]{Institute for Particle Physics Phenomenology, Department of Physics, Durham University, Durham DH1 3LE, United Kingdom}
\affiliation[58]{ARC Centre of Excellence for Particle Physics at the Terascale, School of Physics, The University of Melbourne, Victoria 3010, Australia}
\affiliation[59]{INFN, sezione di Napoli, Complesso di Monte Sant'Angelo, via  Cintia, I-80126, Napoli, Italy}
\affiliation[60]{Universit\'a di Napoli ``Federico II'', Complesso di Monte Sant'Angelo, via  Cintia, I-80126, Napoli, Italy}
\affiliation[61]{Department of Physics and Astronomy, York University 4700 Keele Street, Toronto, Ontario, M3J 1P3, Canada}
\affiliation[62]{INFN, Laboratori Nazionali del Gran Sasso, Assergi, L'Aquila, Italy}
\affiliation[63]{Bethe Center for Theoretical Physics and Physikalisches Institut der Universit\"at Bonn Nussallee 12, 53115 Bonn, Germany}
\affiliation[64]{Theory Group Lawrence Berkeley National Laboratory, Berkeley, CA 94709, USA}
\affiliation[65]{Berkeley Center for Theoretical Physics University of California, Berkeley, CA 94709, USA}
\renewcommand{\thefootnote}{$\bm\star$} %
\footnotetext{Editor of the paper}
\renewcommand{\thefootnote}{$\bm\S$} %
\footnotetext{Convener of the Chapter}
\renewcommand{\thefootnote}{\arabic{footnote}} %

\preprint{CERN-SPSC-2015-017 (SPSC-P-350-ADD-1)}
\maketitle
\flushbottom

\section*{Acknowledgement}

We are extremely grateful to Walter Bonivento, Annarita Buonaura, Geraldine
Conti, Hans Dijkstra, Giovanni De Lellis, Antonia Di Crescenzo, Andrei
Golutvin, Elena Graverini, Richard Jacobsson, Gaia Lanfranchi, Thomas Ruf,
Nicola Serra, Barbara Storaci, Daniel Treille for their invaluable
contributions during all stages of work on this document and in particular for
their help with models' sensitivity estimates.

We are grateful to all the members of \ship for their dedicated work to make
this experiment possible.

\InsertSection{ship_introduction.tex} %
\InsertSection{ship_vector_portal.tex} %
\InsertSection{ship_scalar_portal.tex} %
\InsertSection{ship_neutrino_portal.tex} %
\InsertSection{ship_axion_portal.tex} %
\InsertSection{ship_susy.tex} %
\InsertSection{ship_tau_neutrino.tex} %
\InsertSection{ship_tau3mu.tex} %
\InsertSection{ship_conclusion.tex} %

\appendix
\InsertSection{ship_appendix_SHiP.tex}
\InsertSection{ship_appendix_notations.tex}
\InsertSection{ship_acknowledgements.tex}

\renewcommand{\bibsep}{3pt}%

\providecommand{\href}[2]{#2}\begingroup\raggedright\endgroup

\end{document}


\chapter{The \ship experiment}
\label{cha:ship-experiment}

In this chapter we provide a brief description of the \ship
experiment. Although not intended to be exhaustive it allows to do the
estimates of the sensitivity towards the detection of new physics. The details
may be found at~\cite{Bonivento:2013jag,TP} and at
\url{http://ship.web.cern.ch}.
	\section*{Experimental setup}
	A dedicated beam line extracted from the SPS will convey a 400~GeV/$c$ proton beam at the SHiP facility~\cite{Bonivento:2013jag,TP}. The beam will be stopped in a Molybdenum and Tungsten target, at a center-of-mass energy $E_{CM} =\sqrt{2 E_b m_p} \simeq 27$~GeV.
	Approximately $2\times 10^{20}$ proton-target collisions ($PoT$) are foreseen in 5 years of operation.
	
	The target will be followed by a hadron stopper, intended to stop all $\pi^\pm$ and $K$ mesons before they decay, and by a system of shielding magnets to sweep muons away from the fiducial decay volume.
	
A neutrino detector consisting of OPERA-like bricks of laminated lead and
emulsions, placed in a magnetic field downstream of the muon shield,
will allow to measure and identify charged particles produced in
charged current neutrino interactions. It is followed by a tracking
system and muon magnetic spectrometer.

	An upstream tagger, together with the muon spectrometer of the
        neutrino detector, will allow to detect and veto charged particles produced outside the main decay volume. The fiducial decay volume begins approximately 63.8~m downstream of the primary target, and is contained in a 60~m long cylindrical vacuum tank with elliptical section of $x$ and $y$ semiaxes 2.5~m and 5~m long, respectively.
	
	An straw tagger is placed in vacuum 5~m downstream of the entrance lid of the vacuum tank. Its purpose is to help reducing background arising from interactions in the material upstream of the decay volume.
	
	An additional background tagger surrounds the fiducial decay volume, which walls enclose 30~cm of liquid scintillator.
	
	The tracking system aimed at measuring the decay products of hidden particles is located at the end of the decay volume. It will consist of 5~m long straw tubes organized in 4 stations, with a magnetic field of 1~Tm between the first and the second pair of stations.
	The high-accuracy timing information provided by a dedicated detector following the straw tracker will be used to discriminate combinatorial background.
	
	The particle identification system is placed outside the vacuum tank, and it features an electromagnetic and an hadronic calorimeter, followed by a muon system made of four active layers interlaced with iron.

	\section*{Number of mesons and adopted cross-sections}
	The number of charm and beauty mesons produced at the SHiP target can be estimated as
	\begin{equation}
		N_{\text{mesons}} = 2\times X_{q\overline{q}} \times N_{PoT}
	\end{equation}
	where $X_{q\overline{q}}$ represents the $q\bar{q}$ production rate. The following cross sections have been used for the estimates:
	\begin{itemize}
		\item the proton-nucleon cross section is $\sigma(pN)\simeq 10.7$~mbarn. 
		\item $\sigma(cc)= 18$~$\mu$barn~\cite{Abt:2007zg} and the fraction $X_{cc}=1.7\times 10^{-3}$
		\item $\sigma(bb)= 1.7$~nbarn~\cite{Lourenco:2006vw} and the fraction $X_{bb}=1.6\times 10^{-7}$
	\end{itemize}

	The relative abundances of charmed mesons are approximately as follows (\textsc{Pythia8} simulations):
	\begin{itemize}
		\item [$D^\pm$]  30\%
		\item [$D^0$, $\bar{D}^0$]  62\%
		\item [$D_s^\pm$]  8\%
		\item [$J/\psi$]  1\%
	\end{itemize}
	
	The expected number of $\tau$ leptons for $N_{PoT}=2\times 10^{20}$ is $N_{\tau} = 3 \times 10^{15}$.

	\section*{Number of events in the detector}
	In order to estimate the number long-lived particles produced
        in decays of heavy mesons ($c$- and $b$-mesons) we used the following formula:
	\begin{align}\label{eq:nX}
		&n_{prod} \simeq N_{PoT} \times \chi(pp \to X) \times \epsilon_{geom} \times \frac{60~\text{m}}{\gamma c \tau}
	\end{align}
	where
	\begin{align}
		&\chi(pp \to X) = (N_{\text{mesons}} / N_{PoT})\times BR(\text{meson}\to X).
	\end{align}
	In Eqn.~\ref{eq:nX}, $\epsilon_{geom}$ is the geometric acceptance, computed
        as ratio between the solid angle covered by the detector and
        the average divergence of new particles $X$, and the factor
        $\gamma = <E>/m_X$ is the average energy of $X$ divided by its
        mass $m_X$. The last factor of Eqn.~\ref{eq:nX} approximates the longitudinal
        acceptance if the average energy and lifetime are such that
        $\gamma c \tau \gg 60$~m (the length of the decay
        volume). This formula must be further corrected for the acceptance
        of the final state in the detector, in addition to
        reconstruction and selection efficiency.
	
	Finally, we consider as ``detectable'' the final states with two or more charged particles.


\chapter{Notations}
\label{sec:notations}

\begin{description}
\item{}Throughout the document we use the following conventions:
  \begin{itemize}
\item Index $\alpha = \{e,\mu,\tau\}$ is the flavour index
\item Charged leptons ($e^\pm,\mu^\pm,\tau^\pm$) collectively are
  $\ell_\alpha$
\item Neutrinos $\nu_\alpha$
\item Left lepton doublet $L_\alpha =
  \begin{pmatrix}
    \nu_\alpha\\
    \ell_\alpha
  \end{pmatrix}$
\item $\Phi$ denotes Higss field (two-component SU(2) doublet); $\tilde \Phi_i
  \equiv \epsilon_{ij} \Phi_j$
\item VEV of the Higgs field $ \langle \Phi\rangle = \frac{v}{\sqrt 2}$, where
  \fbox{$v = \unit[246]{GeV}$}

\end{itemize}

\item{}
Chapter~\ref{sec:neutrino-portal} uses the following additional notations:
\begin{itemize}
\item $N_I$ denotes right-handed (two component) gauge singlet neutrinos

\item Index $I = 1,2,\dots \mathcal{N}$ runs over HNL species ($\mathcal{N}
  \ge 2$)
\item Active-sterile $F_{\alpha I}$ -- matrix $3\times \mathcal{N}$
\item Majorana mass of HNL: $M_I$
\item Dirac mass $(m_D)_{\alpha I}$
\item Sterile neutrino mixing angle 
  \begin{equation}
    U^2_{\alpha I} = \frac{v^2 |F_{\alpha I}|^2}{M_I^2}
  \end{equation}
\item Experimental constraints are put either on $U^2_\alpha$, defined as
  \begin{equation}
    \label{notations:eq:1}
    U^2_\alpha = \sum_I U^2_{\alpha I}
  \end{equation}
  the total mixing angle $U^2$ is defined as\footnote{The notation
    $\sin^2(2\theta)$ is sometimes used instead of $U^2$. The correspondence
    is $\sin^2(2\theta) = U^2$ with no extra factors.}
\begin{equation}
  \label{notations:eq:2}
  U^2 = \sum_\alpha U^2_\alpha
\end{equation}
\end{itemize}

\item{}Cosmological notations

  \begin{itemize}

\item $n_B$ -- baryonic number density
\item Baryon asymmetry of the Universe:
  \begin{equation}
    \eta_B \equiv \frac{n_B}s\label{notations:eq:3}
\end{equation}
\item $s$ -- entropy density

\end{itemize}
\end{description}
\section{Abbreviations:}
\label{sec:abbreviations}

\begin{tabular}[t]{ll}
  Axion-like particle                         & ALP              \\
  Baryon asymmetry of the Universe            & BAU              \\
  Brout–Englert–Higgs (field, particle,\dots) & BEH              \\
  Beyond the Standard Model                   & BSM              \\
  Big Bang Nucleosynthesis                    & BBN              \\
  Charged (sector) lepton flavour violation   & CLFV             \\
  Closed time path (formalism)                & CTP              \\
  Confidence Level                            & C.L. or CL       \\
  Dark Matter                                 & DM               \\
  Deep-inelastic scattering                   & DIS              \\
  Electroweak                                 & EW               \\
  Electroweak phase transition                & EWPT             \\
  Electroweak symmetry breaking               & EWSB             \\
  Future Circular Collider                    & FCC              \\
  Grand Unified Theory                        & GUT              \\
  Gross-Llewellyn Smith (sum rule)            & GLS              \\
  Neutrinoless double beta decay              & $0\nu\beta\beta$ \\
  Heavy neutral lepton                        & HNL              \\
  Left-right symmetric model                  & LRSM             \\
  lepton flavour violation                    & LFV              \\
  lepton number violation                     & LNV              \\
  Right-handed (neutrino/current)             & RH               \\
  Super Proton Synchrotron                    & SPS              \\
  Pseudo-Nambu-Goldstone boson                & PNGB             \\
  Pontecorvo-Maki-Nakagawa-Sakata             & PMNS             \\
  protons-on-target                           & PoT or p.o.t.    \\
  R-parity violating/violation                & RPV              \\
  Standard Model                              & SM               \\
\end{tabular}


\chapter{Vector portal}
\label{sec:vector-portal}

We review the main features and physics motivations behind vector portals between the Standard Model (SM) and dark sectors. Several case studies for the SHiP experiment are presented.

\section{Classification of vector portals}

The gauge structure of the Standard Model, the celebrated $SU(3)\times
SU(2)\times U(1)$ combination, is a minimal choice compatible with the chiral
nature of fermions and spontaneous electroweak symmetry breaking that renders
the weak bosons and fermions massive. Establishing this structure, through a
combination of theoretical and experimental efforts, is one of the major
achievements of $20^{\rm th}$ century physics.

It is possible that the gauge structure of the SM descends from a larger gauge group, as is the case in 
Grand Unified Theories, or GUTs. In that case one expects that at least several of the new vector states are very heavy, 
{\em e.g.} $m_V \sim 10^{16}$~GeV, well beyond the direct reach of accelerators. It is also possible that the
SM is accompanied by additional gauge structures that allow for (sub-)TeV
gauge bosons, as is the case of multiple $U(1)$'s, $SU(3)\times SU(2)\times [U(1)]^n$ \cite{Langacker:2008yv}. 
The high-energy LHC experiments place very strong bounds on the possible existence of new
vector states associated with new $U(1)$ gauge groups, provided that the coupling of
such vector states to the SM is sizable \cite{Aad:2014cka,Khachatryan:2014fba}. An alternative possibility, relatively light
vector states ({\em e.g.} in the GeV mass range) with small couplings to the SM, is poorly constrained by the LHC experiments and 
represents instead an attractive physics target for many experiments at the 
intensity frontier \cite{Essig:2013lka}.

The fixed target experiment SHiP proposed at the CERN SPS \cite{Bonivento:2013jag} is a powerful tool for studying extensions of the SM based on new 
vector particles. This chapter discusses the physics motivations and phenomenology for such particles, provides a classification of vector portals, and points to several promising models that can be probed with proton fixed target (and beam dump) experiments.
The specific examples identified as attractive physics opportunities for  SHiP to be discussed in this chapter are: 
\begin{itemize}
\item
Kinetically mixed dark photons in the GeV mass range with mixing angles $\epsilon \sim O(10^{-6})$, 
as well as gauge bosons coupled to baryons, $V^{(B)}$.
\item Signatures of ``dark Higgs'' states produced through the vector portal and the dark Higgsstralhung process.
\item Signatures of heavy neutral leptons (HNL) produced via the vector portal.
\item Signatures of sub-GeV (meta)stable states ({\em e.g.} light dark matter) via its production and subsequent scattering via the vector portal in a neutrino-like detector.
\end{itemize}

Throghout this chapter, we denote the new vector state as $V_\mu$, or simply $V$. We will also often employ a superscript to indicate the SM current coupled to the new vector state, {\em e.g.} $V^{(B-L)},~V^{(B)}$ etc.
Furthermore, the new (beyond the SM) states that couple to $V_\mu$ will be generically called $\chi$, while the new dark Higgs states associated with 
$U(1)'$ group will be called $h'$. 

\subsection{Kinetic mixing}

Consider a QED-like theory with one (or several) extra vector particle(s), coupled to the 
electromagnetic current. A mass term, or in general a mass matrix for the vector states,
is protected against additive renormalization due to the conservation of the electromagnetic current. 
If the mass matrix for such vector states has a zero determinant, ${\rm det}(M^2_V) =0$, then 
the theory contains one massless vector, to be identified with a photon,  and several massive vector states.

This is the model of  `para-photons', introduced by Okun in early 1980s \cite{Okun:1982xi}, that can be reformulated in an equivalent language 
using the kinetic mixing portal. Following Holdom \cite{Holdom:1985ag}, one writes a QED-like theory with two $U(1)$ groups, 
supplemented by the cross term in the kinetic Lagrangian and a mass term for one of the vector fields,
\be
{\cal L} = {\cal L}_{\psi,A} + {\cal L}_{\chi,A'} -\frac{\epsilon}{2} F_{\mu\nu} F'_{\mu\nu} + \frac{1}{2} m^2_{A'}( A'_{\mu})^2. 
\label{kinmix}
\ee
${\cal L}_{\psi,A}$ and $ {\cal L}_{\chi,A'}$ are the standard QED-type Lagrangians, 
\begin{eqnarray}
{\cal L}_{\psi,A} = -\frac{1}{4} F_{\mu\nu}^2 + \bar \psi[ \gamma_\mu (i\partial_\mu - e A_\mu)  -  m_\psi ] \psi, \nonumber\\
{\cal L}_{\chi,A'} = -\frac{1}{4} (F'_{\mu\nu})^2 + \bar \chi[ \gamma_\mu (i\partial_\mu - g' A'_\mu)  -  m_\chi ] \chi,
\label{QED}
\end{eqnarray}
with $F_{\mu\nu}$ and $F'_{\mu\nu}$ standing for the field strength tensors. 
States $\psi$ represent the QED electron fields while states $\chi$ are charged under the "dark" $U(1)'$. 
In the limit of $\epsilon \to 0$, the two sectors become completely decoupled. 
In eq. (\ref{kinmix}), the mass term for $A'$ explicitly breaks the second $U(1)$ but is protected from additive 
renormalization and hence is technically natural\footnote{When breaking of $U(1)'$ is triggered by a Higgs mechanism, 
there can be an additional "gauge hierarchy" issue related to the naturalness of the $h'$ mass term.}. Using the equations of motion, 
$\partial_\mu F_{\mu\nu} = e J_\nu^{\rm EM}$, the interaction term can be rewritten 
as 
\be
-\frac{\epsilon}{2} F_{\mu\nu} F'_{\mu\nu} =A'_\mu \times ( e \epsilon ) J_\mu^{\rm EM},
\ee
showing that the new vector particle couples to the electromagnetic current with strength, reduced by a small 
factor $\epsilon$. The generalization of (\ref{kinmix}) to the SM is straightforward, by subsituting the QED $U(1)$ 
with the hypercharge $U(1)$ of the SM. 

There are a multitude of notations and names referring to 
one and the same model. We shall refer to the $A'$ state as the "dark photon". It can also be denoted as $V^{(Y)}$, a vector state coupled to the 
hypercharge current.  We choose to call the mixing angle $\epsilon$, and throughout this chapter assume $\epsilon \ll 1$. In contrast, 
one does not have to assume a smallness of $g'$ coupling, which can be comparable to the gauge couplings of the SM, 
$g'\sim g_{\rm SM}$.

\begin{figure}
\centerline{\includegraphics[width=0.55\textwidth]{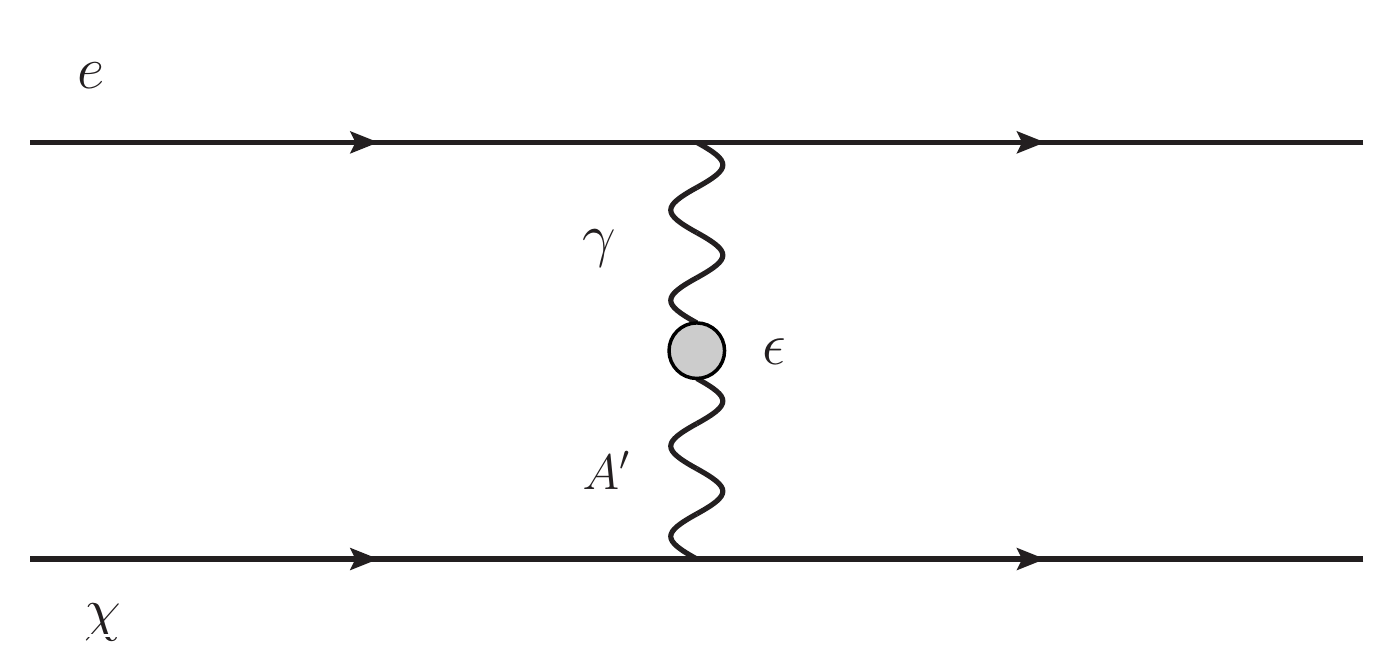}}
\caption{The interaction through the exchange by a mixed $\gamma-A'$ propagator between the SM particles and particles $\chi$ charged under new 
$U(1)'$ group. In the limit of $m_{A'} \to 0$ the apparent electromagentic charge of $\chi$ is $e\epsilon$. }
\label{fig:VPfig1}
\end{figure} 

Although the model of this type is an exceedingly simple and minimal extension of the SM,  one can already learn a number of instructive features:
\begin{enumerate} 

\item The mixing parameter $\epsilon$ is dimensionless, and therefore can retain information about loops of charged particles
at some heavy scale $M$ without power-like decoupling. In the simplest example, a new fermionic field charged under both $U(1)$'s
will generate an additional contribution to the mixing angle that scales as $\Delta \epsilon \sim g'e/(12 \pi^2)\times \log(\Lambda_{UV}^2/M)^2$.  
Alternatively, the mixing $\epsilon$ can be generated at two or higher loop order,
so that one can entertain a wide range of mixing angles. 

\item If both groups are unbroken, $m_V \to 0$, the states $\chi$ are  ``millicharged particles'' with electric charge $q_\chi = e\epsilon$. 
For $m_V \neq 0$, at $|q^2| < m_V^2$,  the particles $\chi$ can be thought of as neutral particles with a non-vanishing electric charge radius, $r_\chi^2 \simeq 6 \epsilon  m_V^{-2}$. The diagram describing the basic interaction between the two sectors is shown in Fig. \ref{fig:VPfig1}. 

\item If there are no states charged under $U(1)'$ (or they are very heavy), and $m_V$ is taken to be zero, then the two sectors 
decouple even at non-zero $\epsilon$. This leads to the suppression of all interactions for a dark photon inside a medium. If 
$m_V$ becomes smaller than the characteristic plasma frequency all processes with emission or absorption of dark photons decouple as 
$\sim m_V^2$ \cite{An:2013yfc}. 

\item The new vector boson interacting with the SM via the electromagnetic current conserves several approximate symmetries of the SM, including parity, 
flavour, and $CP$. Moreover, $A'$ does not couple directly to neutrinos. 
As a consequence of these two features, the interaction strength due to the exchange of $A'$ can be taken to be stronger than that of 
weak interactions, $(e \epsilon)^2/m_{A'}^2;~  (e \epsilon g')/m_{A'}^2\gg G_F$. This property proves very useful in constructing 
light dark matter models with the use of vector portal. 

\end{enumerate} 

Although this model was known to theorists and well-studied over the years ({\em e.g.} Refs. \cite{Galison:1983pa,Foot:1991kb}), there has been a revival of scenarios involving a kinetically-mixed $A'$ during the last decade. Much of this activity has been in response to various astrophysical anomalies which can be interpreted as a sign of dark matter interacting with the SM through a kinetically mixing vector. 
Renewed interest in  dark photons has triggered new analyses of past or existing experiments~\cite{Borodatchenkova:2005ct,Fayet:2006sp,Fayet:2007ua,Bjorken:2009mm,Andreas:2012mt,Gninenko:2013sr,Adare:2014mgk,Lees:2014xha,Babusci:2014sta,Goudzovski:2014rwa}, and generated proposals for new dedicated 
experiments, which are currently at various stages of implementation~\cite{Merkel:2011ze,Abrahamyan:2011gv,Battaglieri:2014hga,Balewski:2014pxa}. 
In this chapter, we will demonstrate that the SHiP proposal
is capable of probing new domains of the parameter space for this model, 
with and without light dark matter.

\subsection{Anomaly-free gauge groups ($B-L$, $L_\mu - L_\tau$ etc)}

The kinetically-mixed portal described above represents the simplest way to couple a new 
vector particle to the SM, without charging any of the SM fields under the 
new gauge group. There is also an interesting alternative route in which certain 
combinations of the SM fields are charged under the new $U(1)'$.

The most prominent example of this type is $V^{(B-L)}$, which, provided the SM is supplemented with three right-handed neutrinos, is anomaly free. While the multitude of scales for the mass of $B-L$ gauge boson is possible, the value of the 
coupling is quite constrained by the fact that neutrinos acquire a new interaction with electrons and baryons, so that broadly speaking we require $(g_{(B-L)})^2/m_{V^{(B-L)}}^2 < G_F$.

The only model-building complication that arises in this construction is related to neutrino 
masses. One can consider Dirac masses for the neutrinos, in which case this problem does not exist. 
On the other hand, the Majorana masses of right-handed (RH) neutrinos, and more generally the effective operator of dimension five that generates light neutrino masses, $(LH)^2$, are incompatible with $B-L$ gauge symmetry. 
One can solve this problem with additional model ingredients.
For example, the right-handed neutrino mass can be associated with the condensation of an additional Higgs field, $\Phi$, that has a charge $-2$ under this gauge group. so that the mass term in the Lagrangian $y_N NN\Phi$ is gauge-invariant. Then the ratio of masses of the $U(1)_{B-L}$ gauge boson and the right-handed neutrino mass would scale as
\be
\frac{m_{V^{(B-L)}}}{m_N} \sim \frac{g^{(B-L)}}{y_N}.
\ee
It is possible then that the masses of the RH neutrinos and the $B-L$ gauge boson can be comparable, and the lightness of 
$V^{(B-L)}$ may in turn imply a relative lightness of $N$. This will be important for the phenomenological signals of the $B-L$ interaction in fixed target experiments. 

Individual lepton flavours can also be gauged in specific anomaly-free combinations. Thus, the $B-3L_i$ combinations, where $i$ is 
an individual flavour, have been considered in the past, and their phenomenology is very similar to the $B-L$ case discussed above. 
One specific group, based on the $L_\mu-L_\tau$ combination, is an exception: since neither electrons nor quarks are charged under this 
group it is very difficult to constrain experimentally, 
and the strength of the new interaction can be comparable with the weak strength. For example, $(g_{(L_\mu-L_\tau)})^2/m_V^2 \sim G_F$
is not excluded. There is an extensive theoretical literature dedicated to such symmetries, see {\em e.g.} Refs. \cite{He:1990pn,He:1991qd,Baek:2001kca,Ma:2001md,Salvioni:2009jp,Heeck:2011wj}.

\subsection{Other froms of vector portals. }

The examples of vector portals described above are special in that they are UV complete and do not require new physics at the weak scale.
In this subsection we discuss other possibilities that require additional steps for UV completion. 

One example with a distinct phenomenology is based on gauged baryon number, $U(1)_B$, \cite{Pospelov:2011ha,Dobrescu:2014fca,Tulin:2014tya,Batell:2014yra}. 
This symmetry is anomalous (in particular there are mixed electroweak-$B$ anomalies)  and therefore a gauged $U(1)_B$ construction requires a UV completion. Such a completion can be obtained, for instance, with new electroweak charged chiral fermions that cancel the anomaly and obtain a weak scale mass through spontaneous symmetry breaking. Therefore, from the low-energy point of view, the $U(1)_B$ portal is legitimate, leading to rather distinct ``leptophobic'' phenomenology. 

Individual quark flavours can also be gauged in anomaly-free combinations. For example, gauging two quark flavour $q_1-q_2$ in a manner similar to $L_\mu-L_\tau$ is possible. The difficulty lies with quark mass sector, where additional 
Higgses must be introduced in order to generate mass terms consistent with the gauge groups of the SM and of the new $U(1)$. 
While this may be possible from a model-building point of view, one has to address the rather severe phenomenological challenge of new tree-level flavour-changing neutral 
currents.

Finally, even anomalous gauge groups can be considered as low energy effective field theories. In general, even in the presence of gauge anomalies gauge invariance can be restored at the expense of non-renormalizability of the theory~\cite{Preskill:1990fr}. This leads to an associated energy scale at which one loses calculability and requires a UV completion. 
For our purposes, it is sufficient to assume that this scale is at the weak scale or above and then consider the effective theory of vector particles,
under the assumption that some UV completion may be found. 

On pure phenomenological grounds, one can classify the  possible vector couplings to the SM fields using the 
framework of the effective Lagrangian where for simplicity we assume the minimal flavour violation (MFV) ansatz. 
The vector interactions of the MFV type \cite{Kamenik:2012hn} are given by the following combinations of the SM fields, Yukawa matrices, and 
unknown coefficients $a_I$ and $b_{IJ}$:
\begin{eqnarray}
{\cal L}_{\rm int} = &V_\mu& \bar Q \gamma_\mu (a_Q{\bf 1} + b_{QU} Y^\dagger_UY_U + b_{QD} Y_D^\dagger Y_D +... ) Q \nonumber \\
+&V_\mu& \bar U \gamma_\mu( a_U{\bf 1} + b_{UU} Y_UY_U^\dagger +...) U  \nonumber \\
\label{stuckelberg_portal} 
+&V_\mu& \bar D\gamma_\mu ( a_D{\bf 1} + b_{DD} Y_DY_D^\dagger +... ) D \\
+&V_\mu &\bar L \gamma_\mu(a_L {\bf 1} + b_{LE} Y_E^\dagger Y_E +...) L  \nonumber \\
+&V_\mu &\bar E \gamma_\mu(a_E {\bf 1} +b_{EE} Y_E Y_E^\dagger+... ) E.
\nonumber
\end{eqnarray} 
At this point, it is more appropriate to think of $V_\mu$ as of  ``ordinary'' Maxwell-Proca field, rather than 
a gauge boson. Written in this form, it is easy to see that the mass $m_V$ is not protected at loop level, and indeed 
will receive additive corrections proportional to the cutoff scale. Within 
effective field theories, the coefficients $a_I$ and $b_{IJ}$ cannot be fixed from first principles, but instead can be constrained directly from experiment. 
Their smallness guarantees that the cutoff can be taken at the TeV scale or higher.

So far we have primarily discussed dimension four vector portals, but the non-renormalizable 
part of (\ref{stuckelberg_portal}) essentially descends from higher-dimensional operators.
It is clear that at dim$>4$, one can construct  many new forms of the higher-dimensional operators. For example, 
one can have a fully gauge invariant dipole portal, 
\be
{\cal L}_{\rm dipole} = \sum_{ij} \bar f_i  (1\times \mu_{ij} + \gamma_5\times d_{ij}) \sigma_{\alpha\beta}  f_j~ V_{\alpha\beta} +(h.c.).
\label{dipole}
\ee
Here, $f_i$ represents  different SM fermions, $V_{\alpha\beta}$ is the field strength of the 
exotic vector state, and $\mu_{ij}$ and $d_{ij}$ are complex-valued Wilson coefficients. This generalizes the dipole operator
discussed in \cite{Dobrescu:2004wz}. Due to the explicit gauge invariance, the mass of the vector boson $V$ does not receive corrections 
from (\ref{dipole}) and can be small. Also, the $SU(2)\times U(1)$ properties of the SM fermions implies that the actual dimension of such operators is six, 
with 
$\mu $ and $d $ scaling as $\sim v_{EW}/\Lambda^2_{\rm UV}$. The phenomenology of this portal, including the 
production and decay of dipole-coupled vectors, has not been studied in any detail in the literature. 
We note in passing that the production and decay of $V$ via (\ref{dipole}) can look similar to the case of the 
axion-like particle, $(\partial_\mu a) \bar f_i \gamma_\mu f$, due to the same scaling with the momentum of the exotic particle. 

\InsertSection{ship_vector_portal_CS.tex} %

\section{Matter states charged under new $U(1)$ }
\label{sec:matter-U1-prime}

The new vector portal is likely to be accompanied by new states charged under $U(1)'$. In this section we review 
different theoretical constructions behind the finite mass for the new vector bosons and discuss the possibility that 
the new $U(1)'$ sector is supersymmetric.

\subsection{Higgs mechanism in the dark sector}

There are several theoretical ways of breaking the $U(1)'$ symmetry. In all UV-complete models, 
the new vector particle couples to the conserved current, which allows for the introduction of a "hard" or "Stuckelberg" mass term, ${\cal L}_m = \frac12 m_V^2 V_\mu^2$. This represents the most minimal possibility. 
If, however, the additional gauge symmetry is broken in the same way as in the SM, a new Higgs field in the dark sector will be introduced. 

The ``Higgsing'' of the $U(1)'$ introduces extra interactions in the dark sector.  Consider the simplest possibility in which the 
new scalar field is charged under $U(1)'$ \cite{Batell:2009yf}, 
\be
{\cal L}_\phi = |D_\mu \phi|^2- \lambda |\phi|^4 + \mu^2 |\phi|^2,
\ee
where $D_\mu$ is the covariant derivative.
After the spontaneous symmetry breaking, $\phi = (v'+h')/\sqrt{2}$, the vector field becomes massive with $m_V = g' v'$. Besides the mass terms, several interaction terms appear:
\be
{\cal L}_{h'} = \frac{1}{2}m_V^2 V_\mu^2 +\frac{1}{2}(\partial_\mu h')^2-\frac{1}{2}m_{h'}^2 h'^2 +\frac{m_V^2}{v'} h' V_\mu^2 +\frac{m_V^2}{v'^2}\,h'^2 V_\mu^2 - \frac{m_{h'}^2 }{2 v'} h'^3 - \frac{m_{h'}^2}{8 v'^2} h'^4,
\ee
where $m_{h'}$ is the mass for the dark Higgs particle. 
Depending on the relative sizes of the quartic coupling $\lambda$ and the square of the gauge coupling $g'^2$ the dark Higgs can be lighter or heavier than 
the vector. This will become important for the potential signals of dark sector as $m_{h'} \leq m_V$ opens the possibility for a macroscopic decay length 
of the dark Higgs \cite{Batell:2009yf,Pospelov:2008zw}. 

A more complicated Higgs sector can lead to an additional important effect for a dark photon beyond 
the kinetic mixing. In particular, a mass mixing between $V$ and the SM $Z$ boson becomes possible \cite{Davoudiasl:2012ag}, 
opening a host of additional phenomenological consequences such as  parity violation mediated by $V$ exchange and the
coupling of $V$ to the SM neutrinos.

\subsection{Supersymmetric $U(1)'$ models}
\label{sec:susy-u1-models}
Supersymmetric extensions of vector portal models are also well-studied.
For a SUSY version of the dark photon model, see {\em e.g.} \cite{Dienes:1996zr,Hooper:2008im,Baumgart:2009tn,Morrissey:2009ur}.
One can imagine a variety of scenarios vis-a-vis the origin of supersymmetry breaking that could lead to different mass patterns between the
observable and dark sectors. One of the most interesting possibilities includes a scenario 
with an approximately supersymmetric dark sector in which the breaking of supersymmetry in the $U(1)'$ sector is mediated by the vector portal coupling from the SM. In the simplest scenarios of this kind, one expects the following relations to hold true:
\be
m_{A'}^2 \sim \epsilon^{1/2}M_Z;~~ m_{A'} = m_{h'}.
\ee
The first relation arises because of the induced $D$-term in the $U(1)'$ sector that scales as $\sim \epsilon$ \cite{Baumgart:2009tn}. 
The second relation is the consequence of the minimal Higgs sector in the SUSY version of the dark photon model \cite{Morrissey:2009ur}.
Equal masses for $h'$ and $A'$ particles forbid the decays of $h'$ to $A'$ and prolong the lifetime of $h'$, thus making supersymmetric dark Higgses  an interesting 
target for the studies at SHiP. More details on supersymmetric scenarios in conjunction with the light vector portal can be found in the 
SUSY chapter of this white paper, Chapter~\ref{sec:susy}.

\section{Physics motivation for light mass (less than weak scale) vector particles}
\label{sec:motivation-light-vector-particles}

The interest in vector portals comes from several sources. The appearance of such light vector portals is quite common in top-down constructions, such string-inspired models and GUT theories. 
At the same time, several variants of the vector portal models, especially with relatively light mediators, have been invoked as a remedy for a number of observational 
anomalies, both in particle physics and astrophysics. This section reviews a subset of the problems that vector portal models help to resolve.

\subsection{Putative solution to the muon $g-2$ discrepancy}

The persistent discrepancy between the measured muon anomalous magnetic moment
and the SM prediction at the level of $\sim$3$\sigma$~\cite{Bennett:2006fi}
has generated significant experimental and theoretical activity aiming for a
possible explanation. The intense scrutiny of the SM contributions has not
produced any obvious candidate for an extra contribution $\Delta a_\mu \sim +3
\times 10^{-9}$ that would cover a theoretical shortfall and match the
observed value.  Among the new physics explanations for this discrepancy are
weak scale solutions~\cite{Czarnecki:2001pv}, as well as possible new
contributions from light and very weakly coupled new particles (see,
e.g.,~\cite{Gninenko:2001hx,Fayet:2007ua,Pospelov:2008zw}).  With the LHC
continuously squeezing the available parameter space for the weak-scale
$g-2$-relevant new physics, solutions with light particles appear as an
attractive alternative.

\begin{figure}
\centerline{\includegraphics[width=0.55\textwidth]{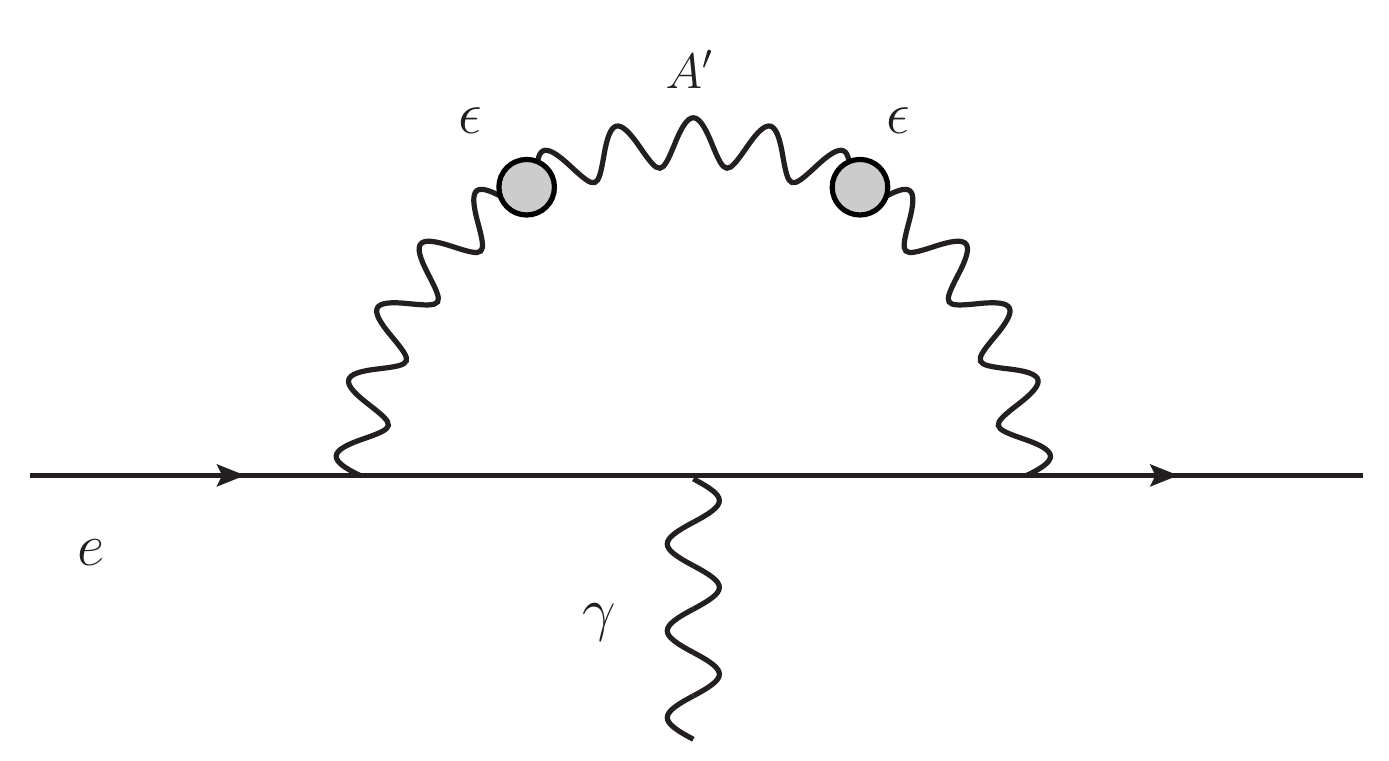}}
\caption{One-loop correction to the muon anomalous magnetic moment due to the exchange of the dark photon. }
\label{fig:VPfig2}
\end{figure} 

It is easy to see that a light vector particle coupled to muons via the vector portal provides an upward correction to $g-2$. In most models the new vector particle does not have an axial-vector coupling to charged 
leptons, and a simple one-loop diagram, Fig. \ref{fig:VPfig2}, gives a positive 
correction to the magnetic anomaly,
\be
\label{g-2}
a_l^{V}  = \frac{\alpha}{2\pi} \left(\frac{g'}{e}\right)^2 \times \int_{0}^{1}dz\frac{2m_l^2z(1-z)^2}{m_l^2(1-z)^2+m_V^2z}
=\frac{\alpha}{2\pi} \left(\frac{g'}{e}\right)^2\times \left\{\begin{array}{c}1 ~~{\rm for} ~~ m_l\gg m_V,\\
										2m_l^2/(3m_V^2) ~~{\rm for} ~~ m_l\ll m_V.\end{array}
\right. 
\ee
In this expression $g'/e$ is the strength of the $V_\mu$ coupling to the muon vector current in units of electric charge. 
For the kinetically-mixed dark photon $A'$, we have $g'/e = \epsilon$. For the choice of $\epsilon \sim{\rm few} \times 10^{-3}$ at $m_V \sim m_\mu$, 
the new contribution brings theory and experiment into agreement. Since 2008, significant  experimental advances have been made towards testing this possibility, while at the same time various theoretical extensions of the simplest kinetic mixing portal explanation have been proposed. The following picture has emerged: 
\begin{itemize} 

\item The minimal dark photon model, with no light particles charged under $U(1)'$ is excluded (or at least very close to being excluded) by a complementary array of experiments. The most difficult part of the parameter space, the vicinity of $m_{A'} \sim 30$ MeV, has finally been ruled out only recently as a solution to the $g-2$ puzzle,
\cite{Lees:2014xha,Goudzovski:2014rwa}. 

\item A slightly extended model of the dark photon can still offer a solution to the $g-2$ discrepancy. For example, if $A' \rightarrow \chi \bar\chi$ in the dark sector the visible $A'\to e^-e^+$ decays will be diluted. In any case, it appears that $m_{A'} < 200$ MeV is required \cite{Essig:2013vha}.  

\item Finally, the least constrained model is based on the gauged $L_\mu-L_\tau$ vector portal 
\cite{Baek:2001kca,Ma:2001md,Heeck:2011wj}, and the vector mass below $m_V \sim 400$ MeV can still be considered as a potential solution to the muon $g-2$ discrepancy  \cite{Altmannshofer:2014cfa,Altmannshofer:2014pba}. 

\end{itemize}

To summarize, a light vector particle remains an attractive solution to the muon $g-2$ discrepancy, and more experimental work is required to 
exclude this possibility in a model-independent way.

\subsection{Mediator of interaction with DM and possible connection to astrophysical positron excess }
\label{sec:light-DM-mediator}

Vector portals offer a means to connect the SM to dark matter. In the last few
years, the direct searches for dark matter have intensified, paralleled by
broad investigations of the theoretical opportunities for dark matter. The
weakly interacting massive particle (WIMP) paradigm offers perhaps the largest
number of opportunities for the experimental discovery of dark matter via its
non-gravitational interaction. In the standard WIMP paradigm, known from 1970s
\cite{Lee:1977ua,Vysotsky:1977pe}, the correct cosmological abundance of dark
matter is achieved via its self annihilation at high temperatures, $T\sim
m_\chi$, where $m_\chi$ is the WIMP mass. Simple calculations show that the
required WIMP abundance is achieved if \be \sigma_{\rm annih} (v/c) \sim
1~{\rm pbn} ~\Longrightarrow~\Omega_{\rm DM} \simeq 0.25, \ee where $v/c$ is
the approximate relative velocity at the time of annihilation. The nature of
the interaction responsible for the self-annihilation of WIMPs to the SM
states is important. It sets the size of the self-annihilation cross section
and ultimately the abundance of WIMP dark matter. If the interactions are
mediated by forces that have weak strength and operate through the exchange of
weak scale particles, then for small and large masses one expects the
following scaling with the WIMP mass, \be
\sigma (v/c) \propto \left\{\begin{array}{c}G_F^2m_\chi^2 ~~{\rm for} ~~ m_\chi \ll M_W,\\
    1/m_\chi^2 ~~{\rm for} ~~ m_\chi\gg
    M_W.\end{array}\right. ~~\Longrightarrow ~~{\rm few~GeV} < m_\chi < {\rm
  few~TeV} \ee This famously determines the so-called "Lee-Weinberg window",
or the mass range for the DM under the assumption of weak-scale mediators.
According to this logic thermal relic MeV-GeV scale dark matter is disfavored.

The crucial  assumption in the argument above is the link between the weak scale and the mass of the mediator particles. As was argued in the previous sections certain vector portals do allow interaction strengths in excess of $G_F$. This in turn opens the door to the construction of rather natural  models of light dark matter, 
which can be made as light as a few MeV \cite{Boehm:2003hm}. 
It is important to note that such light WIMPs are challenging to probe via direct scattering of galactic DM particles on atoms \cite{Essig:2011nj}, and therefore 
alternative strategies aimed covering this mass range must be developed.

\begin{figure}
\centerline{\includegraphics[width=0.55\textwidth]{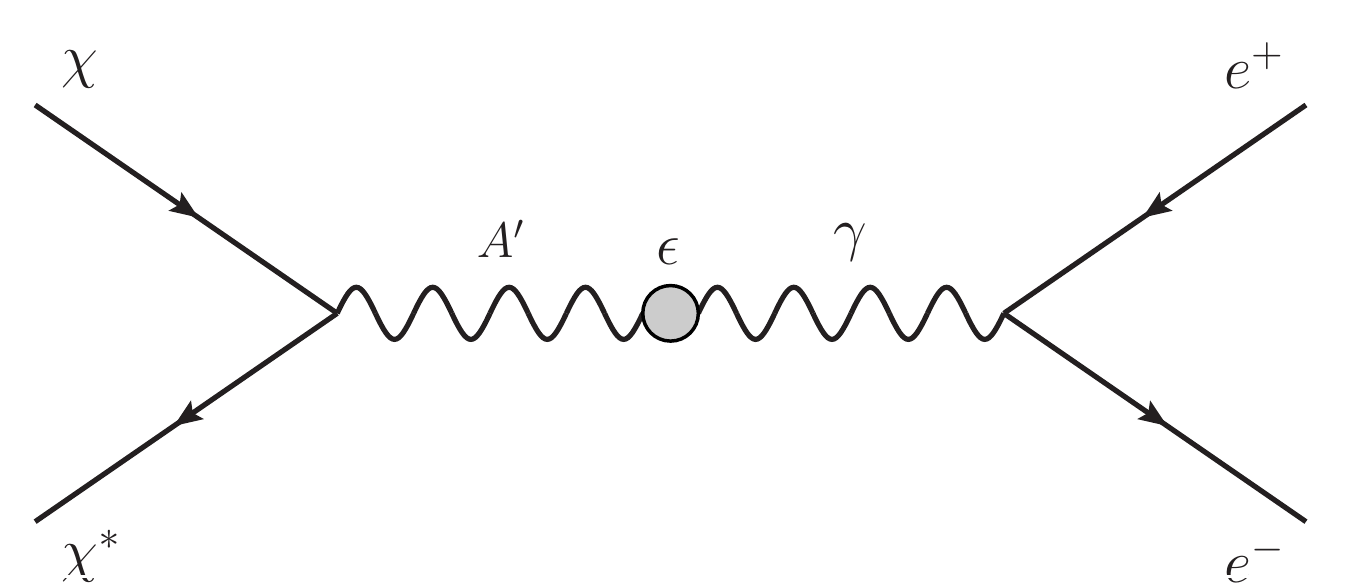}}
\caption{Light ($m_\chi \sim {\rm few}$~MeV) scalar dark matter annihilating to electron-positron pairs due to the mixed $\gamma-A'$ propagator. 
The annihilation occurs in the $p$-wave. }
\label{fig:VPfig3}
\end{figure} 

On the phenomenological side, the light dark matter can be behind an unexpectedly strong emission of 511 keV photons from the galactic bulge, 
as observed by SPI/INTEGRAL \cite{Jean:2003ci}. 
It is presently unclear whether new physics needs to be invoked for the explanation of such emission, and we refer readers to the on-going discussions in the literature \cite{Prantzos:2010wi}. Nonetheless, a dark-matter-related origin of 511 keV excess, in which the non-relativistic or semi-relativistic positrons are products of DM annihilation or decay, is possible \cite{Boehm:2003bt}. 
For example, scalar dark matter charged under the new $U(1)'$ with a mass in the 
$m_\chi \sim $few~MeV range can pass all the existing constraints \cite{Boehm:2003hm}, and supply the requisite source for positrons. 
Direct calculations in the model where mediation of the SM-DM interaction occurs due to the dark photon, as in Fig. \ref{fig:VPfig3}, gives  the annihilation cross-section
in the form
\be
\sigma_{\rm annih} (v/c)\simeq  
\frac{  8 \pi \alpha \alpha_D  \epsilon^2  ( m_\chi^2  + 2 m_e^2  )  v^2     }{   3(m_{A'}^2 -4m_\chi^2)^2    }  \sqrt{ 1 -  m_e^2 / m_\chi^2    }.
\ee
Here $\alpha_D = (g')^2/(4\pi)$, and $m_\chi \gg m_e$ is assumed.
The extra factor of the squared relative velocity in this formula is indicative of the $p$-wave annihilation, and is what ultimately allows this model to escape strong constraints 
on light dark matter annihilation imposed by the accurate measurements of CMB anisotropies. 
The least constrained region of the parameter space corresponds to very light mediators, $m_{A'} < 100$ MeV, and $2 m_\chi < m_{A'}$. 
With this choice of parameters, $\sigma_{\rm annih} (v/c)$ can be significantly larger than 1 pbn, making MeV-scale dark matter possible.  

Another prominent astrophysical anomaly that has generated a great deal of speculation as a possible sign of dark matter is the observation of the rise in the fractional positron flux as a function of energy.  In 2008, the results of PAMELA satellite experiment \cite{Adriani:2008zr,Adriani:2010rc} indicated that the 
fractions of galactic anti-proton flux, $n_{\bar p}/(n_p+ n_{\bar p})$, as a function of energy, behaves according to the fiducial expectations from the astrophysical modelling of the cosmic ray origin and propagation. In contrast, the corresponding fraction of 
positrons, $n_{\bar e}/(n_e+ n_{\bar e})$, exhibits a significant upturn above $E>10$~GeV, prompting speculations about the necessity of 
additional primary sources of energetic positrons. This measurement was independently confirmed through FERMI-LAT observations \cite{FermiLAT:2011ab}, and 
brought to the new level of accuracy by the AMS-2 experiment \cite{Aguilar:2013qda}. The annihilation of heavy dark matter with $m_\chi > M_W$ could be a theoretically attractive source of such positrons. 
Yet the simplest WIMP models do not fit the positron excess because of two problems. The required annihilation rate capable of supplying 
the positron excess is above the WIMP freeze-out annihilation rate by $\sim $ two orders of magnitude. In addition, models where the final state 
annihilation products are heavy SM particles ($b,~t,~W,~Z,~h$) will necessarily produce antiprotons and are therefore are tightly constrained 
by $n_{\bar p}/(n_p+ n_{\bar p})$. 

\begin{figure}
\centerline{\includegraphics[width=0.75\textwidth]{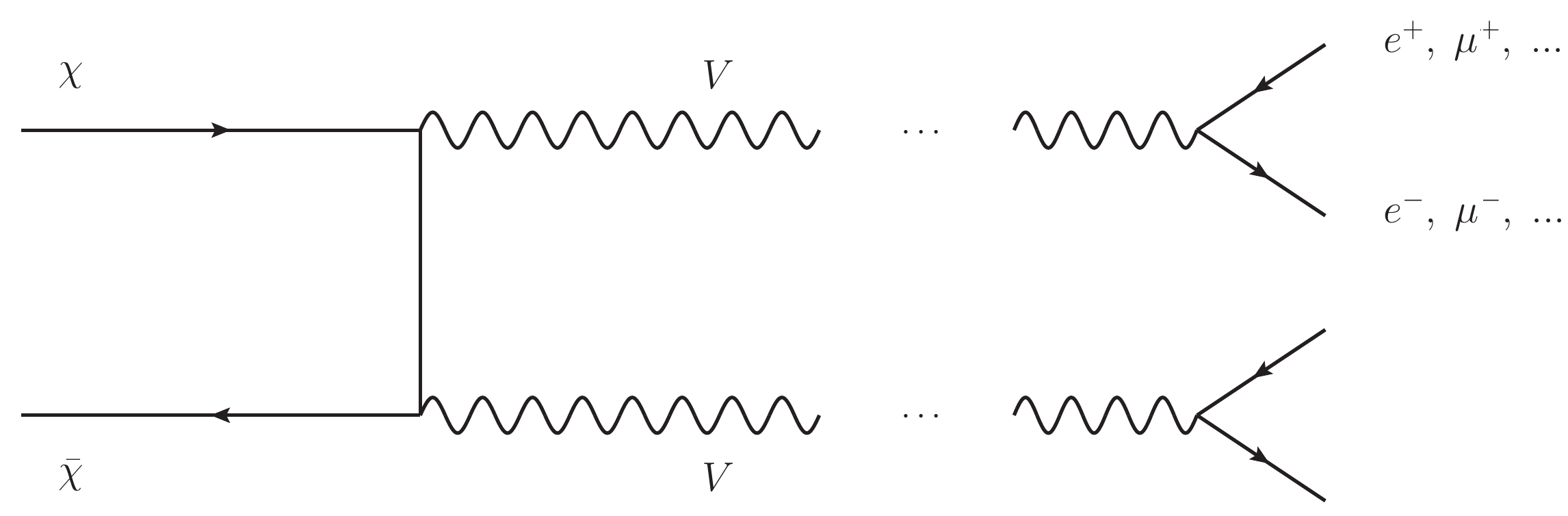}}
\caption{Sequential annihilation of heavy WIMP  dark matter, first to light metastable mediators, 
 with their subsequent decay to light SM states.  }
\label{fig:VPfig4}
\end{figure} 

It was soon realized that these problems can be rather efficiently circumvented if the heavy WIMP dark matter is 
interacting with the SM via relatively light mediators \cite{ArkaniHamed:2008qn,Pospelov:2008jd}, 
and the DM$\to$SM annihilation occurs via an intermediate stage of light mediators as in Fig. \ref{fig:VPfig4}. 
In particular, for the light vector mediator one finds that: 
\begin{itemize}  

\item The WIMP dark matter abundance is regulated via the $\chi \bar \chi \to VV \to {\rm SM} $ annihilation process. If $m_V$ is sufficiently 
light, then the $v\sim 0.3 c$ and $v\sim 10^{-3}c$ annihilation regimes (freeze-out vs galactic environment) can be markedly 
different. The existence of the dark-force-induced attraction between WIMP and anti-WIMP particles creates a Coulomb (or Sommerfeld) type
enhancement of the cross-section, providing an extra factor $S \sim \pi \alpha_D/ v \gg 1$ at small velocities. The 
galactic annihilation can be further enhanced by the resonant states in the $\chi-\bar\chi$ system. This solves the problem for required 
enhancement of the annihilation cross section in the galactic environment. 

\item The fraction of leptons in the final states can be significanly enhanced. Indeed, for $m_V < 2 m_p$, the antiprotons cannot be generated
in the decay of mediators $V$. The annihilation branching to photons is also significantly reduced if $\pi^0$ cannot be easily generated in the decay products. 

\item Finally, the two-stage annihilation process, $\chi \bar \chi \to VV$, followed by decay $V \to {\rm SM}$, allows hiding the 
WIMP dark matter from direct searches.  This is because the coupling to the SM ({\em e.g.} $\epsilon$ for dark photons) is not constrained 
by the annihilation  cross section, and can be taken almost arbitrarily small \cite{Pospelov:2007mp}. 

\end{itemize} 

This theoretical construction proves that galactic positron excess {\em could} come from the annihilation of WIMP dark matter
(there are no show stoppers), although at this point it is impossible to argue in favor of the dark matter explanation vs competing astrophysical 
mechanisms (see, {\em e.g.} \cite{Profumo:2008ms,Hooper:2008kg}). These
models do come under additional constraints from the absence of broad excesses in $\gamma$-ray fluxes \cite{Ackermann:2012qk}, and more recently from 
the absence of DM-induced distortion of the CMB anisotropies as reported by Planck collaboration (Planck collaboration, {\em to appear}). 
While these constraints are very 
important, they cannot rule out the theoretical construction above in a model-independent way. Thus, it is important to constrain the 
DM-SM mediators directly from terrestrial experiments, and for the broad ranges of possible  couplings. 

Vector portals can also open a window to non-thermal dark matter, particularly in the case of sterile neutrino dark matter, $N$ \cite{Shuve:2014doa}. Sterile neutrinos mix with Standard Model (SM) neutrinos, and therefore any new vector interactions involving SM neutrinos, such as gauged lepton number symmetries, indirectly couple to sterile neutrinos as well. This can modify the sterile neutrinos' cosmological abundance, and the existence of new vector interactions coupled to SM leptons also provides a new phenomenological probe of theories with sterile neutrino dark matter. In the minimal model where the sterile neutrinos interact with the SM only via mixing (the neutrino minimal SM, $\nu$MSM), $N$ production is entirely determined by the $N$ mass and its mixing with a lepton doublet $L_\alpha$, given by $\sin\theta_\alpha$.  Production of $N$ proceeds by electroweak production of SM neutrinos, which subsequently oscillate into $N$; this is known as the Dodelson-Widrow mechanism \cite{Dodelson:1993je}. Currently, the combination of constraints on warm DM and X-rays completely rules out the Dodelson-Widrow mechanism as being the sole explanation for the DM abundance. 
 Alternate mechanisms have also been proposed to enhance the production rate of $N$, such as via large lepton asymmetry and/or 
the decays of new singlet scalars in the inflation or Higgs sectors \cite{Shi:1998km,Abazajian:2001nj,Shaposhnikov:2006xi,Kusenko:2006rh,Merle:2013wta}.

New vector mediators $V$ coupled to SM neutrinos also produce a larger abundance of $N$ by analogy with the Dodelson-Widrow mechanism \cite{Shuve:2014doa}:~the $V$ particle results in a higher scattering rate of $N_1$ in the early universe, and consequently a larger production rate of $N$ through oscillation. This is true even if the $V$ has no coupling to $N$ prior to electroweak symmetry breaking.  If the new interaction occurs in the neutral current only, it does not contribute to $N\rightarrow \nu_\alpha\gamma$ decay and can be safe from X-ray constraints. Since the new interactions exclusively couple to SM particles, the vector portal provides a direct connection between sterile neutrino cosmology and the phenomenology of new vector states.
The sterile neutrino abundance as a function of $M_{N}$, mixing angle, $M_{V}$, and $g'$ is computed in Ref.~\cite{Shuve:2014doa}, taking into account all finite-temperature corrections and oscillation damping effects associated with both the electroweak interactions and the new vector state $V$. New vector portal interactions with $M_{V}\sim$ MeV-GeV and $g'\sim10^{-6}-10^{-2}$ are consistent with dark matter with masses 1-100 keV $N$, as well as all cosmological constraints and X-ray bounds on $\sin\theta_\alpha$. This gives a concrete window for phenomenological probes of $V$.

\subsection{Self-intereaction of dark matter via light mediators}

Finally a noteworthy motivation for the light vector portal particles come from the host of observational questions 
related to the dark matter sub-structure and their possible resolution via {\em self-interaction} of DM particles. 
The astrophysical structure of galaxies and clusters of galaxies can be dramatically affected by scattering interactions between 
dark matter particles~\cite{Spergel:1999mh,Dave:2000ar}.  
This effect, known as {\it dark matter self-interactions}, is motivated based on two outstanding issues in our understanding of structure formation.  
\begin{itemize}
\item {\it Core-cusp problem:} Observations indicate that dwarf galaxies exhibit reduced central densities with flat profiles (cores)~\cite{Moore:1994yx,Flores:1994gz,Simon:2004sr,Oh:2010ea,Walker:2011zu}, as opposed to steeper density profiles (cusps) predicted by collisionless DM-only simulations~\cite{Navarro:1996gj}.  More recently, dark matter cores have been found in larger systems as well, including low surface brightness spiral galaxies~\cite{deBlok:2001fe,KuziodeNaray:2007qi,deNaray:2011hy,deBlok:2002tg} and clusters~\cite{Newman:2012nw}.  
\item {\it Too-big-to-fail problem:} Collisionless DM-only simulations predict that MW-like halos should contain a population of $\mathcal{O}(10)$ subhalos that are too massive to host any of the known most massive classical dwarfs, as inferred by stellar kinematics~\cite{BoylanKolchin:2011de,BoylanKolchin:2011dk}.  Satellite populations of the Andromeda Galaxy and the field of the Local Group also exhibit a dearth of massive satellites expected from collisionless DM-only simulations~\cite{Tollerud:2014zha,Garrison-Kimmel:2014vqa}.
\end{itemize}
These discrepancies may be explained if DM particles within halos are not collisionless, as is usually assumed, but instead undergo elastic scattering with other DM particles.  DM self-interactions lead to thermodynamic heat transport from the hotter outer halo to the colder inner halo, thereby reducing the central halo density as DM particles get heated and escape.  Additionally, the formation of cores in satellite galaxies can modify their stellar kinematics, resolving the mismatch between the most massive satellites observed in the MW and those found in simulations.  Recent simulations of self-interacting dark matter have shown that a self-interaction cross section per unit mass of
\begin{equation}
\sigma/m_\chi \gtrsim 1 \; {\rm cm}^2/{\rm g} \sim 2 \times 10^{-24} \; {\rm cm}^2/{\rm GeV}
\end{equation} 
can resolve the core-cusp and too-big-to-fail problems on dwarf scales~\cite{Rocha:2012jg,Peter:2012jh,Vogelsberger:2012ku,Zavala:2012us}.  

On the other hand, these discrepancies may simply point to the inadequacy of using DM-only simulations to make inferences about the real Universe containing both DM and baryons.   A variety of mechanisms have been proposed to generate cores in DM halos through gravitational feedback from baryons, including supernovae in gas-rich dwarfs~\cite{Pontzen:2011ty,Governato:2012fa}, environmental interactions between satellite galaxies and the baryonic disk of the larger host galaxy~\cite{Zolotov:2012xd}, and active galactic nuclei in clusters~\cite{Teyssier:2010dp}.  

Although the origin of these discrepancies remains an open and hotly-debated question, self-interacting dark matter remains an appealing alternative to the WIMP paradigm that can be tested in both astrophysical and terrestrial contexts.  Self-interactions are present for WIMP dark matter.  However, the self-interaction cross section for WIMPs is set by the weak scale, with $\sigma \sim 10^{-36} \; {\rm cm}^2$ and $m_\chi \sim 100$ GeV, giving the figure of merit $\sigma/m_\chi \sim 10^{-38} \; {\rm cm}^2/{\rm GeV}$ that is too small by fourteen orders of magnitude.  Therefore, self-interacting dark matter requires the existence of new dark mediator particles that are much lighter than the weak scale.

The minimal model for self-interacting dark matter consists of a dark matter particle $\chi$ (taken to be a fermion) and a mediator particle, scalar $\phi$ or vector $V$, with interaction~\cite{Ackerman:mha,Feng:2009hw,Buckley:2009in,Loeb:2010gj,Lin:2011gj,Aarssen:2012fx,Tulin:2012wi,Tulin:2013teo}
\begin{equation}
{\cal L}_{\rm int} = \left\{ \begin{array} {ll} g_\chi \bar{\chi} \gamma^\mu \chi V_\mu & {\rm vector \; mediator} \\
g_\chi \bar \chi \chi \phi & {\rm scalar\; mediator} \end{array} \right. \, .
\end{equation} 
The coupling $g_\chi$ can be related to a ``dark'' fine structure constant $\alpha_D = g_\chi^2/(4\pi)$.  In the small velocity limit, the Born self-interaction cross section is\footnote{In general, Eq.~\eqref{born-cross-section} breaks down for $\alpha_\chi m_\chi/m_\phi \gtrsim 1$, where nonperturbative Sommerfeld-type effects become important, or for $m_\chi v/m_V \gtrsim 1$, where $\sigma$ can no longer be represented by a contact interaction.  In these regimes, a numerical solution for $\sigma$ has been performed~\cite{Buckley:2009in,Tulin:2013teo}.}
\begin{equation}
\sigma = 4\pi \alpha_D^2 m_\chi^2 /m_V^4 \approx 6 \times 10^{-23} \, {\rm cm}^2 \left(\frac{\alpha_D}{0.01}\right)^2 \left(\frac{m_\chi}{10 \: {\rm GeV}}\right)^2 \left(\frac{30 \: {\rm MeV}}{m_V}\right)^4 \, 
\label{born-cross-section}
\end{equation} 
independent of whether mediator is scalar or vector.  If the self-interaction cross section is large enough to explain 
small-scale astrophysical observations, the mediator particle must have mass much lighter than the weak scale, typically in the range~\cite{Tulin:2013teo}
\begin{equation}
m_V \sim {\rm MeV} - {\rm GeV} \, ,
\end{equation} 
over a wide range of DM mass ($m_\chi \sim {\rm MeV} - {\rm TeV}$) and coupling ($\alpha_D \sim 10^{-4} - 10^{-1}$).

The existence of new force in the dark sector at the MeV--GeV scale provides a new motivation for direct searches at high-luminosity colliders.  Self-interactions and small scale structure anomalies do not by themselves require that light mediator couples directly to the Standard Model.  However, without such a coupling, and  in the absence of additional light states
there will be a population of mediator particles that dominates the DM mass budget over 
self-interacting $\chi$ particles~\cite{Kaplinghat:2013yxa}.  Thus, they must decay to SM states, and, requiring that decay occurs before Big Bang Nucleosynthesis to avoid excessive entropy injection, there is minimal value on the coupling between $V$ (or $\phi$) and SM particles that depends on specific model realization.  
Typically, a vector mediator coupling through the kinetic mixing portal must have $\epsilon \gtrsim 10^{-10}$, while a scalar mediator $\phi$ coupling by mixing with the Higgs boson must have a mixing angle larger than around $10^{-5}$. On the other hand, these mixing parameters can be much larger and indeed close to the current experimental limits.

\section{Main features of vector portal phenomenology. }

In this section we review the main phenomenological features of light vector states coupled to the SM vector 
currents and light matter states (dark Higgses, HNLs etc) that may be associated with them. 

\subsection{Decay rates, modes, branchings, $c\tau$ for dark photon}

The phenomenology of massive dark photons with $m_V > 2 m_e$ is quite straightforward. The decay rates of the 
$A'$ to the SM states can be easily calculated for any point in the parameter space $\{m_{A'}, \epsilon\}$. 
More generally, the biggest uncertainty in this scenario is related to the question of the existence of  new states $\chi$ charged under $U(1)'$, which would open additional dark decay channels. In general, the decay width is given by 
\be
\Gamma_{A'} = \sum_l \Gamma_{A'\to l^+l^-} + \sum_{\rm hadrons} \Gamma_{A'\to {\rm hadrons}} + \sum_{\chi} \Gamma_{A'\to \chi\bar\chi}. 
\ee
The decays to leptons and new "dark states" $\chi$ are elementary to handle, 
\begin{eqnarray}
\label{Gammal}
\Gamma_{A'\to l^+l^-} =  \frac13 \epsilon^2 \alpha m_V\left(1+\frac{2m_l^2}{m_V^2}\right)\sqrt{1-\frac{4m_l^2}{m_V^2}},\\
\Gamma_{A'\to \chi\bar\chi} =  \frac13  \alpha_D m_V\left(1+\frac{2m_\chi^2}{m_V^2}\right)\sqrt{1-\frac{4m_\chi^2}{m_V^2}},
\end{eqnarray}
while the expressions for the decays to hadrons can be extracted from the well-known spectral function of virtual photon transition, $\gamma^*\to {\rm hadrons}$,
\begin{equation}
\Gamma_{ V \rightarrow {\rm hadrons} }=\frac{1}{3} \alpha \epsilon^2 m_V \sqrt{1-\frac{4 m_\mu^2}{m_V^2}} \left(1+\frac{2 m_\mu^2}{m_V^2}\right) R(s= m_V^2).
\end{equation}
The $R$-ratio can be found in \cite{Agashe:2014kda}. If dark states are kinematically accessible, then "visible" vs "dark" decay widths scale as 
$\sim \alpha\epsilon^2/\alpha_D$. 

If the decays to dark states are non-existent or subdominant, $\Gamma_{A' \to SM} \gg \Gamma_{A'\to \chi \bar\chi} $, 
then the SM decay channels are not diluted by dark decays. This is the most studied and most constrained case. 
Making a specific choice of parameters, one can see that in the absence of dark decays, (boosted) dark photons
can propagate over macroscopic distances. Choosing $m_{A'} =200$ MeV for simplicity, one determines that 
\be
c\tau_{A'} \gamma \sim   40~{\rm m} \times \left(\frac{10^{-6} }{ \epsilon} \right)^2 \times \frac{\gamma}{100} .
\ee
Thus, this range of mixing angles is of special interest for the SHiP-like experiments.

The production of vector states is a more complicated process than their decays, especially if they 
are produced in $p$ collisions with a target. So far, several production channels have been identified: 
\begin{enumerate} 

\item The dark photons, or more generically, light vectors states can be produced in meson decays,
which in turn are copiously produced in $p$-on-target collisions. 
For light vector masses, by far the most important process is $\pi \to \gamma V$, which is suppressed 
by $\epsilon^2$, but not $\alpha$, since the main decay of $\pi^0$ is electromagnetic. In the literature, 
$\eta$, $\rho/\omega$, and $\Delta$ sources of $A'$ have been discussed \cite{Batell:2009di}. The weak decays of $K,~D$
mesons play a subdominant role in the production of dark photons

\item The (quasi)elestic scattering of incident protons on nucleons in the target can lead to the production 
of vector states via bremsstrahlung process $ pp  \to  p p V$ \cite{Blumlein:2013cua,Gorbunov:2014wqa}. 
This type of process is quite similar to the production of vector states from the 
electrons in collision with nuclei, but is more difficult to handle due to the non-elementary nature of nucleons. 

\item For large $m_V$, it is expected that the direct perturbative QCD production of vector states via the underlying 
$q +\bar q \to V;~ q+g \to q+V $ processes should become dominant. Some of the QCD production cross sections 
have been calculated in Ref. \cite{deNiverville:2012ij}.

\end{enumerate} 

It is fair to say that at this point there are no universally accepted calculational tools available for the computation of production rates and distributions of 
extra vector states in $pp$ and $pn$ collisions over a wide range of incident energies and for all values of masses $m_{V}$.
We expect that in the future more theoretical work will be done to fill the needs of the SHiP experiment.

The opposite  regime in which dark decays are dominant, which occurs for 
$2 m_\chi < m_{A'}$ and $\alpha_D \gg \alpha \epsilon^2$, 
is also of considerable interest. In that case, the only model-independent signature is the 
missing energy decays of $A'$.
If, however, $\chi$-states do not fragment further into additional light dark states, they can participate in the scattering on the SM particles via the exchange of the mixed $A-A'$ propagator, Fig. \ref{fig:VPfig1}. 
The rate for the scattering on electrons \cite{Batell:2009di} is given by 
\be
\frac{d\sigma_{e\chi\to e\chi}}{dE_f} = \frac{\alpha_D \epsilon^2}{\alpha} \times \frac{8\pi \alpha^2 m_e(1-E_f/E)}{(m_{A'}^2+2m_eE_f)^2},
\ee 
where $E_f$ is the electron recoil energy, and assuming $E_f \gg m_e$.
This scattering process, as well as scattering on nucleons and nuclei, can be viewed as the search channels for light dark matter 
\cite{Batell:2009di,deNiverville:2011it,Izaguirre:2013uxa}. The treatment of the light dark matter-nucleon scattering can be
complicated as it may occur in a variety of dynamical regimes, as evident from recent literature \cite{Jeong:2012fv,Soper:2014ska,Dobrescu:2014ita}. 

\subsection{Other vector candidates}
\label{sec:other-vector-candidates}
The phenomenology of other $U(1)'$ gauge symmetries and their associated  vector bosons  are in principle straightforward to handle. Many gross features are similar to the case of the
kinetically mixed dark photon. In addition, if the gauge symmetry involves 
lepton number or any individual lepton flavors $L_i$ it is guaranteed that there
exists a new decay channel to neutrinos, $V  \to \nu\bar \nu$. 
When the right-handed neutrinos are lighter than the vector boson an additional decay channel $V\to N \bar N$ opens up, where $N$ denotes the right-handed neutrino, or more generally a HNL, 
\be
\Gamma_{V\to N\bar N} = \frac16 \frac{(g')^{2}}{4\pi} m_V\left(1-\frac{m_N^2}{m_V^2}\right)\sqrt{1-\frac{4m_N^2}{m_V^2}}.
\ee
Thus, the vector portals based on leptonic symmetries may provide a new source of right-handed neutrinos in the SHiP experiment.
Compared to the previous chapters, the signal of HNL charged under light $U(1)'$ will 
scale as ${\rm production\times decay} \sim (g')^2\times |U|^2$, as opposed to the minimal HNL case with $|U|^4$ 
scaling, where $U$ is the HNL mixing angle.  

Another interesting candidate that can give distinct signatures at SHiP is the vector boson of the baryonic $U(1)^{B}$.
If the kinetic mixing parameter for the $V^{(B)}$ gauge boson is small the decays of the baryonic vector may be 
dominated by the underlying $\frac{1}{3}V_\mu^{(B)}g_{(B)}\bar q \gamma_\mu q$ interaction. In the mass range 
$m_B <$1~GeV the main decay channels are hadronic, and because of the symmetries of the underlying current 
they are $V\to \gamma \pi^0$ and $V\to \pi^+\pi^-\pi^0$. These decay channels are additionally suppressed because of their
loop origin ($\gamma \pi^0$) and multiplicity of the final state ($\pi^+\pi^-\pi^0$). Using results of Refs. \cite{Tulin:2014tya,Batell:2014yra},
one can estimate the propagation length of baryonic vector with $m_V=200$~MeV as 
\be
\left.c\tau_{V^{(B)}}\gamma\right|_{\epsilon=0} \sim 70 ~{\rm m}\times \left(\frac{10^{-4} }{ g_{(B)}} \right)^2 \times \frac{\gamma}{100} .
\ee
(A non-zero kinetic mixing angle in excess of $10^{-6} $ would shorten this decay length.)

\subsection{Higgsstrahlung process for $U(1)'$ and delayed decays of $h'$. }
If the mixing angle is in ballpark of $\epsilon\sim 10^{-3}$ the lifetime of dark photon is very short 
and the decay length is microscopic as long as $m_V > 2 m_e$. However, it does not mean that such a theory cannot have long-lived particles. One candidate is the long-lived dark Higgs particle that can be produced in association with $V$ \cite{Batell:2009yf}. 
In the dark photon model, if $m_{h'} \leq m_{A'}$ the decays of $h'$ are suppressed, as 
$h'\to A'_{\rm on-shell}+X_{\rm SM} $ channels become kinematically forbidden. 
As a consequence the decays occur via two off-shell $A'$, either via loop mechanism  or at tree level \cite{Batell:2009yf}, with the decay width scaling as $\Gamma_{h'} \sim \epsilon^4$. 
Taking for concreteness   $m_{A'}\sim m_{h'} \sim 2$~GeV, one 
can estimate the decay rate of $h'$ to two vector mesons ($\rho$ or $\omega$), and 
deduce the ensuing decay length,
\be
\label{lifetimeh'}
c\tau_{h'} \gamma \sim 10^{7} ~{\rm m}\times \left(\frac{10^{-3} }{ \epsilon } \right)^4 \times \left(\frac{\alpha }{ \alpha_D } \right)\times \frac{\gamma}{10} .
\ee
One can see that the decay length can be very long compard to the size of the beam dump experiments, leading to the additional 
suppression of the signal. 

The production of energetic $h'$ can be described by the underlying $q\bar q \to h'A'$ annihilation 
process that has the cross section \cite{Batell:2009yf},
\begin{eqnarray}
\sigma_{q\bar q \rightarrow h'A'}\simeq   0.2 \, {\rm fb}  \times Q_q^2\left(\frac{\alpha_D}{\alpha}\right) \left(\frac{\epsilon^2}{10^{-6}} \right) \frac{(10~{\rm GeV})^2}{\hat s},
\end{eqnarray}
where $Q_q$ is the quark charge in units of $e$. In case of a  more complicated Higgs sector, the rates can be enhanced by 
the production of the on-shell $A'$ decaying to a pair of non-identical $h'$ bosons \cite{Morrissey:2014yma,Schuster:2009au}.

\section{Summary of the existing constraints on light vector and light DM states }

In the last few years, there has been  a concerted effort to explore the parameter space of the 
dark photon model, both with and without dark matter. There are also  a number of experimental studies searching for 
other particles coupled to the SM via vector portal. Here we summarize these developments.

\subsection{Current status of experimental constraints on exotic vector states}

\begin{figure}
\centerline{\includegraphics[width=0.55\textwidth]{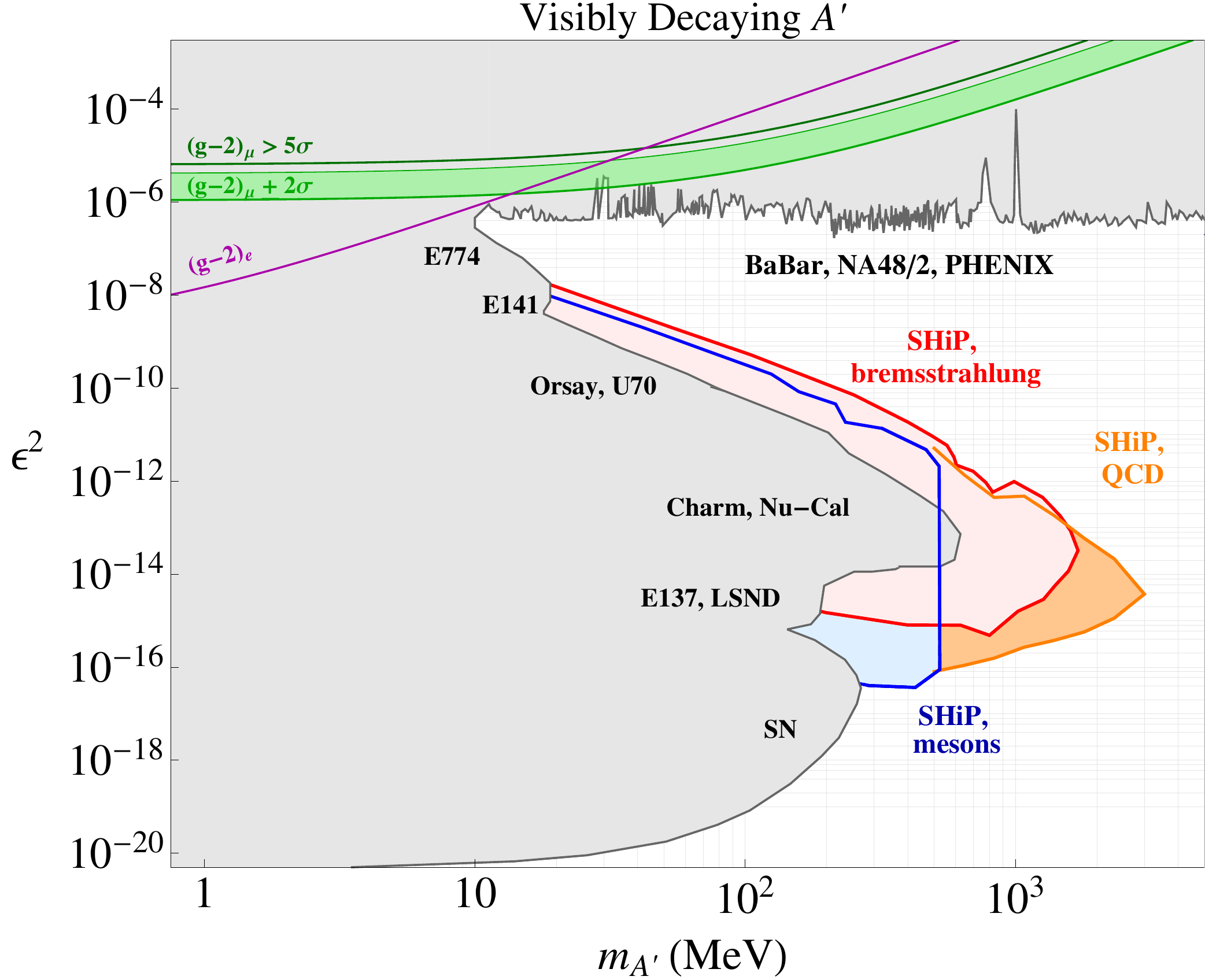}}
\caption{Summary of constraints on the 
dark photon model. The limits at $\epsilon \sim 10^{-7};~m_{A'} > 200$ MeV range come from old experiments, and
 can be improved with SHiP. The $g-2$ region of interest is shown as a green band. The projected 
SHiP sensitivity contour is derived using three modes of production: mesons, bremsstrahlung, and QCD 
production. }
\label{fig:VPfig5}
\end{figure}

Fig. \ref{fig:VPfig5} summarizes the existing constraints on the dark photon model, under the assumption that the dark decays 
are absent or subdominant. One can clearly see that the direct constraints can be subdivided into two main categories:
those that come from the prompt decays of the dark photons, and those that are derived from the delayed 
decays of dark photons inside the detectors placed some distance behind beam dumps.  The combination of 
all constraints covers the region of parameter space motivated by the solution to the muon $g-2$ discrepancy.
Future electron-on-target experiments such as HPS \cite{Essig:2013lka} will add sensitivity in  $\epsilon \sim 10^{-3}-10^{-5}$ and $m_{A'}\leq$ few 100 MeV 
window that is comparatively less constrained.

It is important to note that the excluded regions corresponding to $\epsilon \sim 10^{-6} $ and masses 
in excess of $m_\pi$ come from relatively old experiments, CHARM and E137, which are proton and 
electron beam dump experiments, respectively. (The shape of these regions can be understood as follows: 
above the upper boundary of that region the lifetime $c\tau_{A'} \gamma$ is too short for the sufficient number of 
dark photons to reach the detector. Below the lower boundary, the size of $\epsilon$ is too small to produce and 
detect a sufficient number of $A'$.)  It is clear that the improvement on the number of protons-on-target and shorter 
distance to the detector, such as in the SHiP proposal, will improve the sensitivity to $A'$ around the CHARM/E137 region.

Other models of light vectors have received comparatively less attention. The constraints on $B-L$ vector portal have been 
assessed in \cite{Heeck:2014zfa}, with additional constraint that the RH neutrinos are very light.
The sensitivity of $\phi,\eta,\eta'$ meson decays to the baryonic vector $V^{(B)}$ was derived in \cite{Tulin:2014tya}. 
The full analysis of constraints on $\{\alpha_{(B)},m_{V}\}$ plane has not been performed yet. 

Some cases of other exotic particles produced in association with $V$ have been constrained in experiment.
BaBar studies have placed limits on dark Higgsstrahlung \cite{Lees:2012ra}, by exploiting $A'h'$ production with subsequent 
decays of $h'$ to $2A'$ and eventually to pairs of charged SM particles. The ensuing constraints are quite strong (reaching down to $\epsilon\sim{\rm few}\times  10^{-4}$ at $\alpha_D\sim \alpha$),
but applicable only to $m_{h'} > 2m_{A'}$ region of parameter space. 
Another study at KLOE \cite{Babusci:2015zda}
have searched for missing energy signature from $h'$ decays outside of the detector, and reached the 
constraints at the level of $\epsilon\sim{\rm few}\times  10^{-3}$. 
Constraints on the most motivated case, $m_{h'} \simeq m_{A'}$, are more difficult to obtain 
because they involve stable $h'$ on the scale of the detector.

\subsection{Production and detection of light vector portal DM}

New constraints on vector portals occur when direct production of 
light dark matter states $\chi$ opens up. The missing energy constraints on 
dark photons derived from $e^+e^-$ colliders were analyzed in \cite{Essig:2013vha}. 
Invisible decays of $A'$ are usually harder to detect,
except $K^+\to \pi^+A'\to \pi^+$+missing energy, where  the competing 
SM process, $K^+\to \pi^+\nu\bar\nu $ is extremely suppressed \cite{Pospelov:2008zw}. 
Also, fixed targets experiments sensitive to the missing energy decays of vector states 
have been proposed recently \cite{Andreas:2013lya,Izaguirre:2014bca}. 

\begin{figure}
\centerline{\includegraphics[width=0.55\textwidth]{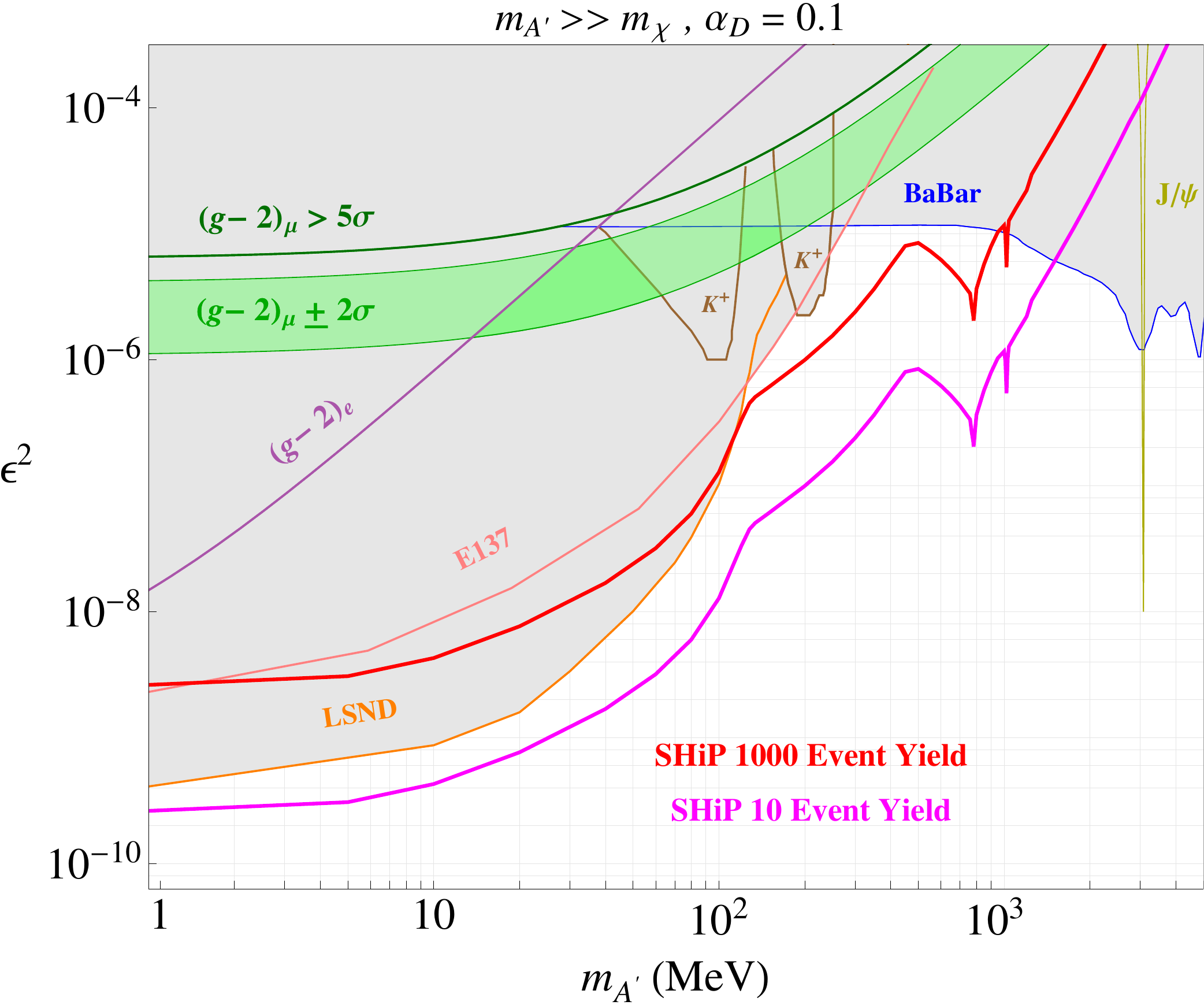}}
\centerline{\includegraphics[width=0.55\textwidth]{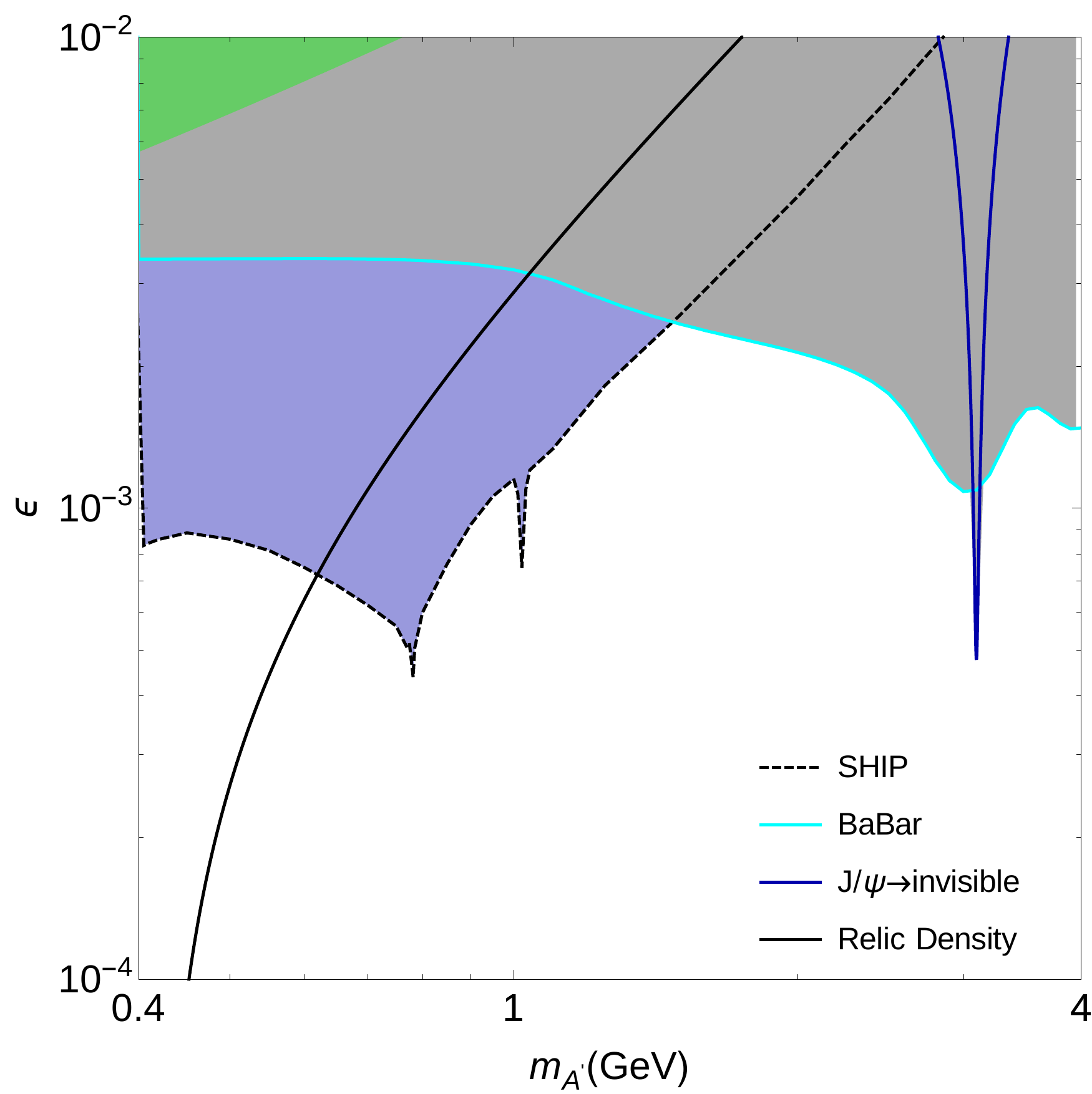}}
\caption{Summary of the constraints on the scattering of the 
light dark matter produced in $e$ and $p$ collisions with fixed targets. Upper pannel,  small dark matter mass $m_\chi$ and $\alpha_D= 0.1$, with SHiP sensitivity
contours assuming 10 and 1000 electron-scattering events; 
Lower pannel is for $m_\chi = 200$ MeV.  The SHiP sensitivty with 10 events is shown as a blue region.}
\label{fig:VPfig6}
\end{figure}

A rather systematic study of the detection of light dark matter produced via the dark photon portal 
has been performed in a number of papers \cite{Batell:2009di,Izaguirre:2013uxa,Batell:2014mga,Kahn:2014sra}.
The most stringent constraints follow from the highest POT experiment, LSND,  provided that the 
dark matter is within kinematic reach. A typical detection scheme in the proton beam dump experiments 
is built on the following chain of events: 
\be
pp \to \pi^0 + X, ~~ \pi^0  \to V \gamma, ~~ V \to \chi\bar\chi, ~~ \chi ~ {\rm scattering~on~electrons/nuclei}
\ee
These results significantly constrain, but do not fully rule out, MeV-scale 
dark matter models, suggested as a candidate explanation of the 511 keV excess from the galactic bulge. 
Currently, the MiniBooNE collaboration is conducting a dedicated search for such states in a beam dump mode run~\cite{Dharmapalan:2012xp}. 
The summary of the existing constraints on light dark matter 
produced via vector portal is given in Fig. \ref{fig:VPfig6}. Similar constraints were also derived for light dark 
matter coupled to the SM via the baryonic vector portal \cite{Batell:2014yra}. It is important to emphasize that these constraints 
cover the low mass region of parameter space inaccessible to traditional underground direct detection experiments. 

\subsection{Cosmological and astrophysical constraints on vector portals }

The question of the astrophysical and cosmological constraints on light particles associated with the vector portal is very important. 
The consistency of deuterium and helium abundance predictions with observations require that the new particles decay 
or significantly reduce their energy density at the time prior to BBN, or more specifically, before the neutron-proton interconversion freeze-out. If coupling constants in the dark sector  ($\epsilon,~\alpha_D,...$)
are sizable,  this requirement translates into $m_{V,\chi} >$ few MeV \cite{Serpico:2004nm,Nollett:2013pwa}. 
Recent measurements of the radiation energy density at the time of the CMB, parametrized by $N_{eff}$, also 
agrees very well with the SM prediction of $3.04$, reinforcing the BBN bound. 
Models that have extra forces acting on neutrinos, such as $B-L$, are also constrained by cosmology. A stronger-than-weak 
force between SM neutrinos and electrons would keep neutrinos in equilibrium for longer, and raise their overall energy density. 
If RH neutrinos are very light and also charged under $B-L$, this would impose very stringent constraints on mediators of 
$B-L$ force \cite{Heeck:2014zfa}. However, if the $B-L$ is broken via a suitable Higgs mechanism, the RH neutrinos can be 
made heavy and unstable, which would significantly weaken constraints on $\{ m_{V},g_{(B-L)}\}$ parameter space. 

Late annihilation of WIMP dark matter into the SM channels other than neutrinos is constrained by the distortions of the 
observed pattern in the CMB anisotropies. This rules out the $s$-wave annihilating WIMPs with masses less than $\sim 10$ GeV. 
However, the scalar dark matter annihilates via a vector mediator in the $p$-wave which significantly lessens all constraints 
from the CMB. Alternatively, light fermionic dark matter may exist in a particle-antiparticle asymmetric state, and be prevented from annihilation 
at the CMB decoupling epoch that way. 

Finally, significant restrictions on parameter space of light particles, including those discussed in this chapter, 
are imposed by stellar astrophysics. If the masses are in excess of few MeV, only supernovae will provide significant bounds 
\cite{Raffelt:1987yt}. The dark photon parameter space is limited by the possibility of it being produced in the SN, and decaying outside of the 
explosive zone, providing a disruptive energy transport mechanism. This imposes strong bounds on the kinetic mixing parameter
$\epsilon$ and vector masses below $200$ MeV. The constraint from supernovae, Fig. \ref{fig:VPfig5}, partially overlaps with the 
beam dump constraints. Light dark matter is also constrained by the same mechanism in SN, if it has a possibility to escape the explosive zone. 
In addition, light particles are constrained if they possess a significant coupling to neutrinos, thus degrading their energy below the ones 
observed in the explosion of SN 1987a. The latter rules out MeV-scale WIMPs with significant coupling to neutrinos \cite{Fayet:2006sa}.

To conclude, astrophysics and cosmology cannot resolutely rule out the existence of light -- few MeV and heavier -- particles 
coupled to the SM via the vector portal. 
Specifically, SHiP has unique sensitivity to a variety of new particles  with masses on 
the order of a few 100 MeV and heavier, and lifetimes below 1 s, which are not constrained by cosmological or astrophysical considerations.

\section{Case studies for SHiP}

Here we describe preliminary sensitivity studies for several benchmark vector portal models. 
Estimates performed for this chapter use basic assumptions about SHiP geometry and normalize the event yield on $10^{20}$ protons on target (POT). 

\subsection{Production and detection of kinetically mixed dark photons and baryonic vectors.}

As stated previously, there exist only partial calculations of the production of vector particles 
at SHiP. 
To derive sensitivity to dark photons, we use the results of the three channels identified earlier 
(meson decays, bremsstrahlung, and direct QCD production). Note that bremsstrahlung production must be 
accompanied by a form factor suppression, if the mass of the dark photon is above the typical QCD scale. 
Using the results of recent estimates in Ref.~\cite{Gorbunov:2014wqa}, and supplementing it with our own estimates of the QCD production 
(for $m_V \gtrsim 1 $ GeV region)
 in the contour shown in red, in Fig. \ref{fig:VPfig5}. 
 The QCD production is simulated using MadGraph5~\cite{Alwall:2014hca}, with the renormalization and factorization scales fixed to $m_{A'}$.
We have extended the QCD estimate to $m_{A'}$ as low as 500 MeV, although we note that in this range there are significant renormalization and factorization scale uncertainties. The detectable level of signal events was set to $N=3$. 

 As one can see, new territories on $\{m_{A'},\epsilon\}$ 
parameter space are going to be covered by SHiP, up to $m_{A'} \sim 3$ GeV. 
If the values of $\epsilon \sim 10^{-7}$ are thought to be generated by loops 
of charged particles, this region corresponds to the two- or three-loop mediation mechanism.

These estimates can be extended to the case of the baryonic vector, when the coupling to the baryon current dominates over 
the kinetic mixing effect. 
In the $m_V >$ GeV mass domain there is no accidental extra longevity of $V^{(B)}$, 
and one can use the approximate translation from the dark photon case, $\{m_{A'},\epsilon\} \to\{V^{(B)},\alpha^{(B)}/\alpha\}$.
Below GeV masses there are no efficient hadronic decays compared to the dark photon case, and as a result the SHiP 
experiment will be sensitive to {\em larger} values of the parameter space relative to the dark photon case. A detailed exploration of this parameter space and comparison to the SHiP sensitivity, with the derivation of bounds from past experiments should be performed in future work.

\subsection{Production and detection of other unstable particles ($h'$, HNL) through the  vector portal }

The parameter space of these models is much wider, $\{ m_{A'}, m_{h'},
\epsilon, \alpha_D \}$ for the dark Higgs, and $\{ m_{V}, m_{N}, \theta,
\alpha_D \}$ for the HNL (Here $\theta$ is the value of the HNL mixing angle,
also called $U$ in this report.) Scanning the parameter space in this case
represents a significant challenge. Instead, we opt for exploring the
sensitivity of SHiP to some representative points on the parameter space.

For the case of dark Higgs-strahlung production of $h'$ with its subsequent decay registered by SHiP detector, 
we  choose the point $m_{h'}=m_{A'} = 2$~GeV. In that case $h'$ is fairly long-lived (eq. (\ref{lifetimeh'})) while the production occurs at 
minimum virtuality of $4 {\rm GeV}^2$, where perturbative QCD can be employed. 
The total number of events, seen as multiple $\pi$ mesons from the decay of $h'$, can be estimated for SHiP as 
\be
N_{\rm events} \sim O(10) \times \left( \frac{\epsilon}{10^{-3} } \right)^6 \times \frac{\alpha'}{\alpha}.
\ee 
This limit is competitive with other constraints ({\em e.g.} the missing energy signal at B-factories), and is indeed 
much better than sensitivity that can be achieved at MINOS or T2K. 

The production of HNLs charged under the vector force is another interesting case to consider. Here we take 
$q \bar q \to V \to NN \to N~{\rm decay}$ sequence of processes, where the last chain is the weak decay of 
$N$, suppressed by the small mixing angle $\theta$. Taking for definitiveness the $B-L$ portal with $m_V=2$ GeV mass mediator, and 
$m_N =0.5$ GeV,
we arrive at the following estimate of the total event rate, 
\be
N_{\rm events} \sim O(10^3)\times \left( \frac{\alpha_{B-L}}{10^{-8} } \right) \times \frac{10^{-1}~{\rm s}}{\tau_N}.
\ee
As it is easy to see, this is a very large rate, reaching down to extremely small values of $\alpha_{B-L}$, and see-saw ballpark for the 
lifetimes of HNLs. Thus, vector portals provide a very efficient mechanism for producing and detecting HNL at SHiP.

\subsection{Possible sensitivity to light DM through SHiP neutrino detector.}
\label{sec:light-DM-SHiP}
Due to the planned neutrino detector at the SHiP facility, the experiment will potentially have a significant reach 
to light dark matter production in the target, with its subsequent scattering in the neutrino detector. 
To estimate the sensitivity we assume a model of kinetic mixing mediation, and calculate the 
number of electron scattering events in a 10 ton detector. A proper calculation would have to include the 
hadronic contributions from deep-inelastic scattering analogous to Ref. \cite{Soper:2014ska}, 
which is expected to significantly increase the total event rate. For now, we take into account the scattering on electrons. 
We normalize the sensitivty to 10 and 1000 candidate events. It remains to be seen if ten events can be separated from the weak interaction background 
of regular neutrinos, but it can be attempted using kinematic differences between neutrinos scattering (via heavy mediator), and 
light dark matter scattering (via a relatively light $A'$.) The resulting sensitivity contour are shown in Fig. 6, bottom pannel,
for a 200 MeV mass of dark matter particle. Even if the backrounds would not be reduced below 1000 events, the sensitivty is quite considerable 
and compares well with previous experiments. Moreover, if one switches from $A'$  to a leptophobic mediator, such as $V^{(B)}$, 
the constraints from lepton-driven experiments (BaBar and E137) will be significantly relaxed, and SHiP will produce a strong additional 
reach to LSND and MiniBooNE. 

\section{Conclusions}

The vector portals from SM to dark sectors are well-motivated, and have been in the focus of both experimental and theoretical communities 
in the last five-to-ten years. We have summarized the main features of vector portal phenomenology, exemplified (but not limited to!) 
by the case of the dark photon $A'$. Preliminary assessment of the sensitivity  shows that the planned SHiP facility will increase the reach 
in terms of the probed  parameter space to many models of vector portals, both through the decay and scattering signatures.


\chapter{SUSY}
\label{sec:susy}

\section{Introduction}\label{susyintro}
Supersymmetry is one of the most popular, but also one of the best motivated extensions of the Standard Model. Connecting
bosons with fermions, by giving each boson a fermionic superpartner and vice versa, its structure is theoretically appealing but it also helps to address or at least alleviate some of the most significant challenges to the Standard Model, e.g. the hierarchy problem or grand unification.

Of course, observation (e.g. the absence of a selectron with mass of 511 keV) tells us that supersymmetry must be broken. Indeed supersymmetry softly broken at or slightly above the scale of electroweak symmetry breaking would provide a natural way to address the hierarchy problem of the Standard Model. In this situation one expects superpartners with masses in the 100 GeV to few TeV range.
This directly leads one to search for them at high energy colliders such as the LHC~\cite{ATLASSUSY,CMSSUSY,Feng:2013pwa,Mitsou:2013uia,Craig:2013cxa}.
Alternatively, heavy particles can also be detected via their effects in loop processes and searches, in particular for flavor violating effects, have been
performed using for example $B$-mesons (see, e.g.,~\cite{Domingo:2007dx}).
While these searches are ongoing and hopes are high, in particular for Run II of the LHC, one has to face the fact that existing searches pose significant constraints on this scenario. Especially the idea of a very high degree of naturalness seems under pressure.

In this situation it seems prudent to rethink our ``natural'' expectations of supersymmetry, go beyond the simplest models assumptions (such as, e.g. the CMSSM, or R-parity conservation), and explore suitable complementary search strategies.
For example in the Standard Model the electron is quite light compared to the electroweak scale. Therefore it may be sensible to also search for SUSY particles a couple of orders of magnitude lighter than the mass of the W and Z bosons.
Indeed such lightness could be ``accidental'' or it could be due to some mechanism. 
In this chapter we explore and survey some possibilities for light superpartners. Interestingly we find that there are quite a variety of opportunities for light superpartners that to date have avoided detection.

The next step is, of course detection of these particles. As this is a physics case study for the SHiP experiment we focus on this proposal and illuminate the potential of SHiP to search for SUSY particles.

We note that actually there is a history of searching for SUSY at fixed target facilities \cite{Ball:1984qc}. While the required light gluinos for those searches are excluded by now, we show in the following that there are still interesting opportunities for finding SUSY at fixed target facilities such as SHiP.

In this chapter we discuss a variety of SUSY particles and models, review their individual motivations and discuss the discovery potential at SHiP. For convenience we highlight at the end of each section the relevant search channel at SHiP. 

\section{A Very Light Supersymmetric Neutralino and R-Parity Violation}
\label{Neutralinos}
\subsection{Motivation for a very light neutralino}
Within the CMSSM the mass of the lightest neutralino, $\neu$, is constrained to be \cite{Agashe:2014kda}
\begin{equation}
m_{\neu}>46\,\mathrm{GeV}\,.
\end{equation}
However this assumes a supersymmetric GUT relation among the electroweak gaugino masses
\begin{equation}
M_1=\frac{5}{3} {\tan}^2\theta_{\mathrm{W}} M_2 \approx \frac{1}{2} M_2\,,
\label{susy-gut}
\end{equation}
and then uses the LEP II chargino search. If the relation Eq.~(\ref{susy-gut}) is dropped, and $M_1$ and $M_2$
are left as free parameters, there is no laboratory bound on the neutralino mass and even a massless neutralino
is allowed \cite{Choudhury:1999tn,Dreiner:2009ic}. This can be understood as follows. Setting the determinant of
the MSSM neutralino mass matrix to zero and solving for $M_1$ we obtain \cite{Dreiner:2009ic}
\begin{equation}
M_1=\frac{M_2 M_Z^2\sin(2\beta)\sin^2\theta_{\mathrm{W}}}
{\mu M_2-M_Z^2\sin(2\beta)\cos^2\theta_{\mathrm{W}}}\,.
\label{massless-cond}
\end{equation}
Thus for every choice of the supersymmetric parameters $M_2,\,\mu,\,\tan\beta$, there is always an $M_1$
resulting in at least one massless neutralino.\footnote{Note that we assumed here that $M_1,\,M_2,\,\mu\in\mathbb{R}$; 
for complex parameters a solution need not exist \cite{Dreiner:2009ic}.}  The resulting $M_1$ is typically a 
factor 50 or so smaller than $\mu,\,M_2$, and the lightest neutralino is almost pure bino. It therefore does not 
couple directly to the $Z^0$, evading LEP I bounds \cite{Choudhury:1999tn}. The pair production of neutralinos
at a collider \cite{Dreiner:2006sb} then necessarily involves virtual squarks or sleptons. If these are heavier than
about 300 GeV the rate is sufficiently small to avoid bounds.

The main bounds on the neutralino are then astrophysical from supernova \cite{Dreiner:2003wh} and white dwarf 
cooling~\cite{Dreiner:2013tja}, or cosmological from the resulting dark matter energy density 
\cite{Dreiner:2009ic,Calibbi:2013poa,Calibbi:2014lga}. The supernova bounds \cite{Dreiner:2003wh} are evaded for neutralino 
masses above about 250 MeV. A massless neutralino is allowed for selectron masses above 1275 GeV or {\it 
below} 320 GeV. For $m_{\tilde e}>1275\,$GeV too few neutralinos are produced. For $m_{\tilde e}<320\,$GeV 
the neutralinos are trapped in the supernova, similar to neutrinos and must be included in the full supernova 
simulation. Since this has not been done to-date the supernova does not give a reliable bound in this region.

Cosmologically the Cowsik-McClelland bound \cite{Cowsik:1972gh,Gershtein:1966gg} on a very light neutrino translates into the 
{\it upper} neutralino mass bound \cite{Dreiner:2009ic}
\begin{equation}
m_{\neu} < 0.7\,\mathrm{eV}\,.
\end{equation}
The neutralino in this case provides hot dark matter, but not enough to negatively affect structure formation. The 
observed dark matter energy density must then originate elsewhere.

If one requires the lightest neutralino to provide the correct dark matter energy density one obtains a {\it lower} 
mass bound, corresponding to the Lee--Weinberg bound \cite{Lee:1977ua,Hut:1977zn,Sato:1977ye}. The proper bound is obtained by
scanning over the supersymmetric parameter space, while dropping the relation in Eq.~(\ref{susy-gut})
and taking into account the latest collider constraints on the supersymmetric parameters.\footnote{For a 
semi--analytical treatment see \cite{Dreiner:2009ic}.} This is thus an on--going process \cite{Hooper:2002nq,Belanger:2002nr,AlbornozVasquez:2011yq,Belanger:2013pna,Scopel:2013bba}. 
The most recent bound including the Higgs--discovery data and also constraints from stau searches gives
\cite{Lindert:2011td,Calibbi:2013poa,Calibbi:2014lga}
\begin{equation}
m_{\neu}>24\,\mathrm{GeV}\,.
\end{equation}
Therefore the mass range
\begin{equation}
0.7\,\mathrm{eV}<m_{\neu}<24\,\mathrm{GeV}\,,
\end{equation}
is excluded for a stable neutralino LSP as it gives too much dark matter. A neutralino LSP in
this mass range is only allowed if it decays, \textit{i.e.} R-parity is violated.

Note that there is no particular motivation to expect $M_1$ to approximately satisfy Eq.~(\ref{massless-cond}),
resulting in a light neutralino. However, there is also no reason the neutralino must be heavy. Since we have no 
convincing theory of supersymmetry breaking, which would predict $M_1,\,M_2,\,\mu$, and $\tan\beta$, experimental 
searches are our best bet.

\subsection{R-parity Violation}
When supersymmetrizing the Standard Model (SM), an extra Higgs doublet must be introduced and then the particle content 
must be doubled. With this minimal particle content the most general renormalizable superpotential, which governs the
non-gauge interactions, is fixed by the SU(3)xSU(2)xU(1) gauge symmetry to contain two sets of terms $W_{MSSM}=
W_{RPC}+W_{RPV}$ with
\begin{eqnarray}
W_{RPC}&=& (Y_e)_{ij} L_i H_1 \bar E_j + (Y_d)_{ij} Q_i H_1 \bar D_j + (Y_u)_{ij} Q_i H_2 \bar U_j
+\mu H_1 H_2\,, \label{superpot1}
\\
W_{RPV}&=& \lambda_{ijk} L_iL_j \bar E_k + \lambda'_{ijk} L_iQ_j \bar D_k + \lambda''_{ijk} \bar U_i \bar D_j \bar D_k
+\kappa_i L_i H_2\,, \label{superpot2}
\end{eqnarray} 
in the notation of \cite{Allanach:2003eb}. The terms in Eq.~(\ref{superpot1}) correspond to the known Yukawa couplings 
in the SM. The terms in Eq.~(\ref{superpot2}) have no correspondence and violate lepton- or baryon number. If all the latter terms are present and have significant couplings there is rapid proton decay which is experimentally excluded. A further symmetry
must be added to forbid at least a subset of the terms $W_{RPV}$. One standard approach is to use a discrete multiplicative symmetry R-parity,
which forbids all the terms in $W_{RPV}$, this is the widely considered MSSM. An alternative symmetry, e.g. baryon-triality,
see for example \cite{Dreiner:2006xw}, just forbids the terms proportional to $\lambda''_{ijk}$. Baryon-number is then 
conserved and the proton is also stable. Both discrete symmetries are discrete gauge anomaly-free \cite{Dreiner:2005rd,Ibanez:1991pr},
and are thus theoretically equally well motivated.

The main features of an additional R-parity violating (RPV) term \cite{Dreiner:1997uz} are that the lightest supersymmetric particle is no longer 
stable.  It is thus not a dark matter candidate. This could instead be the axino, or the axion \cite{Chun:1999cq} (cf. also Chapter~\ref{chapALP}). At colliders we 
would also not expect to have significant amount of missing transverse momentum as a result of supersymmetric production. 
Furthermore at colliders one can have resonant single production of supersymmetric particles. As long as the nature of the dark matter
particle is not known, and supersymmetry has not been discovered at the LHC, R-parity conservation and RPV are equally well 
motivated.

\subsection{Finding Neutralinos at \ship{} via R--Parity violation}
\begin{figure}
\centering
\includegraphics[width=0.2\textwidth]{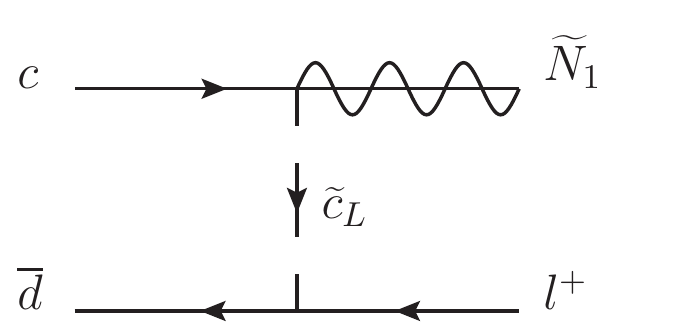} \qquad
\includegraphics[width=0.2\textwidth]{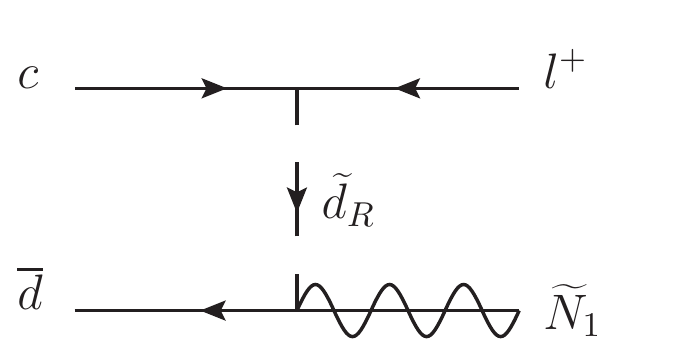} \qquad
\includegraphics[width=0.2\textwidth]{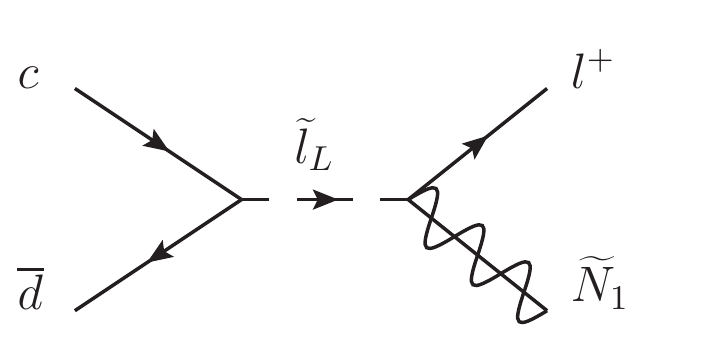}
\caption{Relevant Feynman Diagrams for $D^+ \rightarrow \neu + \ell^+$. }
\label{fig:Ddecay}
\end{figure}
In the following, we assume that there exists an R--parity violating interaction $\Delta W = \lami (L_i \cdot Q_2) \bar{D}_1.$ in the 
superpotential. Moreover we focus on production via the more numerous $D$ mesons. Production via $B$ mesons is also possible and allows to extend the mass reach. We leave this option for future more detailed studies.

\begin{figure}
\begin{flushleft}
\includegraphics[width=0.48\textwidth]{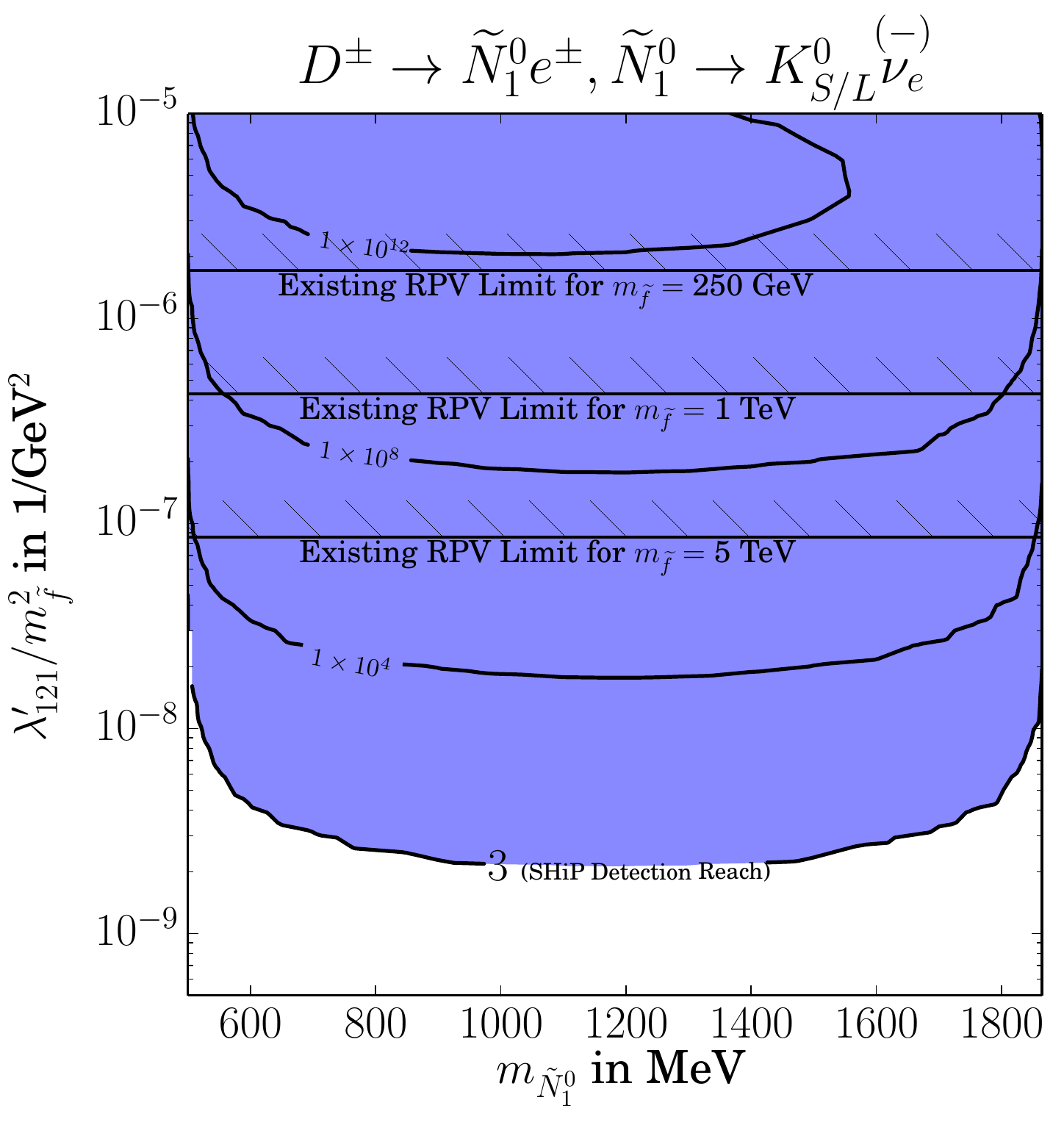} \quad
\includegraphics[width=0.48\textwidth]{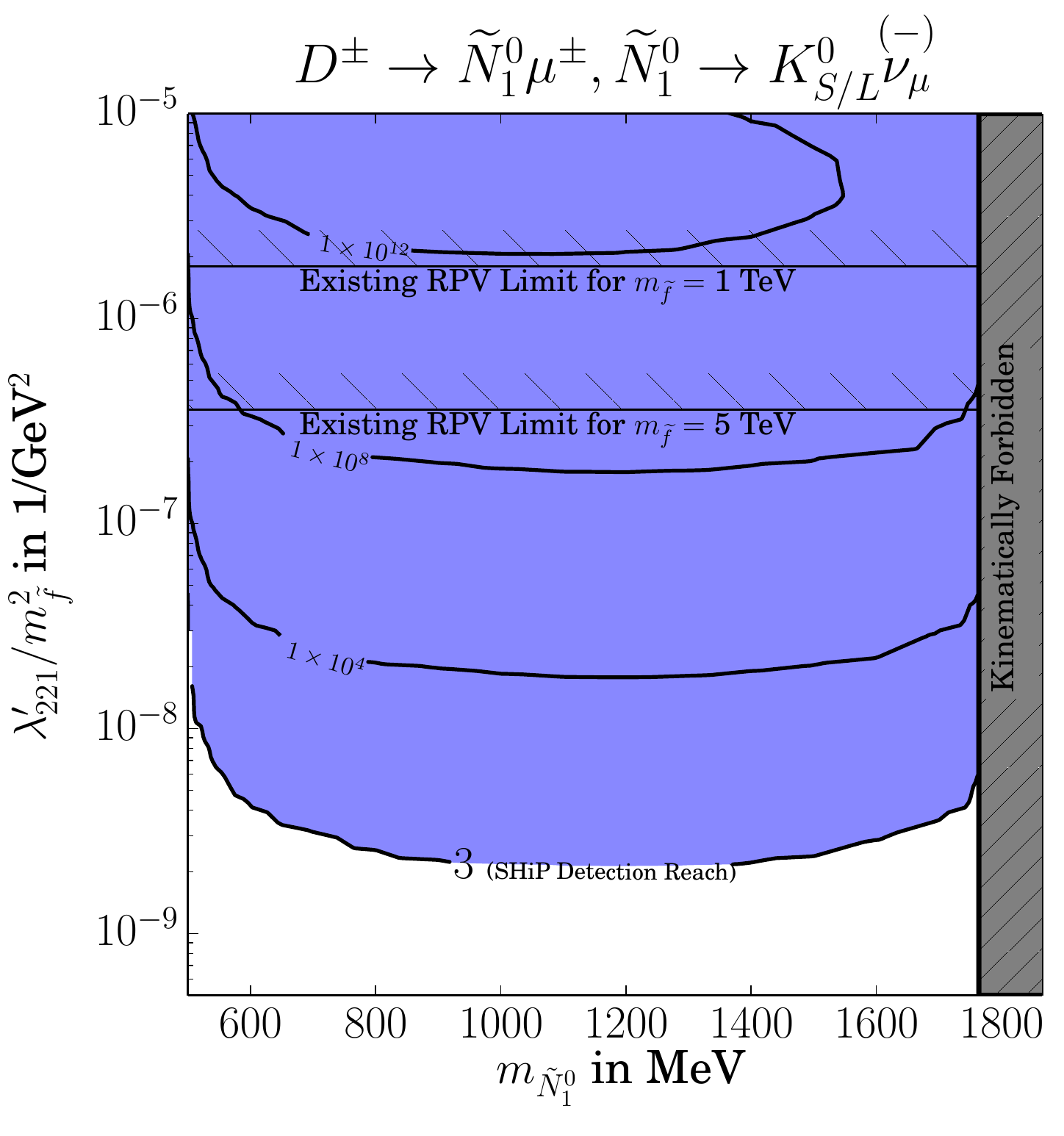} \\
\includegraphics[width=0.48\textwidth]{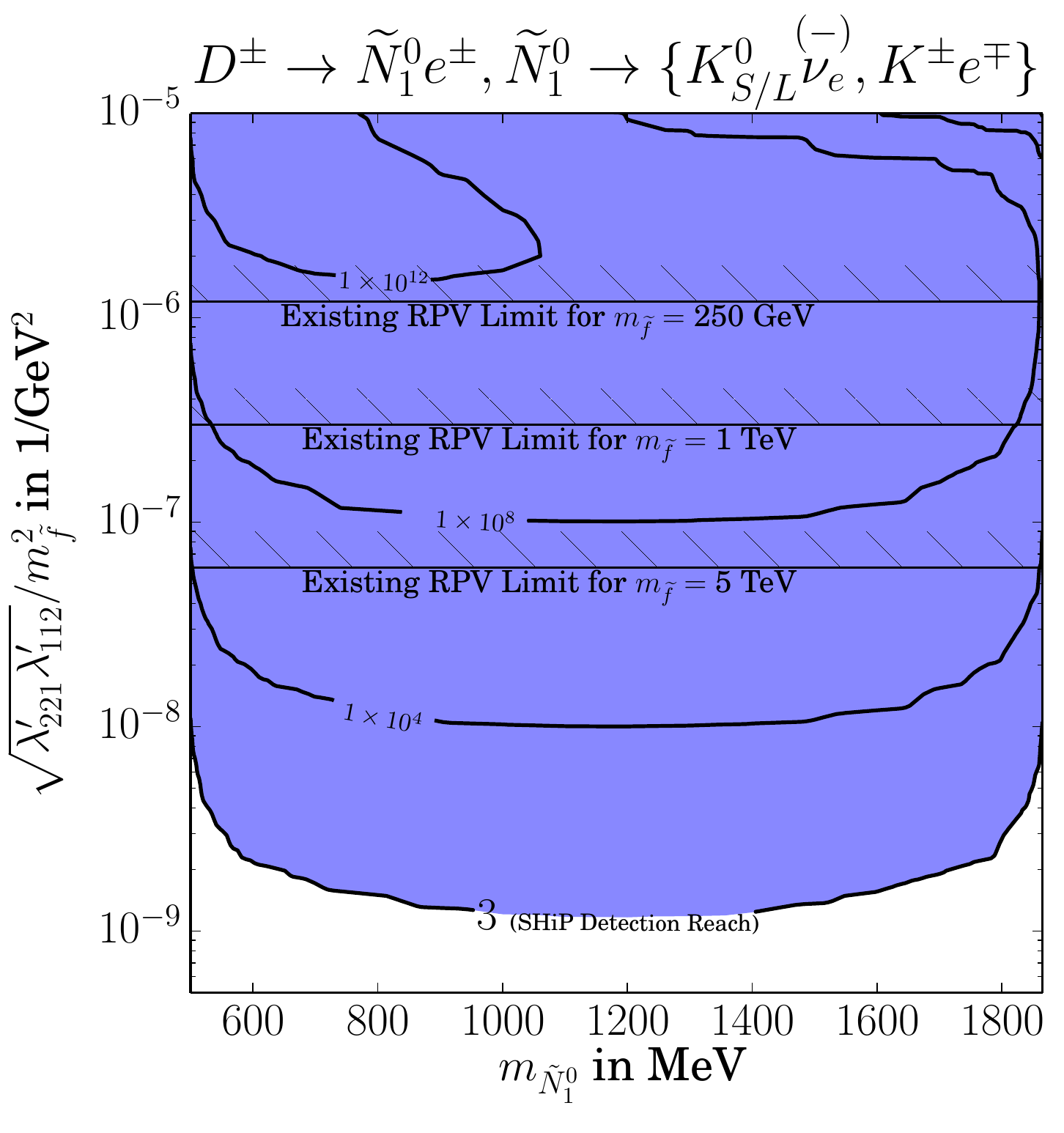}  \quad
\includegraphics[width=0.48\textwidth]{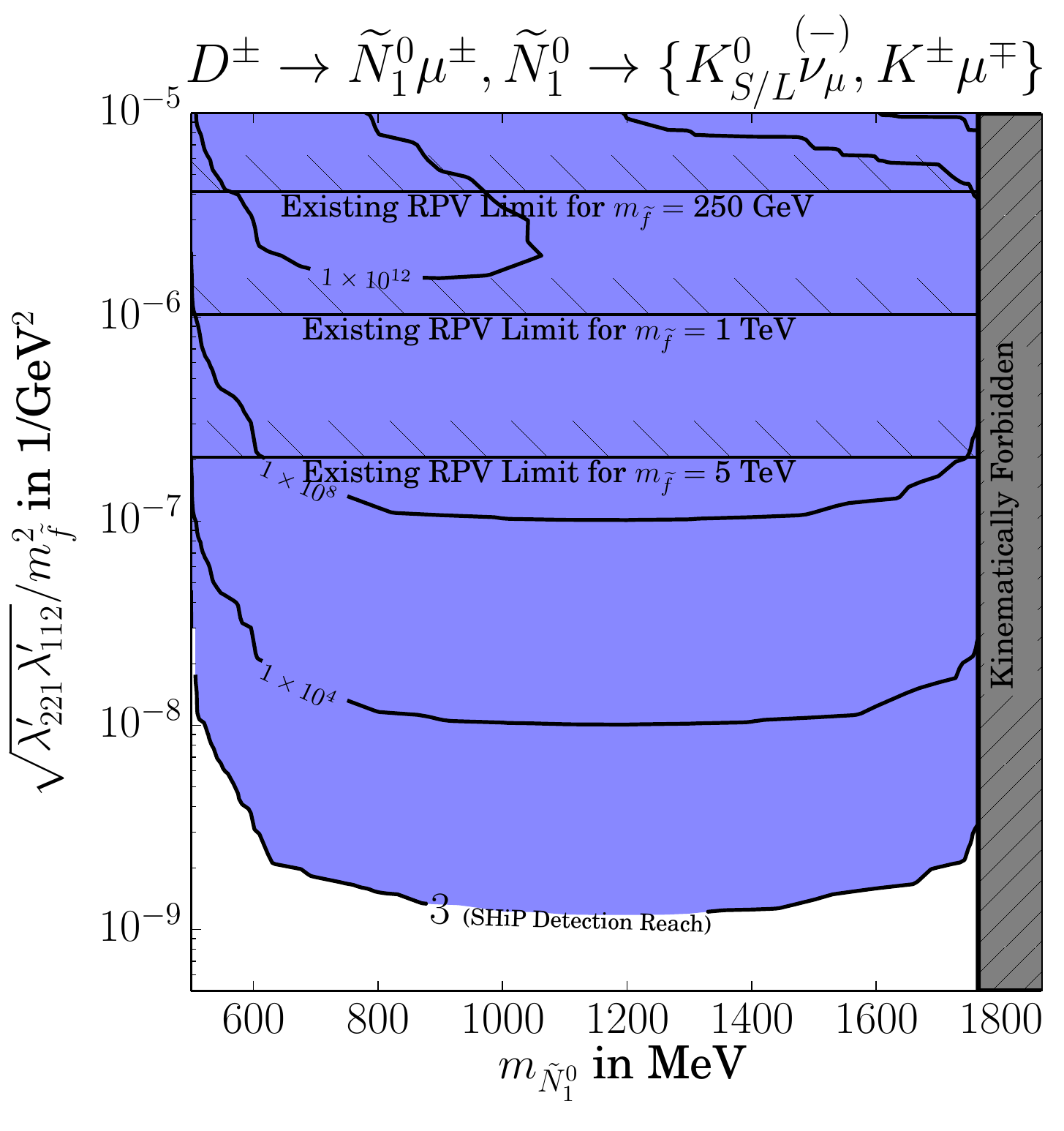} 
\caption{Expected event yields after a runtime of 5 years for RPV-produced
  neutralinos decaying into kaons and leptons with the decay vertex being
  inside the detector. Shown contour lines are determined by interpolating
  results of $28 \times 29$ points. The horizontal lines correspond to the
  existing bounds in Eq.~\eqref{bounds} for fixed sfermion mass, as
  labelled. Previously excluded is always above the horizontal line, below is
  the potentially new region tested by SHiP.}
\label{fig:results}
\end{flushleft}
\end{figure}

For the case at hand, a $D^\pm$ meson produced at \ship{} could decay into a neutralino and a charged lepton via the 
diagrams in Fig.~\ref{fig:Ddecay}, if the neutralino is light enough. The respective partial width for this decay reads
\begin{align}
\Gamma(D^\pm \rightarrow \neu + \lipm) = \frac{9 g_2^2}{256 \pi}  \left(Z_{12} + \tan \theta_W Z_{11} \right)^2 \frac{|\lami|^2}
{m_{\widetilde{f}}^4}  f^2_D  \frac{m_{D^\pm}^2  (m_{D^\pm}^2 - m_{\neu}^2 - m_{\li}^2)}{(m_c+m_d)^2}  |\vec{p}_{\neu}|,
\end{align}
with $\vec{p}_{\neu} = \lambda^{\frac{1}{2}}(m_{D^\pm}^2, m^2_{\neu}, m^2_{\li})/(2 m_{D^\pm})$ being the 3-momentum of the 
neutralino in the $D^\pm$ meson's rest frame\footnote{$ \displaystyle \lambda^{\frac{1}{2}}(a,b,c) \equiv
\sqrt{a^2+b^2+c^2-2ab-2ac-2bc}$}.  For a pure bino-like lightest neutralino, the neutralino mixing matrix elements $Z_{11} = 1, 
Z_{12} = 0$, and vice versa for a pure wino lightest neutralino. We have not considered higgsino admixtures, as the
corresponding couplings are very small for the first two quark/lepton generations we consider here. $f_D$ is the decay 
constant of the $D$ meson which we set to \unit[207]{\MeV{}} according to \cite{Agashe:2014kda}. For the sake of 
simplicity we here assume all sfermion masses to have a common value $m_{\widetilde{f}}$. Comparing this partial width to the 
total width of the charged D--meson, $\Gamma^{D^\pm}_{\text{tot}} = \unit[6.33 \times 10^{-13}]{\GeV{}}$ \cite{Agashe:2014kda}, 
one can find the branching ratio $\text{Br}(D^\pm \rightarrow \neu \lipm)$ for the above decay of interest, which can range between 
$10^{-10}$ and $10^{-2}$ for the considered range of parameters shown in Fig.~\ref{fig:results}.

The same R-parity violating operator yields a coupling which could induce subsequent decays $\neu \rightarrow \overline{K^0_S} \vi, \overline{K^0_L} 
\vi,$ $K^0_S  \overline{\vi}, K^0_L  \overline{\vi}$, which all have equal decay probability in the limit of negligible CP--violating 
effects. Decays 
into charged Kaons exist in addition if the operator $\Delta W^\prime = \lamTi (L_i \cdot Q_1) \bar{D}_2$ (note the different order of the indices) is non-vanishing. The 
diagrams are similar to those shown in Fig.~\ref{fig:Ddecay}. The corresponding decay widths for the lightest neutralino are then as 
follows
\begin{align}
\Gamma(\neu \rightarrow K^0_{S/L} + \hspace{-4pt} \stackrel{(-)}{{\nu}^{i}}) &= \frac{g_2^2}{512 \pi}  \left(3 Z_{12} - 
\tan \theta_W Z_{11} \right)^2 \frac{|\lami|^2}{m_{\widetilde{f}}^4}  f^2_K  \frac{m_{K_0}^4  (m_{\neu}^2 - m_{K^0}^2)}{m_{\neu}
^2(m_s+m_d)^2}  |\vec{p}_{\vi}|, \label{eq:neuwidth1}\\
\Gamma(\neu \rightarrow K^\mp + \lipm) &= \frac{9 g_2^2}{256 \pi}  \left(Z_{12} + \tan \theta_W Z_{11} \right)^2 \frac{|\lamTi|^2}
{m_{\widetilde{f}}^4}  f^2_K  \frac{m_{K^+}^4  (m_{\neu}^2 - m_{K^+}^2 - m_{\li}^2)}{m_{\neu}^2 (m_s+m_u)^2}  |\vec{p}_{\li}|, 
\label{eq:neuwidth2}
\end{align}
with $\vec{p}_{\vi} = (m_{\neu}^2 - m^2_{K_0})/(2 m_{\neu})$, $\vec{p}_{\li} = \lambda^{\frac{1}{2}}(m^2_{K^+}, m^2_{\neu}, 
m^2_{\li})/(2 m_{\neu})$ and the decay constant of the Kaon $f_K$ with value \unit[156]{\MeV{}} \cite{Agashe:2014kda}. 
Depending on the size of the supersymmetric parameters, the resulting proper decay length of the neutralino can range from $\mathcal{O}($mm$)$ 
for $\lami/m^2_{\widetilde{f}} = \unit[10^{-5}]{\GeV}^{-2}$ up to  $\mathcal{O}($\unit[1000]{km}$)$, for $\lami/m^2_{\widetilde{f}} = 
\unit[10^{-9}]{\GeV}^{-2}$.

To estimate the kinematic properties of the neutralino, we use \verb@Pythia 8.1.7.5@ \cite{Sjostrand:2007gs}. We generate 
proton--proton--collisions in the lab frame by setting $E(p_1) = \unit[400]{\GeV}$ and $E(p_2) = m_p$, the proton mass. Since we 
are here only interested in the properties of $D^\pm$ mesons, we exclusively generate \verb@HardQCD:hardccbar@ matrix 
elements, which include the partonic interactions $q \bar{q}, g g \rightarrow c \bar{c}$. We find that 1 $c \bar{c}$ event on average 
creates 0.53 $D^\pm$ mesons. We demand exclusive decays $D^\pm \rightarrow \neu \lipm$, $\neu \rightarrow K^\mp \lipm$ to 
increase the Monte Carlo statistics\footnote{Since the kinematics of the neutralino decay products are not important, we use the 
same single neutralino decay mode for all scenarios and just change the total width formula according to the scenario.}\!\!. The 
lifetime of the $D^\pm$ is kept fixed at the experimental value, whereas we set the lifetime of the neutralino to the numerical value 
calculated from Eqs.~(\ref{eq:neuwidth1}) and (\ref{eq:neuwidth2}).

We are interested in the probability that a given neutralino candidate decays within the detector at \ship. We assume the distance 
target--detector to be $\Ltd = \unit[68.8]{m}$, the total length of the detector to be $\Ld = \unit[55]{m}$ and an elliptic face with semiaxes 5~m and 2.5~m. For each neutralino candidate, we determine the velocity $\beta^z_i$ of the neutralino in the beam direction, the 
boost factor $\gamma_i$, which relates the neutralino--rest frame to the lab frame, and the angle $\theta_i$ between the 
neutralino's momentum and the beam axis. The average values of $\bar{\gamma} \approx 30$ and $\bar{\theta} \approx \unit[60]
{mrad}$ show that their consideration is important; the boost significantly increases the observed decay length in the lab frame and 
the angles are such that the neutralinos can easily leave the detector in radial direction before they decay, as it only covers polar 
angles in the range \unit[25]{mrad} to \unit[100]{mrad}. Using this information, we can find the probability for each such candidate 
to have a decay vertex with coordinates $z_i \in [\Ltd, \Ltd + \Ld]$ and $\rho_i \in [0, R_{d(\phi_{\widetilde{N}_1})}]$.
From averaging the probabilities of $5,000$ individual Monte Carlo Events one finds an estimate for the overall fraction of 
neutralinos that decay within the detector, $r^{\text{detector}}_{\widetilde{N}}$, with statistical uncertainty of less than $5 \%$. From 
\cite{Bonivento:2013jag} we expect $\approx 9 \times 10^{16}$ produced $c \bar{c}$ pairs, which according to our \verb@Pythia@ 
study corresponds to roughly $N_{D^\pm}^{\text{produced}} = 5 \times 10^{16}$ produced $D^\pm$ mesons in 5 years of operation. Using the 
results from above, we find the expected number of observable neutralino decays to be
\begin{align}
N_{\widetilde{N}}^{\text{observed}} = N_{D^\pm}^{\text{produced}}  \times \text{Br}(D^\pm \rightarrow \neu \lipm) \times r^{\text{detector}}_
{\widetilde{N}}\,.
\end{align}
The resulting event yields for different combinations of $m_{\neu}$ and  $\lambda^\prime / m^2_{\widetilde{f}}$ can 
be found in Fig.~\ref{fig:results}, assuming the neutralino to be a pure bino.

\subsection{Comparison with Previous Bounds}
The best bounds we are currently aware of are from \cite{Allanach:1999ic}. These bounds are at the 2$\sigma$
level
\begin{eqnarray}
\lambda'_{121}&<&0.043\frac{m_{\tilde d_R}}{100\,\mathrm{GeV}}\,,\qquad
\lambda'_{221}<0.18\frac{m_{\tilde s_R}}{100\,\mathrm{GeV}}\,,\qquad
\\\nonumber
\lambda'_{112}&<&0.021\frac{m_{\tilde s_R}}{100\,\mathrm{GeV}}\,,\qquad
\lambda'_{212}<0.059\frac{m_{\tilde s_R}}{100\,\mathrm{GeV}}\,. \label{susy:bounds}
\end{eqnarray}
As can be seen, these bounds scale differently, than our event rates at \ship. We have thus inserted into
Fig.~\ref{fig:results} the bounds for fixed sfermion mass. To the left of the line is the new regime which can
be explored by \ship.

\subsection{Concluding remarks}
Using the channels\footnote{Similar to the photon channels discussed in Chapter~\ref{chapALP} the first, neutral channel is quite challenging experimentally.We note however, that it probes different couplings and that the experimental options are being investigated.}
\begin{eqnarray}
pp&\to& D+X\to \neu +X, \neu \xrightarrow{\rm long\,\,lived} K^{0}_{S/L}+ \hspace{-4pt} \stackrel{(-)}{\nu}_{\hspace{-4pt}i}
\\\nonumber
pp&\to& D+X\to \neu +X, \neu \xrightarrow{\rm long\,\,lived} K^{\pm}+\ell^{\mp}
\end{eqnarray}
and similar channels for $B$ meson decays, allows SHIP to significantly extend the discovery reach for light neutralinos in a scenario with R-parity violation. This direct probe is complementary to SUSY searches at the LHC and allows to (indirectly) probe sfermions in the region of a few 10s of TeV.

\section{Light particles from the SUSY breaking sector}
\label{sec:sgoldstinos}
\subsection{Origin of light sgoldstinos}
Supersymmetry, if it exists, must be spontaneously broken. This happens
when an auxiliary component of some chiral superfield gains nonzero
vacuum expectation value (vev). In a particular model the situation 
may be more complicated, involving several chiral and/or vector
superfields, but for the purposes here it is sufficient to
consider the simplest setup. The chiral superfield $\Phi$ is called the
Goldstone supermultiplet. It includes the auxiliary component $F_\Phi$,
with non-zero vev
\begin{equation}
\label{auxiliary-vev} 
\langle F_\phi \rangle \equiv  F
\end{equation}
responsible for the
SUSY breaking, the Majorana fermion---massless 
Goldstone particle---called {\bf goldstino} $\psi$, and the spin-0
component $\phi$ forming scalar $S$ and pseudoscalar $P$ {\bf
  sgoldstinos} \cite{Perazzi:2000id}. Introducing the superspace with 
Grassmanian coordinates $\theta$ and $\bar\theta$, one can write the Goldstone 
superfield\footnote{In any case a chiral superfield must participate in
 the supersymmetry breaking \cite{ArkaniHamed:1998nu}, 
and if many, $\Phi$ is replaced by a linear combination of the
chiral superfields, which plays the same role that we attribute to the
Goldstone supermultiplet. In particular, the fermionic component of
this linear combination is massless, that is the goldstino, 
and the scalar components are sgoldstinos, if no strong mixing exists
in the spin-0 sector.} as 
\begin{equation}
\label{Goldstone-superfield}
\Phi=\phi+\sqrt{2}\theta{\psi}+F_\phi\theta\theta\,.
\end{equation}

If supersymmetry is promoted to a local symmetry, supergravity, the gravitino $\tilde G_\mu$ (spin-3/2 superpartner of graviton) 
acquires a mass $m_{\tilde G}$ as a result of the super Higgs
  mechanism~\cite{Cremmer:1982en} and 
the goldstino disappears from the spectrum becoming 
the longitudinal component of the gravitino, 
\begin{equation}
\label{gravitino-longitudinal}
\tilde G_\mu \to  \tilde G_\mu + \frac{i}{m_{\tilde G}}\,
\sqrt{\frac{2}{3}}\,\partial_\mu{\psi}.
\end{equation}
The gravitino mass is then proportional to the supersymmetry breaking
parameter,   
\[
m_{\tilde G}=\sqrt{\frac{8\,\pi}{3}}\,\frac{F}{M_{{\rm Pl}}}\,.
\]
In many simple cases, like O'Raifeartaigh and Fayet--Iliopolous
models, 
scalar and pseudoscalar sgoldstinos 
\[
S \equiv \frac{1}{\sqrt{2}}\left(\phi+\phi^\dagger\right) \;,
\;\;\;\;
P \equiv\frac{1}{i\sqrt{2}}\left(\phi-\phi^\dagger\right)
\]
remain massless at tree level. Higher order corrections to the scalar
effective potential contribute to their masses. Their values are largely model
dependent, and we keep them as free parameters $m_S$ and $m_P$ in what
follows. For example they can be of the order of the gravitino
mass \cite{Ellis:1984xe} 
or much heavier. For the SHiP experiment the interesting mass
range is $m_{S(P)}\sim 1$\,MeV -- 10\,GeV.  

\subsection{Sgoldstino couplings and phenomenology}
To the leading order in $1/F$, the goldstino $\psi$ couples to all the 
fields through the Goldberger--Treiman formula 
\begin{equation}
\label{Goldberger--Treiman}
{\cal L}_{\psi} \propto \frac{1}{F}\,J^{\mu}_{{\rm SUSY}}
\partial_\mu\psi\;, 
\end{equation}
where $J^{\mu}_{{\rm SUSY}}$ is the supercurrent. 
Using the transformation \eqref{gravitino-longitudinal} 
induces interactions between the gravitino and other fields suppressed
not by the inversed Planck mass, but by the inverse SUSY-violating
parameter $1/F$. Therefore, for low scale of SUSY breaking
$\Lambda_{SUSY}\sim \sqrt{F}$, the gravitino couples to other fields 
mostly through its longitudinal component, i.e. the goldstino. 

The effective interactions of the sgoldstino can be obtained by making use 
of a supersymmetric transformation of
\eqref{Goldberger--Treiman}. Alternatively, following 
Ref.\,\cite{Brignole:1996fn}, one can
use the spurion technique for the Goldstone superfield $\Phi$ to
derive its coupling to the SM particles and their 
superpartners.\footnote{This procedure is transparent and provides the 
  sgoldstino interaction terms to the
leading order in $1/F$ valid in many classes of supersymmetric models.
In other cases, e.g.\,\cite{Lopez:1996fs}, 
the relation between sgoldstino couplings
and the soft supersymmetry breaking MSSM parameters may be more complicated
as compared to what we obtain following this procedure.} 
Namely, the effective interaction 
\[
\frac{M_{\lambda}}{F}\int \!\!d\theta^2\, \Phi W^\alpha W_\alpha
 + {\rm h.c.} 
\] 
yields SUSY violating gaugino mass terms $M_{\lambda}\lambda\lambda+h.c.$,
when the auxiliary component of $\Phi$ \eqref{Goldstone-superfield} is
substituted by a nonzero vev \eqref{auxiliary-vev}. For sgoldstinos it
provides couplings to the SM gauge bosons. In particular,
the interaction terms with photons and gluons 
read\,\footnote{See Refs.\,\cite{Gorbunov:2000th,Gorbunov:2001pd} 
for complete set of interaction terms including all the MSSM soft
parameters, and goldstino-sgoldstino leading order couplings.}  
\begin{align}
\label{sgoldstino-photons}
{\cal L}_{eff} & =  
-\frac{1}{2\sqrt{2}}\frac{M_{\gamma\gamma}}{F}SF^{\mu\nu}F_{\mu\nu}+
\frac{1}{4\sqrt{2}}\frac{M_{\gamma\gamma}}{F} 
P\epsilon^{\mu\nu\rho\sigma}F_{\mu\nu}F_{\rho\sigma}\\
\label{sgoldstino-gluons}
& -\frac{1}{2\sqrt{2}}\frac{M_3}{F}SG^{\mu\nu~a}G_{\mu\nu}^a+
\frac{1}{4\sqrt{2}}\frac{M_3}{F}
P\epsilon^{\mu\nu\rho\sigma}G_{\mu\nu}^\alpha G_{\rho\sigma}^\alpha,
\end{align}
where $F_{\mu\nu}$ is the electromagnetic field strength,
$G_{\mu\nu}^a$, $a=1,\dots,8$ refers to gluons, and $M_{\gamma\gamma}
= M_1 \sin^2\theta_W +M_2 \cos^2\theta_W$ with $M_1$ and $M_2$ 
denoting $U(1)_Y$  and $SU(2)_W$ gaugino
SUSY breaking mass terms, and $M_{3}$ stands for the gluino mass.

 Similarly, the effective interaction 
\begin{equation}
\int\!\! d\theta^2 \l \frac{ A^u_{ij}}{
F}\int\Phi H_uQ_iU^C_j  + \frac{ A^d_{ij}}{
F}\int\Phi H_dQ_iD^C_j  +\frac{ A^l_{ij}}{
F}\int\Phi H_dL_iE^C_j \r + {\rm h.c.} 
\end{equation}
leads for \eqref{Goldstone-superfield}, \eqref{auxiliary-vev} to the
soft supersymmetry breaking trilinear terms. After electroweak
symmetry breaking it provides with
sgoldstino Yukawa-like couplings to the SM fermions, 
\begin{align}
-\frac{\tilde{m}_{D_{ij}}^{LR~2}}{\sqrt{2}F}S\bar{\psi}_{D_i}\psi_{D_j}-
i\,\frac{\tilde{m}_{D_{ij}}^{LR~2}}{\sqrt{2}F}P\bar{\psi}_{D_i}\gamma_5\psi_{D_j} 
-\frac{\tilde{m}_{U_{ij}}^{LR~2}}{\sqrt{2}F}S\bar{\psi}_{U_i}\psi_{U_j}-
i\,\frac{\tilde{m}_{U_{ij}}^{LR~2}}{\sqrt{2}F}P\bar{\psi}_{U_i}
\gamma_5\psi_{U_j}\label{qqS}\\
-\frac{\tilde{m}_{L_{ij}}^{LR~2}}{\sqrt{2}F}S\bar{\psi}_{L_i}\psi_{L_j}-
i\,\frac{\tilde{m}_{L_{ij}}^{LR~2}}{\sqrt{2}F}P\bar{\psi}_{L_i}\gamma_5\psi_{L_j}\;,
\label{llS}
\end{align}
where $\psi_{L_i}$ are leptons, $\psi_{U_i}$ and
$\psi_{D_i}$, are up- and down-type quarks and $i$ runs over
generations; the SUSY violating entries in the slepton, up-squark and
down-squark left-right squared mass matrices are introduced as follows  
\begin{equation}
\label{mass-trilinear}
\tilde{m}_{L_{ij}}^{LR~2} \equiv A^l_{ij} v_d \;, \;\;\;\; 
\tilde{m}_{D_{ij}}^{LR~2} \equiv A^d_{ij} v_d \;, \;\;\;\; 
\tilde{m}_{U_{ij}}^{LR~2} \equiv A^u_{ij} v_u \;, 
\end{equation}
where $v_d$ and $v_u$ are vevs of the Brout--Englert--Higgs fields 
$H_d$ and $H_u$, respectively.

While the gravitino (goldstino) is $R$-odd, sgoldstinos are
  $R$-even, and hence (if kinematically allowed) may be singly
produced in collisions and decays of the SM particles and may decay 
into pairs of the SM particles. Interaction Lagrangians
\eqref{sgoldstino-photons}-\eqref{llS} 
ensure that the rates of all these processes are
proportional to $1/F^2$. 
This defines the sgoldstino phenomenology, whose main
attraction is the unique possibility to determine the scale of
supersymmetry breaking $\Lambda_{\rm SUSY}\sim\sqrt{F}$ in the whole
theory. The sgoldstino couplings are proportional to the SUSY breaking
MSSM soft terms and inversely proportional to $F$. Thus the higher the
SM superpartner scale and the lower the scale of SUSY breaking in the
whole theory, the larger the sgoldstino couplings to the SM
particles. Among the models with relatively 
low scale of SUSY breaking are models
with no-scale supergravity \cite{Ellis:1984kd,Ellis:1984xe}
and models with gauge mediation of
supersymmetry breaking \cite{Giudice:1998bp,Dubovsky:1999xc}. 
 
The list of processes where light sgoldstinos may show up is 
long, since both flavor conserving and flavor violating
couplings are present in the sgoldstino interaction sector. The latter
\eqref{qqS},\eqref{llS} 
are proportional to the off-diagonal SUSY-violating entries in the
slepton and squark mass matrices, which are fixed by the pattern of the
trilinear soft terms \eqref{mass-trilinear}. The off-diagonal mass
terms and soft trilinear couplings give rise to the FCNC processes
through the squark and slepton virtual contribution to the SM rare
processes. Their absence places limits on these couplings, and hence
the numerators in the sgoldstino coupling constants
\eqref{qqS},\eqref{llS} (which are the
ratios of the MSSM soft parameters and SUSY breaking parameter $F$)
are strongly constrained. 
As to the flavor conserving sgoldstino couplings to vector bosons,
e.g. \eqref{sgoldstino-photons},\eqref{sgoldstino-gluons}, 
the largest are those to the particles whose superpartners are the
heaviest. Flavor conserving couplings to fermions are proportional to
the diagonal trilinear soft parameters \eqref{mass-trilinear}. For the
universal (or close to universal) conditions for the soft SUSY breaking
parameters, the trilinear terms are proportional to the SM fermion
Yukawas $Y^{l(d,u)}_i$, and hence 
\begin{equation}
\label{diagonal-fermions}
\tilde{m}_{L_{ii}}^{LR~2} =A^l Y^l_{i} v_d = A^l m^l_{i} \;, \;\;\; 
\tilde{m}_{D_{ii}}^{LR~2} = A^d Y^d_{i} v_d = A^d m^d_i \;, \;\;\; 
\tilde{m}_{U_{ii}}^{LR~2} = A^u Y^u_{i} v_u = A^u m^u_i\;, 
\end{equation}
where $m^{l(d,u)}_i$ denotes the corresponding fermion mass. If the
trilinear parameters are of the same order, 
\[
A^l\sim A^d\sim A^u\,,
\]
the strongest are couplings to the heaviest fermions.

The sgoldstino phenomenology depends on the sgoldstino mass scale and is
governed by the pattern of the MSSM soft terms. It has been thoroughly
studied in the literature.  Very light sgoldstinos can play important
roles in astrophysics and cosmology, and the absence of a visible imprint left by the sgoldstino
places strong limits on its couplings to the SM fields
\cite{Nowakowski:1994ag,Gorbunov:2000th}.  At the same time,  
relic sgoldstinos of 7\, keV mass have been suggested
\cite{Demidov:2014hka} to explain the excess of 3.5\,keV $X$-rays
\cite{Bulbul:2014sua,Boyarsky:2014jta} observed by the orbital
telescopes.

Heavier sgoldstinos can contribute to rare processes including flavor
violation \cite{Brignole:2000wd,Gorbunov:2000th}. 
They can appear in lepton, meson and
baryon decays \cite{Gorbunov:2000th}, as well as $Z$-boson decays
\cite{Dicus:1990dy}. Sgoldstino pair production in $B$-meson decays
\cite{Demidov:2011rd} has been recently searched for at the LHCb
experiment \cite{Aaij:2013lla}.  Note that the HyperCP anomalous
events \cite{Park:2005eka} in $\Sigma^+\to p \mu^+ \mu^-$ with dimuon
invariant mass of $214.3\pm 0.5$\,MeV can be explained as pseudoscalar
sgoldstino production in the $\Sigma$-hyperon decay through the flavor
violating coupling, with subsequent sgoldstino decay into
$\mu^+\mu^-$. The suggested tests of the model in kaon
\cite{Gorbunov:2005nu} and $B$-mesons \cite{Demidov:2006pt} physics
have been performed with negative results
\cite{Hyun:2010an,Abouzaid:2011mi}, closing a part of the model
parameter space.

Sgoldstinos can be directly produced in high energy collision
processes.  Sgoldstino phenomenology at $e^+e^-$-colliders has been
addressed in \cite{Dicus:1990su,Dicus:1990dy,Perazzi:2000id}.  The DELPHI
Collaboration has performed searches for a sgoldstino signal in the
collected data revealing no hints \cite{Abreu:2000ij}.  Signatures
of massive sgoldstinos at the TeVatron have been considered in
\cite{Dicus:1989gg,Dicus:1996ua,Perazzi:2000ty}. Heavy sgoldstinos are
produced mostly through gluon fusion
\cite{Perazzi:2000ty}. Prospects of sgoldstino searches at LHC have
been discussed in \cite{Gorbunov:2002er,Demidov:2004qt}, including
tests of possible sgoldstino mixing with the Higgs boson
\cite{Astapov:2014mea} and of flavor violatings interaction with
top-quark \cite{Gorbunov:2000ht}.

\subsection{Sgoldstinos at SHiP}
If kinematically allowed, sgoldstinos may be 
directly produced by 400 GeV protons scattering on a target and
indirectly through the decays of mesons and baryons emerging from
proton scattering and the subsequent hadron shower in the dump. The direct
sgoldstino production here is naturally dominated by the gluon fusion 
$gg\to S$, $gg\to P$. The coupling to gluons is expected to be the
strongest among the flavor conserving, since in many particular
supersymmetric extensions of the SM, the gluino 
is naturally the heaviest gaugino (as, e.g. in case of universal boundary
conditions for the soft terms, in models with gauge mediation of 
SUSY breaking, etc). The production cross section can be obtained by
simple rescaling of the known results for the fusion of the SM Higgs boson of
the same mass, sgoldstino coupling to gluons is proportional to
$M_3/F$ \eqref{sgoldstino-gluons}. 
The results for scalar and pseudoscalar sgoldstinos of the same mass
coincide. 

\begin{figure}[!t]
  \centering
  \includegraphics[width=.5\textwidth]{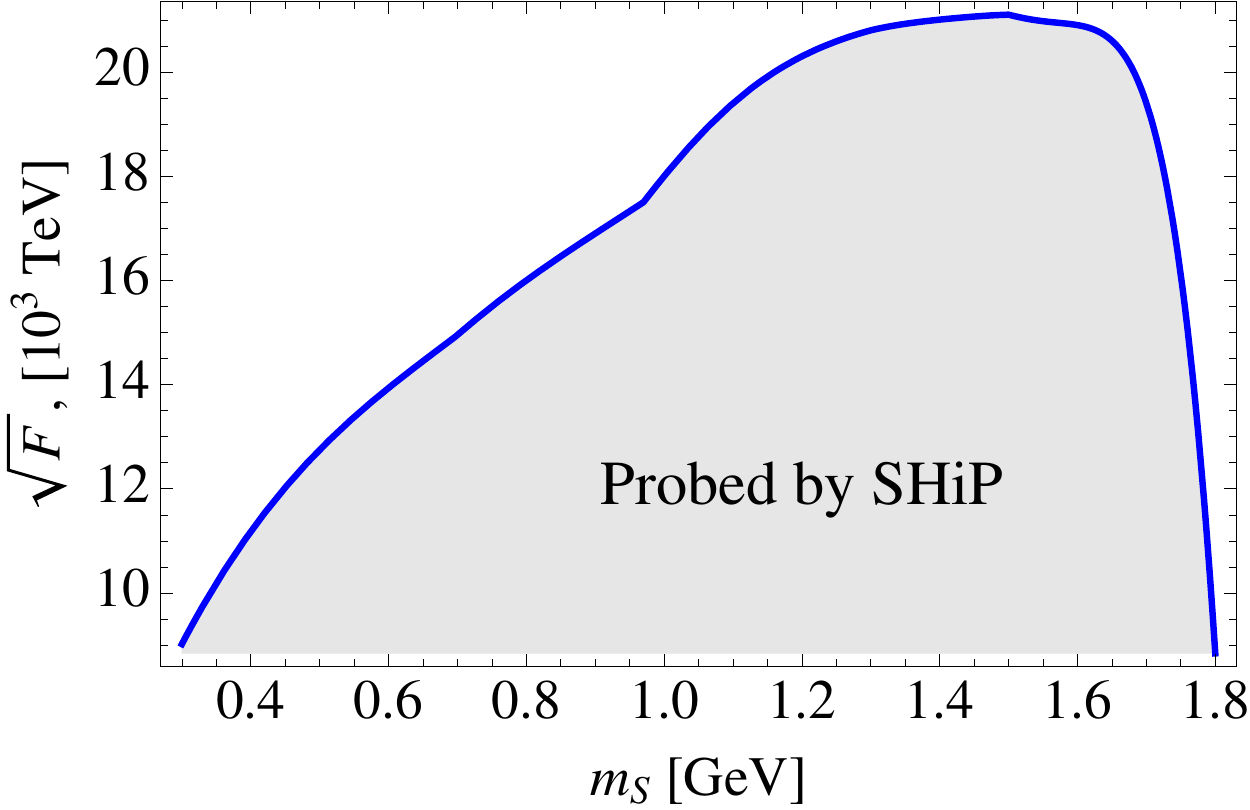}
  \caption{SUSY breaking scale $\sqrt F$ probed by SHiP as a function of
    the sgoldstino mass $m_S < m_D$ in the lepton flavour violating case,
    Eq.~\protect\eqref{qqS}. The down-squark left-right mass matrix element
    was chosen $\tilde m^{LR\,2}_{D_{12}} = (\unit[100]{GeV})^2$ and the gaugino mass
    $M_3 = \unit[3]{TeV}$. The result scales with $\tilde m^{LR\, 2}_{D_{12}}$ and
    $ M_3$ as shown in Eq.~\protect\eqref{charged-pions-from-sgoldstinos}. }
  \label{fig:sgoldstino}
\end{figure}

Indirect sgoldstino production may go through the flavor conserving and flavor
violating couplings, constrained by the absence of rare processes and
superpartners at the LHC. For the flavor conserving case the dominant channel for
${\cal O}(1)$\,GeV mass sgoldstino is $B\to K S(P)$, the same one-loop process
as would dominate the SM Higgs boson production of the same mass. For lighter
sgoldstinos $K\to \pi S(P)$ dominates. In the flavor violating case the result
depends on the chosen pattern of the flavor violating model parameters. In
particular, sgoldstinos of (sub)GeV mass may be produced in the two body $D$-meson
decays like $D_s\to K +S(P)$, $D^0\to \pi^0 +S(P)$, and $D^\pm\to \pi^\pm+
S(P)$.

Sgoldstinos decay into the SM particles mostly through the flavor
conserving couplings. Here the final states are 
\[
S(P) \to \gamma\gamma\,,
\;\;\;\mu^+\mu^-\,,\;\;\;\pi^+\pi^-\,,\;\;\;  
\pi^0\pi^0\,,\;\;\; e^+e^-\,,\dots
\]
with hadronic channels (if open) presumably dominant (provided by
heavy gluino mass and large number of gluons). Indeed, for the ratio of
sgoldstino decay rates into photons and gluons one obtains from
eqs.\,\eqref{sgoldstino-photons},\eqref{sgoldstino-gluons}, 
\[
\frac{\Gamma\l S(P)\to gg\r}{\Gamma\l S(P)\to \gamma\gamma\r}
= (N_c^2-1)\frac{M_{3}^2}{M_{\gamma\gamma}^2}\,.
\]
For the leptonic channels one finds with the universal diagonal trilinear
terms \eqref{diagonal-fermions}
\[
\frac{\Gamma\l S(P)\to e^+e^-\r}{\Gamma\l S(P)\to \mu^+\mu^-\r}
= \frac{m_e^2}{m_\mu^2}\ll 1\,,
\]
therefore the heavy fermions dominate. For the rates to fermions and
gluons one obtains \cite{Perazzi:2000id,Gorbunov:2000th}
\[
\frac{\Gamma\l S(P)\to \mu^+\mu^-\r}{\Gamma\l S(P)\to gg\r}
=\frac{A^{l\,2}}{4\,M_{3}^2}\, \frac{m_\mu^2}{m_{S(P)}^2}\,. 
\]
If the sgoldstino decay rates are saturated by hadronic channels, sgoldstino
lifetimes may be estimated as (for sgoldstino masses $2m_\pi < m_S < 2 m_K$
the decay $S\to\pi^+\pi^-$ dominates)
\begin{equation}
\label{sgoldstino-lifetime}
\tau_{S(P)}=2\times10^{-4}\,\text{s}\,\times 
{\l\frac{\sqrt{F}}{10000\,\text{TeV}}\r^{\!\!4}} \, 
{\l\frac{3\,\text{TeV}}{M_{3}}\r^{\!\!2}} \,
{\l\frac{1\,\text{GeV}}{m_{S(P)}}\r^{\!\!3}} 
\;\;\;\; 2m_\pi < m_{S(P)} < 2 m_K
\end{equation}
(for $m_S > 2 m_K$ the lifetime becomes shorter by 40\% due to additional
channels $S(P) \to K^0\bar K^0$ and $S(P)\to K^+ K^-$).  Here the difference between
scalar and pseudoscalar sgoldstino is only due to their masses, which are
usually different. The same conclusion is true for all the processes, if
parity is violated by sgoldstino couplings, so both scalar and pseudoscalar
sgoldstinos couple through the parity odd and parity even interaction
terms. If parity is conserved by sgoldstino couplings \cite{Gorbunov:2000cz},
the phenomenology of scalar and pseudoscalar sgoldstinos, especially those
involving meson decays, differ.

For the reference values of the model parameters adopted in
Eq.~\eqref{sgoldstino-lifetime} sgoldstinos produced by 400\,GeV protons on
target cover a distance of several kilometers before they decay. Hence the
corresponding signature---a couple of charged particles emerging at a single
point with invariant mass equal to that of the sgoldstino---may be searched for in
the SHiP detector. For a 60\,m length detector with $2\times 10^{20}$ PoTs one
expects the following number of charged pion pairs appearing from the light
sgoldstino decays,
\begin{equation}
\label{charged-pions-from-sgoldstinos}
N_{\pi^+\pi^-}\simeq 4\times 10^4{
\l \frac{\tilde{m}_{D_{ii}}^{LR~2}}{(100\,\text{GeV})^2} \r^2
\l\frac{10000\,\text{TeV}}{\sqrt{F}}\r^{\!\!\!8}}{
\l\frac{M_{\lambda_g}}{3\,\text{TeV}}\r^{\!\!\!2}}{ 
\l\frac{m_{S,P}}{1\,\text{GeV}}\r^{\!\!4}}\;,\;\;\; 
2m_\pi < m_{S(P)} < 2 m_K
\end{equation}
Here we take into account the Lorentz $\gamma$-factor for sgoldstino, produced
in decays of $6.8 \times 10^{17}$ $D$-mesons (see
Appendix~\ref{cha:ship-experiment}) via lepton flavour violating coupling
$\tilde{m}^{LR\,2}_{D_{12}} $. 
The results of the detailed MC simulations, taking
into account the acceptance of the detector, are presented in
Fig.~\ref{fig:sgoldstino}. The scaling of the sensitivity plot with
$\tilde{m}^{LR\,2}_{D_{12}} $ and $M_3$ remains the same as in
Eq.~\eqref{charged-pions-from-sgoldstinos} also for $m_S > 2 m_K$.
We note that the reference numbers in (6.3.13) are all in agreement
with present experimental limits on MSSM parameters.

\subsection{Concluding remarks}
Sgoldstinos are a natural consequence of F-term SUSY breaking and their
typical interaction strength are directly linked to the scale at which SUSY is
broken.  Phenomenologically they share features with scalars from Higgs portal models and pseudoscalar axion-like particles
(ALPs) and similar search strategies can be applied
(cf. Chapters~\ref{sec:scalar-portal} and~\ref{chapALP}).  Promising search channels at SHiP include,
\begin{eqnarray}
pp&\to& S \rm{(gluon\,\, fusion)},\,\, S\xrightarrow{\rm long\,\,lived} \ell^{+}\ell^{-}, \pi^{+}\pi^{-}, \pi^{0}\pi^{0}
\\\nonumber
pp&\to& D +X\to S+X',\,\, S\xrightarrow{\rm long\,\,lived} \ell^{+}\ell^{-}, \pi^{+}\pi^{-}, \pi^{0}\pi^{0},
\end{eqnarray}
and similar for the pseudoscalar $P$.

\section{Light Dirac gauginos}
\label{dgauginos}
\subsection{Origins of Pseudo-Dirac fermions}

A pseudo-Dirac fermion, where an originally Dirac fermion is split into two Majorana components, is an interesting candidate for dark matter, since it allows co-annihilation between the eigenstates but would appear as a Majorana particle for the purposes of direct and indirect detection experiments -- provided that the mass splitting is larger than the typical kinetic energy of the particles, of order $100$ keV, but less than the energy of the thermal bath during freeze-out, typically a few GeV. 

Pseudo-Dirac fermions can occur quite naturally in supersymmetric theories. 
For example in the MSSM we could have a Higgsino which is pseudo-Dirac.
However this is excluded for consideration at the masses $\lesssim 20\,{\rm GeV}$ which are of interest to SHiP as it would be excluded by the Z-width measurement. 

A better candidate is a Dirac gaugino~\cite{Hsieh:2007wq,Carpenter:2005tz,Belanger:2009wf,DeSimone:2010tf,Buckley:2013sca} which is an
extension of the MSSM which is well motivated by the existence of an (approximate) R-symmetry.  

If the Dirac particle is predominantly bino-like then it will have weak enough interactions to generate sufficient dark matter, and will also evade the Z-width constraint for models accessible by SHiP. 
There are typically two classes of Dirac gaugino models: those which preserve an R-symmetry \cite{Hall:1990hq,Kribs:2007ac,Kribs:2012gx,Frugiuele:2012kp,Morita:2012kh,Chakraborty:2013gea,Bertuzzo:2014bwa,Diessner:2014ksa,Chakraborty:2014sda} and those which allow a small breaking \cite{Fox:2002bu,Nelson:2002ca,Benakli:2011kz,Heikinheimo:2011fk,Benakli:2012cy,Benakli:2014cia,Ipek:2014moa,Goodsell:2014dia}. 
In the latter case, a minimal extension of the Higgs sector is allowed, and the neutralino is split into (at least) six majorana fermions. The splitting between the bino eigenstates induced by electroweak symmetry breaking is \cite{Belanger:2009wf} 
\begin{align}
\Delta m \simeq& - \sin 2 \beta \frac{M_Z^2 s_W^2}{\mu} \left(\frac{2 \lambda_S^2 - g_Y^2}{g_Y^2}\right).
\end{align}
Hence if $\mu \sim $TeV we have a mass splitting of the order of $2$ GeV. 

Typically even in models that do not have an explicit R-symmetry breaking on the field theoretical level such a breaking is naturally induced by gravitational effects. For example a splitting $\Delta m \gtrsim $ 100 keV, required to have a genuine pseudo-Dirac candidate as regards direct-detection experiments, is induced through anomaly mediation if the gravitino mass $m_{\tilde G}$ is given by
\begin{align}
m_{\tilde G} \simeq& \frac{16\pi^2}{g_Y^2} \Delta m \simeq 0.1\ \mathrm{GeV} \times \left(\frac{\Delta m}{100\ \mathrm{keV}}\right).
\end{align}

There are several possibilities how pseudo-Dirac particles could arise in supersymmetric models with Dirac gauginos.
Due to the variety of model-building possibilities we shall adopt in the following an effective action approach along the lines of~\cite{DeSimone:2010tf} and leave detailed investigation of specific examples to future work.

\subsection{Effective model, phenomenological features}

For models of interest to SHiP, the pseudo-Dirac particle must be much lighter than other supersymmetric particles -- and also the electroweak gauge bosons. We can therefore integrate them out and parametrise the leading couplings of the intially Dirac fermion $\Psi$ to standard model fermions $f$ as \cite{DeSimone:2010tf}
\begin{align}
\mathcal{L}_{int} =& \frac{1}{\Lambda^2} \bigg[   \ov{\Psi} \gamma^\mu ( c_L P_L + c_R P_R) \Psi\bigg] \bigg[   \ov{f} \gamma_\mu ( c_L^f P_L + c_R^f P_R) f\bigg].
\label{PD:GeneralEffective}
\end{align}
The LHC has already put limits on the strength of these interactions; from the point of view of the LHC the splitting of the Dirac fermion into two components is irrelevant. From \cite{CMS:rwa,Buchmueller:2014yoa} we find that for light dark matter, 
\begin{align}
 \frac{c_{L,R}^{u,d}}{\Lambda} \lesssim 900\ \mathrm{GeV}.
\end{align}
These are also the important bounds for the production of the particles in SHiP. On the other hand, there is no reason that the couplings to up and down quarks must be the important couplings for the dark matter relic abundance. In
\cite{DeSimone:2010tf} an approximate formula for the relic density was given as
\begin{align}
\frac{\Lambda}{C_{DM}} \simeq& 0.8\ \mathrm{TeV} \left(\frac{\Omega h^2}{0.11} \right)^{1/4} \left( \frac{m_1}{100\ \mathrm{GeV}}\right)^{1/2} \exp \left[ -6 \frac{\Delta m}{m} \right].
\end{align}
Here
\begin{align}
C_{DM}^4 \equiv& \frac{1}{4}  | c_L + c_R|^2\sum_{f} ( |c_L^f|^2 + |c_R^f|^2)
\end{align}
where the sum is over all fermions relativistic at freezeout. Clearly for a light dark matter particle, given the above collider bounds, either we should expect to have an overabundance of dark matter or the annihilations should proceed to leptons rather than quarks. Both cases are possible. In the first case a notional excess of dark matter is acceptable because a non-standard thermal history of the universe -- such as with a late-decaying particle that injects additional entropy and dilutes the density to the observed value, e.g. in the ``non-thermal WIMP miracle'' \cite{Acharya:2009zt} -- would then render such a scenario viable. %

\subsection{Origin of the effective model}
Let us now consider the origin of the couplings present in the effective model in the context of TeV-scale SUSY Dirac gaugino models (as opposed to Split-SUSY variants, such as \cite{Antoniadis:2005em,Benakli:2013msa,Dudas:2013gga,Fox:2014moa}, where the dark matter properties are very different), where they can come through the Z, sfermion and Higgs portals. We shall imagine that our dark matter candidate consists of neutral components from $SU(2)$ singlets $b$ (which include the bino), triplets $w$ (including the wino) and doublets $r$ (such as the higgsinos or R-higgsinos) so that we can write schematically
\begin{align}
P_L b =& L_s P_L \Psi + ..., \qquad  P_L w = L_t P_L \Psi+ ..., \qquad  P_L r = L_{r_i} P_L \Psi + ...\,.
\label{PD:fermionparam}\end{align}

We note that for the purposes of SHiP it is perfectly acceptable to integrate out particles with masses of the order of the $Z$ or the Higgs mass and use the effective description (for LHC this may not always be appropriate). 

\subsubsection{Z portal}

Given the above interaction is vector-like and our Dirac particle is neutral, it is natural to consider first the Z-portal for interactions between the dark matter and the visible sector. However, this is strongly constrained by the Z-width; it is not possible to have a light pseudo-Dirac particle coming from an $SU(2)$ doublet such as a Higgsino. On the other hand, if there is some mixing between a bino and Higgsino, then we can write the effective coupling between the Dirac fermion (before splitting the eigenvalues) as
\begin{align}
\mathcal{L}_{Z\ DM} =&  \frac{e}{2s_W c_W} c_Z  Z_\mu \ov{\Psi} \gamma^\mu P_L \Psi.
\end{align}
In the decomposition of our Dirac particle in (\ref{PD:fermionparam}) we have $c_Z = 2\sum_i Y_i L_{r_i}^2$.
The Z-width then constrains 
\begin{align}
|c_Z| \lesssim& 0.1 .
\end{align}
For pure Z-portal dark matter with interactions of the above type, \cite{Arcadi:2014lta} found that dark matter would always be overproduced for $m_\Psi \lesssim 22$ GeV when we satisfy the Z-width constraint. As mentioned above, however, from the perspective of SHiP this merely means that the thermal history of the universe in this case would be non-thermal.   

\subsubsection{Higgs portal} 

If the Pseudo-Dirac fermion couples to the Higgs then it annihilate via Higgs exchange, which will determine the pattern of couplings to Standard Model fermions to be proportional to their masses. If the Higgs sector preserves R-symmetry, then we can write the coupling of the Dirac fermion prior to the mass splitting to be
\begin{align}
\mathcal{L}_{Higgs\ DM} =& H \ov{\Psi} ( c_H P_L + c_H^* P_R) \Psi .
\end{align}
Here the couplings can have gauge current and superpotential origin. 

The Higgs portal is constrained by the invisible Higgs width, from monojet searches \cite{Djouadi:2012zc} and $t \ov{t}H$ production \cite{Zhou:2014dba}; the latter gives a bound
\begin{align}
\mathrm{Br}( H \rightarrow \mathrm{invisible}) < 0.4,
\end{align}
which, since the width of the Standard Model Higgs with a mass of $125$ GeV is $4.04 \times 10^{-3} $ GeV, translates into a bound (in the limit of the Dirac particle much lighter than the Higgs mass) of
\begin{align}
|c_H|^2 <& 8\pi \times 0.4/0.6 \times \frac{\Gamma_H}{m_H} \nn\\
<& 5 \times 10^{-4}.
\end{align}
This leads to a massive overproduction of dark matter; for $m_\Psi = 1$ GeV we find
\begin{align}
\Omega h^2 \simeq& 2.5\times 10^6 \times \left(\frac{10^{-4}}{|c_H|^2}\right).
\end{align}
As in the Z-portal case, this could still be interesting if the thermal history of the universe is non-standard. However, the amount of entropy injection required here is much greater.

\subsubsection{Sfermion portal}

The natural possibility to consider is the scalar portal, which corresponds in SUSY theories to squark or slepton exchange. In the Dirac basis, there are direct couplings of the sfermions $\tilde{f}$ to the gauginos of the form
\begin{align}
\mathcal{L}_{\mathrm{Scalar\ DM}} =& \tilde{f}^*_{i} ( a_{L,i} \ov{f^c} P_L \Psi + a_{R,i} \ov{\Psi} P_R f )  + h.c.
\end{align}
Let us consider an approximately R-symmetric theory. Then there is no mixing between left- and right-handed sfermions, and in the decomposition of our Dirac particle in (\ref{PD:fermionparam}) we have either $a_{L,i} = 0$ or $a_{R,i}=0$. We then have $a_{(L,R),i} = \sqrt{2} g_Y Y L_s + \sqrt{2} g_2 T L_t$. 

The sfermion portal is the most natural way of obtaining a reasonable relic density while evading other constraints: the sleptons can be light (with masses of $\mathcal{O}(100)$ GeV), in contrast to the squarks -- as claimed in \cite{Buckley:2013sca}, the dark matter could be as light as $10$ GeV without being overproduced for a pure bino.

\subsection{Detection at SHiP}

\subsubsection{Decays}
One of the essential requirements for an experiment like SHiP is a particle with a macroscopic decay length.
If we consider the decays of our pseudo-Dirac particle, we shall assume that the dark matter density is dominated by lepton interactions and therefore pairs of leptons will be produced. The decay length is \cite{DeSimone:2010tf}
\begin{align}
L ( \chi_2 \rightarrow \ell^{+}\ell^{-} \chi_1) \simeq 250\ \mathrm{m} \times \frac{E_2}{m_2} \times\left( \frac{\Lambda/C_{DM}}{100\ \mathrm{GeV}} \right)^4 \left(\frac{50\ \mathrm{MeV}}{\Delta m}\right)^5.
\end{align}
where $\chi_{1}$ and $\chi_{2}$ denote the heavier and lighter components of the pseudo-Dirac fermion, respectively.

Since the decay length is very sensitive to the mass splitting of the states -- which then does not affect any of our other observables -- it is not possible to place definitive exclusions; but we may make a detection!

\subsubsection{Production}

\begin{figure}[!ht]
\begin{center}
\includegraphics[width=0.6\textwidth]{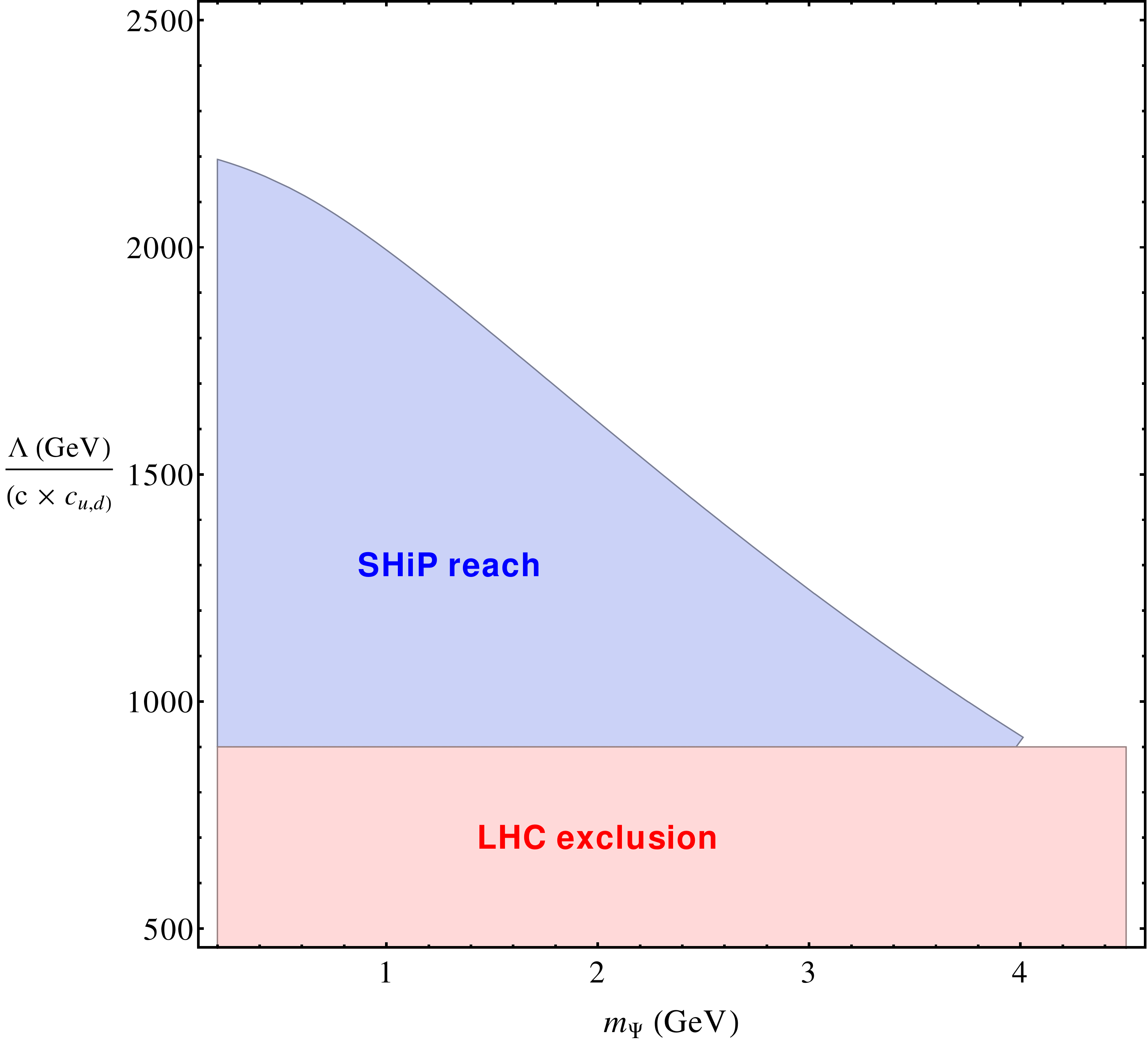}
\end{center}
\caption{Reach of SHiP for a pseudo-Dirac fermion $\Psi$ and compared to existing constraints from LHC.}
\label{PD:FIG:reach}\end{figure}

The production of light fermions coupled to the standard model by the effective operator (\ref{PD:GeneralEffective}) is suppressed by the fraction of anti-up and anti-down quarks in the proton at low momenta. Taking for sake of argument $c_L = c_R \equiv c, c_u^L = c_u^R = c_d^L = c_d^R \equiv c_{u,d}$, $ \frac{\Lambda^2}{c c_{u,d}}  = (2 \mathrm{TeV})^2$ with equal couplings to up and down quarks, and $m_\Psi = 1$  GeV; we find 
\begin{align}
\sigma (p + p \rightarrow \Psi + \ov{\Psi}) \simeq& 2 \times 10^{-6} \mathrm{pb} \times \left( \frac{(2\ \mathrm{TeV})^2}{\Lambda^2/(c \times c_{u,d})}\right)^2.
\end{align} 
This should be compared to the proton cross-section, which at the centre-of-mass energy of $\sqrt{s}=27$ GeV is $40$ mb (cf.~\cite{Cartiglia:2013jya}). Hence for $2\times 10^{20}$ protons on target we expect roughly $10^4$ raw events. We must then take the geometric factors and decays into account; we do this via a rough estimate with a geometric factor of $8\times 10^{-4}$, and taking the decay length in the lab frame to be $90$ m for reference, we have about $2$ expected events -- just enough to make a detection, and with much weaker couplings than the direct LHC bounds. The prospects improve as we decrease the particle mass; we plot in figure~\ref{PD:FIG:reach} the estimated optimistic SHiP reach in this approach over a range of masses and couplings, and find good prospects for detection.

\subsection{Concluding remarks}
For pseudo-Dirac particles suitable for dark matter models the channel,
\begin{eqnarray}
pp&\to& \chi_{2}+\chi_{1,2}, \chi_{2} \xrightarrow{\rm long\,\,lived} \chi_{1}+ \ell^{+}\ell^{-}
\end{eqnarray}
offers significant discovery potential beyond current LHC limits.

\section{SUSY vector portal I: Hidden Photinos}
\label{photinos}
In this and the following section we will consider extensions of the MSSM by extra U(1) gauge groups.
Here, we will consider the case where the extra U(1) gauge boson is {\emph{massless}} (and not accompanied by any hidden matter). In the next section we will then turn to the case where the gauge boson is {\emph{massive}} an accompanied by Higgses.
As we will discuss below, in the first case the observable signal arises from the fermionic superpartner of the massless gauge boson, the hidden photino, whereas in the second case interesting signals arise from the decay of the U(1) gauge boson into the extra Higgses or extra neutralinos.

\subsection{Motivation}
A novel possibility for new light fermionic states coupled to the Standard Model arises in supersymmetric extensions of the Standard Model with additional massless $U(1)$ gauge sectors (cf. also the Chapter~\ref{sec:vector-portal} on dark vector particles). The gauge bosons of these additional $U(1)$ factors may kinetically mix with hypercharge, while the gauge fermions (hidden photinos) may mix with the bino through both kinetic mixing and mass mixing. 
If there is light matter charged under these $U(1)$ factors, then the light matter inherits fractional electromagnetic charge and provides the primary means for probing the new sector. However, if there is no light matter charged under the additional $U(1)$ sectors, then the kinetic mixing between abelian gauge bosons is unphysical and the only portal is the kinetic and mass mixing of the photinos \cite{Ibarra:2008kn, Arvanitaki:2009hb}. 
Hidden photinos may be comparable in mass to, or significantly lighter than, MSSM neutralinos and couple through renormalizable couplings with small mixing angles, making them ideal candidates for exploration at a fixed target experiment.

Beyond their phenomenological novelty, hidden photinos have a natural realization in string theory.
Realistic string constructions typically involve the compactification of extra dimensions on manifolds with non-trivial topology, which is then reflected by the spectrum of light states after compactification. In particular~\cite{Polchinski:1995mt,Grimm:2008dq}, dimensional reduction of Ramond-Ramond (RR) $n$-forms gives rise to massless $U(1)$ fields labeled by the independent cycles of dimension $(n-1)$. For example, dimensional reduction of the RR four-form in Type IIB string theory gives rise to a $U(1)$ gauge group for every 3-cycle of the compactification manifold; the multitude of such cycles in realistic compactifications leads to a potential plethora of RR $U(1)$'s. There are generically no light states charged under these RR $U(1)$s, making them ideal sources of hidden photinos.

\subsection{Features}

Hidden photinos couple to the MSSM through kinetic and mass mixing with the bino. In a theory with $n$ additional $U(1)$s, the mass spectrum consists of four mostly-MSSM-like neutralinos $\tilde N_i$ ($i =1,\dots 4$) that are principally admixtures of the bino, wino, and neutral higgsinos and $n$ mostly-photino neutralinos $\tilde \gamma_a$ ($a = 1,\dots n)$. These photinos couple to the MSSM through small mixings $\kappa$ that are parametrically of the order of the kinetic mixing $\epsilon$ times the ratio of photino mass to bino mass, 
\begin{equation}
\kappa \sim \epsilon \frac{m_i}{M_1}.
\end{equation} 

Photino pairs couple to the Standard Model through the $s$-channel exchange of $Z$ or Higgs bosons and through the $t$-channel exchange of squarks and sleptons. These couplings are suppressed by powers of $\kappa$ as well as possibly small mixings $\eta \sim M_Z/\mu, M_Z/M_{1,2}$ corresponding to further suppression of photino-higgsino and photino-wino mixing.

If $R$-parity is conserved the lightest photino is stable. Heavier photinos decay into the lightest photino via three-body decays primarily through an off-shell $Z$ boson with lifetime
\begin{equation} \label{eq:lifetime}
\tau(\tilde \gamma_a \to \tilde \gamma_1+\ell^{+}\ell^{-}) \simeq 30\,{\rm cm}  \times \left( \frac{10^{-2}}{\kappa} \right)^4 \left( \frac{1}{\eta} \right)^4 \left( \frac{10 \, {\rm GeV}}{\delta m} \right)^5
\end{equation}
where $\delta m$ is the inter-photino mass splitting. If $R$-parity is violated then even the lightest photino can decay directly to the Standard Model. The lifetime depends on the details of the $R$-parity-violating interaction, but in general the lifetime is enhanced by a factor of $\kappa^{-2}$ relative to the equivalent three-body decay of a bino.

\subsection{Existing bounds}

\subsubsection{Direct bounds}
Direct bounds on hidden photinos come from precision measurements of the $Z$ width and potentially from direct production. LEP direct search limits on processes such as $e^+ e^- \to \tilde N_1 \tilde N_2$ \cite{Abbiendi:2003sc} place a negligible constraint on photino masses and mixings. There are no meaningful bounds from LHC Run 1, though photinos appearing in sparticle cascade decays may weaken the sensitivity of existing searches for supersymmetry \cite{Baryakhtar:2012rz}. The invisible $Z$ width, in contrast, provides a sensitive probe of light photinos. In the presence of $N$ photinos lighter than $M_Z/2$, the contribution to the decay width is
\begin{equation}
\delta \Gamma_Z \sim \frac{G_F}{6 \sqrt{2} \pi} N^2 \kappa^4 \eta^4 M_Z^3 \sim 0.03 \, {\rm MeV} \times N^2 \left( \frac{\kappa}{0.1} \right)^4 \left( \frac{\eta}{1} \right)^4
\end{equation}
Given LEP bounds on the invisible $Z$ width \cite{ALEPH:2005ab} this corresponds to a constraint on the mixing parameters $\kappa \eta \lesssim 0.3/\sqrt{N}$ and represents the strongest constraint on photinos from precision electroweak observables. Similarly, the contribution to the invisible Higgs width is
\begin{equation}
\delta \Gamma_h \sim \frac{G_F }{2 \sqrt{2} \pi} N^2 \kappa^4 \eta^2 M_W^2 m_h  \sim 0.1 \, {\rm MeV} \times N^2 \left( \frac{\kappa}{0.1} \right)^4 \left( \frac{\eta}{1} \right)^2.
\end{equation}
Bounds on the Higgs invisible width on the order of $20\%$ from Higgs coupling fits (e.g. \cite{CMS:2014ega}) provide the somewhat stronger constraint $\kappa \sqrt{\eta} \lesssim 0.17/\sqrt{N}$.

\subsubsection{Cosmological bounds}
If a hidden photino is stable and constitutes the LSP, then there are relatively strong constraints coming from the observed dark matter relic abundance \cite{Ibarra:2008kn, Arvanitaki:2009hb}. Photinos are thermalized by MSSM interactions provided $\kappa \gtrsim 10^{-6}$, in which case the thermal relic abundance of such a photino dark matter candidate is typically too large by a factor of $\kappa^{-4}$.
Even in the case that photinos do not reach thermal equilibrium with the MSSM for $\kappa \lesssim 10^{-6}$, out-of-equilibrium production may overclose the universe unless $\kappa \lesssim 10^{-11}$. Of course, these constraints may be ameliorated by a low reheating temperature, a phase of thermal inflation at low energies, or decay of photinos either into new hidden degrees of freedom or back into the Standard Model via $R$-parity-violating couplings, and place no irreducible bound on hidden photinos. Another potential cosmological bound on exceptionally light photinos with masses $\lesssim 30$ MeV comes from supernova cooling via photino-strahlung, but this does not place a significant constraint on the parameter space for hidden photinos.  

\subsection{What SHiP can do}

The production of light photinos at low-energy experiments is much akin to the production of light neutralinos, albeit with extra $\kappa^4$ suppression due to mixing. Light photinos could be produced in pairs at SHiP either through the two-body decay of mesons or through three-body decays such as $B \to K \tilde \gamma \tilde \gamma$; the precise meson production modes depend on the flavor structure of the MSSM sfermion sector, which determines the flavor structure of $q q \tilde \gamma \tilde \gamma$ couplings via squark exchange. The rates for production are akin to those for light neutralinos~\cite{Dreiner:2009er} with an additional suppression factor $\sim \kappa^4$ for pair production (this can be reduced to $\sim\kappa^2$ if there is also a light ordinary neutralino and one produces $\neu\widetilde{\gamma}$). The signatures of photino pair production then depend on whether or not $R$-parity is conserved.

\subsubsection{R-parity conserving photinos}
If $R$-parity is conserved then the lightest photino $\tilde \gamma_1$ is generically stable and we can have only pair production of $\tilde \gamma_1$ in meson decays. However, it is also possible for mesons to decay first to heavier photinos which in turn undergo three-body decays into lighter photinos, as in pair production of $\tilde \gamma_1 \tilde \gamma_2$ in meson decays followed by (possibly displaced) decay of the $\tilde \gamma_2$ via $\tilde \gamma_2 \to q \bar q \tilde \gamma_1$ or $\tilde \gamma_2 \to \ell \bar \ell \tilde \gamma_1$. 

To obtain a (very rough) estimate for the potential sensitivity at SHiP, let us consider the case with several light photinos produced from the decay of charged $B$ mesons (one could also consider the even more numerous $D$ mesons at a slight expense of a reduced mass reach). The production of light neutralinos can be quite high if there exist flavor changing terms from SUSY breaking (even in absence
of these production is possible via loop diagrams)~\cite{Dreiner:2009er}. 
Indeed, high gluino masses as required by the LHC measurements, weaken constraints on flavor changing terms.
Optimistically those could then be such that they maximize the production of photinos.
As a guide let us take the maximal allowed branching ratio for one of the $B$ decay channels~\cite{Chen:2007zk,Aubert:2004ws}
\begin{equation}
BR(B^{-}\to K^{-}+\nu\bar{\nu})=1.4\,\, 10^{-5}\sim 10^{-5},
\end{equation}
which is currently within the allowed regime.
Analog to Section~\ref{fermionproduction} we estimate that $\sim 7\times 10^{13}$ $B$ mesons being produced at SHiP. Using a geometric factor of $\sim 10^{-3}$ for the produced hidden photinos to be in the suitable detector region we would have about $\sim 10^5-10^6$
hidden photinos inside the detector region. With $\kappa$ in the $10^{-2.5}$ region a significant number of those would decay inside the detector volume, giving an intriguing possibility for a discovery.

It should be said that this is, of course, a very optimistic estimate. The true branching fraction and production rate depends on the detailed model parameters (superpartner masses and flavor changing terms) and can and will in general be suppressed by $\kappa^4$ ($\sim \kappa^2$ in the case of a light neutralino). Yet, at the same time before suppression by $\kappa^4$ there are no hard constraints that the branching fraction into hidden photinos could not even be bigger and only the $\kappa^4$ suppression brings it into the experimentally allowed region.

\subsubsection{R-parity-violating photinos}
$R$-parity violation is a compelling setting for hidden photinos, as cosmological bounds on photinos are naturally avoided and LHC bounds on the scale of MSSM states may be lowered. In this case, the phenomenology is much like that of R-parity violating neutralino decays. Production of the lightest photino proceeds either in pair production or single production through meson decays followed by three-body decay to Standard Model final states.
In particular the latter proceeds along very similar lines as the RPV scenario discussed in Section~\ref{Neutralinos} and from there we are optimistic that it yields a significant step into untested parameter regions.

\subsection{Concluding remarks}
Searches for the gauginos of extra ``hidden'' U(1) gauge symmetries are a well motivated setup offering interesting possibilities at SHiP.
Of particular interest are channels such as,
\begin{eqnarray}
pp&\to& B^{-}\to K^{-}+\widetilde{\gamma}\widetilde{\gamma},\,\,\widetilde{\gamma}_{i} \xrightarrow{\rm long\,\,lived} \widetilde{\gamma}_{j}+ \ell^{+}\ell^{-}
\\\nonumber
pp&\to& B+X\to \widetilde{\gamma} +X', \,\,\widetilde{\gamma} \xrightarrow{\rm long\,\,lived} D^{\pm}+\ell^{\mp}
\end{eqnarray}
and similar channels with $D$ mesons instead of the $B$ mesons.
While further more detailed studies are clearly needed, estimates indicate that SHiP will have interesting sensitivity.

\section{SUSY vector portal II: Novel Hidden Photon decays}
\label{photinos2}
In the previous section we have considered the case of an extra ``hidden'' U(1) gauge group with a massless gauge boson. 
In this case the signals originate directly from the additional hidden photinos.
We will now turn to the alternative option where the extra U(1) gauge symmetry is broken and the corresponding gauge boson is 
massive and therefore directly observable. In this case SUSY offers interesting new channels and signatures compared to the case of a simple dark vector scenario as diskussed in Chapter~\ref{sec:vector-portal}.
 
\subsection{Setup}
In much the same way that supersymmetry can stabilize the electroweak
scale against large quantum corrections, it can also stabilize the mass
of a light Abelian vector boson~\cite{ArkaniHamed:2008qp}.  
To do so, there must exist additional light particles beyond just 
the dark vector.  These new particles can produce new signals 
in lower-energy experiments at the intensity frontier~\cite{Morrissey:2014yma}.

 Extending the vector portal to be supersymmetric implies that the dark
vector $X^{\mu}$ of the exotic $U(1)_x$ gauge invariance
must be accompanied by a dark gaugino fermion $\widetilde{\gamma}^{\alpha}$.
Supersymmetry also implies that the kinetic mixing interaction must 
be extended to include new interactions involving the dark gaugino
and any scalars charged under the dark gauge group or SM hypercharge:
\beq
\mathscr{L} ~\supset~ -\frac{\epsilon}{2}B_{\mu\nu}X^{\mu\nu}
+ \epsilon\,\left[
\widetilde{B}^{\dagger}i\bar{\sigma}\ccdot\del\widetilde{\gamma} + (h.c.)\right]
+ \epsilon\,D_xD_Y \ ,
\label{eq:susyvec}
\eeq
where $D_i = g_i\sum_aq_a^i|\phi_a|^2$ sums over all the scalar fields 
in the theory weighted by their charge $q_a^i$ under the corresponding 
$U(1)_i$ factor, $i=x,\,Y$.  The first term in Eq.~(\ref{eq:susyvec}) 
is the familiar vector portal interaction, the second is a kinetic mixing between
the dark gaugino and the hypercharge gaugino $\widetilde{B}$,
and the third term is an extension of the Higgs portal interaction
to the new scalar fields required by supersymmetry.

In this section we will focus on a scenario where the dark photon gets a mass
$m_x$ from a dark Higgs mechanism~\cite{Morrissey:2009ur}.  
This occurs when a scalar dark Higgs field charged under $U(1)_x$ develops 
a vacuum expectation value~(VEV).  A minimal supersymmetric realization
of this consists of two dark Higgs supermultiplets
$H$ and $H'$ with equal and opposite $U(1)_x$ charges $\pm x_H$.  
Each supermultiplet contains a complex scalar and a Weyl fermion.
Scalar VEVs can be induced by the $D$-term interaction of Eq.~\eqref{eq:susyvec}
or by soft supersymmetry breaking terms in the dark sector.

When both dark Higgs scalars develop VEVs, the Weyl fermion Higgsinos
mix with the dark gaugino, one of the dark Higgs scalars is eaten by dark vector,
and the remaining Higgs scalars mix amongst themselves.  The physical
states in this minimal Higgsed dark vector realization consist
of a single massive dark vector $Z^x$, two real dark scalars $h_{1,2}^x$,
one dark pseudoscalar $A^x$, and three Majorana dark neutralinos
$\widetilde{N}_{1,2,3}^x$.  If $R$-parity is conserved, the lightest dark neutralino
will be stable and contribute to the dark matter density.

The natural size of the dark vector mass is of the same order as the soft
supersymmetry breaking terms in the dark sector.  These can be much smaller
than the corresponding terms in the visible sector (which must be larger than
about a TeV) if supersymmetry breaking is communicated less efficiently to the
dark sector~\cite{ArkaniHamed:2008qp}.  However, the soft terms in the dark
sector should be larger than at least $m_{soft}^x \gtrsim
\min\{\sqrt{\epsilon}M_Z,\,\epsilon\,m_{soft}^{vis}\}$ due to the kinetic
mixing interaction~\cite{Morrissey:2009ur,Cheung:2009qd}.

\subsection{SHiP Sensitivity}

The signals of this minimal supersymmetric dark vector theory 
in fixed-target experiments depend on how the dark vector decays.  
There are four main cases~\cite{Morrissey:2014yma}:
\begin{enumerate}
\item[A.] $Z^x\to SM+SM$: signals are similar to the non-supersymmetric
dark vector (cf. Chapter~\ref{sec:vector-portal}).\vspace{-0.3cm}
\item[B.] $Z^x \to \widetilde{N}_1^x\widetilde{N}_1^x$: vector decays are invisible.
\vspace{-0.3cm}
\item[C.] $Z^x \to h_1^xA^x$: scalars decay slowly to the SM
and can give rise to displaced signals.
\vspace{-0.3cm}
\item[D.] $Z^x\to \widetilde{N}_1^x\widetilde{N}_2^x$: can give displaced decays
when the 3-body mode $\widetilde{N}_2^x\to\widetilde{N}_1^xZ^{x*}$ is slow. 
\end{enumerate}%
Cases A and B are similar to minimal dark vector and minimal dark
matter theories respectively, so we will concentrate on cases C and D.
For this, we will set $\alpha_x = g_x^2/4\pi = \alpha$ and scale all
the dimensionful parameters of the theory as fixed ratios of the
dark vector mass $m_x$ as in Ref.~\cite{Morrissey:2014yma}.

\begin{figure}[ttt]
  \centering
  \hspace*{-1.8cm}\includegraphics[scale=0.35]{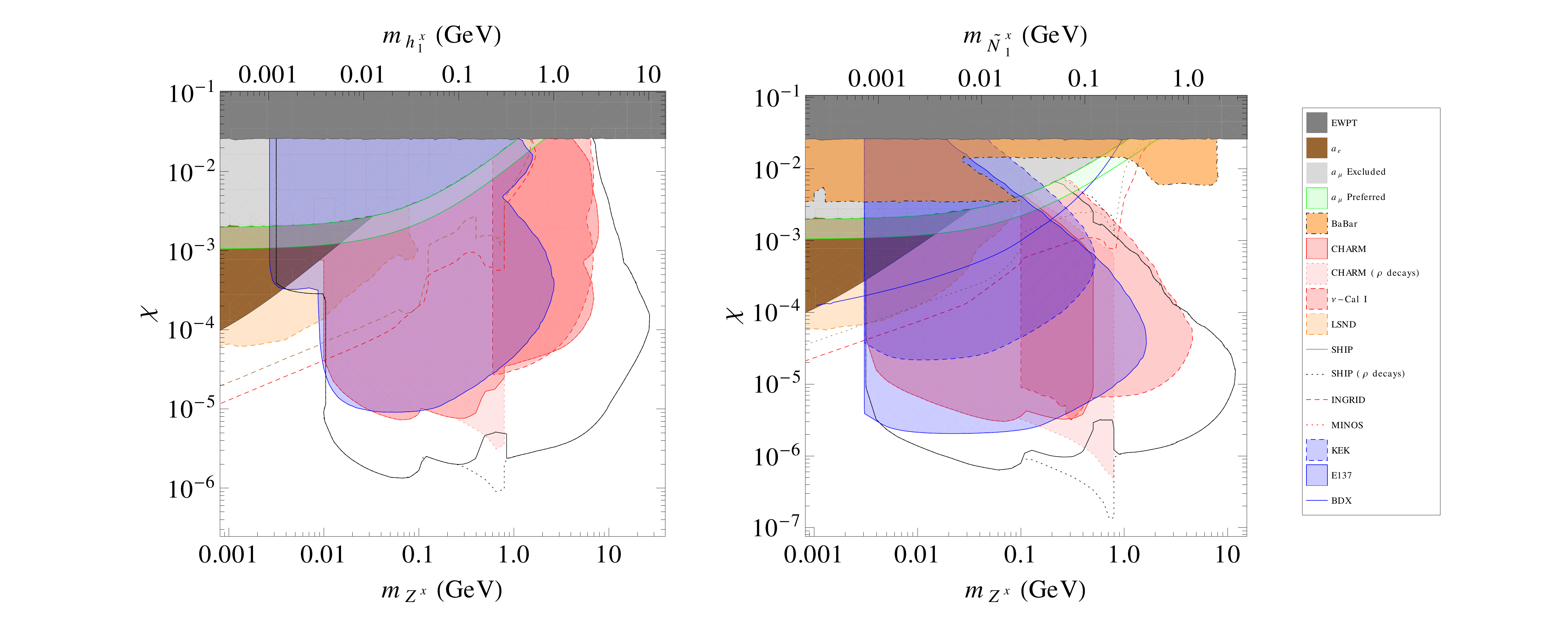}
  \caption{Sensitivity estimates for Cases C~(left) and D~(right) 
of the minimal supersymmetric dark vector theory discussed in the text. Benchmark slopes are as in~\cite{Morrissey:2014yma}.
  }\label{fig:ship}
\end{figure}

Adapting the same strategy and techniques as in~\cite{Morrissey:2014yma} to the geometry of SHiP 
in the left and right panels of 
Fig.~\ref{fig:ship} we show the reach of the SHiP experiment to cases
C and D of this minimal supersymmetric dark photon theory relative to 
other searches.  A significant improvement is achievable.

\subsection{Concluding remarks}
For SUSY models with an extra, spontaneously broken U(1) gauge symmetry SHiP offers significant improvement over existing limits
based on channels such as, e.g.
\begin{eqnarray}
pp&\to& Z^{x}+X\to h^{x}_{1}\,A^{x}_{1} +X,\,\,A^{x}_{1} \xrightarrow{\rm long\,\,lived} h^{x}_{1}+ \ell^{+}\ell^{-}
\\\nonumber
pp&\to& Z^{x}+X\to \widetilde{N}_{2}^{x}\widetilde{N}_{1}^{x} +X, \,\,\widetilde{N}^{x}_{2} \xrightarrow{\rm long\,\,lived} \widetilde{N}^{x}_{2}+\ell^{+}\ell^{-}.
\end{eqnarray}

\section{Axinos and saxions, ALPinos and sALPs}
\label{Axinos}
\subsection{Motivation}
As discussed in Chapter~\ref{chapALP} axions and axion-like particles (ALP) are well motivated in string theory but also
in field theoretic settings with spontaneously broken global approximate symmetries. 
In a supersymmetric version of axion models\footnote{For notational simplicity we will from now on talk about axions, axinos and saxions but unless specified otherwise this includes ALPs, ALPinos and sALPs.} (cf. Chapter~\ref{chapALP}), the axion $A$ forms a supermultiplet
with a fermonic partner, axino $\axino$, and a scalar partner, saxion $s$. 

The interactions for these partner particles are typically very weak (often suppressed by the large scale of spontaneous symmetry breaking $f_{A}$) and therefore testing them requires a high precision/intensity experiment such as SHiP.

\subsection{Phenomenology of saxions and axinos and possibilities at SHiP}
The most direct constraints typically arise from searching for the axion itself (cf.~Chapter~\ref{chapALP} for those constraints).
However, if the axino and/or the saxion are relatively light it is interesting to search for these particles directly. 

Masses of axino and saxion are naively considered to be of the order of the gravitino mass, however those can be lower in specific cases and take values between MeV and GeV~\cite{Chun:1992zk,Chun:1995hc,Kim:2012bb}. 

\subsubsection{Saxions}
Since saxions are just the scalar partners of the corresponding pseudo-scalar axions they feature interactions very similar to that of the axion itself.
The interactions typically include couplings to two axions as well as to two gauge bosons, e.g. photons and gluons,
\begin{equation}
\begin{split}
  \label{susy:eq:2}
 {\mathcal L} = \frac{s}{\sqrt{2}f_A} (\partial_\mu A)^2+  C \frac{\alpha}{8\pi f_A} s F^{a\mu\nu} F^a_{\mu\nu} + C_3\frac{\alpha_3}{8\pi f_A} s G^{a\mu\nu} G^a_{\mu\nu},
\end{split}
\end{equation}
with some constants $C$ and $C_{3}$. 
This is essentially the scalar equivalent to the pseudo-scalar interactions of ALPs discussed in Chapter~\ref{chapALP} plus an additional interaction with two axions.
In addition we can also have (strongly model dependent) interactions with Standard Model fermions.

Let us briefly discuss the two main differences: the parity and the interaction with two axions.
For most direct experimental tests such as SHiP the difference in parity only leads to minor modifications (see, e.g., Figs.~\ref{fig:scalar:yuk_SHIP} and~\ref{fig:yuk_SHIP} where we consider scalars and pseudoscalars interacting with SM fermions; we expect this level of similarity also for other couplings such as one to two gauge bosons).
The interaction with two axions is more important. If the axion mass is less than half the saxion mass, the saxion can dominantly decay via $s\to AA$.
Depending on the value of $f_{A}$ and the mass of the axion this can be an invisible decay, or the axion $A$ itself can decay into something visible.
If the decay is invisible this can lead to a significant reduction in the observed number of events and a corresponding reduction in sensitivity.
One can always consider this as the worst case scenario for the sensitivity. Decays of the produced axions can only lead to additional events.
Aside from this modification we expect for saxions at SHiP very similar signatures and quantitatively also similar improvements to those for ALPs, and we refer the reader to Chapter~\ref{chapALP} for details.

\subsubsection{Axino LSP with R-parity breaking}
If R-parity is broken, an axino LSP can decay and produce signatures in the SHiP experiment. 
In bilinear R-parity violation models,
 the light axinos dominantly decay to photon and neutrino with a lifetime\footnote{Here we consider parameter values that are more appropriate for ALP models and not for the QCD axion. Nevertheless we model the interactions (also for the superpartners) on those of the KSVZ axion. However, one should take this with a grain of salt, as strong model dependencies are possible.}~\cite{Covi:2009pq} 
\begin{equation}
\tau_{\tilde A} \sim1{\rm m}\left( \frac{|U_{\tilde{\gamma}\tilde{Z}}| \xi_i}{10^{-1}}\right)^{-2}  \left(\frac{\maxino}{10\gev}\right)^{-3}\left(\frac{f_A}{10^{5}\gev}\right)^{2} ,
\end{equation}
with $\xi_i=\VEV{\tilde{\nu}_i}/v$ and $\VEV{\tilde{\nu}_i}$ being the sneutrino vev which parametrizes the size of R-parity violation. 
In models with DSFZ type interactions axinos can additionally decay into $e^- e^+\nu$ with similar decay rate.
The QCD axino with $f_A\gtrsim 10^9\gev$ is found to be hard to decay
 inside the SHiP experiment.
However, for general ALPinos, the corresponding Peccei-Quinn breaking $f_A$ scale might be smaller. 
 
\begin{figure}
\centering
\includegraphics[width=0.38\textwidth]{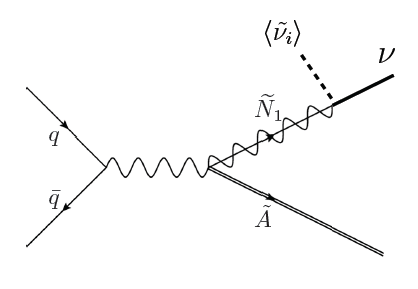}
\caption{Feynman diagram for the R-parity violating production of an axino.}
\label{rviolprod}
\end{figure}

With R-parity violation, a single axino can be produced together with a neutrino, e.g. from two quarks via an s-channel $Z$ boson as shown in Fig.~\ref{rviolprod}.

\subsubsection{Axinos with R-parity conservation}
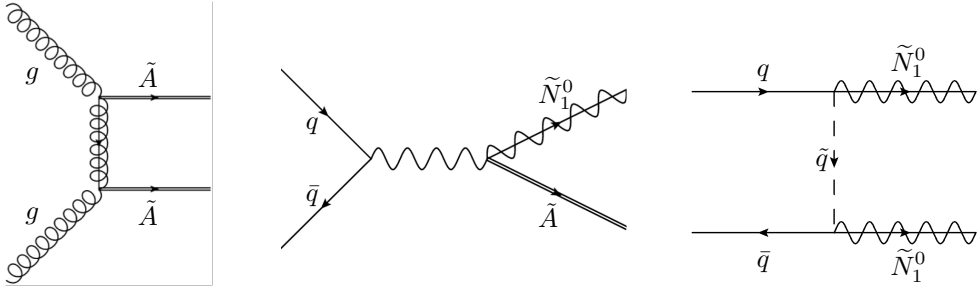
\begin{figure}
\centering
\begin{picture}(130,150)(-20,0)
\includegraphics[width=0.18\textwidth]{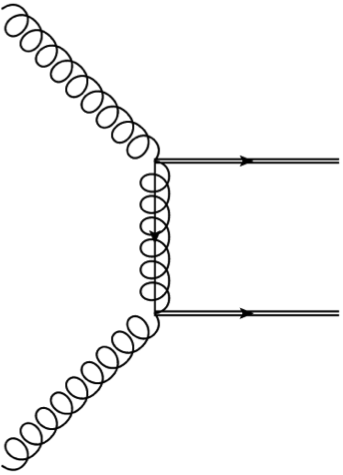}\qquad
\put(-70,25){$\displaystyle g$}
\put(-70,78){$\displaystyle g$}
\put(-28,23){$\displaystyle \tilde{A}$}
\put(-28,75){$\displaystyle \tilde{A}$}
\end{picture}
\begin{picture}(130,150)(10,-12)
\includegraphics[width=0.3\textwidth]{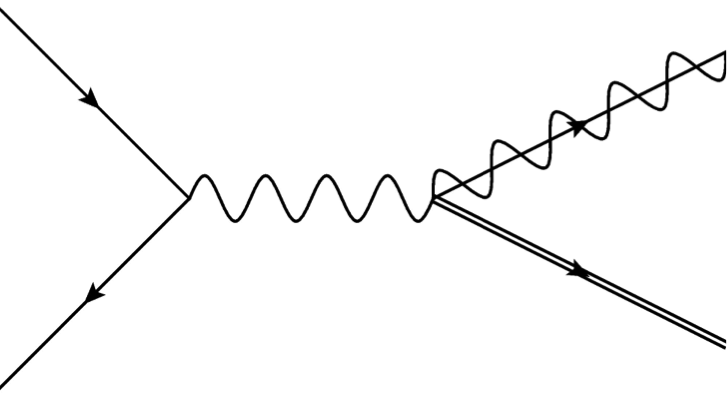}
\put(-120,20){$\displaystyle \bar{q}$}
\put(-120,47){$\displaystyle q$}
\put(-33,10){$\displaystyle \tilde{A}$}
\put(-33,57){$\displaystyle \neu$}
\end{picture}
\begin{picture}(130,80)(-10,-15)
\includegraphics[width=0.25\textwidth]{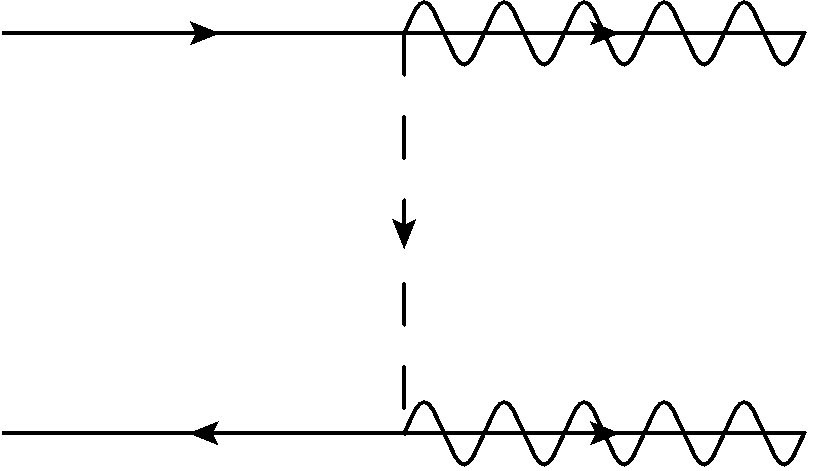}
\put(-33,-11){$\displaystyle \neu$}
\put(-33,66){$\displaystyle \neu$}
\put(-61,30){$\displaystyle \tilde{q}$}
\put(-83,-8){$\displaystyle \bar{q}$}
\put(-83,64){$\displaystyle q$}
\end{picture}
\caption{Typical Feynman diagrams for axino/neutralino pair production from gluons (left) or in association with a light neutralino (middle) or production of two light neutralinos. All of these production channels also work if R-parity is unbroken. According to the number of axinos produced the production amplitude is suppressed by $1/f^{2}_{A}$, $1/f_{A}$ and $1$.}
\label{axinoprod}
\end{figure}

If R parity is conserved, supersymmetric particles must be produced in pairs and this, of course also holds for the axinos.
Typical R-parity conserving production channels are shown in Fig.~\ref{axinoprod}. Since the diagram on the left hand side is suppressed by two factors of $f_{A}$, the corresponding production is likely to be too small to be observed, and a light neutralino is required for direct R-parity conserving production. In this case we can have axino/neutralino production (middle) or neutralino pair production (right).

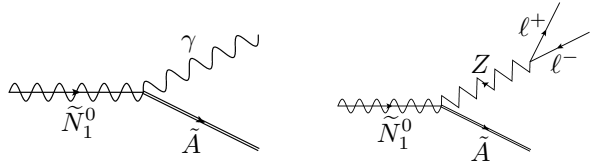
\begin{figure}
\centering
\begin{picture}(140,80)(-40,0)
\includegraphics[width=0.22\textwidth]{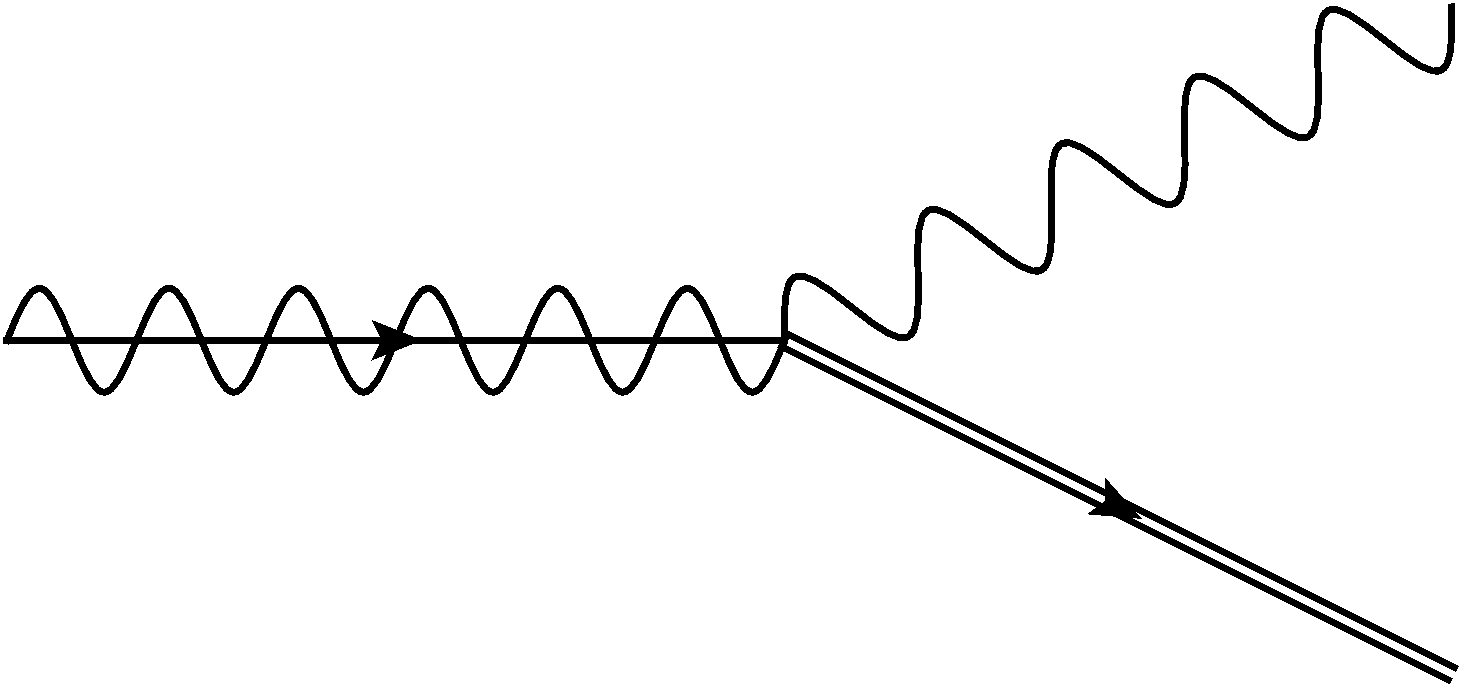}
\put(-30,0){$\displaystyle \tilde{A}$}
\put(-30,40){$\displaystyle \gamma$}
\put(-75,8){$\displaystyle \neu$}
\end{picture}
\begin{picture}(140,80)(-20,0)
\includegraphics[width=0.22\textwidth]{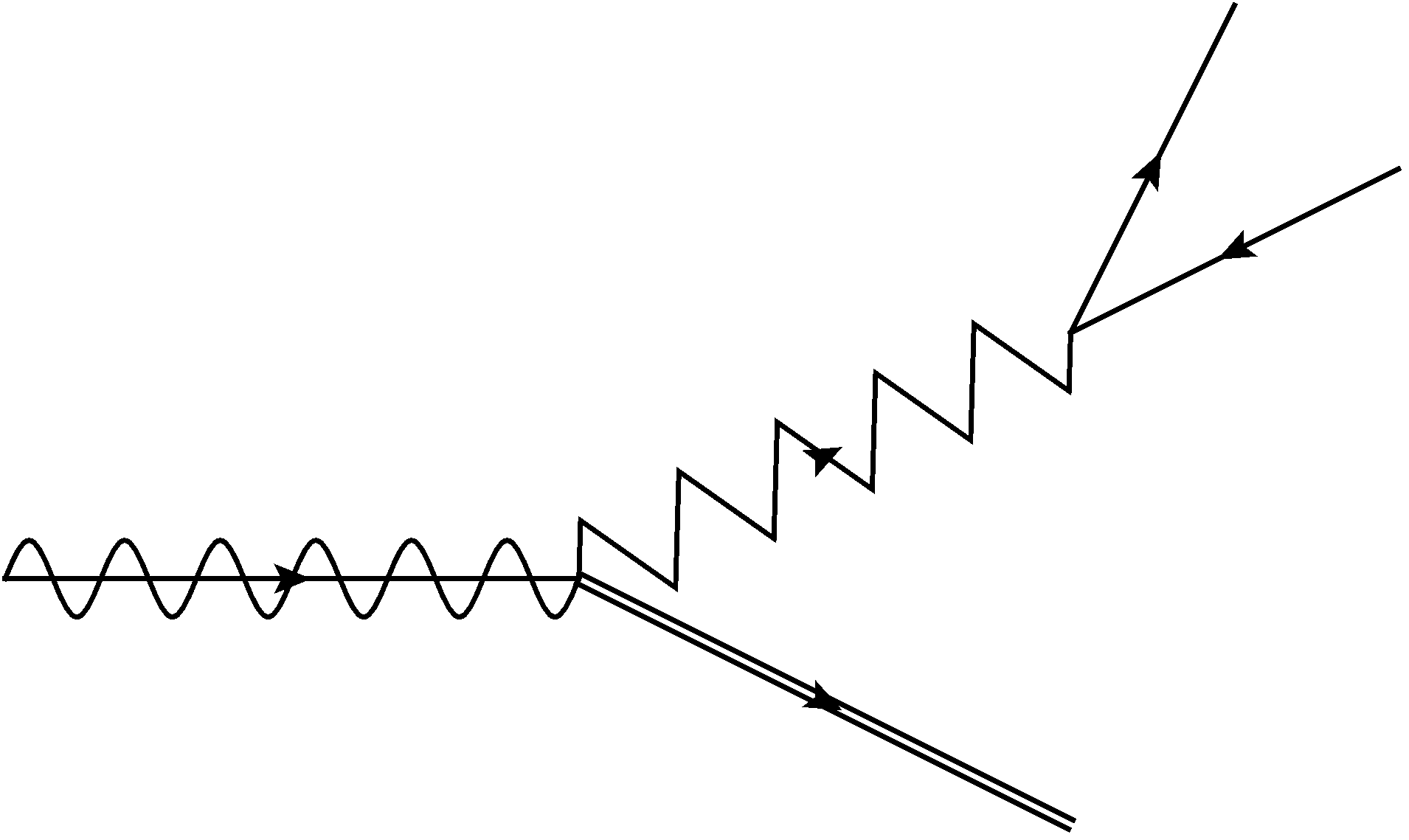}
\put(-45,-3){$\displaystyle \tilde{A}$}
\put(-45,30){$\displaystyle Z$}
\put(-80,4){$\displaystyle \neu$}
\put(-27,44){$\displaystyle \ell^{+}$}
\put(-15,30){$\displaystyle \ell^{-}$}
\end{picture}
\caption{Neutralino to axino decay channels.}
\label{axinodecay}
\end{figure}

An interesting option then also opens up for the decay. The heavier of the two (either the axino or the neutralino) can decay into the lighter one, as shown in Fig.~\ref{axinodecay}, for the case when the neutralino is heavier.
The corresponding lifetime can be quite large  and therefore suitable to the SHiP setup since the relevant interaction is suppressed by $f_{A}$.
The decay shown on the right hand side then leads to an interesting but also challenging mono-photon signature in SHiP. In addition we can also have (somewhat suppressed) decays with two charged tracks as shown on the right hand side of Fig.~\ref{axinodecay}.
If the axino is heavier, we can just reverse the roles of axino and neutralino in the decay diagrams. 

\subsection{Concluding remarks}
If axion/ALP models are embedded in a supersymmetric context two additional particles appear that lend themselves to searches at SHiP.
The first option is the saxion/sALP whose phenomenological features are qualitatively relatively close to that of the axion/ALP and can be searched at SHiP with the strategies discussed in Chapter~\ref{chapALP}. However, if the decay of the saxion into two axions is kinematically allowed, it is typically dominant, which can lead to reductions in the observed number of events with SM particles.

The fermionic axino/ALPino leads to different signatures and strategies. If R-parity is broken possible channels include,
\begin{eqnarray}
pp&\to& \axino\nu,\,\,\axino \xrightarrow{\rm long\,\,lived} \nu+ \ell^{+}\ell^{-},
\\\nonumber
pp&\to& \axino\nu,\,\,\axino \xrightarrow{\rm long\,\,lived} \nu+ \gamma,
\end{eqnarray}
If R-parity is conserved interesting signatures exist if there is an additional light neutralino. In this case one could have mono-photon signatures such as
\begin{eqnarray}
pp&\to& \neu\neu,\,\,\neu \xrightarrow{\rm long\,\,lived} \axino+ \gamma,
\end{eqnarray}
but also signals with charged track such as
\begin{eqnarray}
pp&\to& \neu\neu,\,\,\neu \xrightarrow{\rm long\,\,lived} \axino+ \ell^{+}\ell^{-},
\end{eqnarray}
More studies are needed for quantitative statements on the sensitivity.

\section{Additional Possibilities}
\label{additional}
In this section we briefly mention several other options that could lead to observable signatures at SHiP, but which are in very early stages of discussion and need further investigation to determine the sensitivity at SHiP and for comparison with existing limits.

\subsection{Pair production of light neutralinos and decay to gravitino}
In Section~\ref{Neutralinos} we already discussed the possibility of light neutralinos. The focus was on R-parity violating production and decays.

Even if R-parity is conserved, the neutralino is not necessarily the lightest superpartner. In theories with gauge mediated supersymmetry breaking this is typically the gravitino.
For example the neutralino can decay into a photon and a gravitino (cf.~\cite{Giudice:1998bp}),
\begin{equation}
\tau(\neu\to\tilde{G}\gamma)=\frac{16\pi F^2}{k^2\kappa_{\gamma}m^{5}_{\neu}}=10\,{\rm m}\,\,\frac{1}{k^2\kappa_{\gamma}}\left(\frac{10\,{\rm GeV}}{m_{\neu}}\right)^5\left(\frac{\sqrt{F}}{100\,{\rm TeV}}\right)^4.
\end{equation}
Here $F$ is the scale of supersymmetry breaking, $k$ is a coupling strength to the SUSY breaking sector and $\kappa_{\gamma}$
encodes the relevant neutralino mixing angles for the neutralino-photon-gravitino coupling.

Pair production of such light neutralinos could occur via squark exchange in the t-channel. (On the parton level the diagrams are similar to what we will have in the case of flavor violating production discussed in the next subsection and shown in Fig.~\ref{flavorviol}, but due to the absence of flavor violation, formation of intermediate mesons is not necessarily preferred.)

In this setup the channel of interest would be,
\begin{eqnarray}
pp&\to& \neu\neu,\,\,\neu \xrightarrow{\rm long\,\,lived} \tilde{G}+ \gamma,
\end{eqnarray}
somewhat similar to the axino case discussed above.

\subsection{Flavor violating production of light neutralinos}
\begin{figure}
\centering
\begin{picture}(130,80)(0,-10)
\includegraphics[width=0.25\textwidth]{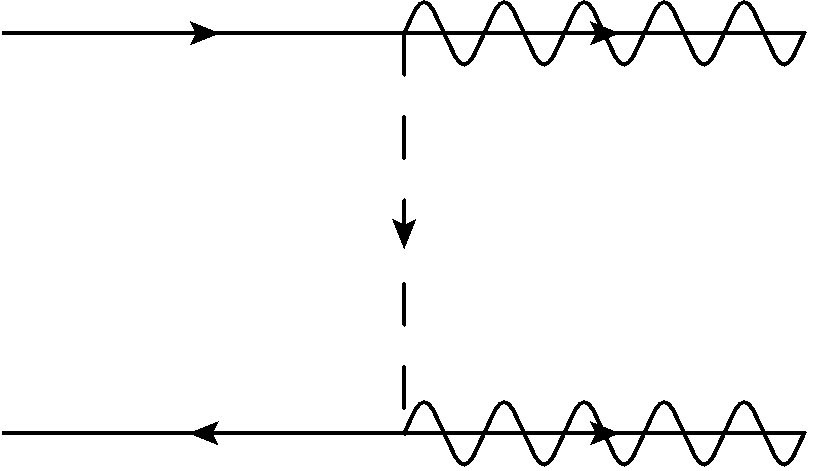}
\put(-33,-11){$\displaystyle \neu$}
\put(-33,66){$\displaystyle \neu$}
\put(-61,30){$\displaystyle \tilde{t}$}
\put(-83,-8){$\displaystyle \bar{u}$}
\put(-83,64){$\displaystyle c$}
\end{picture}
\caption{Neutralino pair production from neutral $D$ mesons via flavor violating processes with t-channel stop exchange.}
\label{flavorviol}
\end{figure}

\begin{figure}
\centering
\begin{picture}(130,80)(0,-10)
\includegraphics[width=0.25\textwidth]{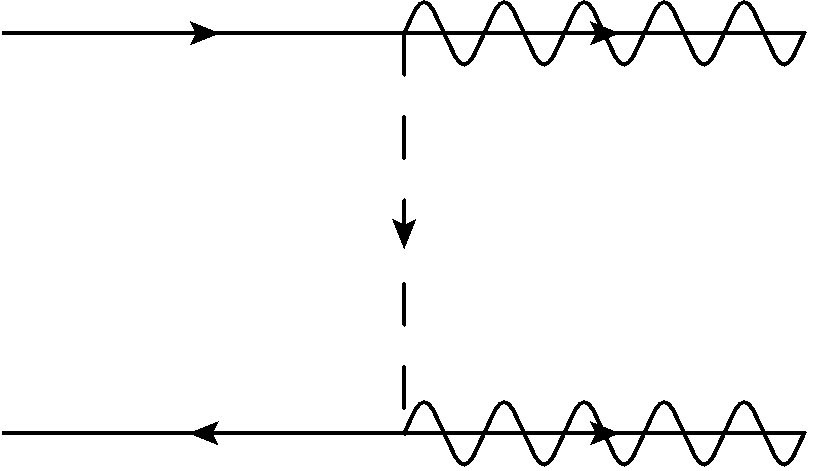}
\put(-33,-11){$\displaystyle \neu$}
\put(-33,66){$\displaystyle \neu$}
\put(-61,30){$\displaystyle \tilde{t}$}
\put(-83,-8){$\displaystyle \bar{u}$}
\put(-83,64){$\displaystyle u$}
\end{picture}
\caption{Direct neutralino pair production in a process with t-channel stop exchange. Compared with production via mesons shown in Fig.~\ref{flavorviol} this can benefit from larger parton distribution functions and from somewhat less constrained mixings between up- and top-squarks~\cite{Dittmaier:2007uw,AranaCatania:2011ak}.}
\label{flavorvioldirect}
\end{figure}

Flavor violating squark terms could lead to a significantly enhanced pair production of light neutralinos.
A possible diagram is shown in Fig.~\ref{flavorviol}, where we have production from the decay of $D$-mesons. 
Alternatively one could also have direct production of neutralino pairs as shown in Fig.~\ref{flavorvioldirect}.
The case of a stop exchange is particularly interesting since this can still be fairly light.

For R-parity conserving setups decay can proceed to a gravitino and a photon as discussed above. If R-parity is violated we could have decays similar to those discussed in Section~\ref{Neutralinos}.

All in all we have potential channels,
\begin{eqnarray}
pp&\to& D^{0}+X\to \neu\neu+X,\,\,\neu \xrightarrow{\rm long\,\,lived} \tilde{G}+ \gamma \quad{\rm(R-parity\,\,conserved)},
\\\nonumber
pp&\to& D^{0}+X\to \neu\neu+X,\,\,\neu \xrightarrow{\rm long\,\,lived} K^{\pm}+\ell^{\mp} \quad{\rm(R-parity\,\,violated)},
\end{eqnarray}
and similar direct processes where the intermediate meson production and decay step is absent.
It will be interesting to see if SHiP can compete with or exceed constraints from B-physics for those couplings.

\section{SUSY at SHiP: Final remarks}
Many supersymmetric models still allow for new light neutral particles in the MeV to few GeV range which is accessible to SHiP.
In these models SHiP often allows for significant discovery potential beyond existing constraints.

Importantly supersymmetry not only provides additional theoretical motivation but also encourages to consider additional interesting but also challenging signatures in the SHiP detector. In particular we can have the following decays of a long lived neutral particle,
\begin{eqnarray}
X&\to& \ell^{+}\ell^{-}
\\\nonumber
X&\to& \pi^{+}\pi^{-}
\\\nonumber
X&\to& K^{\pm}+\ell^{\mp}
\\\nonumber
X&\to& K^{0}+\emiss
\\\nonumber
X&\to& \pi^{0}+\pi^{0}
\\\nonumber
X&\to& \gamma+\gamma
\\\nonumber
X&\to& \gamma+\emiss
\end{eqnarray}
where especially the lower ones may be challenging. The above also shows that
there are significant opportunities in these neutral channels which strongly
motivates the study of their experimental sensitivity, which is currently
under way.


\chapter{Tau neutrino physics and other precision measurements in SHiP}
\label{sec:tau-neutrino-physics}

The high intensity proton beam dump experiment under consideration produces
an high intensity neutrino flux with all three flavors: electron, muon and tau
neutrinos and antineutrinos. In this section we will present a list of observables
that could be measured exploiting the variety of such a neutrino flux. The basic
assumptions are that $2 \times 10^{20}$ protons with $400\,\rm{GeV}$ energy from
the SPS at CERN will be stopped in a target of Molybdenum and Tungsten, and that
on the path of the emerging flux of neutrinos there will be a neutrino detector,
consisting for the most part of lead, that allows to identify the flavor of all the three kinds of charged
leptons produced in charged current neutrino interactions and also the sign of the
charge for tau and muon charged leptons, as well as the other
relevant kinematical variables of the Deep Inelastic Scattering (DIS) processes
for both the charged and the neutral current (CC and NC) interactions. We will
further assume that the neutrino detector will allow to identify the production
and decay of charmed resonances.
As we will discuss later in this section, under these assumptions there will
be the possibility to measure tau neutrino charged current cross sections.
This is currently the least known charged current interaction vertex of the Standard
Model (SM) with 7.5 events detected by the DONUT collaboration~\cite{Kodama:2007aa}
and 4 by the OPERA collaboration~\cite{Agafonova:2014ptn}. As we will see in
section~\ref{sec:nutau}, with the SHiP
experiment the statistics of such events can be increased by three orders of magnitude
allowing to perform new studies of the cross sections both at the inclusive and differential level.
To the extent of performing
detailed measurements of tau neutrino cross sections it is crucial to asses the precision
in the determination of the tau neutrino flux. This is discussed in section~\ref{sec:nutauflux}.
The SHiP experiment covers the DIS kinematics in a typical range of the Bjorken scaling
variable $x$ down to $x \sim 5 \times 10^{-4}$ for tau neutrinos with energy $E_\nu = 50$~GeV
and $x \sim 5 \times 10^{-3}$ for muon neutrinos with energy $E_\nu = 20$~GeV.
Given the high rate of muon neutrinos and anti-neutrinos DIS events expected, in section~\ref{sec:DIS} 
we review the current status of deep-inelastic scattering with emphasis on charged current
interactions off unpolarized fixed targets. We address the most relevant open questions
in our fundamental understanding of the proton and discuss the physics potential of
neutrino-nucleon deep-inelastic scattering.
Finally, we consider other limits that can be improved exploiting
the neutrino flux in SHiP; among them there are the limit on the tau neutrino magnetic
moment (section~\ref{sec:mag}) and the limit on the production cross section of a weakly
decaying charmed baryon with five valence quarks (section~\ref{sec:pq}). We will draw our conclusion
in section~\ref{sec:conc}.

\section{Tau neutrino physics}
\label{sec:nutau} 

Direct measurements of tau neutrino charged-current (CC) interactions are a fairly recent
phenomenon. The DONUT experiment reported 9 tau neutrino events with a background of 1.5
events from their neutrino beam produced with the $800\,\rm{GeV}$ Tevatron beam at
Fermilab~\cite{Kodama:2007aa}.
Their cross section measurement agrees well with the related muon neutrino CC cross section,
however, their statistical and systematic errors are each $33\%$ of the best
fit measurement respectively~\cite{Kodama:2007aa}.
More specifically, in the DONUT experiment it was not possible to identify the charge of the
outgoing tau lepton, for this reason the collaboration quoted the average of $\nu_{\tau}$
and $\bar{\nu}_{\tau}$ charged current cross sections as:
\begin{eqnarray} \nonumber
\sigma_{\nu_l} &=& \sigma_{\nu_l}^{const}\,E\,K(E), ~~~~l=e,\mu,\tau \\
\sigma_{\nu_l}^{const} &=& 0.39 \pm 0.13 \pm 0.13 \times 10^{-38}~\rm{cm}^2~\rm{GeV}^{-1}
\end{eqnarray}
where $K(E)$ describes the kinematical suppression due to the tau mass and $\sigma_{\nu_l}^{const}$
has to be compared with the average of $\nu_{\mu}$ and $\bar{\nu}_{\mu}$ charged current cross sections,
that is $0.51\times10^{-38}~\rm{cm}^2~\rm{GeV}^{-1}$. As a final remark concerning the averaged tau neutrino
cross section measured by the DONUT collaboration, we note that at the time of the analysis
of Ref.~\cite{Kodama:2007aa} the charm quark fragmentation in $D_s$ and the $D_s \to \tau \nu_{\tau}$
decay branching ratio were known with less precision.
Together with the charm production cross section in proton-nucleon collisions,
these are the three basic ingredients to estimate the tau neutrino flux in a beam dump experiment
as they enter directly as overal factors in the normalization of the flux.
The $D_s \to \tau \nu_{\tau}$ branching ratio, for example, in Ref.~\cite{Kodama:2007aa}
was assumed to be $(6.4 \pm 1.5)\%$, as suggested in the PDG of 2006, while recently it has been
measured with increased precision to be $(5.54 \pm 0.24)\%$, as reported by the PDG of
2014~\cite{Agashe:2014kda}.

The OPERA experiment reported 4 tau neutrino events with practically no
background~\cite{Agafonova:2014ptn}.
In OPERA these events were found while searching for $\nu_{\mu} \to \nu_{\tau}$ oscillation
starting from an almost pure muon neutrino flux. For this reason in OPERA only $\tau^-$ leptons
have been detected and not $\tau^+$. Even if the OPERA collaboration could now publish a measurement
of the $\nu_\tau$  cross section for the first time, within SHiP there will be the possibility
to measure separately $\nu_{\tau}$ and $\bar{\nu}_{\tau}$ cross sections in the same experiment and
with higher precision, identifying for the first time also $\bar{\nu}_{\tau}$ interactions through
the detection of the positively charged tau leptons produced.
According to our preliminary estimations, the number of detectable tau neutrino and anti-neutrino
charged current events in the SHiP neutrino detector will be several thousands.
Indeed, such a large sample of events will bring opportunities to make new measurements of the
structure functions in tau neutrino and antineutrino charged current events.
With the usual DIS variables: $x \equiv Q^2/2\,p \cdot q$ and $y \equiv p \cdot q/p \cdot k$,
where the momentum assignments are:
\begin{eqnarray}
\nu_{\tau}/\bar{\nu}_{\tau}(k)\,+\,N & \to & \tau^-/\tau^+(k')\,+\,X\\
q^2 \equiv & (k-k')^2 & = -Q^2,
\end{eqnarray}
the tau neutrino and anti-neutrino charged current cross sections in terms of the structure
functions $F_1,...,F_5$ are \cite{Albright:1974ts}:
\begin{eqnarray} \nonumber
\frac{d^2\sigma^{\nu(\bar{\nu})}}{dx\ dy} &=& \frac{G_F^2 M
E_{\nu}}{\pi(1+Q^2/M_W^2)^2}
\Biggl(
(y^2 x + \frac{m_{\tau}^2 y}{2 E_{\nu} M})
F_1
+ \left[ (1-\frac{m_{\tau}^2}{4 E_{\nu}^2})
-(1+\frac{M x}{2 E_{\nu}}) y\right]
F_2
\\ \nonumber
&\pm& 
\left[x y (1-\frac{y}{2})-\frac{m_{\tau}^2 y}{4 E_{\nu} M}\right]
F_3 
+
\frac{m_{\tau}^2(m_{\tau}^2+Q^2)}{4 E_{\nu}^2 M^2 x} F_4
- \frac{m_{\tau}^2}{E_{\nu} M} F_5
\Biggr)\, ,
\label{eq:nusig}
\end{eqnarray} 
where $+F_3$ applies to neutrino scattering and $-F_3$ to antineutrinos, $M$ and $m_{\tau}$
are the nucleon and $\tau$ lepton masses respectively, $E_{\nu}$ is the initial
neutrino energy and $G_F$ is the Fermi constant. As we will see, at the lower $\nu_\tau$
energies the SHiP experiment offers the first opportunity to measure the 
structure functions $F_4$ and $F_5$. At the Born level, neglecting target mass corrections, the
Albright-Jarlskog relations apply \cite{Albright:1974ts}
\begin{eqnarray}
F_4&=&0\,, \\
F_5 &=&\frac{ F_2 }{2x}\,.
\end{eqnarray}
The QCD predictions for the DIS structure functions $F_4$ and $F_5$ are known up to NLO
accuracy~\cite{Kretzer:2002fr}, including full dependence on heavy-quark masses, though.
The detailed relationships between the five structure functions, including NLO QCD together
with target mass and charm quark mass corrections, are discussed, for example, in
Refs.~\cite{Schienbein:2007gr,Kretzer:2002fr,Kretzer:2003iu,Jeong:2010nt}.

To evaluate neutrino-lead charged current scattering, we find the neutrino cross section per nucleon via
\begin{equation}
\label{sigma}
\sigma_{CC} = \frac{Z\sigma_{\nu p}+(A-Z)\sigma_{\nu n}}{A}\,,
\end{equation}
and similarly for antineutrinos.

In evaluating the cross section, we have extrapolated the low $Q$ structure functions according
to Ref.~\cite{Reno:2006hj} below $Q^2=2$ GeV$^2$.   To avoid double counting with quasi-elastic and few
particle production, a minimum hadronic final state invariant mass $W$ is set. A frequent choice
consist of taking $W_{min}=1.4$ GeV.
A second choice, $W_{min}=m_p+m_\pi$, approximates single and few pion production
with the DIS cross section. We show both choices for $W_{min}$ in the next few figures.
\begin{figure}[tbh]
  \centering 
  \subfloat{ \includegraphics[width=0.479\textwidth]{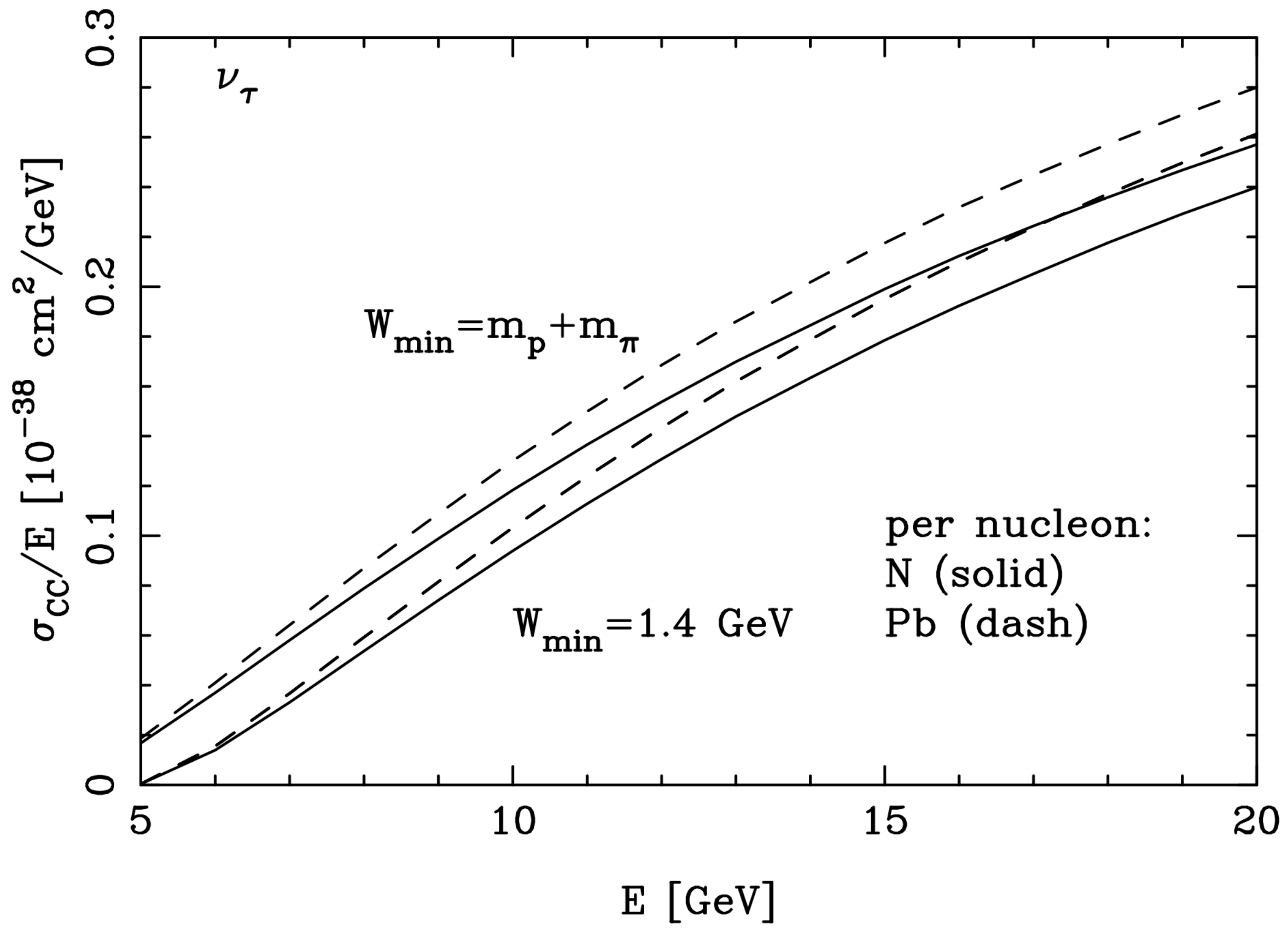}
    \label{fig:sig}
  }
  \subfloat{ \includegraphics[width=0.479\textwidth]{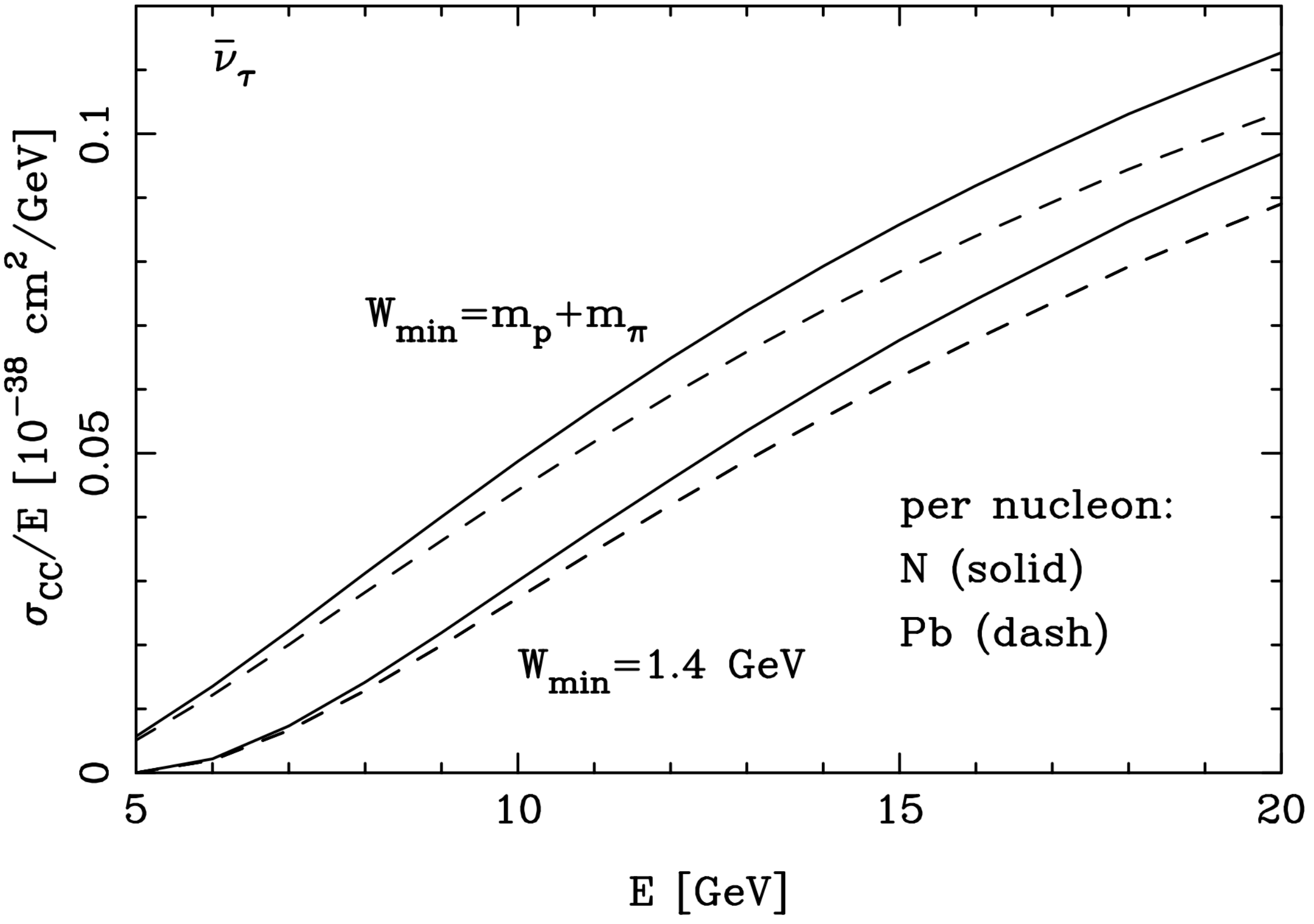}
    \label{fig:asig}  
  }
  \caption{
    The charged current cross section per nucleon, scaled by incident energy for (a) $\nu_\tau$ and
    (b) $\bar{\nu}_\tau$ scattering with isoscalar nucleons (solid) and lead (dashed) targets.
    The upper solid and dashed curves are for $W_{min}=m_p+m_\pi$, while the lower solid and dashed
    curves are for $W_{min}=1.4$ GeV.            
  }
  \label{fig:sigcc}
\end{figure}
Fig. \ref{fig:sigcc} (a) shows the $\nu_\tau$ CC cross section per nucleon per GeV as a function of energy
for two targets: isoscalar nucleons (N, solid curves) and lead (Pb, dashed curves) for two choices of
$W_{min}$. At $E=20$ GeV, the cross section per nucleon on lead is almost 10\% larger than on an isoscalar
target. The neutrino-neutron cross section is larger than the neutrino proton cross section
since the valence contribution $\nu_\tau d\to \tau u$ is large.  Fig. \ref{fig:sigcc}(b) shows that the
cross section per nucleon for $\bar{\nu}_\tau$ scattering with lead is smaller than with isoscalar nucleon
targets since $\bar{\nu}_\tau u \to \bar{\tau} d$ is the dominant valence process.  At low energies, the CC
cross sections for $\nu_\tau p$ and $\bar{\nu}_\tau p$ are comparable.
\begin{figure}[tbh]
  \centering
  \subfloat{ \includegraphics[width=0.479\textwidth]{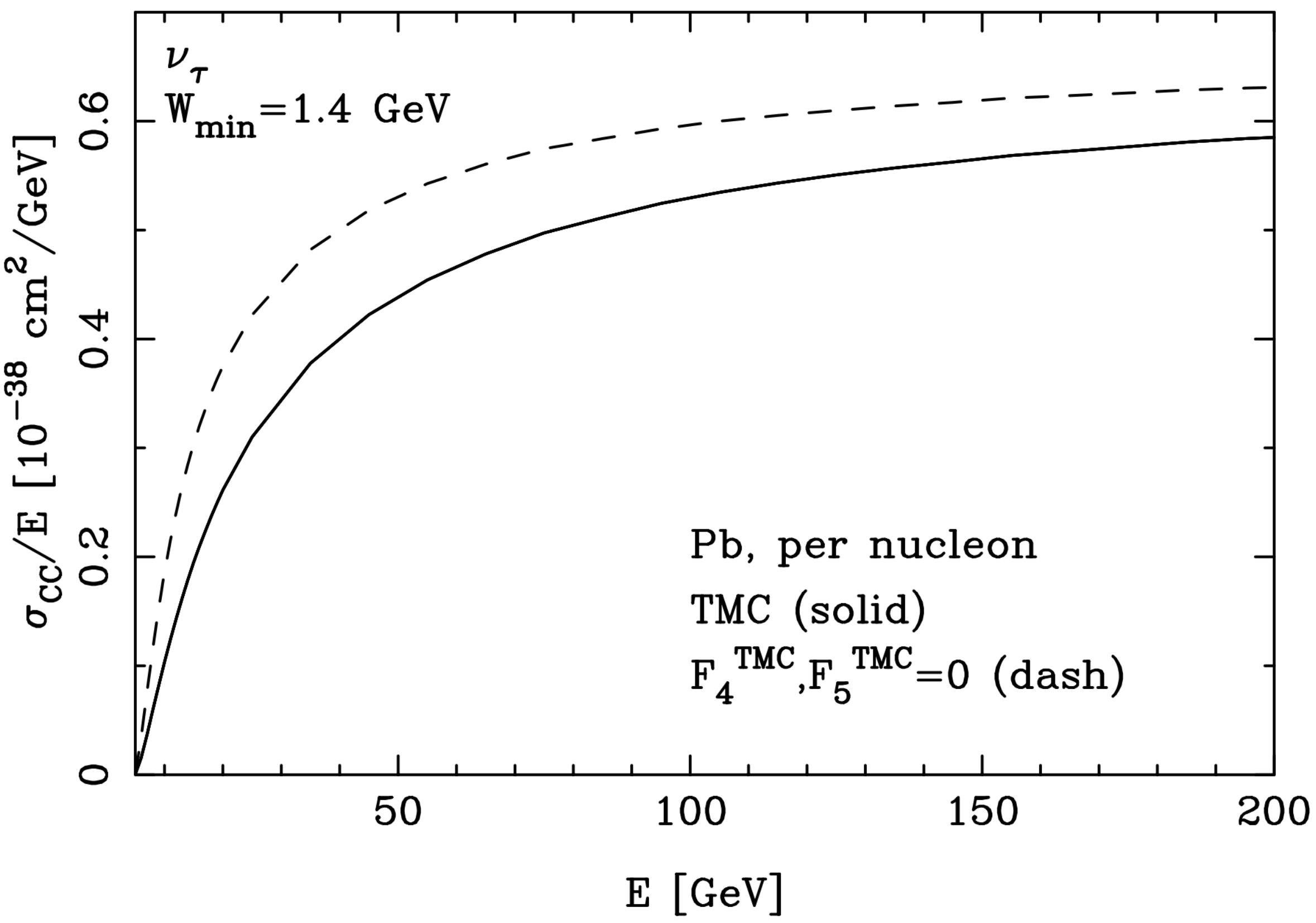}
    \label{fig:f45}
  }
  \subfloat{ \includegraphics[width=0.479\textwidth]{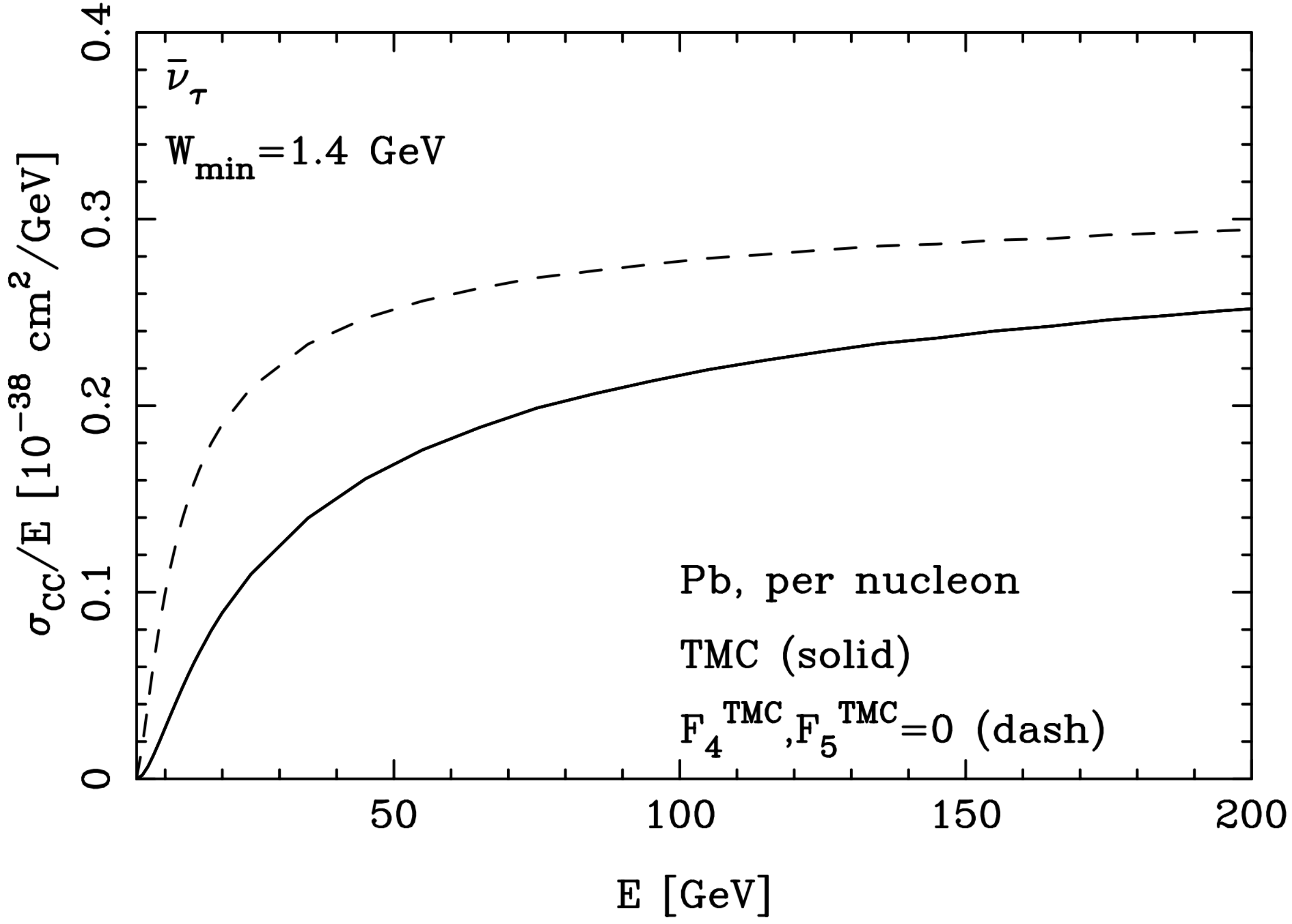}
    \label{fig:af45}  
  }
  \caption{
    The charged current cross section per nucleon, scaled by incident energy for (a) $\nu_\tau$ and (b)
    $\bar{\nu}_\tau$ scattering with a lead  target with $W_{min}=1.4 GeV$. The dashed curve has $F_4$ and
    $F_5$ set to zero, while the solid curve has the full expression for the target mass corrected (TMC)
    cross section.        
  }
  \label{fig:sigf45}
\end{figure}
The additional structure functions $F_4$ and $F_5$ have an impact on the charged current cross section,
as shown in Fig.~\ref{fig:sigf45}(a) and (b).  This is primarily due to $F_5\simeq F_2/(2x)$, reducing the
cross section. At $E=20$ GeV, for tau neutrino CC scattering, the full expression for the cross section
yields a result about 30\%\ lower than when these terms are neglected. At $E=200 $ GeV, the contribution
of these terms reduces the cross section by about 7\%. For $\bar{\nu}_\tau$ CC scattering, the solid line
is about 53\% lower than the dashed line for $E=20$ GeV and  about 14\%\ lower for 200 GeV. Target mass
corrections are included here. They amount to about a 7\%\ correction for $\nu_\tau$ CC scattering at $E=10$ GeV.

\subsection{Flux of tau neutrinos}
\label{sec:nutauflux} 

The precision on the tau neutrino cross section measurement is linked to the precision in the estimation
of the flux of neutrinos.
\begin{figure}[ht]
\centering
\includegraphics[width=0.70\textwidth]{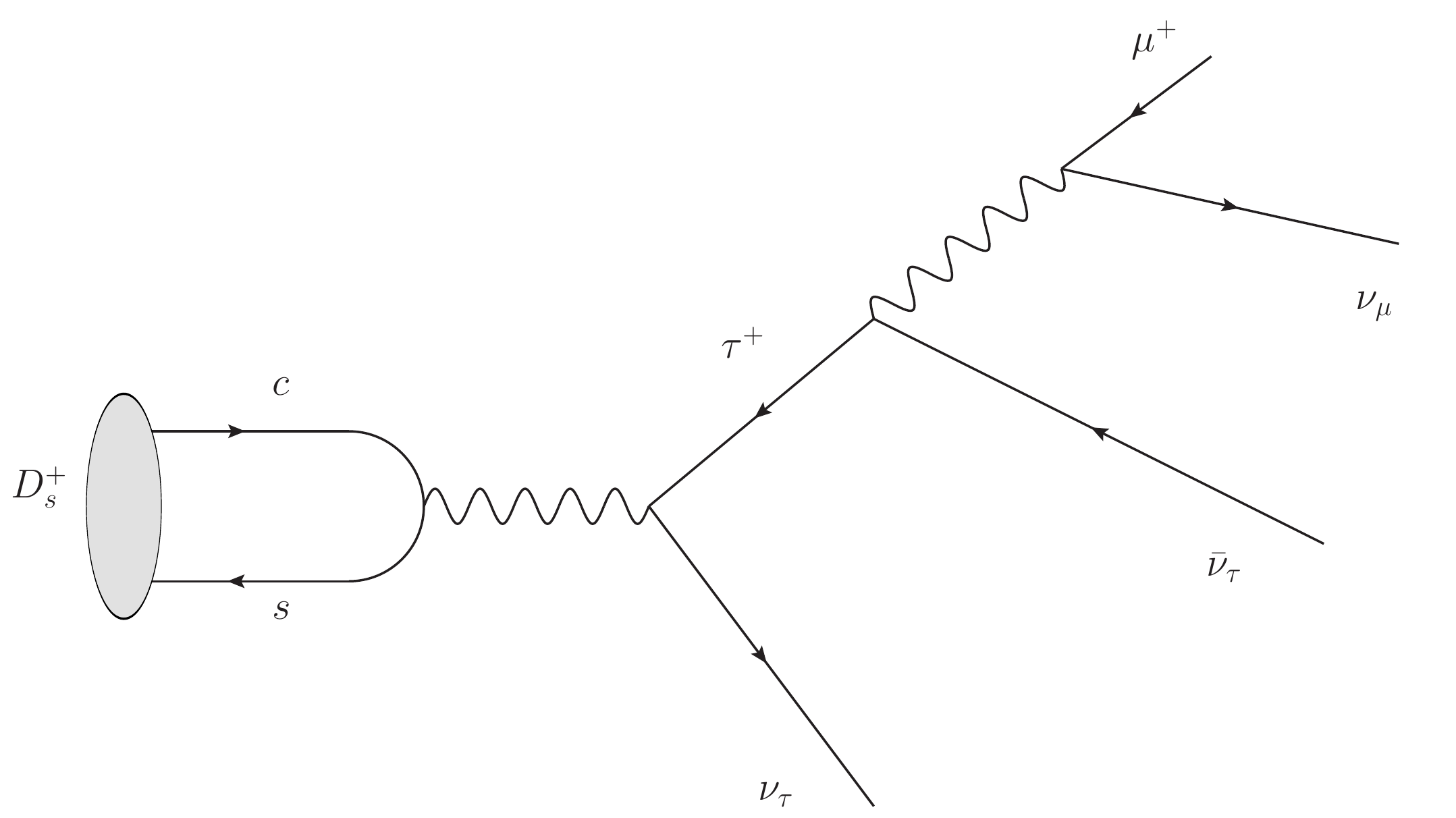}
  \caption{
           Example of $D_s$ decay chain producing a $\nu_{\tau}$ and a $\bar{\nu}_{\tau}$.
 $18\%$ of the tau leptons decay to a muon (see also Table~\ref{table:branching}),
in all tau decay modes another tau neutrino is produced.
}
  \label{fig:nutaus}
\end{figure}
As sketched in Fig.~\ref{fig:nutaus}, the prompt flux of $\nutau+\nutaub$ comes from the
leptonic decay $D_s\to \tau\nutau$, with a decay length of $c\tau_{D_s} = 149.9 \ \mu$m and
branching ratio $B(D_s\to \tau \nutau)=(5.54\pm 0.24)\% $~\cite{Agashe:2014kda}. In this
decay channel, two tau neutrinos are produced: one directly from the decay and the other
from the ``chain decay'' $D_s\to \tau\to \nu_\tau$. For a $D_s$ with energy $E_D$, the lepton
energy distributions are constant, constrained  by
\begin{eqnarray}
0 &\leq& E_\nu \leq (1-R_\tau)E_D\\
R_\tau E_D &\leq &  E_\tau \leq E_D\ ,
\end{eqnarray}
for $R_\tau=m_\tau^2/m_D^2$. With $m_\tau \sim m_{D_s}$, the neutrino energy range is much lower
than the tau energy range: $\langle{E_\nu}\rangle\simeq 0.09 E_D$ while $\langle{E_\tau}\rangle
\simeq 0.91 E_D$.  The tau itself decays promptly, $c\tau_\tau=87.03 \ \mu$m, to $\nu_\tau$
with approximately 1/3 of the tau energy going to the neutrino. Except for the lowest energies,
it is the chain decay $D_s\to \tau\to \nu_\tau$ that dominates the $\nutau+\nutaub$ flux.
Here, we estimate the prompt tau neutrino plus antineutrino fluxes from $D_s$ decays. We show,
for comparison, the prompt muon neutrino plus antineutrino fluxes from charmed hadrons using
the same approximations.
In the illustrative results below, we use two approximations for particles produced, starting with a
$400\,\rm{GeV}$ proton beam incident on  a series of two $58\,\rm{cm}$ targets, first molybdenum,
then tungsten (combined column depth $1,712\,\rm{g/cm^2}$).
First we use the collinear approximation
to get the energy distribution of all of the charmed particles produced in the hadronic
interaction. In a second approximation, we account for the detector geometry of a
$2\rm{m} \times 0.75\rm{m}$ detector downstream $51.5\rm{m}$ from the front of the target.
The fragmentation and decays then proceed in the collinear approximation.
For definiteness, we discuss $D_s$ production, but this applies more generally to charmed
hadrons $h_c$. Given the differential cross section for target $A$, $d\sigma(pA\to D_s X)/dE_{D}$,
the number of neutrinos produced per unit energy is approximately
\begin{equation}
\phi_{\nu_\tau+\bar{\nu}_\tau} \simeq  2\int_{E_\nu}^{E_b} dE_{\Ds}\, \frac{N_p}{\sigma_{pA}} \frac{d\sigma
_{pA\to \Ds X}}{dE_{\Ds}}(E_b,E_{\Ds}) \times \sum_i \frac{dn_i}{dE_\nu}(E_{\Ds},E_\nu)\ .
\label{eq:phinu}
\end{equation}
This approximate expression depends on $N_p$, the number of protons on target, the total $pA$
cross section, and the differential energy distribution of the $D_s$ (and its charge conjugate).
The column depth of either sample target is much larger than the interaction lengths of the incident
proton and the $D_s$, both of which are larger than the decay length of the $D_s$. Regenerated protons,
emerging with lower energies from the first interaction, have been neglected in Eq.~(\ref{eq:phinu}).
As discussed below, the factor of two accounts for the sum of neutrinos and antineutrinos, equal for
prompt decays since the energy distributions of $c$ and $\bar{c}$ are equal, as are the subsequent decay
distributions. The summation of $dn_i/dE_\nu$ is over $i=$direct or $i=$chain decay energy distributions.

Over the full energy range of $E_\nu$ that come from $D_s$ decays, Eq.~(\ref{eq:phinu}) gives
\begin{equation}
N_{\nu_\tau+\bar{\nu}_\tau}(E_\nu^{min}=0)=\int_0^{E_b}\, dE_\nu \phi_{\nu_\tau+\bar{\nu}_\tau}
= 4\times N_p\frac{\sigma_{pA\to c\bar{c} X}}{\sigma_{pA}}f_{D_s} B(D_s\to \tau\nu_\tau).
\end{equation}
Here $f_{\Ds}$ is the fragmentation fraction for $c\to \Ds$. The factor of four accounts for $c+\bar{c}$
and the two neutrinos associated with each $D_s$ produced.

The interaction length of protons on a
molybdenum target is $\lambda_{pMo}=A_{Mo}/(N_A\sigma_{pMo})=156$ g/cm$^2$~\cite{Agashe:2014kda}, giving a cross
section per nucleon of $\sigma_{pN}=\sigma_{pMo}/A_{Mo}=10.7$ mb, corresponding to an approximate $A$ dependence
of $A^{0.71}$.
The $A$ dependence of the charm production cross section is approximated by
$\sigma_{pA\to h_c X}\simeq A\sigma_{pN\to h_c X}$ for 
$A_{Mo}=95.95$ \cite{Abt:2007zg,Leitch:1994vc}. For $D_s$ production, we use NLO
QCD~\cite{Nason:1987xz,Nason:1989zy,Mangano:1991jk} as implemented in the numerical program of Cacciari et
al.~\cite{Cacciari:1998it,Cacciari:2001td}, to calculate $d\sigma_{pN\to h_c X}/{dE_{c}}$ and the Kniehl and Kramer
fragmentation functions (LO at the initial scale $\mu_0$)~\cite{Kniehl:2006mw} to convert to the hadronic energy
distribution. The normalizations of the fragmentation functions are rescaled to match the fragmentation fractions
of Ref.~\cite{Abramowicz:2013eja}. In particular, the fragmentation fraction for $c\to D_s$ is set to $f_{\Ds}=0.077$.

\begin{figure}[ht]
\centering
\subfloat{  \includegraphics[width=0.479\textwidth]{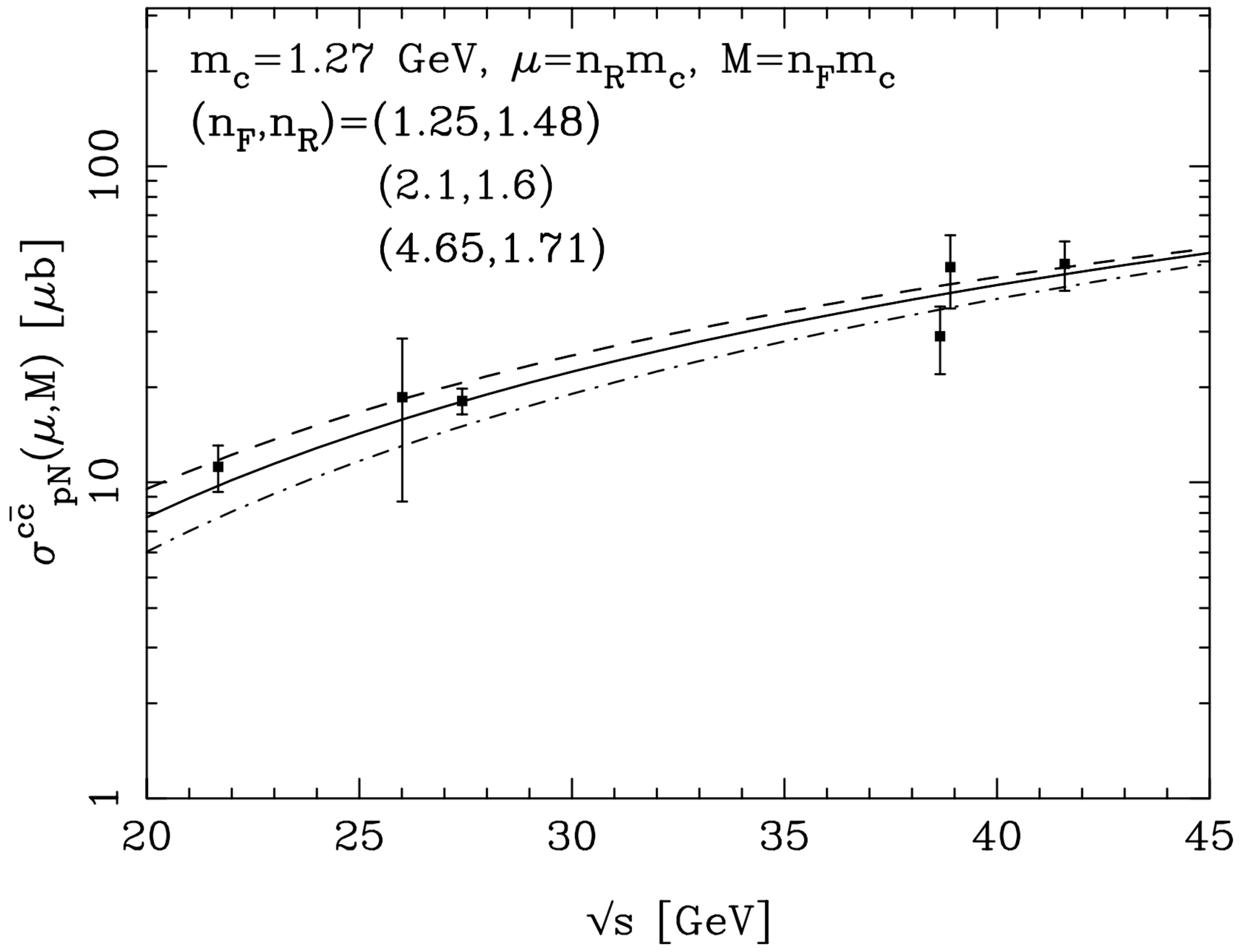}
	  \label{fig:sigccb}}
\subfloat{
	  \includegraphics[width=0.479\textwidth]{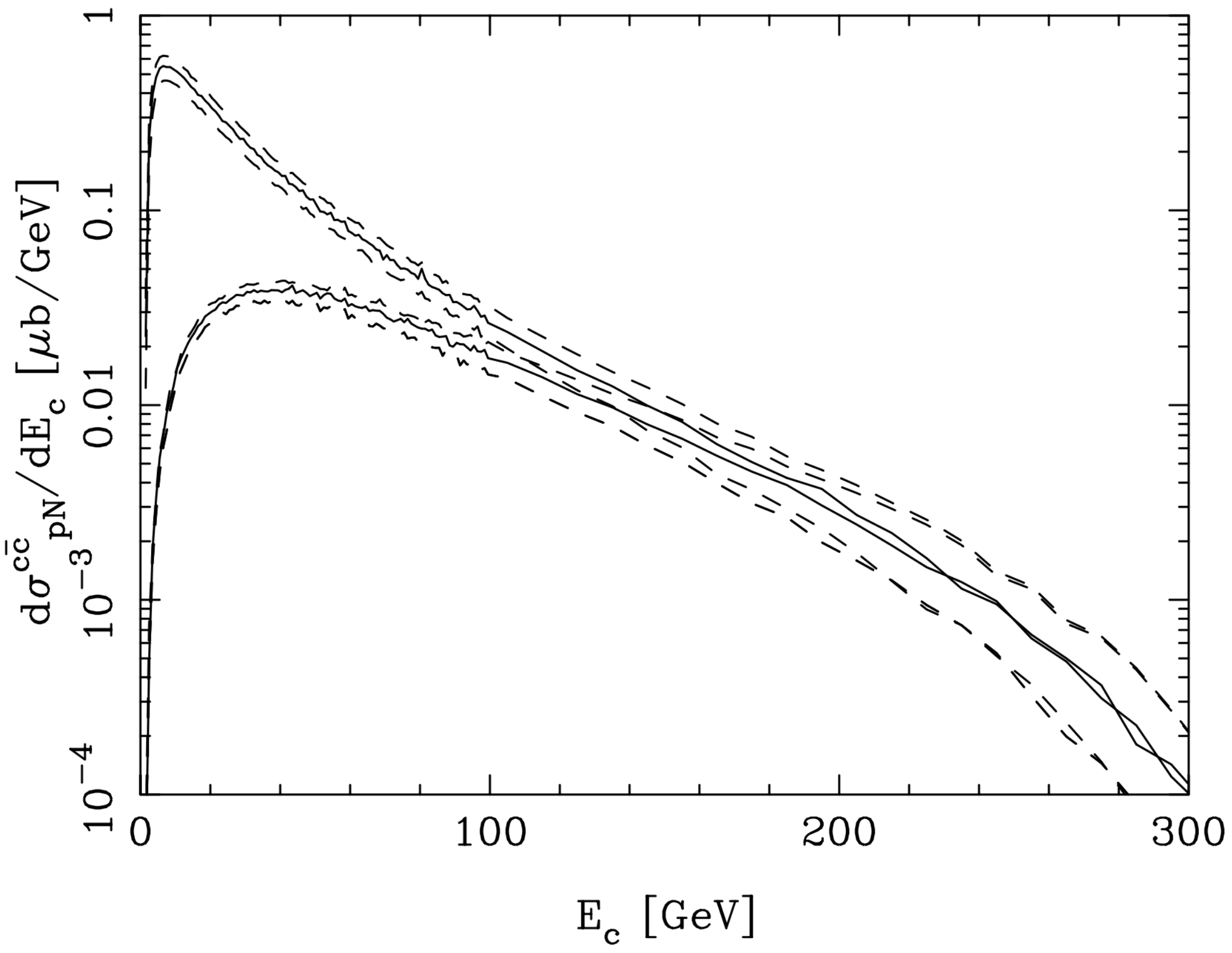}
	  \label{fig:dsdxe}  }
  \caption{
           (a) The charm production cross section
               $\sigma_{pN  \to c\bar{c} +X }$ at NLO with $m_c=1.27$ GeV using the
               CT10 parton distributions, as a function of $\sqrt{s}$. The central curve
               has  factorization
               and renormalization scales $M_F=2.10 m_c$ and $\mu_R=1.6 m_c$,
               respectively. Data points are shown from the compilation of fixed target results collected
in Ref. \cite{Lourenco:2006vw} and from HERA-B \cite{Abt:2007zg}.
(b) The differential cross section $d\sigma(pA\to c\bar{c}X)/dE_c$ for the charmed quark, as a function of
$E_c$ for $E_b=400$ GeV. The upper solid line uses the central scale choices $(M_F,\mu_R)=(2.1,1.6)m_c$
and the collinear approximation, while the lower solid line uses the central scale choices with the
detector direction requirement for the charmed quark momentum vector. The dashed lines are for
$(n_F,n_R)=(1.25,1.48)$ and (4.65,1.71) to represent a range of uncertainty in the NLO perturbative evaluation
with $m_c=1.27$ GeV.
}
  \label{fig:xsec}
\end{figure}

Inputs to the differential cross section are the charm mass, the factorization scale $M_F$, the renormalization
scale $\mu_R$ and the parton distribution functions. We take $m_c=1.27$ GeV~\cite{Agashe:2014kda} and 
$(M_F,\mu_R)=(n_F,n_R)m_c$ with values guided by Ref.~\cite{Nelson:2012bc}. In Ref.~\cite{Nelson:2012bc}, Nelson,
Vogt and Frawley vary the factorization and renormalization scales to determine the range of scales consistent
with fixed target and RHIC data on charm production. We use the scale combinations $(n_F,n_R)=(1.25,1.48)$ (dashed),
$ (2.1,1.6)$ (solid)  and $(4.65,1.71)$ (dot-dashed) that bracket the data, as shown in Fig. \ref{fig:xsec}(a),
for protons incident on an isoscalar target. The data are shown from the compilation in
Ref.~\cite{Lourenco:2006vw} and from HERA-B in Ref. \cite{Abt:2007zg}. With these choices for the
scale factors, the charm production cross section, evaluated with the CT10 NLO parton distribution 
functions \cite{Lai:2010vv} range between 15.1-20.1 $\mu$b for $\sqrt{s}=27.4$ GeV. For $E_b=400$ GeV, using the central
(solid line) factorization scale, the charm production cross section per nucleon is $\sigma_{pN\to c\bar{c}X}=18.1\ \mu$b.
We stress that the procedure outlined above cannot be considered as a genuine QCD prediction in the usual sense.
This is because the overall normalization of
the cross section is an input taken from data that provide it with an uncertainty lower then
the current theoretical uncertainty on the $c\bar{c}$ cross section for high energy hadron collisions.
Nevertheless, we consider NLO QCD corrections reliable for correlating measurements
at different energies first and, second, to correlate the overall normalization of the charm production 
cross section with the tail of the energy distribution of the charmed particles produced.
For the incident proton beam energy of $E_b=400$ GeV ($\sqrt{s}\simeq 27.4$ GeV), the differential energy distributions of
the outgoing charmed quark are shown in Fig. \ref{fig:xsec}(b). The upper curves show the energy distribution in the collinear approximation. The lower solid line is the same distribution with the requirement that the charmed quark momentum vector
intersects the detector 51.5 m downstream from the front of the target. For $E_c\simeq 10$ GeV, the lowest dashed
curve is about $\sim 20\%$ lower than the solid curve. At 100 GeV, there is a wider range, from a few percent higher
than the solid curve, to $\sim 30\%$ lower. The scale dependences of the differential charm production cross section
contribute to the tau neutrino flux uncertainties.
The $D_s$ and leptonic tau decay distributions have simple forms
\cite{Gaisser:1990vg}. For example, $D_s$ decay to $\nu_\tau$ has
\begin{equation}
\frac{dn_{\rm direct}}{dE_\nu}(E_{D_s},E_\nu) =\frac{1}{\Gamma(E_{D_s})}\frac{d\Gamma(E_{D_s},E_\nu)}{dE_\nu} =\frac{1}{E_{D_s}}F_{D_s}^{\nu}\Biggl(\frac{E_\nu}{E_{D_s}}\Biggr)\ ,
\end{equation}
where 
\begin{equation}
F_{D_s}^{\nu}(x)=B(D_s\to \nu_\tau \tau)/(1-R_\tau)\times \theta(1-R_\tau - x)
\end{equation}
for $R_\tau = m_\tau^2/m_{\Ds}^2$. The energy distribution $F_{D_s}^\nu$ is the same for neutrinos and antineutrinos.
For $\tau$ decays to neutrinos, the decay distribution of $\tau^-$ at rest differs from the decay distribution of $\tau^+$. 
Because of the opposite helicities of $\nu_\tau$ and $\bar{\nu}_\tau$ in the $D_s$ decay, the polarizations of $\tau^-$ and
$\tau^+$ are also opposite. The net effect is that in the laboratory frame, the neutrino energy distribution in
the chain decay $D_s^+\to \tau^+\to \bar{\nu}_\tau$ is the same as for $D_s^-\to \tau^-\to \nu_\tau$ in the absence of depolarization
\cite{Gaisser:1990vg,Barr:1989ru}.
For the tau decay, in the energies of interest here,
electromagnetic energy loss is negligible (see, for example,
Ref. \cite{Dutta:2000hh,Bugaev:2003sw}), so the polarization of the produced tau is retained
in the evaluation of the tau decays. Therefore, the $\nu_\tau$ and the $\bar{\nu}_\tau$ fluxes from both the direct $D_s\to \nu_\tau$
and the chain decays are identical.
A summary of the function forms for $\Ds\to \tau\to \nu_\tau$
can be found in Ref.~\cite{Pasquali:1998xf}. We include the leptonic decays of the tau and the two body decays of $\tau\to (\pi,\rho,a_1)\nu)\tau$,
overall accounting for $\sim 87\%$ of the tau decays. The parameters for the charm fragmentation and the PDG central values for the relevant branching ratios of the tau lepton are reported in Appendix~A.

\begin{figure}[ht]
\centering
\includegraphics[width=0.75\textwidth]{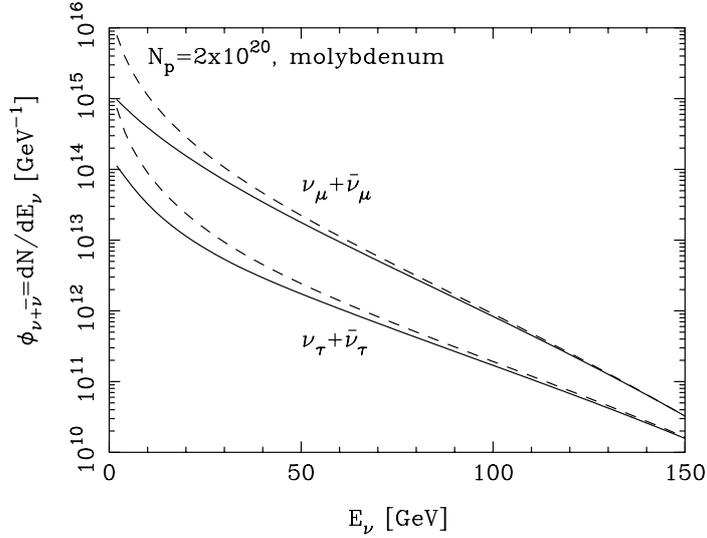}
  \caption{
           The number of neutrinos plus antineutrinos per unit energy, $\phi_{\nu+\bar{\nu}}$, as a function of energy for
$\nu+\bar{\nu}=\nu_\tau+\bar{\nu}_\tau$ (lower curves) and $\nu+\bar{\nu}=\nu_\mu+\bar{\nu}_\mu$ (upper
curves) for $N_p=2\times 10^{20}$ from charmed hadron decays, 
with $M_F=2.10 m_c$ and $\mu_R=1.6 m_c$ for a molybdenum target. The  contribution from $D_s\to\tau\to \nu_\tau$
dominates the tau neutrino flux for most of the energy range. The dashed lines use the collinear approximation, while the solid
lines account for the detector
direction requirement for the charmed quark momentum vector.
}
  \label{fig:phinu}
\end{figure}

Taking $N_p=2 \times 10^{20}$ protons, our results for the number of neutrinos per unit energy are shown in Fig.~\ref{fig:phinu}.
We use Eq.~(\ref{eq:phinu}) to find
$\phi_\nu$ for $\nu=\nu_\tau+\bar{\nu}_\tau$ and $\nu=\nu_\mu+\bar{\nu}_\mu$ from
$h_c=D^0,\bar{D}^0, \ D^\pm,\ D_s^\pm,\ \Lambda_c^\pm$, were we take the relativistic limit.
Following Ref. \cite{Bugaev:1998bi}, we approximate the decays of the charmed mesons to $\nu_\mu$ as approximate three-body decays.
We take the $\nu_\mu$ energy distribution equal to the muon energy distribution parameterized in Ref. \cite{Bugaev:1998bi}. 
For the $\Lambda_c$, we also approximate the decay by a three-body distribution. 
 As indicated above, we use the collinear approximation, and 
with Eq.~(\ref{eq:phinu}), we account only for the first $pA$ interaction to produce $c\bar{c}$. 

\begin{figure}[ht]
\centering
\includegraphics[width=0.75\textwidth]{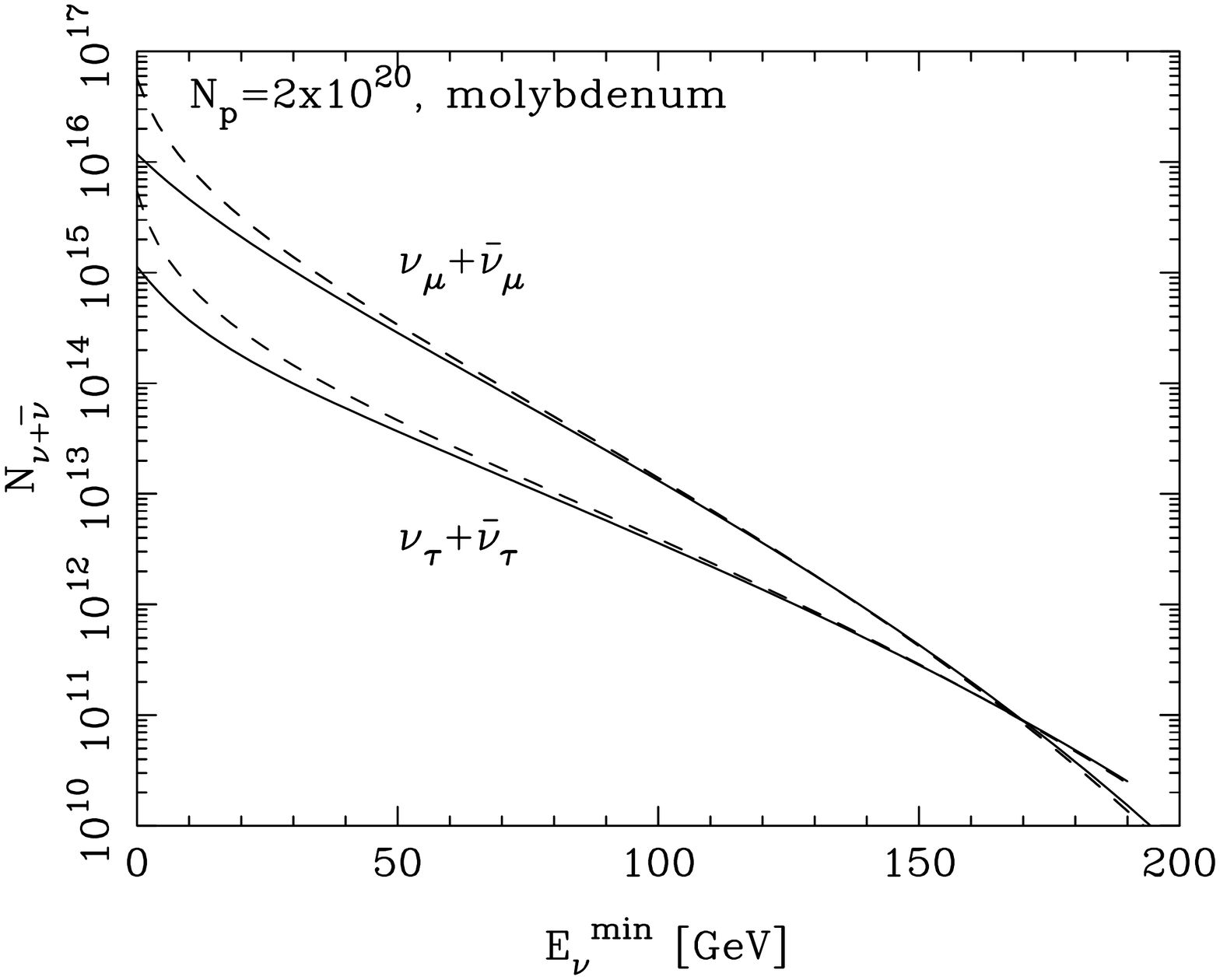}
  \caption{
           The integral of $\phi_{\nu+\bar{\nu}}$ as a function of $E_\nu^{min}$ of the prompt tau and muon neutrino plus
anti-neutrino fluxes for $N_p=2\times 10^{20}$ on a molybdenum target. The upper curves are  $\nu+\bar{\nu}=\nu_\mu+\bar{\nu}_\mu$ and lower curves are $\nu+\bar{\nu}=\nu_\tau+\bar{\nu}_\tau$,
as in Fig. \ref{fig:phinu}.
}
  \label{fig:phinu-int}
\end{figure}
The $\Ds\to\tau\to \nu_\tau$ decay chain dominates the direct $D_s\to \nu_\tau$ contribution
above a few GeV. With $D_s$ decays, the tau neutrino flux equals the tau antineutrino flux.
The charmed hadron contributions to the muon neutrino flux are similar in shape. The $D^\pm$ and the $D^0+\bar{D}^0$ contributions to
$\nu_\mu+\bar{\nu}_\mu$ are nearly equal.

Fig. \ref{fig:phinu-int} shows the integral of $\phi_\nu$ from $E_\nu=E_\nu^{min}\to 400$ GeV, $N_\nu(E_\nu^{min})$, 
for $\nu=\nu_\tau+\bar{\nu}_\tau$ (lower curves)  and $\nu=\nu_\mu+\bar{\nu}_\mu$ (upper
curves) for $(n_F,n_R)=(2.1,1.6)$. Overall, in the collinear approximation, these parameters give $5.7\times 10^{15}$
$\nu_\tau+\bar{\nu}_\tau$ for $N_p= 2 \times10^{20}$ protons for all neutrino energies, and $1.8 \times 10^{15}$ for
$E_\nu\geq 4$ GeV.

\begin{figure}[ht]
\centering
\includegraphics[width=0.75\textwidth]{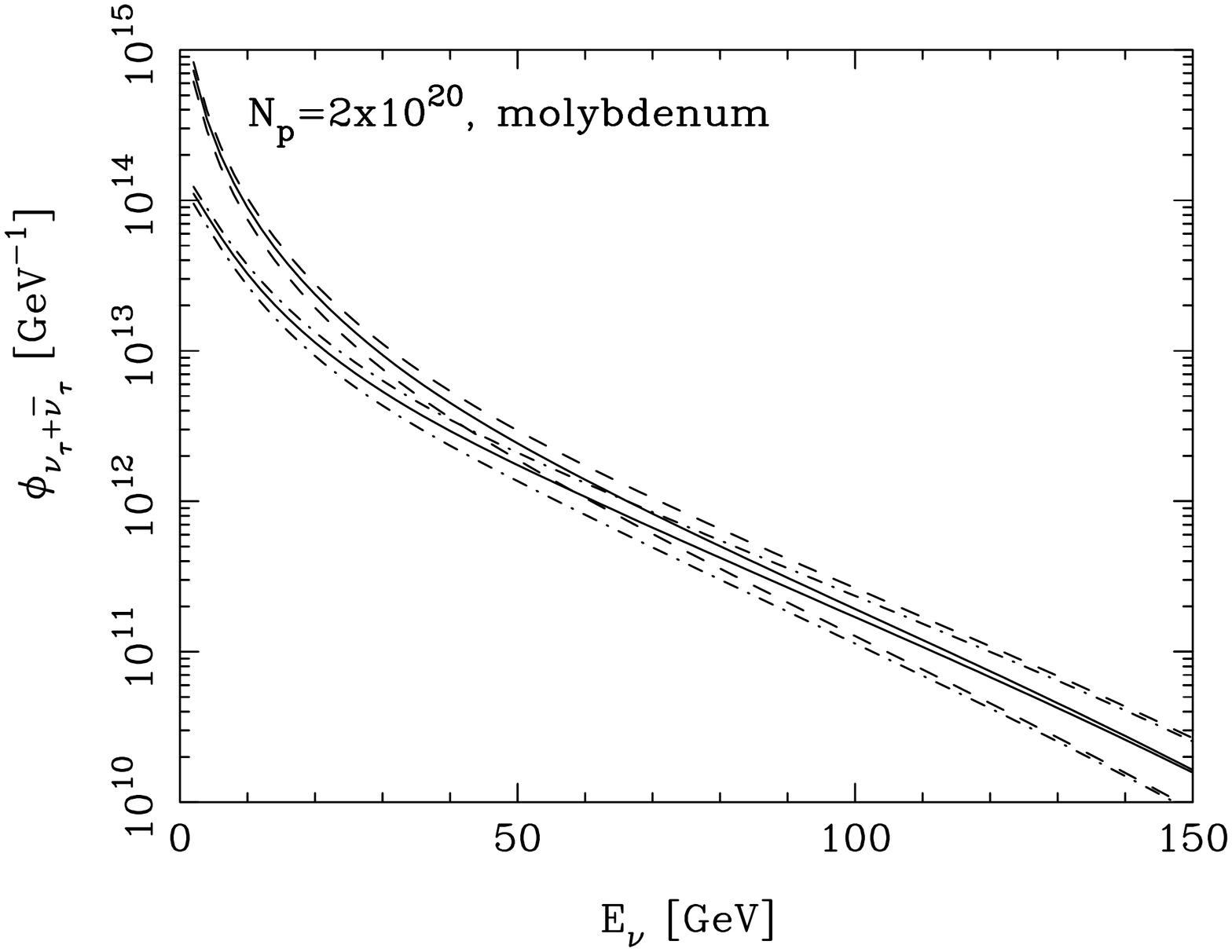}
  \caption{
           The flux $\phi_{\nu+\bar{\nu}}$ as a function of $E_\nu$ for $\nu+\bar{\nu}=\nu_\tau+\bar{\nu}_\tau$,
for the central scale choices,  $M_F=2.10 m_c$ and $\mu_R=1.6 m_c$ (solid line)
for a molybdenum target and $N_p=2\times 10^{20}$ protons. The upper curves
used the collinear approximation with other scale choices shown with the dashed lines, while the lower curves
account for the detector
direction requirement for the charmed quark momentum vector and other scale choices (dot-dashed lines).
}
  \label{fig:phinu-band-150}
\end{figure}

The relative number of neutrinos of the two flavor shifts as a function of energy, and there is a fairly
strong dependence on factorization and renormalization scales. As an estimate of the theoretical error in the flux
predictions, the quantity $\phi_\nu$ as a function of $E_\nu$ for $\nu=\nu_\tau+\bar{\nu}_\tau$,
for the central scale choices $M_F=2.10 m_c$ and the scale variation given above are shown
in Fig.~\ref{fig:phinu-band-150}.
One expects additional uncertainties associated with the choice of charmed quark mass and fragmentation functions,
the latter especially at high neutrino energies where the flux is low. The additional uncertainties from the fragmentation fraction ($< 10\%$) and branching fraction for $D_s\to \tau \nu_\tau$ ($<5\%$) are however small compared to the scale dependence.
The ratio of $\nu_\tau/\nu_\mu$
is much less sensitive to the scale dependence than the absolute flux. For the integrated flux, the ratio differs by just a few percent
for $E_\nu^{min}=10$ GeV and by about 20\% for $E_\nu^{min}=50$ GeV. Therefore, the relative contributions
by neutrino flavor of the prompt fluxes should help pin down the tau neutrino flux.

\subsection{Expected sensitivity}

The theoretical uncertainty associated to the cross sections plotted in Fig.~\ref{fig:sigf45}
are mainly related to the scale choices, to the Parton Distribution Functions (PDF's) uncertainty
and the modeling of the target mass corrections. Furthermore, the disentanglement of the nuclear
effects will also be a source of
uncertainty in the extraction of the cross section from the measurements; we estimate that these
sources will give all together an uncertainty of $\sim 10\%$ or less for a large interval of
relevant neutrino energies. The largest source of uncertainty will be then given by the one on the knowledge
of the fluxes $\sim 20\%$.
To estimate the inclusive number of events that might be collected with the neutrino detector,
we take the convolution of the flux entering the surface of the detector with the cross sections above.

Using the tau neutrino plus antineutrino fluxes of Fig. \ref{fig:phinu-band-150} including the detector
geometry restriction, the
number of events is
\begin{equation}
N_{evts} = \int_{E_{min}}^{E_{max}} \, dE_\nu 
\Bigl( \phi_\nu \frac{L_{Pb}}{\lambda_{\nu Pb}}+\phi_{\bar{\nu}} \frac{L_{Pb}}{\lambda_{\bar{\nu} Pb}}\Bigr)
\label{eq:nevt-tau}
\end{equation}
where $\phi_\nu=\phi_{\bar{\nu}}=\phi_{\nu+\bar{\nu}}/2$ for tau neutrinos. 
For $M=1$ ton$=10^3$ kg, and $\rho_{Pb}=11.35$ g/cm$^3$, the depth of the detector is $d=5.87$ cm for a cross sectional
detector area of 2.0 m by 0.75 m.
The column depth of the lead is $L_{Pb}=d\rho_{Pb}=66.7$ g/cm$^2$.
The interaction length of neutrinos in lead is $\lambda_{\nu Pb}=A/(N_A \sigma_{\nu A})$ for $A=207$ and $Z=82$, with 
$\sigma_{\nu A}/ A$ given in Eq.~(\ref{sigma}).

\begin{figure}[ht]
\centering
\includegraphics[width=0.75\textwidth]{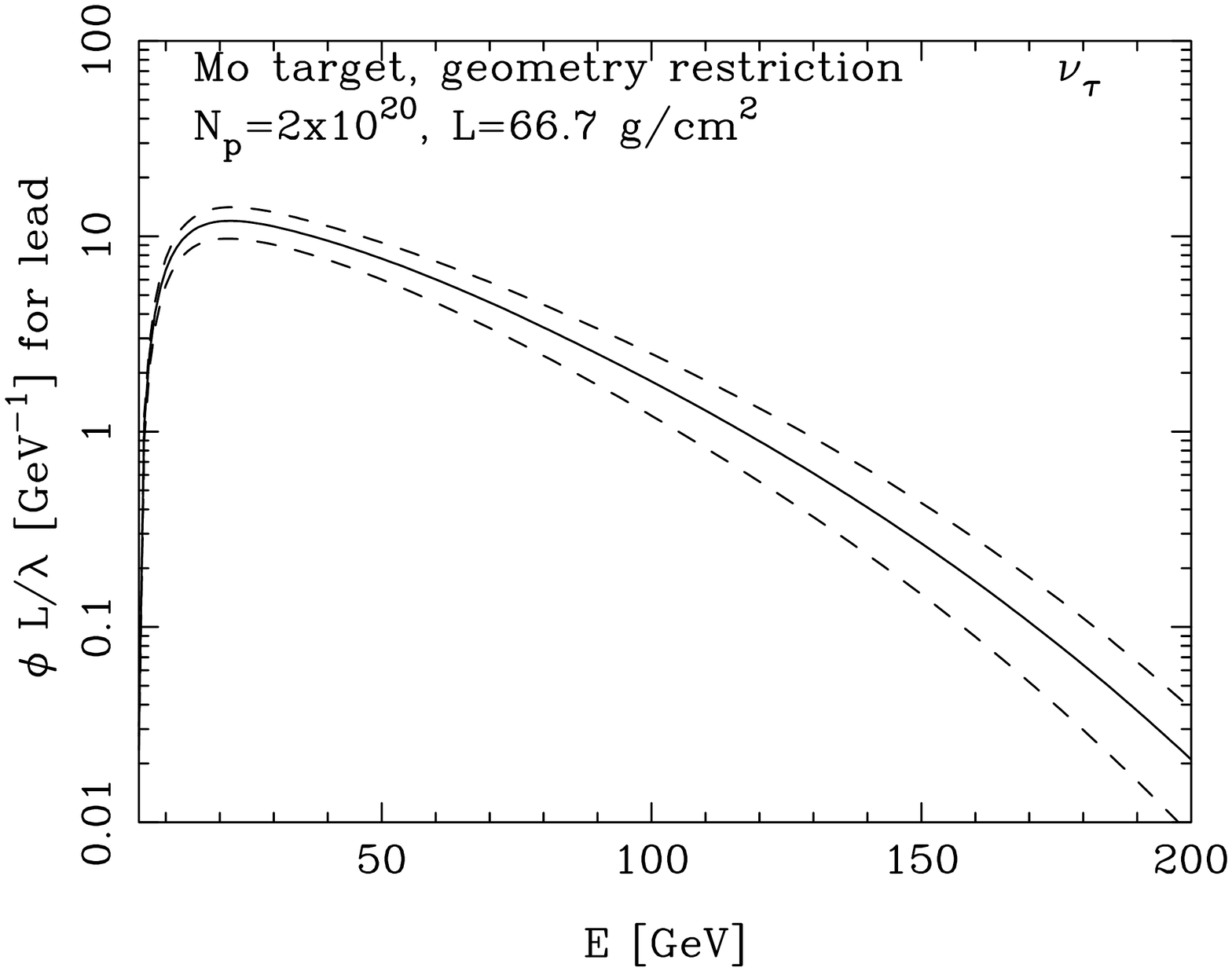}
  \caption{
           The quantity $\phi_{\nu}\times L/\lambda$ as a function of $E_\nu$ for
for the central scale choices,  $M_F=2.10 m_c$ and $\mu_R=1.6 m_c$ (solid line). The scale choices
 $(n_F,n_R)=(1.25,1.48)$ and (4.65,1.71) are shown with the dashed lines. The flux is evaluated 
for a molybdenum target and $N_p=2\times 10^{20}$ protons with the detector geometry restriction on the charmed
quark momentum. The lead detector column depth is taken to be $L=66.7$ g/cm$^2$. }
  \label{fig:tauflcc-v2}
\end{figure}

\begin{figure}[h]
\centering
\includegraphics[width=0.75\textwidth]{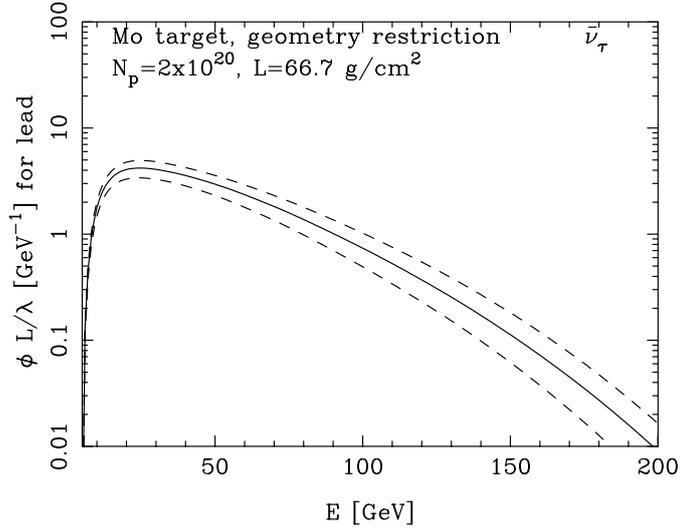}
  \caption{ As in Fig. \ref{fig:tauflcc-v2}, for anti-neutrinos.
           }
  \label{fig:atauflcc-v2}
\end{figure}

\begin{figure}[h]
\centering
\includegraphics[width=0.75\textwidth]{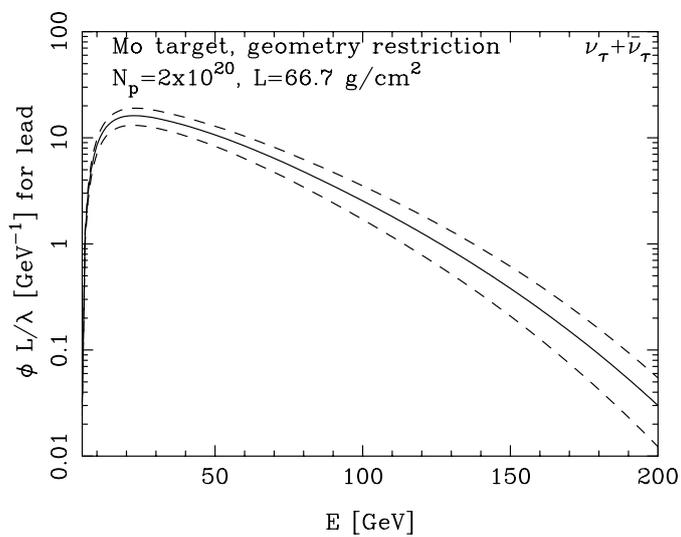}
  \caption{
           The quantity $\phi_{\nu}\times L/\lambda_\nu+\phi_{\bar{\nu}}\times  L/\lambda_{\bar{\nu}}$ as a function of $E_\nu$ for
for the central scale choices,  $M_F=2.10 m_c$ and $\mu_R=1.6 m_c$ (solid line). The scale choices
 $(n_F,n_R)=(1.25,1.48)$ and (4.65,1.71) are shown with the dashed lines. The flux is evaluated 
for a molybdenum target and $N_p=2\times 10^{20}$ protons with the detector geometry restriction on the charmed
quark momentum. The lead detector column depth is taken to be $L=66.7$ g/cm$^2$. }
  \label{fig:tauatauflcc-v2}
\end{figure}

\begin{figure}[h]
\centering
\includegraphics[width=0.75\textwidth]{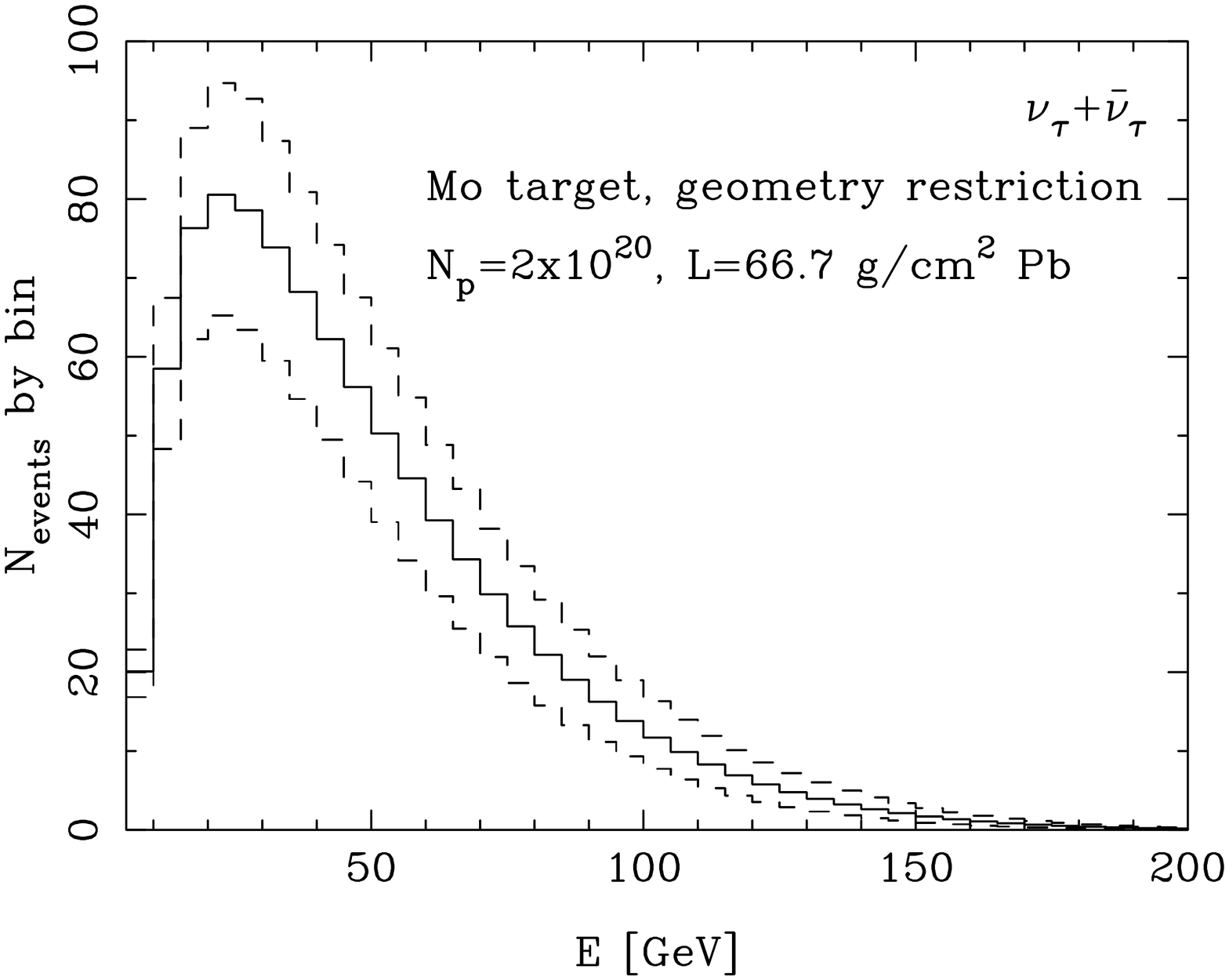}
  \caption{
           The number of events in each 5 GeV energy bin for $E_\nu=5-200$ GeV for
for the central scale choices,  $M_F=2.10 m_c$ and $\mu_R=1.6 m_c$. The scale choices
 $(n_F,n_R)=(1.25,1.48)$ and (4.65,1.71) are shown with the dashed lines.
The flux is evaluated 
for a molybdenum target and $N_p=2\times 10^{20}$ protons with the detector geometry restriction on the charmed
quark momentum. The lead detector column depth is taken to be $L=66.7$ g/cm$^2$. }
  \label{fig:bybin}
\end{figure}

Fig. \ref{fig:tauflcc-v2} and \ref{fig:atauflcc-v2} show the separate neutrino and antineutrino contributions
to the integrand in Eq.~(\ref{eq:nevt-tau}), and Fig.~\ref{fig:tauatauflcc-v2} shows the sum of the two terms.
Fig.~\ref{fig:bybin} shows the number of events for the central scale choice (solid lines in Figs.~\ref{fig:tauflcc-v2}
and~\ref{fig:atauflcc-v2}) integrated in 5 GeV bins from $E=5$ GeV to $E=200$ GeV. The sum of the number of events
is 936 events for the central scale choice (685 tau neutrino interactions and 252 tau antineutrino interactions).
For $(n_F,n_R)=(4.65,1.71)$, there are 723 events (529 tau neutrino and 194 tau antineutrino),
while for $(n_F,n_R)=(1.25,1.48)$ , we find 1151 events (840 tau neutrino and 311 tau antineutrino). The number of
neutrino and antineutrino interactions corresponding to Figs.~\ref{fig:tauflcc-v2} and~\ref{fig:atauflcc-v2},
integrated over energy in 5 GeV intervals, is shown in Fig.~\ref{fig:bybin}. The sum
of the entries in each bin, from 5 GeV through the 195-200 GeV bin, gives the total number of events quoted above.

We note that for a good range of neutrino energies among 10
and 50 GeV, where a significant fraction of CC tau neutrino DIS interactions is expected, the theoretical uncertainty
on the flux, that is the largest one, can be estimated to be uniform of the order of 20\%.
We assume that the current knowledge of the structure functions $F_1$, $F_2$ and $F_3$ from previous precise experimental measurements in
$\nu_\mu$ and $\bar{\nu}_\mu$ CC has a much smaller error, parametrized in terms of scale and PDF uncertainties,
to be lower then 10\%.
By subtracting the contribution of $F_1$, $F_2$ and $F_3$ from the expected number of events observed, that is about
twice the expected contribution of $F_4$ and $F_5$, we deduce that in SHiP there will be the opportunity to asses
the cumulative impact of the structure functions $F_4$ and $F_5$ differentially in the interval of energies among 10 and 50 GeV.
\begin{figure}[h]
\centering
\includegraphics[width=0.46\textwidth]{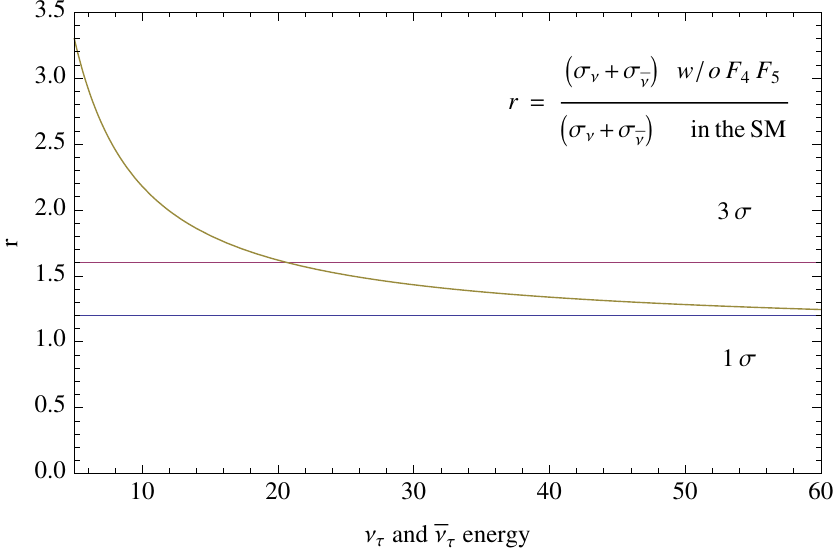}
\includegraphics[width=0.45\textwidth]{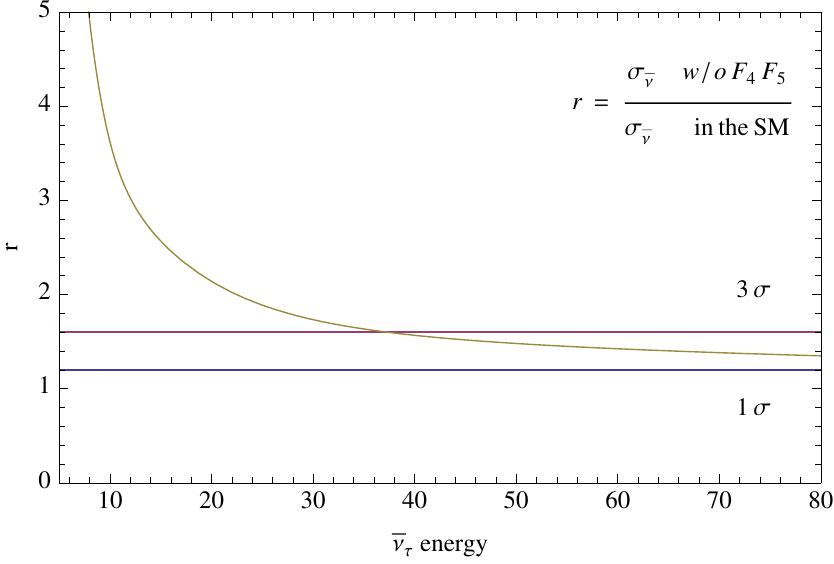}
  \caption{
   Left panel: ratio of the averaged tau neutrino cross sections off lead as a function of the incident neutrino energy.
   Right panel: the same ratio for the anti-neutrino cross section. }
  \label{fig:sensit}
\end{figure}
In the left panel of Fig.~\ref{fig:sensit} we show the ratio of the averaged $\nu_\tau$ and $\nutaub$ cross sections
on lead as a function of the incident neutrino energy, while in the right panel the same ratio is plotted for
$\nutaub$ cross section only. Systematic uncertainties on these ratios are expected
to cancel and, given the relatively large number of events expected, we assume that the theoretical error
is the dominant source of uncertainty.
Under these assumptions in the two plots we also included the lines for one and two standard deviations
from the hypothesis of no contribution from $F_4$ and $F_5$ structure functions.
However, we point out that, even including radiative corrections,
$F_4$ is not expected to give an important contribution for the cross section in the Standard Model,
indeed it contributes $-5\%$ at 5 GeV, but only $1\%$ at 10 GeV and further decreasing with energy.
As a final remark we stress here that the total number of events quoted above is affected by
the geometrical acceptance and the collinear approximation used.
Including a more detailed description of the $D_s \to \tau \to \nutau$ chain would lower the expected
number of events estimated here. On the other hand the collinear approximation that we have adopted here
does not spoil the relative theoretical uncertainties quoted.

\section{Deep inelastic muon and electron neutrino scattering}
\label{sec:DIS} 

Historically, the formulation of QCD as the gauge theory of the strong interactions~\cite{Gross:1973ju,Gross:1974cs}
has been closely related to the early deep-inelastic scattering (DIS) experiments 
and the observed scaling violations in inclusive structure functions, which parametrize the hadronic interaction. 
Over time experiments with lepton- and neutrino scattering off fixed targets at CERN, FNAL and SLAC 
as well as collisions of electrons and positrons on protons at DESY's HERA collider 
have provided unique information on the nucleon structure. 
The available experimental data is of very high precision. 
At the same time, it spans a large range in the kinematics of the Bjorken-$x$ variable and 
of the measured values of momentum transfer $Q^2$ between the scattered 
lepton or neutrino and the nucleon.

The current picture of the proton structure from DIS data is complemented by
measurements at the Large Hadron Collider (LHC) for a number of hadro-production processes. 
These include, for instance, cross sections for the hadro-production of jets or 
differential distributions for the production of $W^\pm$- and $Z$-gauge
bosons, possibly in association with charm quarks.
These data can contribute to our understanding of the flavor composition of
the nucleon, it can constrain the gluon distribution at large values of
Bjorken-$x$ and it is sensitive to the value of the strong coupling constant $\alpha_s(M_Z)$.

Dramatic further improvements in different ranges of kinematics 
can be expected from a number of planned future colliders.
The proposal for a Large Hadron Electron Collider (LHeC) to collide an
electron beam of 60 GeV, up to possibly 140 GeV, 
with the intense hadron beams of the LHC, would allow for coverage 
of a kinematic range extended by a factor of twenty in $Q^2$, and in the inverse
Bjorken-$x$, if compared to HERA,
and with great physics potential~\cite{AbelleiraFernandez:2012cc} regarding high precision DIS measurements.

The design of the planned electron-ion collider EIC aims instead at
center-of-mass energies in the range from 50 to 140 GeV, but allows 
to collide electrons and positrons with polarized protons and with light to heavy nuclei 
at high luminosity.
Major research themes of the EIC are the spin and flavor structure of the
proton as well as three-dimensional images of the nucleon
structure~\cite{Boer:2011fh}.

The planned SHiP experiment~\cite{Bonivento:2013jag} is distinct from these
activities to the extend that it realizes intense neutrino beams.
Beyond establishing for first time the direct observation of tau anti-neutrinos, 
the resulting neutrino-nucleon DIS interactions from intense electron and muon
neutrino beams will allow to study the strangeness of the nucleon.
Besides the tau neutrino events discussed in the previous subsection,
the SHiP experiment is expected to collect ${\cal O}$(2M) charged current
interactions from muon neutrinos and ${\cal O}$(1M) from electron neutrinos,
the latter with an average energy of the interacting neutrinos around 30 GeV. 
Previous studies of the physics potential with neutrino beams have focused on  
experiments at the front-end of a muon storage ring~\cite{Mangano:2001mj} 
and date back to the pre-LHC era, so that an update is mandatory.
In the following we will review the status of perturbative QCD computations and
the most important DIS measurements from muon and electron neutrinos
that can be performed in the SHiP experiment.

\subsection{Status of perturbative QCD calculations}

The key observables for tests of perturbative QCD are either inclusive structure functions or,
in the semi-inclusive case, differential cross sections, which describe the hard hadronic interaction
in the QCD improved parton model.
The particle data group (PDG)~\cite{Agashe:2014kda} 
provides a summary of DIS, including the definitions of kinematic variables.

\begin{figure}[h!]
\begin{center}
\includegraphics[width=0.75\textwidth]{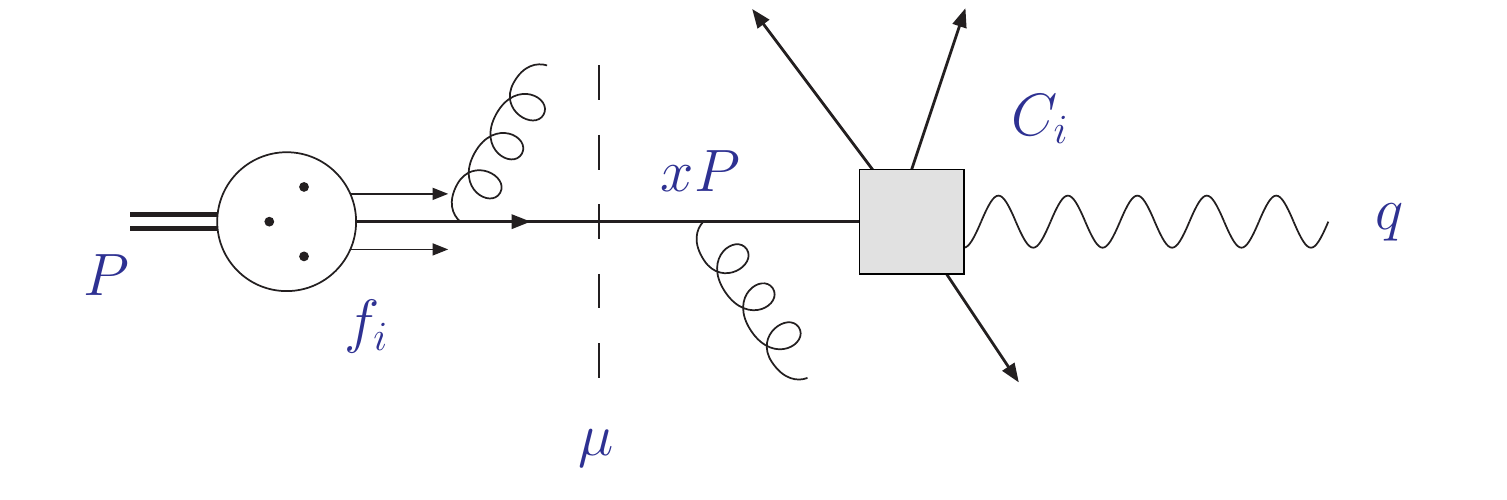}
\end{center}
\vspace*{-3mm}
\caption{\label{fig:ep-rest} 
  {QCD factorization of the cross section for the scattering 
    of a deeply virtual gauge-boson with (space-like) momentum $q$ 
    ($-q^2 = Q^2 > 0$) off a proton with momentum $P$ 
    in their center-of-mass frame, see Eq.~(\ref{eq:F2factorisation}). 
}}
\end{figure}

Precision predictions in perturbative QCD rest on factorization. 
Sensitivity to the dynamics at scales of the proton mass 
is separated from hard high-energy scattering at a large scale of the momentum transfer $Q^2$. 
In Fig.~\ref{fig:ep-rest} this is illustrated for lepton-nucleon DIS 
in the one-boson exchange approximation.
With the help of factorization at a scale $\mu$ the structure functions 
$F_k$ ($k=2,3,L$) can be expressed as convolutions 
of parton distributions (PDFs) $f_{i}$ ($i=q,{\bar q},g$) in the nucleon 
and short-distance Wilson coefficient functions $C_{k,i}$,
\begin{eqnarray}
  \label{eq:F2factorisation}
  F_k(x,Q^2) &=& \sum\limits_{i=q,{\bar q},g} \, 
  \int_x^1 dz \, f_{i}\left(\frac{x}{z},\mu^2\right)\, 
  C_{k,i}\left(z,Q^2,\alpha_s(\mu),\mu^2 \right) 
  \, , 
\end{eqnarray}
in so-called leading twist approximation up to power corrections of ${\cal O}(1/Q^2)$. 
Such higher twist contributions are relevant in the analysis of DIS data at
low $Q^2$, see, e.g., \cite{Alekhin:2012ig}.

QCD factorization has also been established for differential distributions in DIS, 
e.g., for semi-inclusive DIS where the cross section $d^2 \sigma/ dx dQ^2$ 
is subject to a decomposition similar to Eq.~(\ref{eq:F2factorisation}).
In the differential case, the hard parton scattering cross sections 
are augmented by an additional prescription for any exclusive final state parton, 
e.g., a jet algorithm or fragmentation functions, in order to ensure infrared finiteness.

The coefficient functions $C_{k,i}$ in Eq.~(\ref{eq:F2factorisation}) can be calculated perturbatively 
in powers of the strong coupling constant $\alpha_s$, 
\begin{eqnarray}
\label{eq:C-exp-alphas}
  C &=& 
  C^{(0)} + \alpha_s\, C^{(1)} + \alpha_s^2\, C^{(2)} + \alpha_s^3\, C^{(3)} + \, \ldots
  \, ,
\end{eqnarray}
where the expansion coefficients $C^{(0)}$ are denoted as the leading order (LO), 
$C^{(1)}$ as the next-to-leading order (NLO) and so on. 
The PDFs $f_i$ are non-perturbative objects 
describing the nucleon momentum fraction $x=Q^2/(2P \cdot q)$ (see Fig.~\ref{fig:ep-rest}) carried 
by the quark or gluon and have to be determined in global fits to experimental data.
Within perturbation theory, PDFs at different scales are related, though, 
by the well-known evolution equations,
\begin{equation}
\label{eq:evolution}
\frac{d}{d \ln \mu^2}\,
  \left( \begin{array}{c} f_{q_i}(x,\mu^2)  \\ f_{g}(x,\mu^2)  \end{array} \right) 
\: =\:
\sum_{j}\, \int\limits_x^1\, \frac{dz}{z}\, 
\left( \begin{array}{cc} P_{q_iq_j}(z) & P_{q_ig}(z) \\ P_{gq_j}(z) & P_{gg}(z) \end{array} \right)\,
  \left( \begin{array}{c} f_{q_j}(x/z,\mu^2)  \\ f_{g}(x/z,\mu^2) \end{array} \right)
\, ,
\end{equation}
where the so-called splitting functions $P_{ij}$ enter 
to account for the different possible parton splittings in the collinear limit.
The $P_{ij}$ are universal quantities in QCD and, like the $C_{k,i}$, 
can be computed in a power series in $\alpha_s$, 
\begin{eqnarray}
\label{eq:P-exp-alphas}
  P & = &
  \alpha_s\, P^{(0)} + \alpha_s^2\, P^{(1)} + \alpha_s^3\, P^{(2)} + \, \ldots
  \, .
\end{eqnarray}
The QCD predictions for DIS observables have reached an unprecedented level of precision. 
The Wilson coefficients $C_{k,i}$ and the splitting functions $P_{ij}$ are
known to high enough orders in perturbation theory, so that the effect of
radiative corrections on structure functions is well under control.
The splitting functions $P_{ij}$ in Eq.~(\ref{eq:P-exp-alphas}) 
are known to next-to-next-to-leading order (NNLO)~\cite{Moch:2004pa,Vogt:2004mw} 
and, likewise, the coefficient functions $C_{k,i}$~\cite{vanNeerven:1991nn,Zijlstra:1992qd,Zijlstra:1992kj,Moch:1999eb}
in Eq.~(\ref{eq:C-exp-alphas}). 
Moreover, for photon and charged current $W^\pm$-boson exchange, the hard corrections at order 
next-to-next-to-next-to-leading order (N$^3$LO) are also available~\cite{Vermaseren:2005qc,Moch:2007rq,Moch:2008fj}.

For the production of heavy quarks in DIS, like charm, mass effects need to included.
The theoretical description of DIS heavy-quark production in the kinematical
range of SHiP can be based on the fixed-flavor-number (FFN) factorization scheme with three light quarks in the 
initial state and the heavy-quarks appearing in the final state only. 
The respective Wilson coefficients for the heavy-quark electro-production 
are known to NLO completely~\cite{Laenen:1992zk,Laenen:1992xs} 
and approximately to~\cite{Kawamura:2012cr}, 
the latter based on the massive operator matrix elements at three loop order~\cite{Bierenbaum:2009mv,Ablinger:2014nga}.
This ensures the exact description of all logarithmically enhanced terms in
the heavy-quark mass $m$ up to $\alpha_s^3 \ln^k(Q^2/m^2)$ with $k=1, \dots, 3$.
Similarly, the heavy-quark coefficient functions for charm quark production in
charged current DIS are known exactly at NLO~\cite{Gottschalk:1980rv,Gluck:1996ve,Blumlein:2011zu} 
and at asymptotic values of $Q^2 \gg m^2$ at NNLO~\cite{Buza:1997mg,Blumlein:2014fqa,Alekhin:2014sya}.
The exact quark mass dependence for charm quark production in charged current
DIS to NNLO is not available at present, though, 
it could, in principle, be extracted from Ref.~\cite{Brucherseifer:2014ama}.
\begin{figure}[h!]
\centering
    \includegraphics[width=12.0cm]{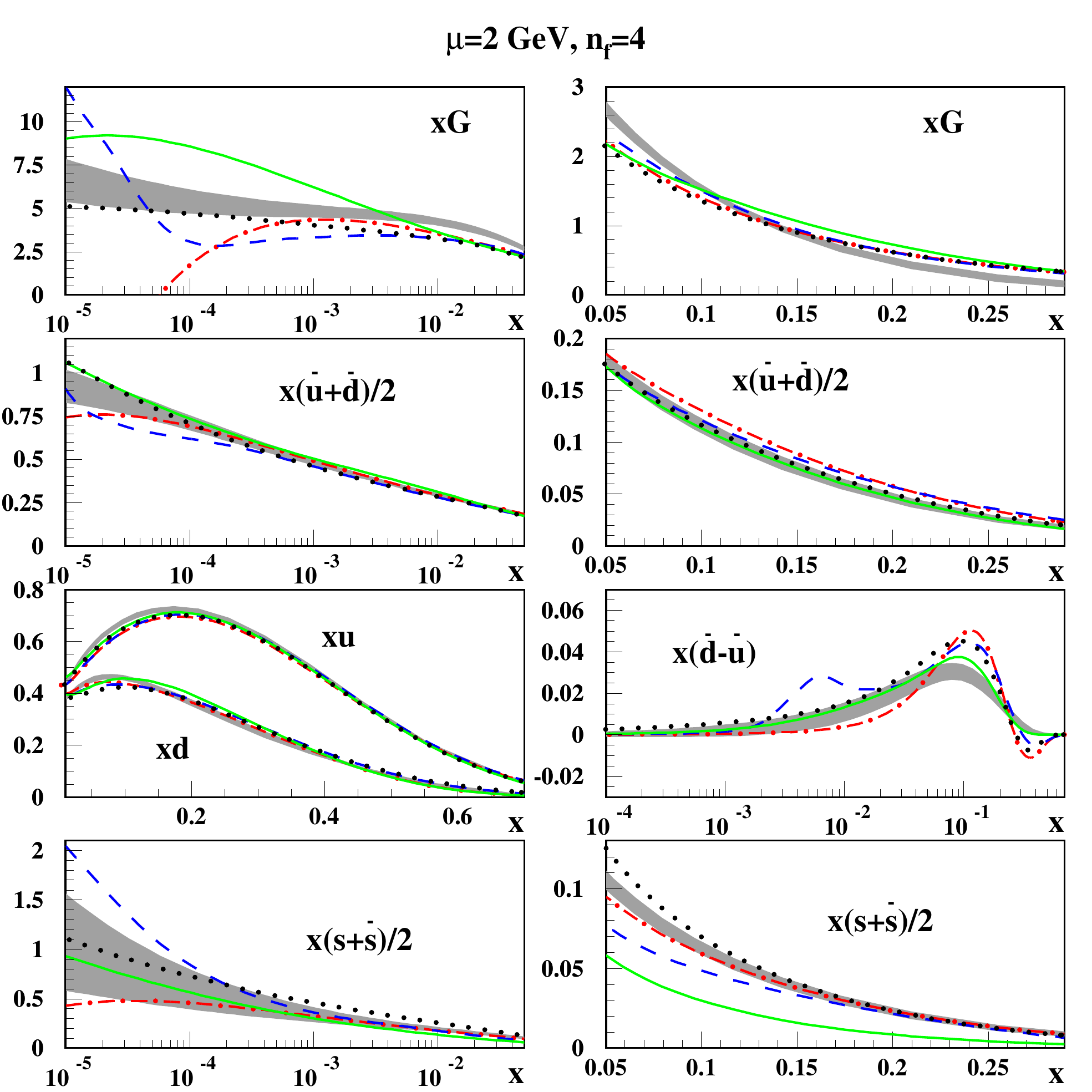}
  \caption{\small
    \label{fig:pdfcomp}
    The 1$\sigma$ band obtained for the 4-flavor NNLO ABM12 PDFs~\cite{Alekhin:2013nda}
    at the scale of $\mu=2~{\rm GeV}$ versus $x$ (shaded area) 
    compared with the central fits obtained by 
    JR09~\cite{JimenezDelgado:2008hf} (solid lines), 
    MSTW~\cite{Martin:2009iq} (dashed dots),  
    NNPDF (v2.3)~\cite{Ball:2012cx} (dashes),
    and CT10~\cite{Gao:2013xoa} (dots).
    Plot taken from~\cite{Alekhin:2013nda}.
  }
\end{figure}
Available QCD theory predictions and experimental data for inclusive DIS allow 
to deduce important information about PDFs or the value of the strong coupling constant $\alpha_s(M_Z)$
as the non-perturbative input to QCD precision predictions.
This is, of course, a situation that, generally, needs to be addressed beyond DIS, since  
experimental data from the (anti-)proton-proton colliders Tevatron at FNAL and the LHC 
provide invaluable further constraints, e..g., on the flavor composition of the proton.
Analyses of the proton structure are, therefore, carried out as global fits, 
which are typically accurate to NNLO in QCD. 
Results for such NNLO PDF sets have been obtained by 
ABM11~\cite{Alekhin:2012ig}, ABM12~\cite{Alekhin:2013nda}, 
CT10~\cite{Gao:2013xoa}, HERAPDF (v1.5)~\cite{herapdf:2009wt,herapdfgrid:2011},
JR09~\cite{JimenezDelgado:2008hf},  
MSTW~\cite{Martin:2009iq}, NNPDF (v2.3)~\cite{Ball:2012cx} 
and the recent updates 
MMHT 2014\cite{Harland-Lang:2014zoa} and NNPDF (v3.0)\cite{Ball:2014uwa}.
Some of the results are displayed in Fig.~\ref{fig:pdfcomp}.
\begin{figure}[h!]
\centering
    \includegraphics[width=14.0cm]{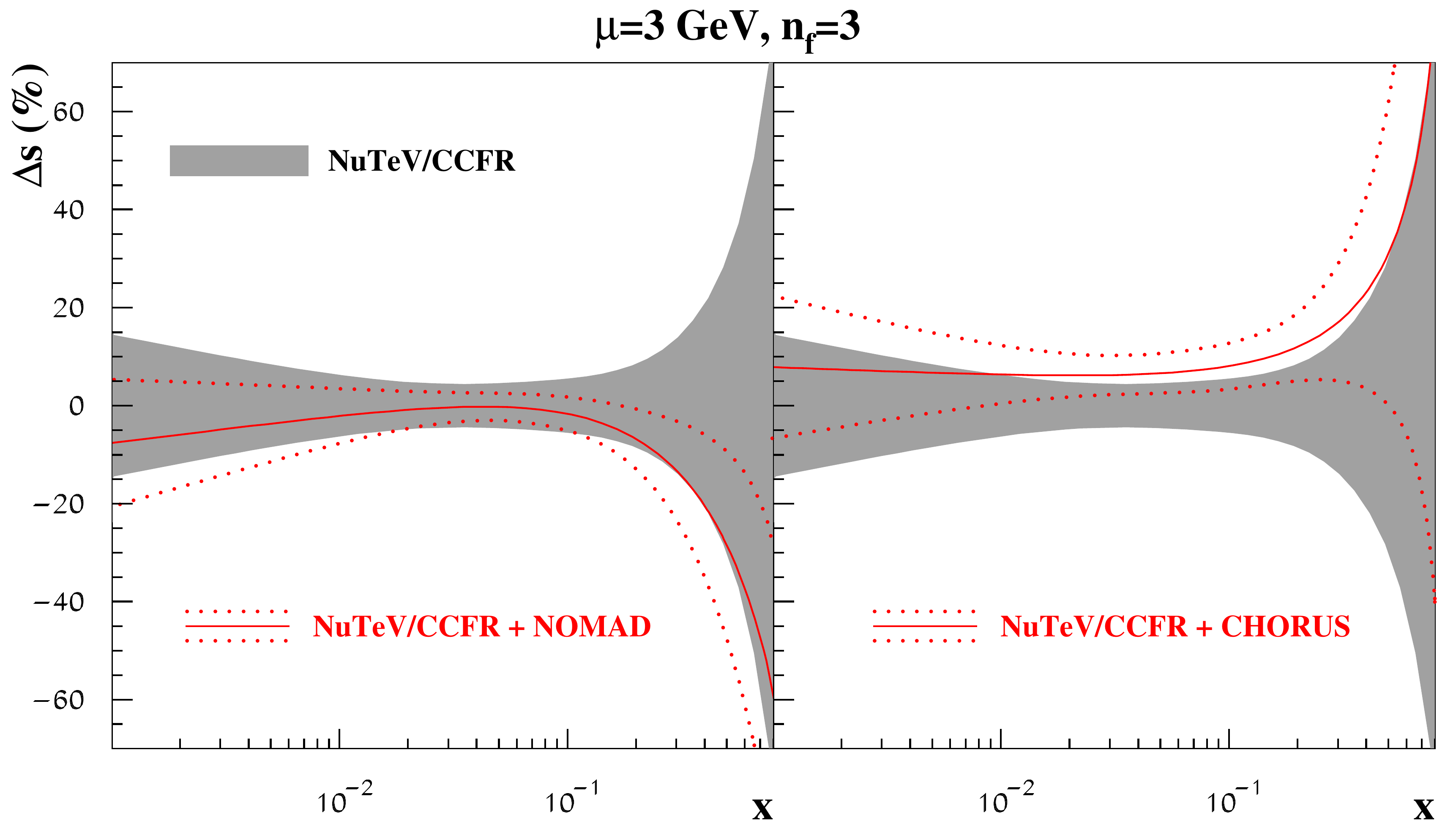}
   \caption{\small
    \label{fig:strange}
    The relative strange sea uncertainty obtained from variants of 
    the ABM12 PDF analysis~\cite{Alekhin:2013nda} 
    with only NuTeV and CCFR data~\cite{Goncharov:2001qe} 
    employed to constrain the strange sea
    (grey area) and with the solid lines displaying
    the relative change in the strange sea due to
    the NOMAD~\cite{Samoylov:2013xoa} (left panel) and 
    CHORUS~\cite{KayisTopaksu:2011mx} (right panel)
    data sets. 
    The dots correspond to the strange sea uncertainty after inclusion of the new data sets. 
    Plot taken from~\cite{Alekhin:2014sya}.
  }
\end{figure}
Given the current status of perturbative QCD and the precision of the experimental data  
a future physics program of neutrino nucleon DIS as in the SHiP experiment 
can help to address a number of very relevant questions.
For PDFs improvements can be made with respect to the flavor asymmetry 
of sea quarks, in particular the strangeness.
The information on the strange sea quark content of the nucleon shown in Fig.~\ref{fig:pdfcomp}
relies almost entirely on the data by the NuTeV and CCFR experiments~\cite{Goncharov:2001qe} 
from charm di-muon production in (anti)neutrino-iron DIS interactions.
The recent publication of new data samples for charm di-muon production in neutrino-iron 
DIS interactions by the NOMAD experiment~\cite{Samoylov:2013xoa}
and for inclusive charm production in nuclear emulsions by the CHORUS experiment~\cite{KayisTopaksu:2011mx} 
shown a clear improvement of the situation as illustrated in Fig.~\ref{fig:strange}.
Note that, however, even if combining NuTeV/CCFR data with the NOMAD or CHORUS data
the bands shrink, the two new best fit represented in Fig.~\ref{fig:strange} appear to
go in opposite directions. For this reason we believe that only with a new and large data sample
we might have a definitive leap in our knowledge of the nucleon strangeness.

\subsection{Strangeness from heavy-quark DIS in CC interactions}

\begin{figure}[ht]
\centering
\includegraphics[width=0.6\textwidth]{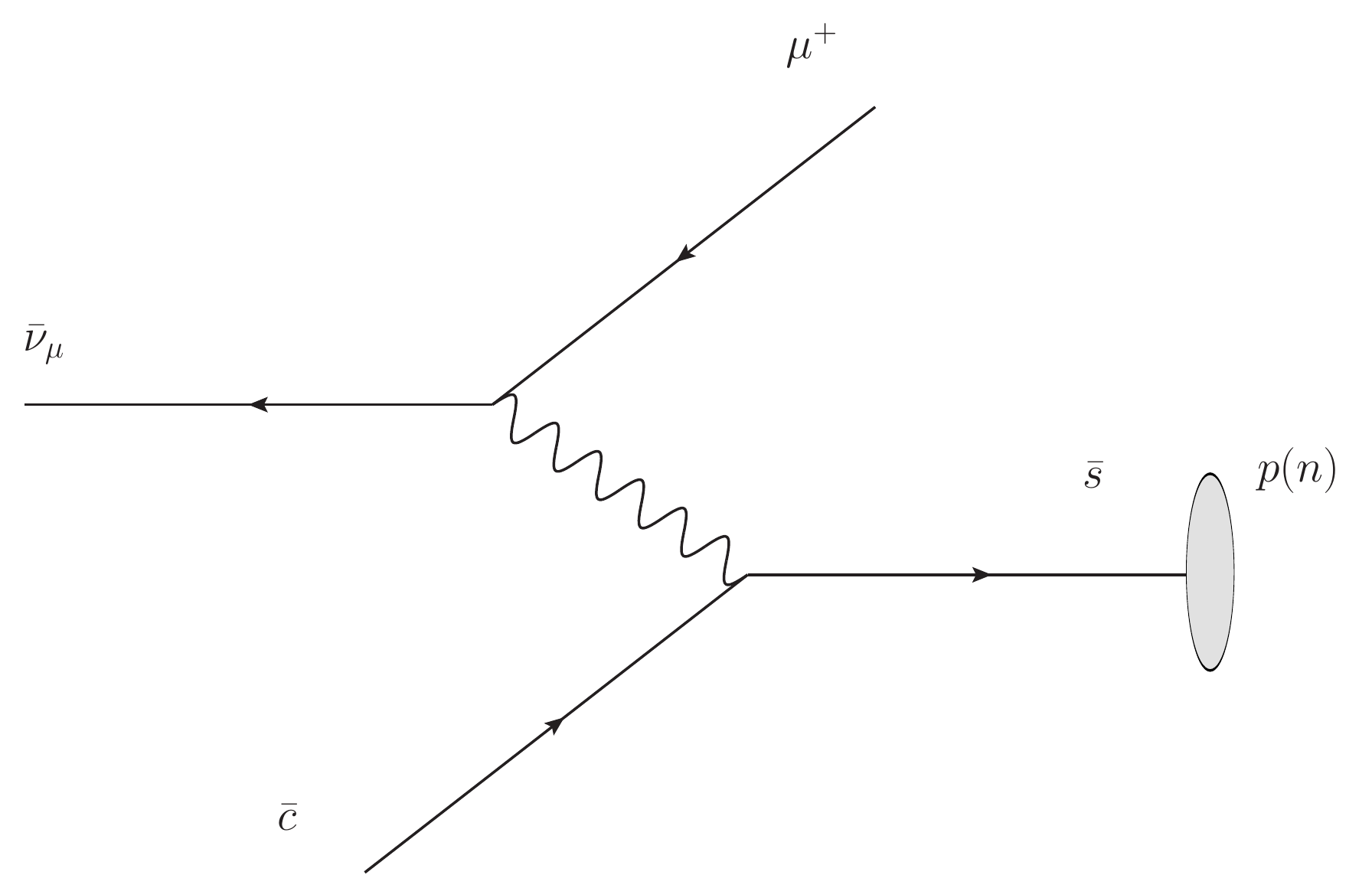}
  \caption{
           Diagram for anti-charm production in anti-neutrino charged current interactions.
}
  \label{fig:antis}
\end{figure}

\begin{figure}[ht]
\centering
\includegraphics[width=0.75\textwidth]{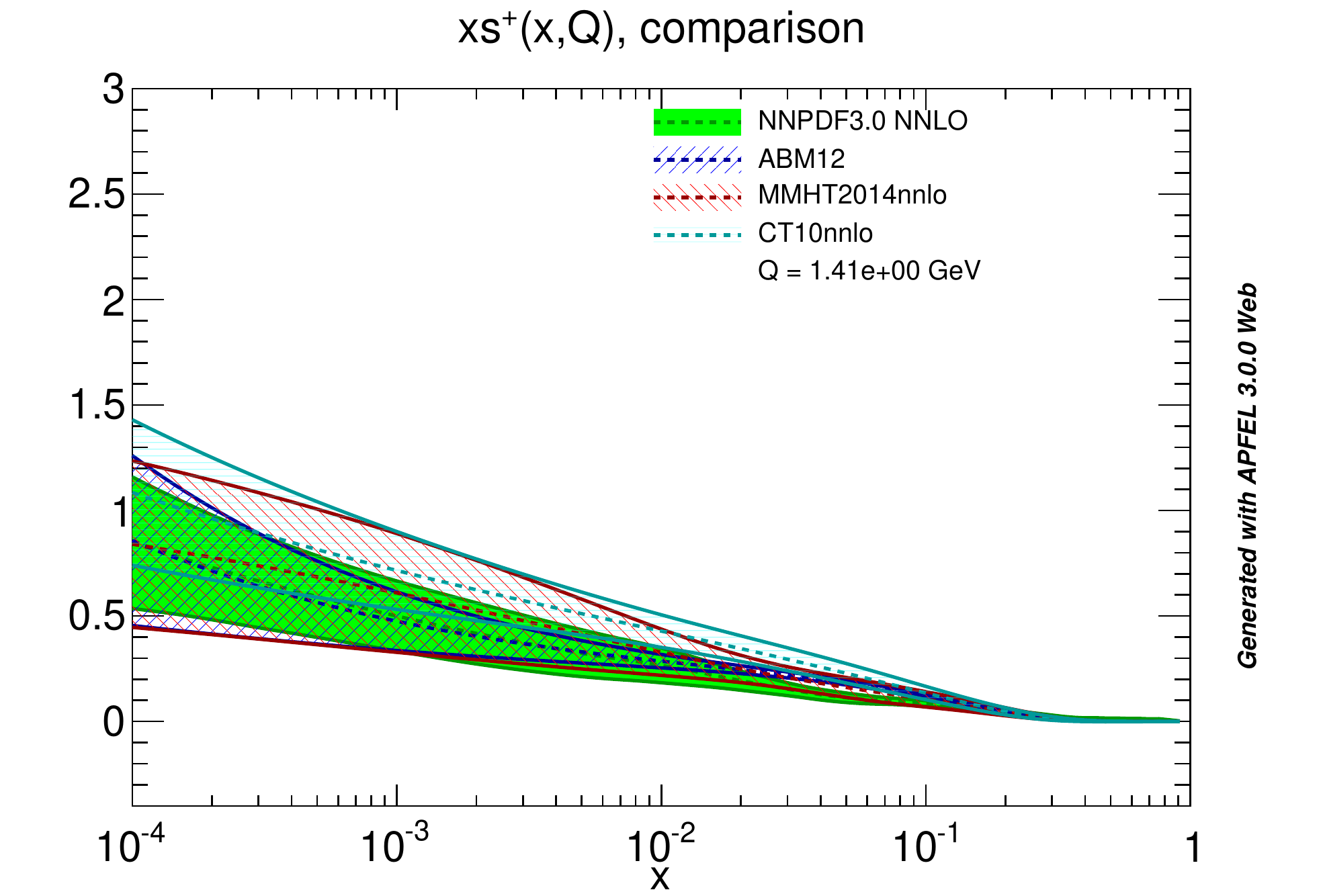}
  \caption{
           $s+\bar{s}$ distribution in the proton.
}
  \label{fig:strangeness}
\end{figure}

Anti-charm production in charged current anti-neutrino interactions selects anti-strange quark in the nucleon.
This is because while $\bar{d}$ quarks are not privileged with respect to the $\bar{s}$ quarks in the proton sea,
the latter are favoured to produce charm by the Cabibbo angle. The process is sketched in Fig.~\ref{fig:antis}.
Precise knowledge of the strangeness is an important information for many precision tests of the SM
as well as for BSM searches at the LHC. The $W$ mass is a fundamental parameter of the model and
at the LHC $W$ boson production proceeds for the 80\% through the $u\bar{d}$ channel and 20\%
through the $c\bar{s}$ channel. For a detailed discussion about the strong impact of the uncertainty
of the strange quark content of the proton on $W$ mass measurements see for example Ref.~\cite{Bozzi:2015hha}.
In Fig.~\ref{fig:strangeness} the current situation for the proton strangeness ($s+\bar{s}$) is
summarized including uncertainty bands. Although the four PDF collaborations: NNPDF3.0, MMHT, CT10
and ABM12, use approximately the same data points, the final results are quite different in the $x$
range $(0.001,0.1)$.
The strange see quark determination by all the PDF groups relies mainly on the
dimuon data collected by the NuTeV/CCFR collaboration.
Despite the high number of charged current $\nu_{\mu}$ and $\bar{\nu}_{\mu}$ events,
$1.280.000$ and $270.000$ respectively, the detection of charm production was limited
by the selection based on the muon decay channel for the charmed resonances produced.
Furthermore a cut of $5\,\rm{GeV}$ was imposed for the muon identification.
For these reasons only $5.102$ $\nu_{\mu}$ and $1.458$ $\bar{\nu}_{\mu}$ events were identified
as charm production in charged current interactions.

The neutrino detector under consideration in SHiP would have a lower cut for the muon
identification and, more important, would allow the detection of any decay of the produced charmed hadrons.
For these reasons, even if the number of expected CC muon neutrino events is comparable
with the ones detected by NuTeV/CCFR the charmed data collection in SHiP will be much larger.
Indeed, in the whole data taking of the SHiP experiment $1.8$ million of $\nu_{\mu}$ and
0.6 million of $\bar{\nu}_{\mu}$ events are expected~\cite{ship-tp}. This would translate into
the order of $70$ thousands charm and $27$ thousands anti-charm produced.
Furthermore, the electron neutrino charged current are also expected to contribute
about $20$ thousands induced charm production. Under these assumptions we expect a
considerable improvement in the knowledge of the nucleon strangeness. In the following
we consider only the contribution of the CC muon neutrino interactions.

To be more specific, we used a simulated data sample normalized to $2 \times 10^{20}$ POT
to estimate number and the distribution of muon neutrino interactions with
charm production. We followed closely the NuTeV analysis~\cite{Goncharov:2001qe}
where the collected data were divided in five bins for the $x$ variable, three for $y$ and three energies
per event sample.
In SHiP the energies will be lower than NuTeV, nevertheless distributing the events in bins with an
equal content we find similar
values for the other two variables $x$ and $y$. As a result we estimate that the statistical uncertainty in each bin
will reduce from $\sim 10-20\%$ in~\cite{Goncharov:2001qe} for both $\nu_\mu$ and $\bar{\nu}_\mu$
interactions, to $\sim 2-3\%$ for charm production and $\sim 4-5\%$ for anti-charm production.
Note that in the NuTeV analysis the systematic error was dominant.
To make a realistic estimation of the level of improvement in our analysis we included also a systematic
uncertainty.
\begin{figure}[ht]
\centering
\includegraphics[width=0.75\textwidth]{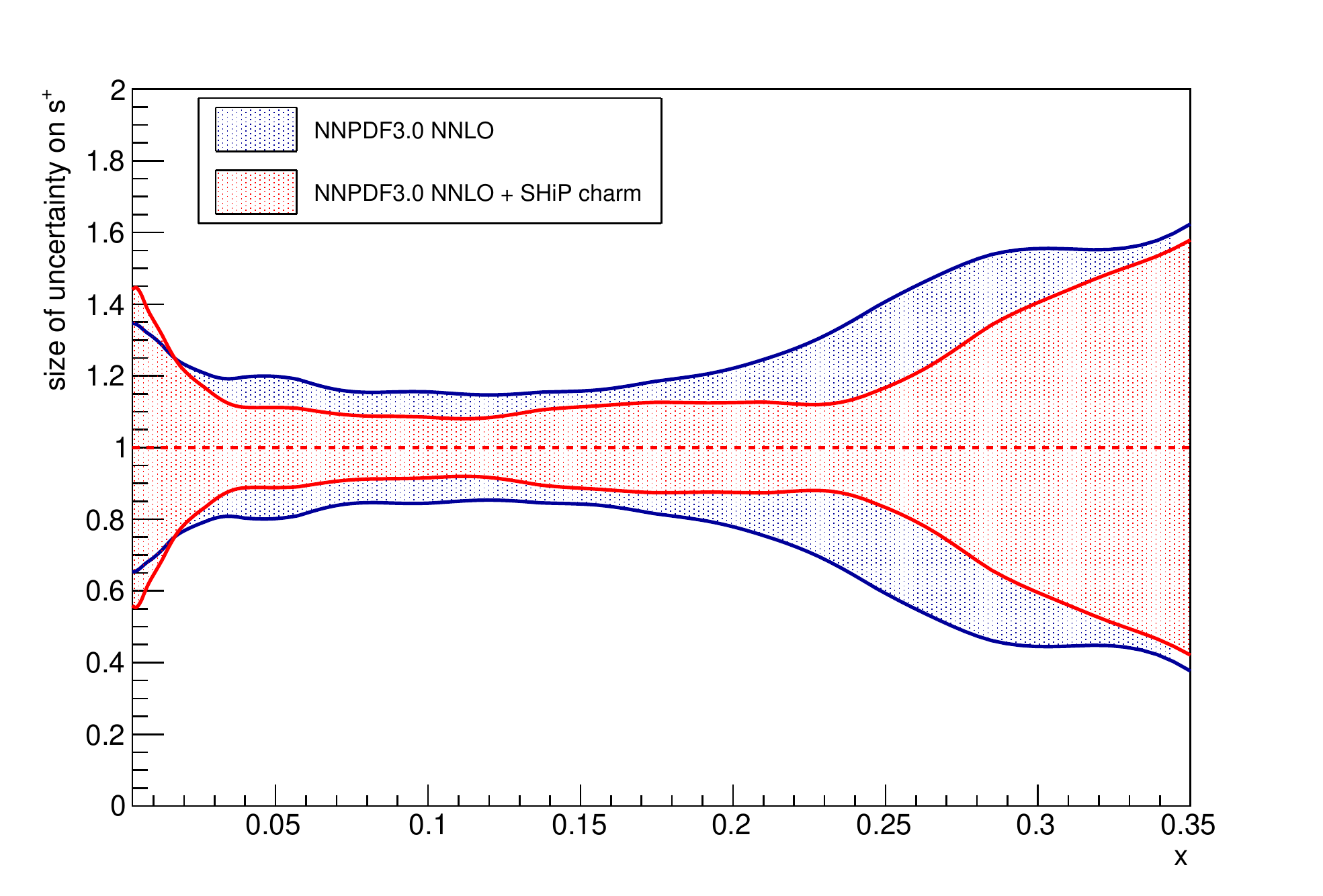}
\includegraphics[width=0.75\textwidth]{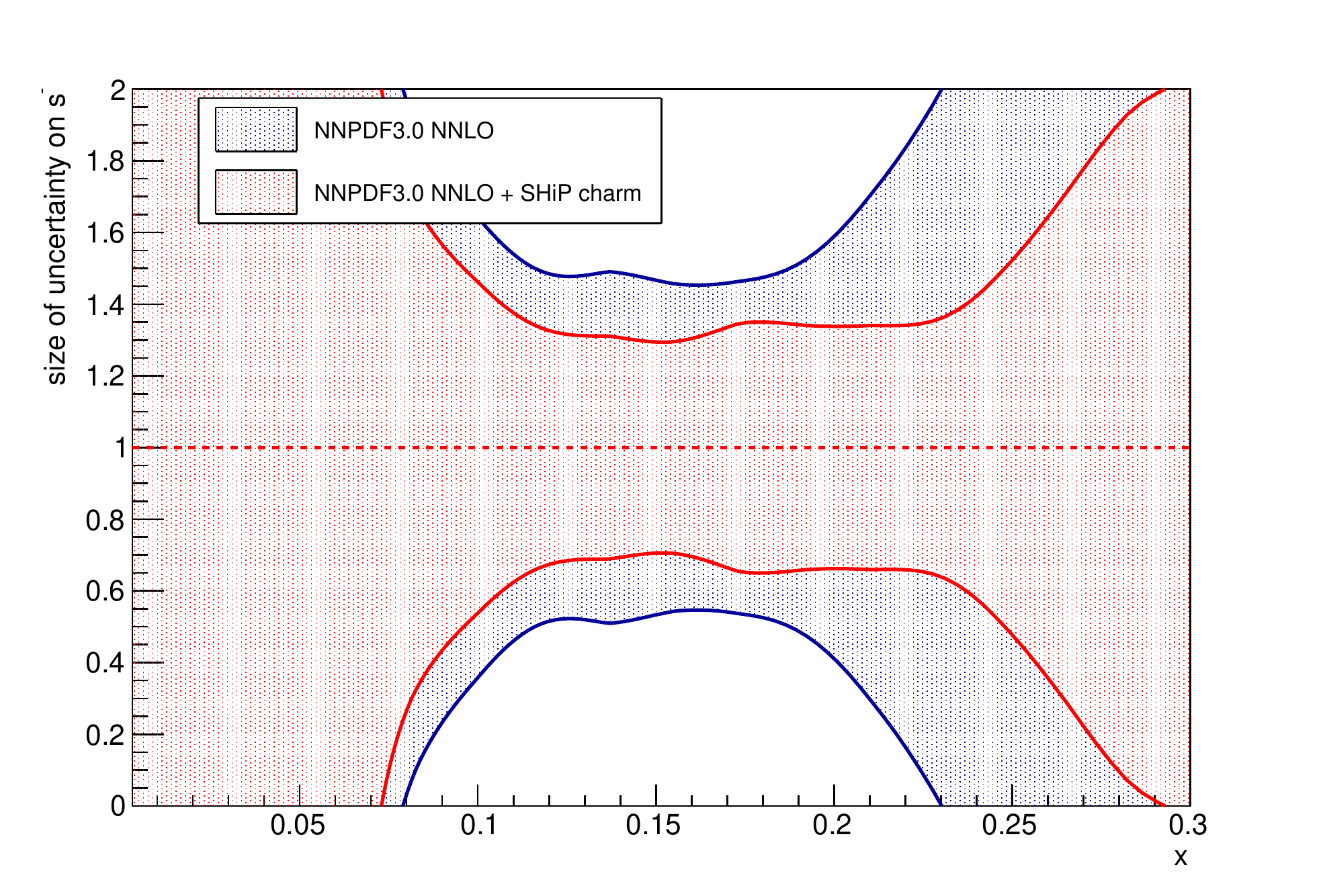}
  \caption{
           $s+\bar{s}$ (top) and $s-\bar{s}$ (bottom) uncertainty bands with and without adding simulated charm data
           from the SHiP experiment to the NNPDF3.0 NNLO fit~\cite{Ball:2014uwa}.
}
  \label{fig:strangeness2}
\end{figure}
We assessed the potential impact of the simulated charm data by adding them to the NNPDF3.0 NNLO
fit~\cite{Ball:2014uwa} using the Bayesian reweighting technique
introduced and described in~\cite{Ball:2010gb,Ball:2011gg}.
Pseudodata are generated fluctuating the central value around the central NNPDF3.0 NNLO predictions.
Their distribution is assumed to be a gaussian with central value 0 and standard deviation equal to the
sum in quadrature of the statistical uncertainty inferred from simulation, and the total systematic uncertainty,
assumed to be uniform 5\% (and uncorrelated) for all pseudodata points.
In the method of Refs.~\cite{Ball:2010gb,Ball:2011gg} one first generate an ensamble
of unweighted replicas representing the probability distribution function under consideration.
It is then possible to attribute a weight expressing the relative probability of the new data for
different PDF's.
Finally one can use the weights to sample the new probability density using the original
replica reweighted.
The constraing power of the SHiP pseudodata is measured by the Shannon entropy
of the reweighted set, which measures the number of effective ($N_\mathrm{eff}$) replicas after reweighting.
In this analysis, starting from a prior set with 1000 replicas, we obtain 
$N_\mathrm{eff}=62$. The reweighted $s^+$ and $s^-$ distributions are showed in Fig.~\ref{fig:strangeness2}.
The reduction of the associated uncertainty is significant in the ranges $0.03-0.3$ for $s^+$ and
$0.08-0.3$ for $s^-$ respectively.

\subsection{Nuclear effects in $\nu N$ DIS and global analyses of nuclear PDFs}

High statistics neutrino deep-inelastic scattering experiments
have generated significant interest in the literature as they provide
crucial information for global fits of parton distribution functions as has been
discussed in the preceding section.
The neutrino DIS data provide the most stringent constraints on the strange quark
distribution in the proton \cite{Kusina:2012vh}, and allow for a flavor decomposition of the PDFs
which is {\em essential} for precise predictions of the benchmark gauge boson production
processes at the LHC.
Moreover, the neutrino experiments have been used to make
precision tests of the Standard Model (SM) in the neutrino sector. 
A prominent example is the extraction of the weak mixing angle $\theta_W$ in
a Paschos--Wolfenstein type analysis.
A good knowledge of the neutrino DIS cross sections is also very important for
long baseline experiments of the next generation which aim at measuring
small parameters of the NMS mixing matrix such as the mixing angle $\theta_{13}$
with more precision and eventually the CP violating phase $\delta$.

Due to the weak nature of neutrino interactions the use of heavy nuclear
targets is unavoidable, and this complicates the analysis of the precision
physics discussed above since model-dependent nuclear corrections must be appplied 
to the data.
Our understanding of the nuclear corrections relies largely on charged-lepton--nucleus ($\ell A$) DIS data. 
In the early 80s, the European Muon Collaboration (EMC)
\cite{Aubert:1983xm} found that the nucleon structure functions
$F_{2}$ for iron and deuterium differ as a function of $x$.
This discovery triggered a vast experimental program to investigate the nuclear modifications
of the ratio $R[F_{2}^{\ell A}] =F_{2}^{\ell A} /(A\, F_{2}^{\ell N})$
for a wide range of nuclear targets with atomic number $A$, see, e.g., Table~1
in \cite{Kovarik:2010uv}.
By now, such modifications have been established in a kinematic range
from relatively small Bjorken $x$ ($x\sim10^{-2}$) to large $x$
($x\sim0.8$) in the deep-inelastic region with squared momentum transfer
$Q^{2}>1\ {\rm GeV}^{2}$. 
The behavior of the ratio $R[F_{2}^{\ell A}]$
can be divided into four regions: (i) $R>1$ for $x\gtrsim0.8$ (Fermi
motion region), (ii) $R<1$ for $0.25\lesssim x\lesssim0.8$ (EMC
region), (iii) $R>1$ for $0.1\lesssim x\lesssim0.25$ (antishadowing
region), and (iv) $R<1$ for $x\lesssim0.1$ (shadowing region).
Various models and physics mechanisms have been proposed
to explain the nuclear modifications in the different kinematic regions.
The shadowing suppression at small $x$ is generally assumed to occur 
due to coherent multiple scattering inside the nucleus of a $q\bar{q}$ pair coming 
from the virtual photon \cite{Armesto:2006ph} with destructive interference of the 
amplitudes \cite{Brodsky:1989qz}.
The antishadowing region is theoretically less well understood but
might be explained by the same mechanism with constructive interference
of the multiple scattering amplitudes \cite{Brodsky:1989qz} or by
the application of momentum, charge, and/or baryon number sum rules.
Conversely, the modifications at medium and large $x$ are usually
explained by nuclear binding and medium effects and the Fermi motion
of the nucleons \cite{Geesaman:1995yd}, however, a complete understanding
is still lacking and subject of ongoing investigations.

The charged-lepton--nucleus ($\ell A$) DIS data form the backbone of global
analyses of nuclear parton distribution functions (NPDFs)
\cite{deFlorian:2011fp,Schienbein:2009kk,Eskola:2009uj,Hirai:2007sx}.
The NPDFs differ from the PDFs inside free nucleons
such that the measured ratios of structure functions is reproduced.
The NPDFs can be used to compute nuclear correction factors 
for different observables in a flexible way.
Furthermore, as in the free nucleon case, neutrino--nucleus DIS data are very useful 
for a better flavor separation of the NPDFs.
In particular, the opposite charge di-muon data from NuTEV and CCFR
allow to constrain the nuclear strange quark distribution in an iron target.

\begin{figure}[t]
\includegraphics[width=0.48\textwidth]{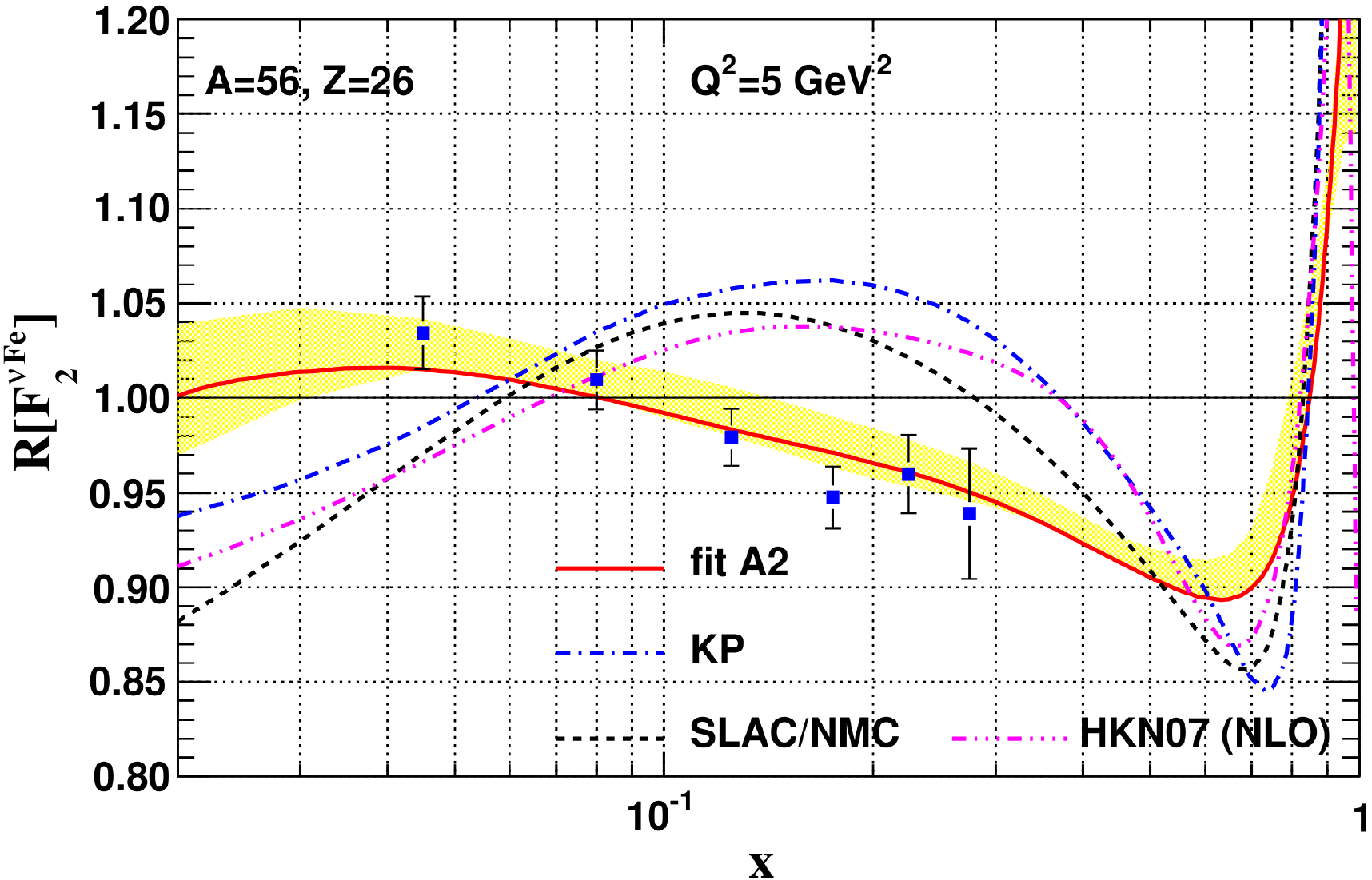}
\includegraphics[width=0.48\textwidth]{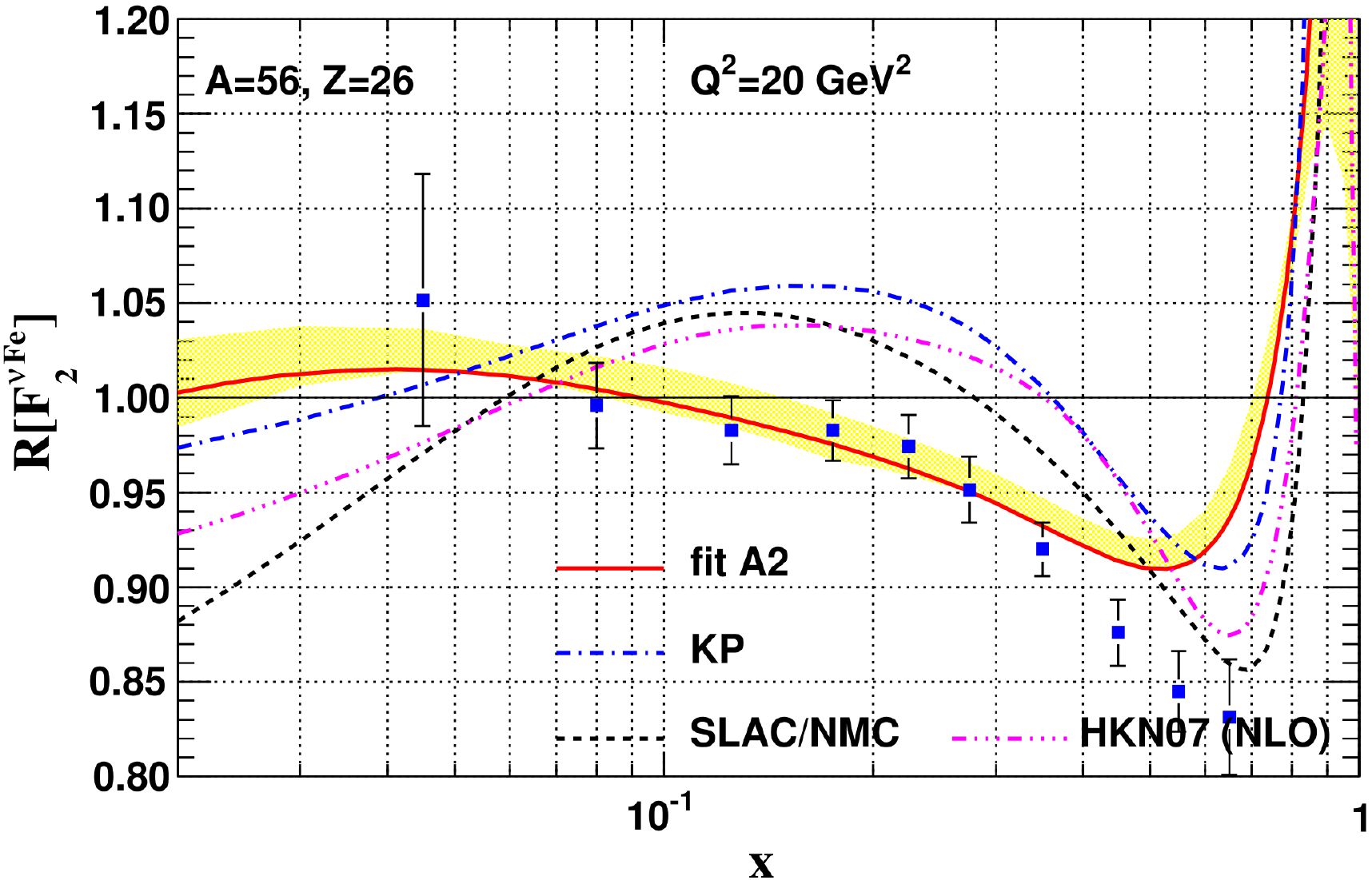}
\caption{The nuclear correction factor 
$R[F_2^{\nu Fe}] \simeq F_2^{\nu Fe}/F_2^{\nu D}$
as a function of $x$ for $Q^{2}=5\,{\rm GeV}^{2}$ (left) and  $Q^{2}=20\,{\rm GeV}^{2}$ (right).
The solid (red) line is the prediction of a fit to $\nu Fe$ DIS
data \protect\cite{Schienbein:2007fs}. 
This fit is compared with predictions based on $\ell^\pm A$ DIS data.
In particular, with the SLAC/NMC parametrization,
the model from Kulagin and Petti (KP) \protect\cite{Kulagin:2004ie},
and the predictions using the HKN'07 NPDFs \protect\cite{Hirai:2007sx}.
}
\label{fig:Rnu1}
\end{figure}

\begin{figure}[t]
\includegraphics[width=0.48\textwidth]{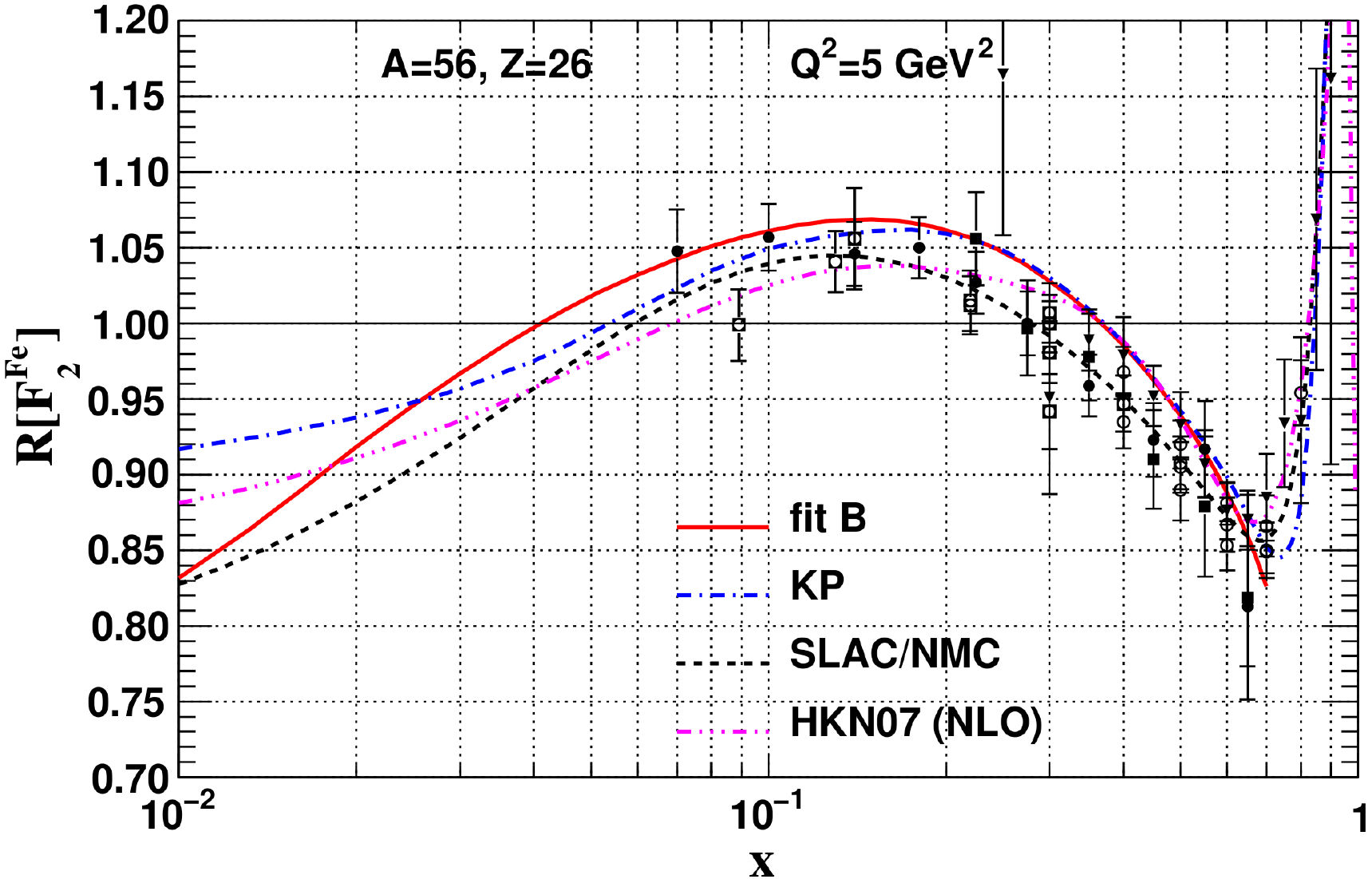}
\includegraphics[width=0.48\textwidth]{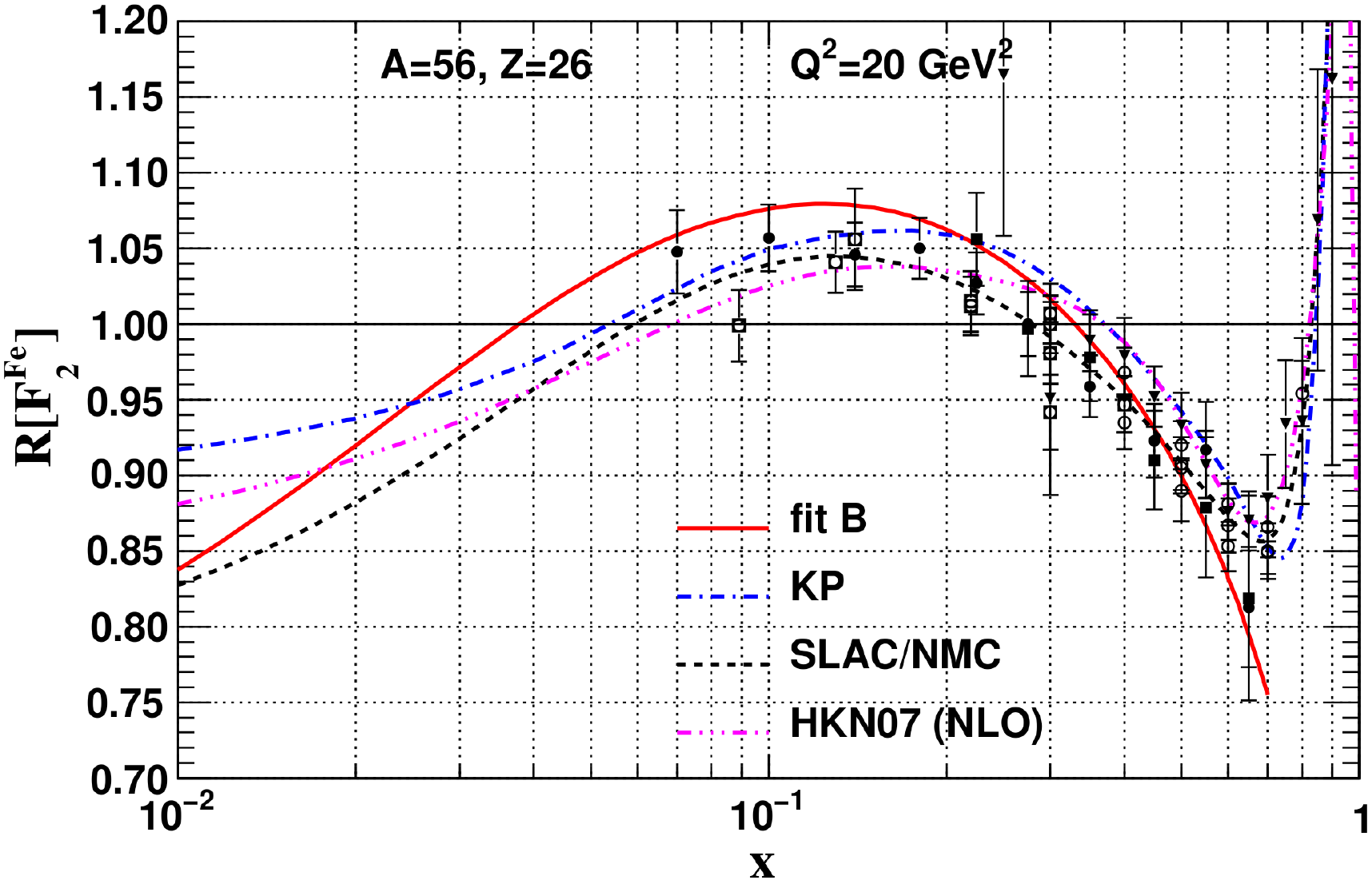}
\caption{The nuclear correction factor 
$R[F_2^{\ell Fe}] \simeq F_2^{\ell Fe}/F_2^{\ell D}$
as a function of $x$ for $Q^{2}=5\,{\rm GeV}^{2}$ (left) and  $Q^{2}=20\,{\rm GeV}^{2}$ (right).
The solid (red) line is the prediction of a fit (fit B =decut3) to $\ell Fe$ DIS
data and Drell-Yan data in the $A$-dependent framework of 
\protect\cite{Schienbein:2009kk}. 
This fit is compared with predictions based on $\ell^\pm A$ DIS data.
In particular, with the SLAC/NMC parametrization,
the model from Kulagin and Petti (KP) \protect\cite{Kulagin:2004ie},
and the predictions using the HKN'07 NPDFs \protect\cite{Hirai:2007sx}.
}
\label{fig:Rl1}
\end{figure}

However, in a recent analysis \cite{Schienbein:2007fs} of $\nu Fe$ DIS data
from the NuTeV collaboration it was found that the nuclear correction factors are
surprisingly different from the predictions based on the $\ell^\pm Fe$ DIS data
with important implications for global analyses of proton PDFs.
This can be seen in Fig.~\ref{fig:Rnu1} where the nuclear correction factor
$R[F_2^{\nu Fe}] \simeq F_2^{\nu Fe}/F_2^{\nu D}$ is shown
as a function of $x$ for $Q^2=5$ GeV$^2$ (left) and $Q^2=20$ GeV$^2$ (right).
Here, the solid line is the prediction of a fit to $\nu Fe$ DIS data from the
NuTeV collaboration \cite{Schienbein:2007fs}.
The other three curves are based on $\ell^\pm A$ data and have quite a different
$x$-shape.
At the same time, the nCTEQ fits 
to $\ell A$ DIS and DY data \cite{Schienbein:2009kk}
are well in agreement with the charged-lepton data as can be seen in Fig.~\ref{fig:Rl1}.

The study in \cite{Schienbein:2007fs} 
left open the question, whether the neutrino DIS data could be reconciled
with the charged-lepton DIS data by a better flavor separation of the NPDFs. 
This question was addressed in Ref.\ \cite{Kovarik:2010uv}
by performing a global $\chi^2$-analysis
of the combined data from $\nu A$ DIS, $\ell A$ DIS and the DY process.
This study confirmed that the neutrino DIS cross-section data are indeed slightly
incompatible with the charged lepton DIS+DY data
if the precision {\em correlated} errors of the neutrino data are taken into account.
In case that the less precise uncorrelated errors are used the tension between the
NuTeV/CCFR and the charged-lepton DIS data sets disappears \cite{Paukkunen:2010hb,deFlorian:2011fp}.

In conclusion, precise measurements of neutrino--lead DIS cross sections with the SHiP experiment
have a great potential to (i) improve the knowledge of NPDFs and
(ii) help resolve the tension between the NuTeV/CCFR data and charged-lepton DIS data.
This would improve our knowledge of the nuclear correction factors needed for the proton global analyses.
Most importantly, for the determination of the strange PDF in the nucleon using di-muon data a solid understanding of 
the nuclear modifications of the individual parton flavors is necessary.
An improved understanding of these questions will then allow to test
theoretical models predicting these nuclear modifications.
An intriguing point is also that a large DIS data sample for lead together with proton--lead data from the LHC
would allow dedicated determinations of lead PDFs without the need to model the nuclear $A$-dependence.

\subsection{$\alpha_S$ measurement via Gross-Llewellin Smith sum rule}

Inclusive quantities in DIS such as several sum rules are known to high
orders in QCD. Of particular interest for neutrino-induced DIS is the 
Gross-Llewellyn Smith (GLS) sum rule, which is well studied in the literature.
The GLS integral is given by 
\begin{eqnarray}
  \label{eq:gls}
  S_{\rm GLS}(Q^2) &=& \frac{1}{2}\, \int\limits_{0}^{1}dx \left( F_3^{\nu p} + F_3^{\nu n} \right)\, ,
\end{eqnarray}
and is known to N$^3$LO in QCD, see~\cite{Larin:1991tj}, up to higher twist
corrections.

New high statistics data from a DIS experiment can also be used for
precision determinations of the strong coupling constant $\alpha_s$
exploiting Eq.~(\ref{eq:gls}).
The current precision is reviewed, e.g., in~\cite{Moch:2014tta}. 

Measurements of $\alpha_s$ with the GLS sum rule need to
account for the correlation of the value of $\alpha_s$ with higher twist
contributions.
A phenomenological parametrization of the higher twist contributions 
to structure functions considers them as independent from the
leading twist ones
\begin{eqnarray}
  \label{eq:htwist}
  \displaystyle
  F_i^{\rm ht}(x,Q^2) &=&
    F_i^{\rm TMC}(x,Q^2)
  +
    \frac{H_i^{\tau=4}(x)}{Q^2}
  +
    \frac{H_i^{\tau=6}(x)}{Q^4}
  + \dots
  \, ,\qquad
\end{eqnarray}
where $F_i^{\rm TMC}$ are the structure functions at leading twist
supplemented by kinematic corrections due to hadron mass effects
i.e., the so-called target mass correction (TMC). 
The coefficients $H_i$ are some polynomial in $x$ and 
any $Q^2$-dependence of the higher twist operators 
due to the QCD evolution is neglected. 

Various studies have quantified the effect of higher twist contributions 
on the value of $\alpha_s$ in global analysis, 
which generally depend on the order of perturbation theory considered.
Twist six $H_i^{\tau=6}$ contributions are generally found to be consistent 
with zero in global fits, while the twist four terms $H_i^{\tau=4}$ can be
significant, see e.g., \cite{Alekhin:2011ey,Alekhin:2012ig}.
For $\alpha_s$ measurements via the GLS sum rule additional uncertainties 
arise from extrapolation into the unmeasured kinematics region at small and
large $x$.

Previous studies (cf. Sec.5 of \cite{Mangano:2001mj}) for physics at the
front-end of a neutrino factory have estimated 
the error on $\alpha_s$ measurements from Eq.~(\ref{eq:gls})
in a kinematical range similar to SHiP, i.e., $1.5 \le Q^2 \le 80$~GeV
to be 
\begin{eqnarray}
  \label{eq:aserr}
  \Delta \alpha_s &=& 0.0035  
\end{eqnarray}
assuming the simultaneous determination of higher twist terms and
${\cal O}(\alpha_s^2)$ QCD corrections.
If higher twist terms are fixed, the uncertainty reduces by roughly one third to $\Delta \alpha_s = 0.0026$.
Currently, even the N$^3$LO coefficient functions are known, 
higher twist contributions are better understood and constrained, so that a
significant further reduction of the uncertainty on $\alpha_s$ in 
Eq.(\ref{eq:aserr}) seems to be feasible. For these reasons the
SHiP experiment could make an important contribution here.

\subsection{Precise Ratios for Neutrino Nucleon Interactions}

The SHiP experiment offers also the possibility to perform
precise measurements of Electroweak parameters.
An open question in particle physics is still the significance of the ``NuTeV anomaly''
which reported a $3\sigma$ deviation in the measument of $\sin^2\theta_W$.
A theoretical
resolution of the anomaly or new experimental measurements will be important.
The SHiP experiment will be using high intensity $\nu$  and $\bar{\nu}$ beams over a
large range of energies and will have the oppportunity to check and improve previous results.

The NuTeV analysis followed a simultaneous fitting of
$R^{\nu(\bar{\nu})}=\sigma^{\nu(\bar{\nu})}_{NC}/\sigma^{\nu(\bar{\nu})}_{CC}$ which is related but not
identical to the PW-relation~\cite{Paschos:1972kj}
\begin{equation}
\label{eq:rminus}
R^- = \frac{\sigma^{\nu}_{NC}-\sigma^{\bar{\nu}}_{NC}}{\sigma^{\nu}_{CC}-\sigma^{\bar{\nu}}_{CC}}
    = \frac{1}{2} - \sin^2\theta_W + \delta R_{A}^- + \delta R_{QCD}^- + \delta R_{EW}^-.
\end{equation}
The original relation was derived for isoscalar targets but in the NuTeV experiment
the target was Iron which brings the corrections included on the rhs of equation~\ref{eq:rminus}.
The corrections have been studied in several articles and we will not repeat them here.
We simply refer to the articles in Refs.~\cite{Davidson:2001ji,McFarland:2003jw,Kretzer:2004tv}.
Other ratios of neutral to charged current reactions are also determined accurately
because the largest contribution is determined by isospin symmetry and the smaller terms
are estimated by using available data  and/or theoretical relations. These ratios can be
used, with sufficient care, either to determine the Weinberg angle or to search for new
effects manifested in the change of the ratios as function of energy or as function of
the distance traveled by the neutrinos.
An attractive ratio is~\cite{Paschos:2002ya}
\begin{equation}
\label{eq:rone}
P = \frac{\sigma^{\nu}_{NC}+\sigma^{\bar{\nu}}_{NC}}{\sigma^{\nu}_{CC}+\sigma^{\bar{\nu}}_{CC}}.
\end{equation}
Before comparing the ratio above with the number of events detected in SHiP we have to
take into account the differences in the $\nu$ and $\bar{\nu}$ fluxes.
Indeed, first estimates based on full Monte Carlo simulations of the events in the
proton target show that above $40\rm{GeV}$
the fluxes of $\nu$ and $\bar{\nu}$ are not far from each other in the absolute values and have
the same shape. We estimate that with the help of a detailed Monte
Carlo simulation tuned to reproduce the number of charged leptons and anti-leptons detected
respectively in the experiment, and the total number of neutral current events, one can correct for the
relative difference in the fluxes and keep the uncertainty on $\sin^2\theta_W$ associated to this procedure
at the level of 1\% or lower. Note that $\sin^2\theta_W$ has been measured with 1\% accuracy
in neutrino nucleon scattering by CDHS~\cite{Blondel:1989ev} and CHARM~\cite{Allaby:1987vr} at CERN.
We note that Monte Carlo uncertainty in signal modeling has also been used by other experiments
to determine the level of purity of a muon neutrino beam with respect to muon antineutrinos and the
electron neutrino contamination in a muon neutrino beam. Here, Monte Carlo simulations
would address the difference in the energy spectra of the sum of muon and electron neutrinos and the
anti-neutrinos. In the assumption that the electrons are identified, they do not represent a limitation.
In general one has to correct the $P$ ratio above for quark mixing, quark sea effects,
$c$-quark threshold effects, non-isoscalarity, $W-Z$ propagator differences, the finite muon mass,
QED and EW radiative corrections.
Indeed, quark sea effects and $c$-quark threshold effects could be measured within the
same experiment with high level of accuracy and we are lead to the conclusion
that after the flux asymmetry mentioned above the non-isoscalarity is the next
higher source of uncertainty.
As we will see below, the correction for non isoscalarity on the  $P$ ratio is
less then 1\% so that the correction for the asymmetry in the fluxes sets the sensitivity
on a possible measurement of $\sin^2\theta_W$.
An important property of this ratio is the disappearance of the interference terms
between vector and axial-vector currents in both numerator and denominator. The weak
hadronic current is given by
\begin{equation}
\label{eq:whcurrent}
J_{\mu} = x V^3_{\mu}\,+\,y A^3_{\mu}\,+\,\gamma V^0_{\mu}\,+\,\delta A^0_{\mu}\,.
\end{equation}
The currents $V^3_{\mu}$ and $A^3_{\mu}$ are the isospin partners of the charged currents
and $V^0_{\mu}$, $A^0_{\mu}$ are iscalar components. The quark content of the currents is
\begin{equation}
\label{eq:currentcontent3}
V^3_{\mu}=\frac{1}{2}\,(\bar{u} \gamma_{\mu} u\,-\,\bar{d} \gamma_{\mu} d)
~~~~~~
A^3_{\mu}=\frac{1}{2}\,(\bar{u} \gamma_{\mu} \gamma_5 u\,-\,\bar{d} \gamma_{\mu} \gamma_5 d)
\end{equation}
and
\begin{equation}
\label{eq:currentcontent0}
V^0_{\mu}=\frac{1}{2}\,(\bar{u} \gamma_{\mu} u\,+\,\bar{d} \gamma_{\mu} d)
~~~~~~
A^0_{\mu}=\frac{1}{2}\,(\bar{u} \gamma_{\mu} \gamma_5 u\,+\,\bar{d} \gamma_{\mu} \gamma_5 d).
\end{equation}
In the Standard Model 
\begin{eqnarray}
\label{eq:smpar}
x &=& 1 -  2 \sin^2 \theta_W \quad\quad\quad\, y = - 1  \\
\gamma &=& - \frac{2}{3} \sin^2 \theta_W \quad\quad\quad\quad \delta = 0.
\end{eqnarray}
As already mentioned the $A^3_{\mu} \otimes V^0_{\mu}$ term in Eq.~(\ref{eq:rone}) is eliminated
for any target (even if the target is not isoscalar).

For the denominator in $P$ we write 
\begin{equation}
\label{eq:AIV}
\sigma(\nu_l\,N \to l^- \, X) = A\,+\,I\,+\,V
\end{equation}
where $V$ comes from the vector current alone, $A$ from the axial current alone and the
interference term $I$ is eliminated by the averaging over neutrinos and antineutrinos.
The expression is written as 
\begin{equation}
\label{eq:rone2}
P = \frac{1}{2}\,\frac{A\,+\,x^2 V\,+\,\gamma^2\,S\,+\,2 x \gamma (V^3 \otimes V^0)}{A\,+\,V}
\end{equation}
where $S$ comes from the current $V^0_{\mu}$ alone and the last term is the interference between
$V^3_{\mu}$ and $V^0_{\mu}$. The last two currents also appear in electromagnetism and their
contributions can be extracted from experimental data. Since the SHiP experiment will cover
high energies we will estimate these terms using deep inelastic distribution functions.
We can mention already that coefficients in front of the terms are smaller and thus reduce
the uncertainties entering the final answer.

We can simplify the right hand side of Eq.~(\ref{eq:rone2}) by appealing to propreties that
we already known for the reactions.

(i)  At high energies deep inelastic scattering dominates, so that the scattering off Nuclei
is the incoherent sum over protons and neutrons (the EMC effect can be included separately).
It is the incoherent sum of parton distribution functions on the quarks. The quark content
of the currents $V^0_{\mu}$ and $V^3_{\mu}$ is the same except for a minus sign which for incoherent
scattering on quarks is irrelevant. Thus
\begin{equation}
\label{eq:veqs}
V = S\,.
\end{equation}
This means that the first three terms in the numerator of Eq.~(\ref{eq:rone2}) see the isoscalar
content of the target. Their contribution over neutrons and protons is proportional to the
atomic number $A$.

(ii) The intereference in the last term of the numerator involves electromagnetic currents
and is related to the difference of electron reactions on protons and neutrons. In fact in the
average over equal numbers of protons and neutrons this term vanishes, thus only the excess of
neutrons in the target nucleus contributes to the ratio making the correction small.

(iii) Defining
\begin{equation}
\frac{d V^{em}}{dx dy}=\frac{G^2}{\pi}\frac{Q^4}{4 \pi \alpha} \frac{\sigma^{em}}{dx dy}
\end{equation}
we find
\begin{equation}
\frac{d V^{em}_{ep}}{dx}=\frac{2 G^2 M E}{\pi}\left[\frac{4}{9}(u + \bar{u})
+\frac{1}{9}(d + \bar{d}) + \frac{1}{9}(s + \bar{s}) \right] \,.
\end{equation}
Then normalizing the denominator in Eq.~(\ref{eq:rone2}) to an isoscalar target with
atomic number $A$ we obtain
\begin{equation}
\label{eq:rone3}
P=\frac{1}{2}\,\Bigg\{ 1 - 2 \sin^2 \theta_W + \frac{20}{9} \sin^4 \theta_W 
-\lambda (1 - 2 \sin^2 \theta_W ) \sin^2 \theta_W  \Bigg\}\,,
\end{equation}
with $\lambda$ originating from the unequal numbers of protons $Z$ and
neutrons $(A-Z)$ in the target. The functional form is
\begin{equation}
\label{eq:lambda}
\lambda = \frac{A - 2 Z}{A}
\frac{V^{em}_p - V^{em}_n}
{\frac{1}{2}\left[(\sigma^{\nu p}_{cc}+\sigma^{\bar{\nu}p}_{cc}) +  (\sigma^{\nu n}_{cc}+\sigma^{\bar{\nu} n}_{cc}) \right]}\,.
\end{equation}
All quantities are known and we can proceed to numerical estimates. For an isoscalar target
$Z=A/2$ and $\lambda=0$, then
\begin{equation}
\label{eq:rone4}
P=R_1=\frac{1}{2}\,\Bigg\{ 1 - 2 \sin^2 \theta_W + \frac{20}{9} \sin^4 \theta_W  \Bigg\}\,.
\end{equation}
If the target is made of lead we estimate $\lambda=0.042$ with an error at the level
of a few \%, showing that the ratio is rather stable and non-isoscalarity correction introduces
a per mille level uncertainty.

\section{Limit on Tau neutrino magnetic moment}
\label{sec:mag}

Neutrinos are electrically neutral fundamental particles that couple to the other particles only through weak
interaction in the Standard Model (SM). Nevertheless, in the minimal extension of the SM where neutrinos
are proposed to be Dirac particles, they can acquire a magnetic moment ($\mu_\nu$) and give rise to
electromagnetic interactions. This is possible because of loop diagrams in which a neutrino radiates a virtual
$W$-boson transforming into a lepton and these combine again to give back the  original
neutrino~\cite{Fujikawa:1980yx}. The intermediate
particles can then couple to an external electromagnetic field. However, the magnetic moment produced in this
manner is extremely small and is proportional to the mass of the neutrino~\cite{Fujikawa:1980yx}:
\begin{equation}
\mu_\nu = \frac{3\,e\,G_F\,m_\nu}{8\,\pi^2\, \sqrt{2}} \simeq 
(3.2 \times 10^{-19}) \, \Big (\frac{m_\nu}{1 \, {\rm eV}} \Big ) \, \mu_B\,.
\end{equation}
The magnetic moment of neutrinos can be enhanced in other new physics models (see, e.g.,~\cite{Shrock:1982sc}).
An increase in the cross section $\sigma_{\nu e}$ for the elastic scattering of neutrino on electron, which 
involves only the neutral current in the SM,  would imply a larger magnetic moment and would be a clear signal of new physics.
The SM contribution to the energy ($T$) distribution of the scattered electron in the laboratory frame is given by
\cite{Amsler:2008zzb,Marciano:2003eq}
\bea
\label{eq:smneutralcurrent}
\frac{
\sigma_{(\nu e, \overline \nu e)}
}
{dT} \Big|_{SM} &=& \frac{G_F^2 m_e}{2 \pi} \left[ (g_V \pm g_A)^2 +(g_V \mp g_A)^2 \times 
\left( 1 -\frac{T}{E_\nu} \right)^2 -(g_V^2 -g_A^2) \frac{m_e T}{E_\nu^2}  \right]\,, \nonumber\\
\eea
where the upper (lower) sign refers to the interaction with $\nu$ ($\overline \nu$).
The contribution arising from magnetic moment of neutrino is given by the additional
term~\cite{Domogatsky:1971tu}:
\bea
\label{eq:magnet}
\frac{
\sigma_{(\nu e, \overline \nu e)}
}
{dT} \Big|_{\mu_\nu} = \frac{\pi \alpha_{em}^2 \mu_\nu^2}{m_e^2} \left( \frac{1}{T} - \frac{1}{E_\nu} \right).
\eea
There is no interference between the SM and magnetic moment amplitudes as the latter involves a flip in the 
spin of the neutrino.

\begin{figure}[ht]
\centering
\includegraphics[width=0.65\textwidth]{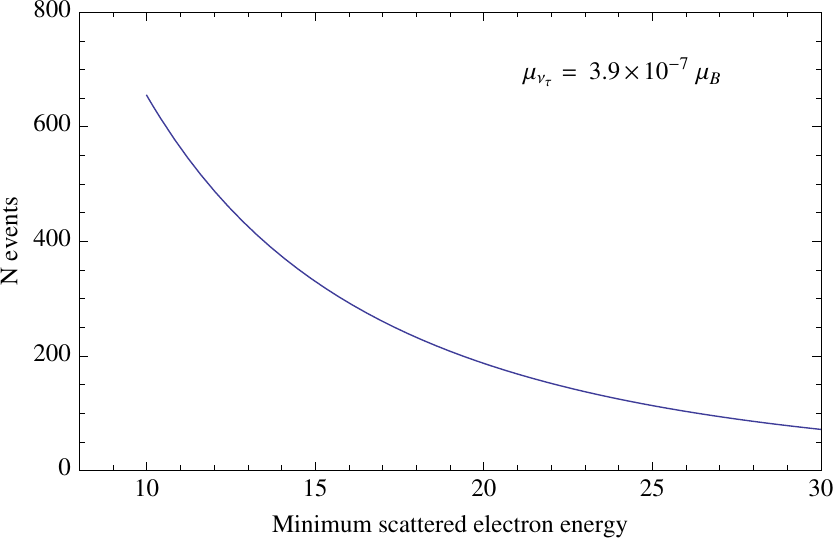}
  \caption{
          Number of expected events in SHiP producing an electron with energy greather then $10\,\rm{GeV}$
          assuming for $\mu_{\nu_{\tau}}$ the upper limit set by the DONUT experiment~\cite{Schwienhorst:2001sj}.
}
  \label{fig:magneticevt}
\end{figure}

In the laboratory frame the kinematical constraint on the angle between the incoming neutrino and the scattered
electron is~\cite{Radel:1993sw} 
\bea
\label{eq:cons}
\theta_{\nu-e}^2 < 2 m_e/E_e
\eea 
and for electron energies of about $1\,\rm{GeV}$, $\theta_{\nu-e}< 30~ {\rm mrad}$.
This constraint can be used as a criterion to suppress the background events arising from $\nu-{\rm nucleon}$ 
scattering.
The actual limits on $\nu_e$ and $\nu_{\mu}$ magnetic moment are $2.9 \times 10^{-11}\mu_B$~\cite{Beda:2013mta} and
$6.8 \times 10^{-10}\mu_B$~\cite{Auerbach:2001wg} respectively, while for the tau neutrino the limit of
$3.9 \times 10^{-7}\mu_B$
has been set by the DONUT experiment~\cite{Schwienhorst:2001sj}. In DONUT one event was found, within the
acceptance cuts, while 2.3 were expected from the background processes.
For this reason the new limit was deduced by comparing the expected rate for a given value of the magnetic moment
with the level of background events.
To get an estimate of the number of events to be found in SHiP for a given value of the
tau neutrino magnetic moment we can integrate the differential cross section in Eq.~(\ref{eq:magnet})
above assuming a minimum energy for the produced electron of $10\,\rm{GeV}$. Convoluting with the spectrum of tau
neutrino
and anti-neutrino given in Fig.~\ref{fig:phinu}, rescaling to $2\times 10^{20}$ POT and 
considering a target of 10 tons, the number of expected events is
\begin{equation}
N_{ev}=4.3  \times  10^{15} \frac{\mu_\nu^2}{\mu_B^2}\,.
\end{equation}
Backgrounds events consist of charged current $\nu_e$ scattering off nuclei for which only the high
energy electron is detected, and neutral current Standard Model neutrino electron scattering of
Eq.~(\ref{eq:smneutralcurrent}).
The number of background events and the relative uncertainly can only be evaluated with a detailed
simulation of the full neutrino flux and of the performance of the detector~\cite{ship-tp}.
In Fig.~\ref{fig:magneticevt} we show the number of events with an outgoing electron with
energy greather then $10\,\rm{GeV}$, that would be produced in the SHiP
experiment assuming a value for the magnetic moment of tau neutrino equal to the upper limit set by the DONUT
experiment of  $3.9 \times 10^{-7}\mu_B$~\cite{Schwienhorst:2001sj}.

\section{Charmed pentaquark searches}
\label{sec:pq}

\begin{figure}[ht]
\centering
\includegraphics[width=0.70\textwidth]{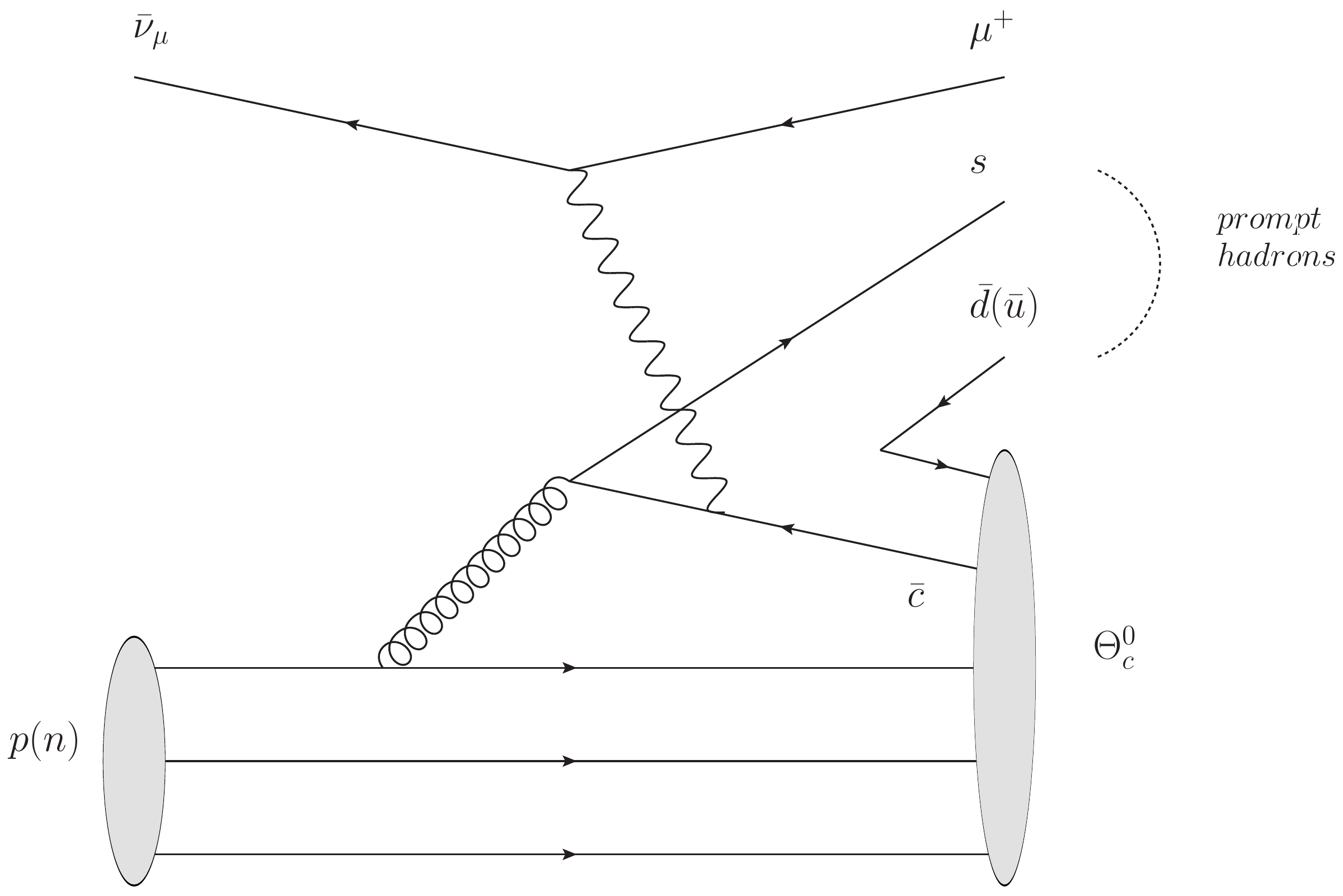}
  \caption{
           $\Theta^0_c$ production in $\bar{\nu}_{\mu}$ interactions.
}
  \label{fig:pentac}
\end{figure}

The recent discovery of charged charmonia and bottomonia states
is a proof of the existence of particles with four valence quarks.
The nature of the new states found is still debated, they could be
systems bound by QCD forces between colored objects (tetraquarks) or by
residual forces between color singlet objects like hadron molecules.
The correlation of the mass of the new states to meson-meson thresholds
favors the hypothesis of these states to be molecules of mesons.
On the other hand the sizeable production of these states at hadron colliders
disfavors extended structures with respect to compact structures.
For a recent status report of the situation see, for example, Ref.~\cite{Maiani:2014ola}.
Once we might consider a violation of the paradigm of hadronic particles being made of
$q\bar{q}$ or $qqq$ states only, we can also consider the existence of five quark states.
Indeed, the neutrino detector of the SHiP experiment provides a unique
opportunity to improve the limits on the production cross section
of a weakly decaying baryon state ($B=1$) with $C=-1$, called $\Theta_c^0$.
Negative searches of such a state have been reported in three experiments.
The Babar collaboration searched for a $\Theta_c^0$ resonance in $e^+e^-$
scattering at the energy of $10.58\,\rm{GeV}$~\cite{Aubert:2006qu},
while the Delphi collaboration used $Z$ boson decays~\cite{Abdallah:2007ad}.
However, such experimental situations could be not optimal for pentaquark generation.
The lack of valence quarks available in these reactions could represent
a detractor for the creation of a five valence quark state.
Finally, in the analysis reported by the CHORUS collaboration, a weakly decaying
charmed pentaquark state was searched among the charged current anti-neutrino
DIS events in~\cite{DeLellis:2007kz}. The situation is sketched in Fig.~\ref{fig:pentac}.
The analysis of Ref.~\cite{DeLellis:2007kz} produced as a result an upper limit
for the production cross section of a $\Theta_c^0$
\begin{equation}
\label{eq:thetaclim}
\frac{\sigma_{\Theta_c^0}}{\sigma_{\bar{\nu}}}<0.039 ~~~~~~ \rm{at}~90\%~C.L.~
\rm{for} ~ \tau_{\Theta_c^0} \sim 0.5 \tau_{D_0}\,,
\end{equation}
and below $0.085$ at 90\% C.L. for longer lifetime of the $\Theta_c^0$.
These limits were obtained with a relatively low statistic. Indeed, 2262 charged current anti-neutrino
events were collected by the CHORUS experiment, with only 32 events with anti-charm production.
This is because the CHORUS detector was using an almost pure $\nu_{\mu}$ beam.
On the other hand, in the SHiP neutrino detector, one expects about 0.6 million of anti
neutrino interactions and, of them, about 50.000 events with anti-charm production. Assuming
that the neutrino detector in SHiP would have a detection efficiency similar to the one of the CHORUS
experiment, the higher statistic would give an improvement of the limit in Eq.~(\ref{eq:thetaclim})
of about three orders of magnitude.

\section{Summary}
\label{sec:conc}

In this section we have discussed a collection of physics observables that could be
measured exploiting the features of an high intense neutrino flux from a 400 GeV proton
beam dump experiment.
We presented results for the tau neutrino physics and the opportunities to identify tau
anti-neutrino interactions and measure the impact of the $F_5$ nucleon structure function on DIS
for the first time.
Furthermore, we have briefly summarized the current status of DIS experiments. 
The theoretical description of DIS data can build on a very mature understanding, which
could be confronted with future experimental data from a new neutrino DIS fixed target
experiments in order to improve our knowledge about the fundamental structure of matter
and the important dynamics of quarks and gluons in nucleons.
In the last part we have considered the actual limits on the neutrino magnetic moment
and the production cross section of five quark states as examples of observables that
might be studied with an high level of accuracy.

\vfil\eject

\subsection*{Appendix A: Parameters used to describe chram quark fragmentation and tau lepton decays}
\label{App:1}
\begin{table}[tb]
\begin{center}
\begin{tabular}{|l|c|c|c|c|}
\hline
Particle & $N_{KK} $& $B_c(2m_c)$& $N$  & $\epsilon$   \\
\hline
$D^0$ & 0.694 &0.728 &  $0.694\times 54.5/72.8$ & 0.101\\
\hline
$D^+$ & 0.282  &0.289 &  $0.282\times 22.2/28.9$ & 0.104\\
\hline
$D_s^+$ & 0.0498  & 0.123 & $0.050\times 7.7/12.3$ & 0.0322 \\
\hline
$\Lambda_c^+$ & 0.00677  &0.0617 & 0.00677 & 0.00418 \\
\hline
\end{tabular}
\caption{\label{table:fragmentation} Parameters for the charm quark fragmentation 
used in the simulation. The numbers $N_{KK}$ and $\epsilon$ come from Kniehl and Kramer's~\cite{Kniehl:2006mw}
Table I. The fragmentation fraction $B_c$ is from their Table III.}
\end{center}
\vspace{-0.6cm}
\end{table}
The parameters for the fragmentation functions used are shown in Table \ref{table:fragmentation},
where the fragmentation function has the form
\begin{equation}
D_c(x,\mu_0^2) = \frac{Nx(1-x)^2}{[(1-x)^2+\epsilon x]^2}\,.
\end{equation}
In Ref.~\cite{Kniehl:2006mw} the authors take $\mu_0=1.5$ GeV. They also report the fragmentation fraction in their
Table III, listed here in the column labeled $B_c(2m_c)$. To account for different fragmentation fractions, rather
than using the $N=N_{KK}$ of Kniehl-Kramer in the fragmentation function, we have used $N$ rescaled by fragmentation
fractions from Ref.~\cite{Abramowicz:2013eja}. Note, we have used the fragmentation function parameterized
at LO with our NLO differential distribution, a fragmentation parameterization where  $m_c=1.5$ GeV.
\begin{table}[tb]
\begin{center}
\begin{tabular}{|l|c|}
\hline
Process & $B $   \\
\hline
$D^0\to \nu_\mu$ & 0.067\\
\hline
$D^+\to \nu_\mu$ & 0.176\\
\hline
$D_s^+\to \nu_\mu$ & 0.065\\
\hline
$\Lambda_c^+\to \nu_\mu$ & 0.045\\
\hline
$D_s^+\to\nu_\tau \tau$ & 0.0554  \\
\hline
$\tau\to e $ &  0.178\\
\hline
$\tau\to \mu $ &  0.178\\
\hline
$\tau\to \pi$ &  0.12\\
\hline
$\tau\to \rho $ &  0.26\\
\hline
$\tau\to a_1 $ &  0.13\\
\hline
\end{tabular}
\caption{\label{table:branching} Parameters for the branching fractions used in this section. For the tau decays,
the formulae are listed in~\cite{Pasquali:1998xf}. }
\end{center}
\vspace{-0.6cm}
\end{table}
Table \ref{table:branching} shows the branching fractions used. Finally, also note the approximations in
$H_c\to \nu_\mu X$ using three body decays, further approximated by the muon energy distribution parameterized
in \cite{Bugaev:2003sw} as discussed above.